\documentclass[a4paper, twoside, 11pt]{book}%,draft

\usepackage[english]{babel} % hyphenation and typographical rules for a language
\usepackage[utf8]{inputenc} % input encoding
\usepackage[T1]{fontenc} % fonts encoding

\usepackage{graphicx} % Include graphics
\usepackage{subfig} % Use small figures inside a larger one
\usepackage{epstopdf}
\usepackage{xcolor} % color managment
\usepackage{wrapfig} % Figures besides text
\usepackage{float} % H comand for figures - put the figure where it appears

\usepackage{amsfonts} % Mathematical fonts
\usepackage{amsmath} % Mathematical structure
\usepackage{amssymb} % Mathematical symbols
\usepackage{amstext} % Mathematical \text command
\usepackage{slashed} % Feynman slashed notation
\usepackage{amsthm} % Typesetting theorems
\usepackage{mathrsfs} % RSFS fonts
\usepackage{booktabs} % Enhance the tables
\usepackage{longtable} % Allow write long tables
\usepackage{bm} % to put bold and italic text in math mode
\usepackage{gensymb} % Generic symbols, as degree, celsiuc, perthousand, etc.

\usepackage{setspace} % space between lines
\usepackage{makeidx} % creation of indexes
\usepackage{pdfpages} % include pdf files
\usepackage[Glenn]{fncychap} % Chapter headings
\usepackage{layout} % layout of the document
\usepackage[a4paper, top=3cm, inner=3cm, outer=2cm, bottom=2cm, includeheadfoot]{geometry} % document dimensions
\usepackage{hyperref} % hypertext links
\usepackage{url} % verbatim for urls
\usepackage{multicol} % single and multiple columns
\usepackage{enumerate} % redefine labels of enumerate
\usepackage{indentfirst} % indent first paragraph after section
\usepackage{enumitem} % control layout of enumerate
\usepackage{mdframed} % frame box around a region
\usepackage{csquotes} % Quotes in the language is used in the text
\usepackage{xspace} % Commands not to eat spaces
\usepackage[numbers, square, merge, sort&compress]{natbib} % Improved bibliography styles
\usepackage{notoccite}

\usepackage[sc, osf]{mathpazo} % Palatino mathematical font
\usepackage{lettrine} % Old german fonts
\usepackage{aurical} % Lukas P fonts
\usepackage{wasysym} % WASY2 fonts (astronomical, phonetic, musical, etc)
\usepackage{calligra} % Calligraphic font
\usepackage{suetterl} % Schulschriften german font
\usepackage{hiero} % Hieroglyphic fons
\usepackage{trajan}%{\trjnfamily ABCDEFGHIJKLMNOPQRSTUVWXYZ} % Old Roman font
\usepackage{CJKutf8} % Chinese-Japanese-Korean Typesetting
\usepackage{kotex} % Korean typesetting

\usepackage{accents} % Creation of accents and placement of scripts

\newcommand{\ket}[1]{|#1\rangle}
\newcommand{\bra}[1]{\langle#1}
\newcommand{\der}[2]{\frac{d#1}{d#2}}

\newcommand{\parc}[2]{\frac{\partial#1}{\partial #2}}

\newcommand{\abs}[1]{\left\vert#1\right\vert}

\newcommand{\esp}[1]{\left[#1\right]}
\newcommand{\corc}[1]{\left(#1\right)}
\newcommand{\llav}[1]{\left\{#1\right\}}
\newcommand{\MCC}{\mathcal{\hat C}}

\DeclareMathOperator{\Tr}{Tr}
\DeclareMathOperator{\tr}{tr}
\DeclareMathOperator{\defm}{:=}

\DeclareMathOperator{\tto}{\longrightarrow}
\DeclareMathAlphabet{\mathpzc}{OT1}{pzc}{m}{it}
\DeclareMathAlphabet{\mathbf}{U}{bf}{m}{n}
\DeclareMathAlphabet{\mathfrak}{U}{frak}{m}{n}

\renewcommand{\S}{\mathcal{S}}

\newcommand{\M}{\mathcal{M}}

\renewcommand{\O}{\mathcal{O}}

\newcommand{\Z}{\mathbb{Z}}
\renewcommand{\Re}{\mathfrak{Re}}
\renewcommand{\Im}{\mathfrak{Im}}

\newcommand{\MeV}{{\rm MeV}}
\newcommand{\GeV}{{\rm GeV}}
\newcommand{\TeV}{{\rm TeV}}

\newcommand{\ds}[1]{{\displaystyle #1 }}

\newcommand*{\pbar}[1]{\accentset{(-)}{#1}}
\newcommand{\CNB}{{\rm C}\nu{\rm B}}

\usepackage{fancyhdr}                          %   Cabeçalhos e rodapés decentes
	\renewcommand{\chaptermark}[1]{%           % - cabeçalhos em minusculas
		\markboth{#1}{}}                       %
	\fancypagestyle{plain}{%                   %
		\fancyhead{}                           % - sem cabeçalhos em páginas limpas
  }

\definecolor{corurl}{RGB}{0,139,139}
\definecolor{corlinks}{RGB}{39,64,139}
\definecolor{corcite}{RGB}{139,0,0}

\hypersetup{
	colorlinks = true,
	urlcolor   = corurl,
	linkcolor  = corlinks,
	citecolor  = corcite,
	pdfborder  = 0 0 0
}

\makeindex

\definecolor{cor2}{rgb}{0,0,0}
\definecolor{cor1}{rgb}{0,0,0}
\definecolor{ultramarine}{RGB}{0,32,96}
\definecolor{brownb}{RGB}{153,51,0}

\definecolor{DarkBlue}{RGB}{47,47,79}%{0,34,102}

\AtBeginDocument{\setlength{\DefaultFindent}{0.5em}}
\graphicspath{{Imagens/nu-MP/}{Imagens/SeeSaw/}{Imagens/nu2HDM/GN/}{Imagens/nu2HDM/DL/}{Imagens/nu2HDM/Flavor/}{Imagens/NF/}{Imagens/RelicNu/}}

\begin{document}

%\layout{}

	\frontmatter
      	
      	\pagestyle{empty}

		% Modelo de páginas-título para teses e dissertações no IFUSP
% Versão 1.0
% Daniel C. Guariento
%
% - INSTRUÇÕES DE USO -
%
% Este modelo deve ser utilizado em substituição aos comandos de
% preâmbulo do documento: \author, \title e \date. O código deve ser
% posicionado no lugar em que estaria o comando \maketitle, logo
% após \begin{document}.
%
% Utilize a classe de documento article ou report, com a opção
% a4paper, sem comandos modificadores de fonte para obter o resultado
% ótimo. Os parâmetros definidos aqui podem funcionar em papel
% letter, mas isso não foi testado.
%
% Siga as orientações nos comentários ao longo do código para
% colocação dos dados da sua tese.
%
% Nota: Se você utiliza algum pacote para edição de cabeçalhos e
% rodapés (como o fancyhdr), então você precisa preencher os campos
% \title e \author no preâmbulo do documento. Os dados que irão
% aparecer nos cabeçalhos ou rodapés serão os que estiverem nesses
% campos.
%
% Enjoy.

{\fontfamily{ptm}\selectfont

\begin{titlepage}

	\setlength{\voffset}{0pt}
	\setlength{\hoffset}{0pt}

	\centering

	\Large{
		Universidade de S\~ao Paulo\\
		Instituto de F\'isica
	}

	\vspace{\stretch{3.5}}

	\LARGE\textbf{
		%	Fundamentos teóricos e fenomenológicos da natureza dos neutrinos.
		Neutrinos Massivos: Consequ\^encias fenomenol\'ogicas e cosmol\'ogicas
	}

	\vspace{\stretch{1.5}}

	\Large{Yuber Ferney Perez Gonzalez}

	\vspace{\stretch{2}}	
	
	\begin{flushright}
	
	\normalsize
		Orientadora: Profa. Dra. Renata Zukanovich Funchal.

		\begin{minipage}{0.5\textwidth}
				
		\vspace{3em}
		%\medskip

			%\rule{\linewidth}{0.5mm}\\
			Tese de doutorado apresentada ao Instituto de Física como requisito parcial para a obtenção do título de 
			Doutor em Ciências.\\
			%\rule{\linewidth}{0.5mm}
			
		\end{minipage}
	\end{flushright}

	\medskip

	\begin{flushleft}

	\normalsize	
		Banca Examinadora: \\~\\
		Profa. Dra. Renata Zukanovich Funchal  (IF/USP)\\
		Prof. Dr.  Gustavo Alberto Burdman (IF/USP) \\
		Prof. Dr.  Hiroshi Nunokawa (PUC/RJ) \\
		Prof. Dr.  André Paniago Lessa (CCNH/UFABC) \\
		Prof. Dr.  Orlando Luis Goulart Peres (IFGW/UNICAMP)
	\end{flushleft}

	\vspace{\stretch{1.5}}	
	\normalsize
	S\~ao Paulo\\
	2017% insira o ano correto aqui

\end{titlepage}

} %\null\newpage 
		\includepdf[pages={1}]{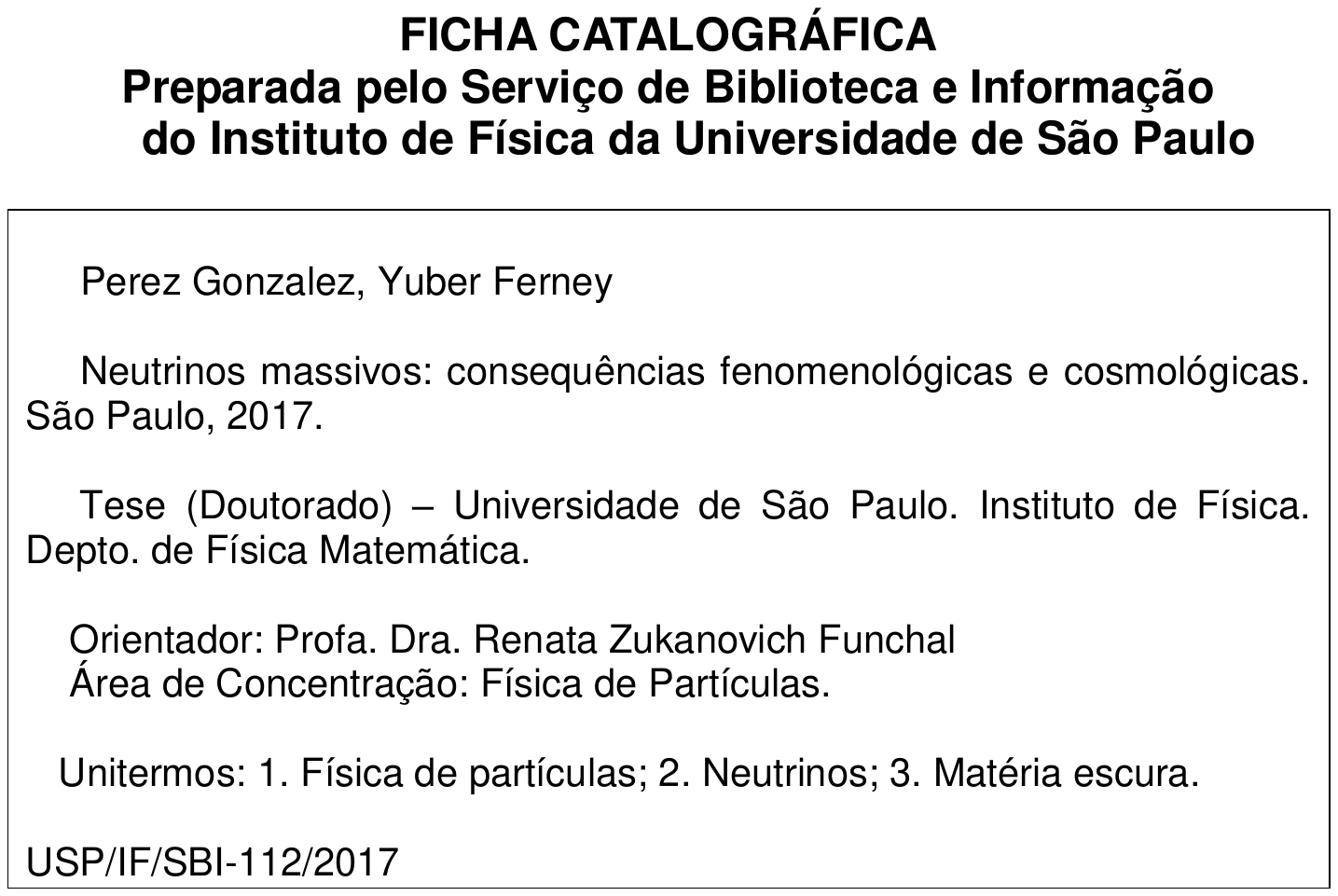}
		% Modelo de páginas-título para teses e dissertações no IFUSP
% Versão 1.0
% Daniel C. Guariento
%
% - INSTRUÇÕES DE USO -
%
% Este modelo deve ser utilizado em substituição aos comandos de
% preâmbulo do documento: \author, \title e \date. O código deve ser
% posicionado no lugar em que estaria o comando \maketitle, logo
% após \begin{document}.
%
% Utilize a classe de documento article ou report, com a opção
% a4paper, sem comandos modificadores de fonte para obter o resultado
% ótimo. Os parâmetros definidos aqui podem funcionar em papel
% letter, mas isso não foi testado.
%
% Siga as orientações nos comentários ao longo do código para
% colocação dos dados da sua tese.
%
% Nota: Se você utiliza algum pacote para edição de cabeçalhos e
% rodapés (como o fancyhdr), então você precisa preencher os campos
% \title e \author no preâmbulo do documento. Os dados que irão
% aparecer nos cabeçalhos ou rodapés serão os que estiverem nesses
% campos.
%
% Enjoy.

{\fontfamily{ptm}\selectfont

\begin{titlepage}

	\setlength{\voffset}{0pt}
	\setlength{\hoffset}{0pt}

	\centering

	\Large{
		University of São Paulo\\
		Physics Institute
	}

	\vspace{\stretch{3.5}}

	\LARGE\textbf{
		%	Fundamentos teóricos e fenomenológicos da natureza dos neutrinos.
		Massive Neutrinos: Phenomenological and Cosmological Consequences
	}

	\vspace{\stretch{1.5}}

	\Large{Yuber Ferney Perez Gonzalez}

	\vspace{\stretch{2}}	
	
	\begin{flushright}
	
	\normalsize

		Supervisor: Prof. Dr. Renata Zukanovich Funchal.	
		
		\begin{minipage}{0.5\textwidth}
			
		\vspace{3em}

			%\rule{\linewidth}{0.5mm}\\
			Thesis submitted to the Physics Institute of the University of São Paulo in partial fulfillment of the
			requirements for the degree of Doctor of Science.\\
			%\rule{\linewidth}{0.5mm}
			
		\end{minipage}
	\end{flushright}

	\medskip

	\begin{flushleft}

	\normalsize	
		Examining Committee: \\~\\
		Prof. Dr. Renata Zukanovich Funchal  (IF/USP)\\
		Prof. Dr.  Gustavo Alberto Burdman (IF/USP) \\
		Prof. Dr.  Hiroshi Nunokawa (PUC/RJ) \\
		Prof. Dr.  André Paniago Lessa (CCNH/UFABC) \\
		Prof. Dr.  Orlando Luis Goulart Peres (IFGW/UNICAMP)
	\end{flushleft}

	\vspace{\stretch{1.5}}	
	\normalsize
	S\~ao Paulo\\
	2017% insira o ano correto aqui

\end{titlepage}

} \null\newpage
		\pagenumbering{arabic}
		%\include{./Preambulo/Dedicatoria} \null\newpage
		%%%%%%%%%%%%%%%%%%%%%%%%%%%%%%%%%%%%%%%%%%%%%%%%%%%%%%%%%%%%%%%%%%%%%%%%%%%%%%%%%%%%%%%%%%%%%%%%%% 
%%%%%%%%%%%%%%%%%%%%%%%%%%%%%%%%%%%%%%%%%%%%%%%%%%%%%%%%%%%%%%%%%%%%%%%%%%%%%%%%%%%%%%%%%%%%%%%%%%
%%%%%%%%%%%%%%%%%%%%%%%%%%%%%%%%%%%%%%%%%%%%%%%%%%%%%%%%%%%%%%%%%%%%%%%%%%%%%%%%%%%%%%%%%%%%%%%%%%

\chapter{Acknowledgments - Agradecim(i)entos}

\lettrine{A}{pós} estes anos de doutorado que, sem lugar a dúvida, foram fundamentais na minha formação, devo agradecer o apoio e ajuda de muitas personas. Devo começar agradecendo à oportunidade que o Instituto de Física da Universidade de São Paulo me brindou ao me permitir fazer meu doutorado aqui. Espero haver retribuído do melhor modo possível essa oportunidade.\\

Debo agradecerle a mi familia que, desde la distancia, siempre me apoyó y me hizo sentir en casa. Mis más sinceros  agradecimientos a mis padres Tulia y Orlando, a mis hermanos Jenny, Yury, William, Gerardo y a mi cuñado Óscar que, junto con mis sobrinos, me motivaron a seguir siempre. ¡Muchas gracias!\\

Agradeço à professora Renata Zukanovich Funchal, minha orientadora, quem não somente me ensinou muitos aspectos da física de partículas, mas também me mostrou como ser um pesquisador íntegro; aprendi dela que devemos sempre ser melhores, dia a dia, não só com relação à física, mas em todos os aspectos da vida.  A sua ampla visão foi fundamental pra evitar que eu tomasse caminhos que não teriam saída. O seu empenho e devoção à pesquisa são uma referência inabalável que sempre levarei comigo. A sua veemente integridade é indubitavelmente um padrão que levarei em diante. Meu mais sinceiro muito obrigado!\\

Agradeço aos professores da banca, Gustavo Burdman, André Lessa, Hiroshi Nunokawa e Orlando Peres, pelas múltiplas sugestões, correções e indicações que sem dúvida fi\-ze\-ram que esta tese fosse muito melhor. \\

Agradeço ao Enrico Bertuzzo, quem considero como um coorientador dessa tese, pelos múltiplos ensinamentos, sugestões e interesse nos distintos projetos que desenvolvemos juntos; sua dedicação e entusiasmo pela física de partículas são um modelo do que significa ser um excelente pesquisador.  Grazie mille!\\

Agradeço ao Pedro A. N. Machado por tantas colaborações, incentivos, observações durante a evolução do meu doutorado. Sua perspectiva abrangente e ímpeto sempre foram, e são, admiráveis e dignas de imitar; com certeza, é um exemplo para todos nós que iniciamos nosso caminho como pesquisadores. Valeu, meu chapa!\\

Agradeço ao Alysson Morais, André Britto, Miguel Ferreira, Rafael Francisco e Walace Elias pela sua amizade e pelos distintos momentos de camaradagem que me ajudaram a criar uma visão muito mais ampla da vida. Tantos momentos juntos jamás serão esquecidos. 
Muito obrigado galera!\\
 
Devo certamente dar um agradecimento especial ao Walace Elias, pois, apesar do curto tempo, sua amizade foi fundamental em numerosos momentos.
Com certeza, as inúmeras histórias que temos ficarão comigo pra sempre. Agradeço porque tive a oportunidade de co\-nhe\-cer várias partes desse Brasil gigante graças a ele. Obrigado mesmo.\\

Também, agradeço à Nayara Fonseca, à Carolina Corrêa e ao Jorgivan Dias, pois desde o momento em que cheguei ao Brasil foram um grandes amigos que me deram seu apoio e me deixaram muitos ensinamentos.\\

Le agradezco a mis amigos coterráneos Hugo Camacho, Antonio Sánchez,  Javier Buitrago y Faiber Alonso, pues fueron siempre una grande compañia y gracias a ellos me sentí más cerca de casa; también, le agradezco a Martín Arteaga, Victor Peralta, Javier Lorca, Pablo Ibieta, quienes me mostraron como 
nosotros latinoamericanos tenemos tanto en común y a la vez existen muchas diferencias. ¡Muchas  gracias!\\

I would like to thank Zahra Tabrizi, Suchita Kulkarni and Frank Deppisch for the collaborations, which improved my knowledge in particle physics, and for having taught me how to perform research in an appropriate manner. Agradeço também ao Olcyr Sumensari, pois é um colaborador, e amigo, esplêndido, com quem tive inúmeras conversas sobre física de partículas e contribuiu com meu desenvolvimento. \\

Agradeço às secretárias do departamento de Física - Matemática, Amélia, Cecília e Simone por sempre estarem prestes a ajudar-me, inclusive com as mais simples dúvidas e coisas insignificantes. Agradeço também aos funcionários da Comissão de Pós-Graduação pela atenção em todos os momentos.\\

Agradeço finalmente ao \textbf{Conselho Nacional de Desenvolvimento Científico e Tecnológico (CNPq)} e
à \textbf{Fundação de Amparo à Pesquisa do Estado de São Paulo (FAPESP)}, através do processo
{\it$2013/03132$-$5$}, pelo apoio financeiro fundamental dado para a realização da presente tese.\\
\vspace{2cm}

{\suetterlin\itshape\LARGE\noindent Yuber Ferney Perez Gonzalez\\}
\vspace{1cm}
\textbf{\\ São Paulo, 6 de Dezembro de 2017}

		%%%%%%%%%%%%%%%%%%%%%%%%%%%%%%%%%%%%%%%%%%%%%%%%%%%%%%%%%%%%%%%%%%%%%%%%%%%%%%%%%%%%%%%%%%%%%%%%%
%%%%%%%%%%%%%%%%%%%%%%%%%%%%%%%%%%%%%%%  EPIGRAFE  %%%%%%%%%%%%%%%%%%%%%%%%%%%%%%%%%%%%%%%%%%%%%%
%%%%%%%%%%%%%%%%%%%%%%%%%%%%%%%%%%%%%%%%%%%%%%%%%%%%%%%%%%%%%%%%%%%%%%%%%%%%%%%%%%%%%%%%%%%%%%%%%
\thispagestyle{empty}

\vspace*{2cm}

\begin{quotation}
  \begin{center}	
  	\begin{center}
		{\DisplayHieroglyphs
		\begin{hieroglyph}{\leavevmode \HquarterSpace \HwordSpace
			\loneSign{\Aca GG/50/}\HinterSignsSpace
			\Cadrat{\CadratLineI{\Aca GO/60/}\CadratLine{\Aca GD/69/}}\HinterSignsSpace
			\loneSign{\Aca GG/32/}\HinterSignsSpace
			\loneSign{\Aca GY/40/}\HinterSignsSpace
			\Cadrat{\CadratLineI{\Aca GF/65/\hfill\Aca GZ/32/}\CadratLine{\Aca GV/62/}}\HinterSignsSpace
			\LoneHorizontalLine{\loneSign{\Aca GD/33/}\HinterSignsSpace
			\loneSign{\Aca GZ/32/}}\HinterSignsSpace
			\Cadrat{\CadratLineI{\Aca GD/52/}\CadratLine{\Aca GAa/32/}}\HinterSignsSpace
			\Cadrat{\CadratLineI{\Aca GY/32/}\CadratLine{\Aca GV/62/}}\HinterSignsSpace
			\HfullSpace \HwordSpace
			\Cadrat{\CadratLineI{\Aca GN/66/}\CadratLine{\HquarterSpace }}\HinterSignsSpace
			\negAROBspace\negAROBspace\negAROBspace\negAROBspace\negAROBspace\negAROBspace\loneSign{{\Hsmaller\Aca GI/42/}}\HinterSignsSpace
			\negAROBspace\negAROBspace\negAROBspace\HquarterSpace \HinterSignsSpace
			\Cadrat{\CadratLineI{\Aca GN/66/}\CadratLine{\HquarterSpace }}\HinterSignsSpace
			\negAROBspace\negAROBspace\negAROBspace\negAROBspace\negAROBspace\negAROBspace\loneSign{{\Hsmaller\Aca GI/42/}}\HinterSignsSpace
			\negAROBspace\negAROBspace\negAROBspace\HquarterSpace \HinterSignsSpace
			\loneSign{\Aca GAa/55/}\HinterSignsSpace
			\loneSign{\Aca GA/33/}\HinterSignsSpace
			\Cadrat{\CadratLineI{\Aca GD/52/}\CadratLine{\Aca GV/62/}}\HinterSignsSpace
			\loneSign{\Aca GV/59/}\HinterSignsSpace
			\Cadrat{\CadratLineI{\Aca GN/66/}\CadratLine{\Aca GD/69/}}\HinterSignsSpace
			\loneSign{\Aca GAa/32/}\HinterSignsSpace
			\loneSign{\Aca GG/50/}\HinterSignsSpace
			\Cadrat{\CadratLineI{\Aca GD/68/}\CadratLine{\Aca GA/32/}}\HinterSignsSpace
			\loneSign{\Aca GW/52/}\HinterSignsSpace
			\loneSign{\Aca GM/48/}\HinterSignsSpace
			\Cadrat{\CadratLineI{\Aca GD/52/}\CadratLine{\Aca GAa/32/}}\HinterSignsSpace
			\Cadrat{\CadratLineI{\Aca GY/32/}\CadratLine{\Aca GA/32/}}\HinterSignsSpace
			\HfullSpace \HwordSpace
			\loneSign{\Aca GD/68/}\HinterSignsSpace
			\loneSign{\Aca GW/58/}\HinterSignsSpace
			\Cadrat{\CadratLineI{\Aca GN/66/}\CadratLine{\Aca GX/32/}}\HinterSignsSpace
			\loneSign{\Aca GG/77/}\HinterSignsSpace
			\Cadrat{\CadratLineI{\Aca GM/67/}\CadratLine{\Aca GD/52/}}\HinterSignsSpace
			\loneSign{\Aca GG/77/}\HinterSignsSpace
			\loneSign{\Aca GN/62/}\HinterSignsSpace
			\loneSign{\Aca GU/56/}\HinterSignsSpace
			\Cadrat{\CadratLineI{\Aca GX/32/\hfill\Aca GZ/32/}\CadratLine{\Aca GY/32/}}\HinterSignsSpace
			\HfullSpace \HwordSpace
			\Cadrat{\CadratLineI{\Aca GD/68/}\CadratLine{\Aca GN/66/}}\HinterSignsSpace
			\loneSign{\Aca GU/56/}\HinterSignsSpace
			\loneSign{\Aca GG/77/}\HinterSignsSpace
			\loneSign{\Aca GG/77/}\HinterSignsSpace
			\loneSign{\Aca GA/32/}\HinterSignsSpace
			\Cadrat{\CadratLineI{\Aca GD/69/}\CadratLine{\InternalCadrat{\CadratLineI{\Aca GQ/34/}\CadratLine{\Aca GD/52/}}}}\HinterSignsSpace
			\loneSign{\Aca GAa/50/}\HinterSignsSpace
			\loneSign{\Aca GY/40/}\HinterSignsSpace
			\loneSign{\Aca GG/58/}\HinterSignsSpace
			\loneSign{\Aca GAa/32/}\HinterSignsSpace
			\loneSign{\Aca GG/77/}\HinterSignsSpace
			\Cadrat{\CadratLineI{\Aca GY/32/}\CadratLine{\InternalCadrat{\CadratLineI{\Aca GZ/33/}\CadratLine{\Aca GI/41/}}}}\HinterSignsSpace
			\HfullSpace \HwordSpace
			\HfullSpace \HinterSignsSpace
			\HfullSpace \HinterSignsSpace
			\HfullSpace \HinterSignsSpace
			\HfullSpace \HinterSignsSpace
			\cartouche{\LoneHorizontalLine{\Cadrat{\CadratLineI{\Aca GQ/34/}\CadratLine{\Aca GX/32/}}\HinterSignsSpace
			\loneSign{\Aca GV/59/}}\HinterSignsSpace
			\Cadrat{\CadratLineI{\Aca GR/35/}\CadratLine{\LoneHorizontalLine{\Aca GX/32/\hfill\Aca GQ/34/}}}\HinterSignsSpace
			\loneSign{\Aca GA/85/}}%
		}\end{hieroglyph}
	  	}
	\end{center}
   \vspace{3.5cm}\large
   \begin{flushleft}
   	{\calligra
   	Do not be arrogant because of your knowledge,\\
   	but discourse with the ignorant man as with the learned.\\
   	For no limits can be set to knowledge,\\
   	and none has yet achieved perfection in it.\\
   	\qquad    Ptahhotep - 2350 B.C.
   	    }
   \end{flushleft}
   \vspace{3.5cm}
    \begin{flushright}
      {\fontfamily{pzc}\selectfont 
      	Não sejas arrogante pelo teu conhecimento,\\
      	Segue o conselho tanto do ignorante quanto do sábio,\\
      	O conhecimento não tem limites,\\
      	E ninguém tem alcançado a perfeição nele.\\
      	\qquad\qquad\qquad Ptahhotep (2350 A.C.)
      	}
    \end{flushright}
  \end{center}
\end{quotation} \null\newpage

		%%%%%%%%%%%%%%%%%%%%%%%%%%%%%%%%%%%%%%%%%%%%%%%%%%%%%%%%%%%%%%%%%%%%%%%%%%%%%%%%
%%%%%%%%%%%%%%%%%%%%%%%%%%%%%%%%%%%%%%%%%%%%%%%%%%%%%%%%%%%%%%%%%%%%%%%%%%%%%%%%
%%%%%%%%%%%%%%%%%%%%%%%%%%%%%%%%%%%%%%%%%%%%%%%%%%%%%%%%%%%%%%%%%%%%%%%%%%%%%%%%

\chapter{Abstract}\label{cha:Abs}

%%%%%%%%%%%%%%%%%%%%%%%%%%%%%%%

\lettrine{T}he XX century witnessed the quantum and relativistic revolutions in physics. The development of these two theories, namely, Quantum Mechanics and Relativity, was the inception of many crucial discoveries and technological advances. Among them, one stands out due to its uniqueness, the neutrino discovery. However, several neutrino properties are still obscure. Neutrinos are the only fundamental particles whose nature is currently unknown. Such fermions can either be different from their antiparticles, i.e., Dirac fermions, or be their own antiparticles, that is, Majorana fermions. On the other hand, the smallness of neutrino masses is a problem seemingly related to the neutrino nature; thus, as essential task consists in addressing the phenomenologically viable models in both cases. Furthermore, it is important to search for other physical process in which the neutrino nature may manifest through different experimental signatures. A rather difficult but promising method corresponds to the detection of the cosmic neutrino background, viz.\ neutrinos which are relics from the Big Bang. Previous works have shown that detection rates for Dirac and Majorana neutrinos can give different results. Nevertheless, this distinction was obtained considering the Standard Model framework only. Therefore, it is important to understand the consequences of having Non-Standard Interactions contributing to the detection of neutrinos from the cosmic background. Another remarkable relic predicted by Cosmology is the unidentified Dark Matter, composing $\sim 25\%$ of the Universe. All searches regarding the Weakly Interacting Massive Particle, one of the principal candidates for Dark Matter, have given negative results; this has compelled experiments to increase their sensitivity. Notwithstanding, neutrinos may stand in the way of such experimental searches given that they may constitute an irreducible background.\\

In this thesis, we will address these three different phenomena, neutrino mass models, detection of the cosmic neutrino background and the neutrino background in Dark Matter searches, by considering the different characteristics in each case. In the study of neutrino mass models, we will consider models for both Majorana and Dirac neutrinos; specifically, we will probe the neutrinophilic two-Higgs-doublet model. Regarding the detection of relic neutrinos, we will analyse the consequences of the existence of the beyond Standard Model physics in the capture rate by tritium. Finally, we will scrutinize the impact of neutrinos in Direct Detection WIMP searches, by considering Standard Model plus additional interactions in the form of simplified models.\\

\noindent{\bf Keywords:} Neutrino Physics; Dirac and Majorana neutrinos; Cosmic Neutrino Background; Dark Matter.
 		%%%%%%%%%%%%%%%%%%%%%%%%%%%%%%%%%%%%%%%%%%%%%%%%%%%%%%%%%%%%%%%%%%%%%%%%%%%%%%%%%%%%%%%%%%%%%%%%%%
%%%%%%%%%%%%%%%%%%%%%%%%%%%%%%%%%%%%%%%   RESUMO  %%%%%%%%%%%%%%%%%%%%%%%%%%%%%%%%%%%%%%%%%%%%%%%%
%%%%%%%%%%%%%%%%%%%%%%%%%%%%%%%%%%%%%%%%%%%%%%%%%%%%%%%%%%%%%%%%%%%%%%%%%%%%%%%%%%%%%%%%%%%%%%%%%%

\chapter{Resumo}\label{cha:Res}

%%%%%%%%%%%%%%%%%%%%%%%%%%%%%%%%%%%%%%%%%%%%

\lettrine{A}o longo do século XX testemunhamos as revoluções quântica e relativista que aconteceram na Física. O desenvolvimento da Mecânica quântica e da teoria da relatividade foi o prelúdio de inúmeras descobertas e avanços tecnológicos fundamentais; em particular, a descoberta dos neutrinos. No entanto, a sua total compreensão ainda é um mistério para a física de partículas. Entendidos como partículas fermiônicas fundamentais, os neutrinos possuem sua natureza desconhecida. Podendo ser diferentes de suas antipartículas, denominadas férmions de Dirac, ou também podendo ser as suas próprias antipartícula, sendo conhecidas como férmions de Majorana. Por outro lado, o valor de sua massa continua sendo um problema em aberto, supostamente relacionado à sua natureza. Portanto, é importante estudarmos modelos fenomenológicos viáveis para as duas naturezas possíves dos neutrinos. Além disso, é necessário procurar outros processos físicos cujos resultados experimentais sejam distintos de acordo com a natureza do neutrino. Um método bastante difícil, mas promissor, corresponde à detecção do fundo de neutrinos cósmicos, isto é, os neutrinos relíquia do Big Bang. Análises prévias mostraram que as taxas de detecção para neutrinos de Dirac e de Majorana resultam em valores distintos. Porém, este resultado foi obtido supondo como base o Modelo Padrão; assim, é crucial entender as possíveis consequências da existência de interações desconhecidas na detecção dos neutrinos da radiação cósmica de fundo. Outra relíquia notável prevista pela Cosmologia é a desconhecida Matéria Escura, que compõe $\sim 25\%$ do Universo. Todas as buscas por WIMPs (do inglês Weakly Interactive Massive Particles), um dos principais candidatos a Matéria Escura, tem dado resultados negativos. Isto tem forçado a criação de experimentos cada vez mais sensíveis. Contudo, os neutrinos poderão ser um obstáculo nessas buscas experimentais, pois estes convertir-se-ão em um fundo irredutível.\\ 

Na presente tese, abordaremos estes três fenômenos diferentes, modelos de massa para os neutrinos, a detecção do fundo de neutrinos cósmicos e o fundo de neutrinos em experimentos de detecção direta de Matéria Escura, considerando as distintas características em cada caso. No estudo dos modelos de massa para os neutrinos consideraremos modelos para neutrinos de Majorana e Dirac; exploraremos modelos neutrinofílicos com dois dubletos de Higgs. Enquanto à detecção dos neutrinos relíquia, analisaremos as consequências da presença de física além do Modelo Padrão na taxa de captura pelo trítio. Finalmente, examinaremos o impacto dos neutrinos em experimentos de detecção direta de WIMPs, supondo as interações do Modelo Padrão junto com interações adicionais na forma de modelos simplificados.\\

\noindent{\bf Palavras Chave:} Física de neutrinos; Neutrinos de Dirac e Majorana; Fundo cósmico de neutrinos; Matéria Escura.
		%%%%%%%%%%%%%%%%%%%%%%%%%%%%%%%%%%%%%%%%%%%%%%%%%%%%%%%%%%%%%%%%%%%%%%%%%%%%%%%%%%%%%%%%%%%%%%%%%%
%%%%%%%%%%%%%%%%%%%%%%%%%%%%%%%%%%%%  Bib Note  %%%%%%%%%%%%%%%%%%%%%%%%%%%%%%%%%%%%%%%%%%%%%%%%%%
%%%%%%%%%%%%%%%%%%%%%%%%%%%%%%%%%%%%%%%%%%%%%%%%%%%%%%%%%%%%%%%%%%%%%%%%%%%%%%%%%%%%%%%%%%%%%%%%%%

\chapter{Bibliographic Note}

The works completed by the author during his Doctoral studies in chronological order are the following

\begin{enumerate}

	\item Pedro A.\ N.\ Machado, Yuber F.\ Perez Gonzalez, Olcyr Sumensari, Zahra Tabrizi and Renata Zukanovich Funchal, {\it On the Viability of Minimal Neutrinophilic Two-Higgs-Doublet Models}, JHEP$\mathbf{12}$ ($2015$) $160$, \href{https://arxiv.org/abs/1507.07550}{[arXiv:$1507.07550$ [hep-ph]]}.
	
	\item Enrico Bertuzzo, Yuber F.\ Perez G., Olcyr Sumensari and Renata Zukanovich Funchal, {\it Limits on Neutrinophilic Two-Higgs-Doublet Models from Flavor Physics}, JHEP$\mathbf{01}$ ($2016$) $018$, \href{https://arxiv.org/abs/1510.04284}{[arXiv:$1510.04284$ [hep-ph]]}.

	\item Enrico Bertuzzo, Frank F.\ Deppisch, Suchita Kulkarni, Yuber F.\ Perez Gonzalez and Renata Zu\-ka\-no\-vi\-ch Funchal, {\it Dark Matter and Exotic Neutrino Interactions in Direct Detection Searches}, JHEP$\mathbf{04}$ ($2017$) $073$, \href{https://arxiv.org/abs/1701.07443}{[arXiv:$1701.07443$ [hep-ph]]}.
	
	\item Enrico Bertuzzo, Pedro A.\ N.\ Machado, Yuber F.\ Perez-Gonzalez and Renata Zu\-ka\-no\-vich Funchal, {\it Constraints from Triple Gauge Couplings on Vectorlike Leptons}, Phys.\ Rev.\ $\mathbf{D 96}$ ($2017$) $035035$, \href{https://arxiv.org/abs/1706.03073}{[arXiv:$1706.03073$ [hep-ph]]}.
	
	\item Mart\'in Arteaga, Enrico Bertuzzo, Yuber F.\ Perez-Gonzalez and Renata Zu\-ka\-no\-vich Funchal, {\it Impact of Beyond Standard Model Physics in the Detection of the Cosmic Neutrino Background}, JHEP$\mathbf{09}$ ($2017$) $124$, \href{https://arxiv.org/abs/1708.07841}{[arXiv:$1708.07841$ [hep-ph]]}.
		
\end{enumerate}

The present document is based in the works related to neutrino phenomenology studies, namely, the publications on neutrinophilic two-Higgs double modes (references $1,2$); the effect of exotic Dark Matter - neutrino interactions in direct detection experiments, reference $3$; and the study on the repercussion of Non-Standard interactions on the detection of the cosmic neutrino background, reference $5$.

		\listoffigures
		\listoftables
		
		\tableofcontents 

	\onehalfspacing
	
	\mainmatter
	
		\pagestyle{fancy}\setcounter{page}{21}

		%%%%%%%%%%%%%%%%%%%%%%%%%%%%%%%%%%%%%%%%%%%%%%%%%%%%%%%%%%%%%%%%%%%%%%%%%%%%%%%%%%%%%%%%%%%%%%%%%
%%%%%%%%%%%%%%%%%%%%%%%%%%%%%%%%%%%%%%%%%%%%%%%%%%%%%%%%%%%%%%%%%%%%%%%%%%%%%%%%%%%%%%%%%%%%%%%%%%
%%%%%%%%%%%%%%%%%%%%%%%%%%%%%%%%%%%%%%%%%%%%%%%%%%%%%%%%%%%%%%%%%%%%%%%%%%%%%%%%%%%%%%%%%%%%%%%%%%
\chapter*{Introduction}\label{cha:Intro}\addcontentsline{toc}{chapter}{Introduction}
\chaptermark{Introduction}

\begin{flushright}
	{\Fontlukas\large \guillemotleft\,Having answered the Count's salutation, I turned to the glass again to see how I had been mistaken.\\
	This time there could be no error, for the man was close to me, and I could see him over my shoulder.\\
	But there was no reflection of him in the mirror!\guillemotright\newline
	From Jonathan Harker's Diary,\\ 
	Bram Stocker's Dracula}
\end{flushright}

%\section*{A Historical Overview}

\vspace{1cm}

\lettrine{I}{f} a neutrino, an exceptionally imaginative one, were able to write a historically accurate novel about some macroscopic beings trying to understand its cha\-rac\-te\-ris\-tics and behaviour, how would it depict such a history? Surely, it would remark about the singularity of the neutrino hypothesis' inception on the science of these non-quantum beings. Such work would also emphasize the efforts of many brave scientists who worked, and are still working, to give a complete insight about these amazing particles. Let us put ourselves on the storyteller's shoes, and try to imagine how a novel about these classical beings attempting to comprehend a neutrino would be. We will base this unpretentious {\it gedanken} experiment on the book from C. Sutton \cite{Sutton:1992qx}.\\

How would such a saga begin? Probably it would consider the origin of the curiosity of the macroscopic creatures. Without going so far, it could begin with the development of subatomic Physics. Particle physics is one of the more recent fields of natural sciences. Its genesis however goes to ancient Greek and Indian philosophers, who thought that nature is composed by indivisible particles. Physics as we know it is an experimental discipline; the atomic hy\-po\-the\-sis could only be confirmed by the end of the XIX century. The birth of Elementary Particle Physics can be traced back to the works of Henri Becquerel \cite{Becquerel:1896zz}, who discovered radioactivity in 1896, and J.\ J.\ Thomson \cite{Thomson:1897cm}, who showed that cathodic rays are composed by particles, i.e. {\it electrons}. The greatest advances came after the advent of Relativity and Quantum Mechanics. Moreover, the development of Quantum Mechanics is intrinsically related to the evolution of Particle Physics.\\

One of the types of radioactivity discovered was named {\it beta-rays}, and the works from Marie Sk\l odowska and Pierre Curie \cite{curie1900charge,curie1903recherches} and Walter Kaufmann  \cite{akademie1902nachrichten} showed that the particles in these rays are actually electrons. Several studies were performed to understand the origin and properties of those electrons. Specially, Lise Meitner and Otto Hahn \cite{hahn1908uber,von1911nachweis,meitner1913magnetische} studied the e\-ner\-gy spectrum of these beta-rays. It was hypothesized that those electrons had a unique e\-ner\-gy. As the initial and final nucleus --nuclei were discovered previously by Ernest Rutherford-- have well known masses and energies, the outgoing electron would have a definite energy. Nonetheless, experiments showed that electrons were emitted with several energies, making the beta spectrum a continuous one \cite{Chadwick:1914zz}. This created a crisis in the scientific community since this continuous spectrum seemed to violate the conservation of energy.\\ 

Niels Bohr went through an extreme path; he proposed that conservation of energy would not be respected in the quantum realm. This however made Wolfgang Pauli uncomfortable.  He thought deeply about this ``problem'', making him to propose that in the beta process not only electrons were produced, but also an additional neutral particle. Pauli would then be named the discoverer of all neutrinos by the ingenious author. In his letter to the ``Radioactive Ladies and Gentlemen'', he speculates with certain hesitation about \texttt{``einen verzweifelten Ausweg''} --a desperate escape-- \cite{Pauli:1930pc}
\begin{quotation}
	\texttt{``... Nämlich die Möglichkeit, es könnten elektrisch neu\-tra\-le Teil\-chen, die ich Neutronen nennen will, in der Kernen existieren, welche den Spin $1/2$ haben und das Ausschliessungsprinzip befolgen ...''}\footnote{``... Namely, the possibility that in the nuclei there could exist electrically neutral particles, which I will call neutrons, that have spin 1/2 and obey the exclusion principle...''}
\end{quotation}
In these few words the neutrino hypothesis was born. Pauli chose the simplest name one could imagine, {\it neutrons}, as there were no other known neutral particles. He further reflects about the properties of such particles and the possibility of their detection \cite{Pauli:1930pc},
\begin{quotation}
	\texttt{``... Die Masse der Neutronen müsste von derselben Grössenordnung wie die Elektronenmasse sein und jedenfalls nicht grösser als 0.01 Pro\-to\-nen\-mas\-se ...''}\footnote{``... The mass of the neutrons should be of the same order of magnitude as the electron mass and in any event not larger than $0.01$ proton mass...''}
\end{quotation}
It is clear that Pauli did not only have doubts about the mass, but also about the origin of the beta rays. He thinks that the {\it neutrons} interacted through its small magnetic dipole, making its detection difficult indeed. These and other reasons made him not to publish his idea; however, the conception of a new neutral particle spread through the community.\\

At this point, the neutrino chronicler would refer to the work of one of the greatest physicists that ever lived, Enrico Fermi. In his work \cite{Fermi:1933jpa}, he takes a step further on the Pauli's hypothesis by proposing the existence of a new interaction, called {\it weak} interaction. He supposes that nucleus are composed by protons and recently discovered neutrons, and the beta process consists in the emission of an electron together with a \textbf{\textit{neutrino}}. He is the one res\-pon\-si\-ble for the name as neutrino comes from Italian meaning {\it the little neutral one}. He also establishes a method to determine the value of the neutrino mass: by studying in detail the end point of the electron's spectrum, it is possible to infer the magnitude of such mass. Taking the experimental data from the time, he found that neutrinos are particles with a mass much smaller than the electron mass. In any case, the neutrino idea was now firmly established.\\

Nevertheless, the magnitude of the neutrino mass is related with an important property of neutrinos. H.\ Weyl \cite{Weyl:1927vd} showed that if a fermion is massless it can be described by a field with a definite chirality (see appendix \ref{ap:Lgroup}). Therefore, if the neutrino has a zero mass, parity symmetry would be violated in decay processes. This was considered completely unrealistic at the time as all the experiments with electromagnetic and strong interactions were compatible with the conservation of parity. Other great physicists in the neutrino saga, T.\ D.\ Lee and C.\ N.\ Yang, emerged at this time. They actually presented a modification of the Fermi theory, and showed that experiments were needed to shed light on the conservation of parity for the specific case of the weak interaction \cite{Lee:1957qr}. Experimentalist such as C.\ S.\ Wu et.\ al.\ \cite{Wu:1957my} and Goldhaber, Grodzins and Sunyar \cite{Goldhaber:1958nb} performed experiments showing that beta decays did not respected the parity symmetry and that neutrinos were always left-helical. Everyone was as astonished as Jonathan Harker when he did not see the reflection of Count Dracula on the mirror. Anyhow, Landau \cite{Landau:1957tp}, Lee and Yang \cite{Lee:1957qr} and Salam \cite{Salam:1957st} did not get frightened by the violation of parity, and they established the, now denominated, two-component theory of a massless neutrino, in which a neutrino is described by a Weyl left-handed fermion.\\

However, the creative neutrino writer would remark, neutrinos are way more diverse than the macroscopic creatures initially thought. Studying cosmic rays, S.\ Neddermeyer and C.\ Anderson, who also discovered the positron \cite{Anderson:1933mb}, found a particle closely related to the electron, the muon ($\mu$) \cite{Neddermeyer:1937md}. Further analysis on the muon's properties showed that it decays weakly into an electron and some invisible particle which was first thought to be the Pauli-Fermi neutrino. Nevertheless, the absence of some kinematically allowed decays, such as
\begin{align*}
	\mu^-\to e^-+\gamma,
\end{align*}
led to the introduction of a conserved quantity which is different for electrons and muons, the leptonic number. Then, in the muon decay, it is necessary to have two {\it distinct} neutrinos which different leptonic numbers
\begin{align*}
	\mu^-\to e^-+\bar{\nu}_e+\nu_\mu.
\end{align*}
Then, the Pauli-Fermi neutrino was actually an electron antineutrino ($\bar{\nu}_e$) and $\nu_\mu$ was a {\it muon neutrino}. Furthermore, after the discovery of a third particle similar to the electron, the tau lepton ($\tau$), it was necessary to introduce a third neutrino, the tau neutrino ($\nu_\tau$). \\ 

Using the Fermi theory, it was initially thought that neutrinos were impossible to detect. This is because the probability of a single neutrino interacting with a detector is tiny. However, technological advances allowed us to detect electron antineutrinos by considering the huge flux of those particles coming from a nuclear reactor. This was achieved by Cowan and Reines in 1956 \cite{Cowan:1992xc}, inaugurating the neutrino experimental age. The $\nu_\mu$ was discovered in 1962 \cite{PhysRevLett.9.36}, and finally the $\nu_\tau$ was found in the year 2000 \cite{Kodama:2000mp}. Thus, we finally settled the current picture of three neutrinos, each belonging to a distinct family; these types of neutrinos are called flavour eigenstates.\\

The ingenious neutrino would now mention that the curious macroscopic beings also discovered how two apparently different interactions, the electromagnetic one, which a neutrino does not experience, and the weak are two facets of a unique interaction, the Electroweak interaction. This led to the establishment of the Standard Model (SM) \cite{Glashow:1961tr,Weinberg:1967tq,Salam:1968}, model which describes the electromagnetic, weak and strong interactions, based in two simple ideas: the gauge principle and the Higgs mechanism \cite{Englert:1964et,Higgs:1964pj,Guralnik:1964eu}. The gauge principle explains how interactions arise in a natural way after imposing that global symmetries present in the model have to be local. The Higgs mechanism describes how gauge bosons associated to weak interactions and fermions acquire mass due to a spontaneous symmetry breaking.  The SM is without any doubt extremely successful. It has predicted neutral charged currents; the existence of a neutral gauge boson $Z^0$; and a neutral scalar $H$. All these have been found experimentally. The latest one was the discovery of a particle that seems to be the Higgs boson in 2012 \cite{Aad:2012tfa,Chatrchyan:2012xdj}. Nevertheless, the SM still has unsolved problems within its construction. The {\it Hierarchy Problem} is one of the difficulties which has been attacked for years without great success. This problem is related to the large difference between the Planck and Weak scales, which makes the Higgs mass unstable after radiative corrections. Other difficulties are related to the explanation of the structure of the model itself. For instance, why are there three families? Why the fermion masses are so different? Unfortunately, the SM does not address these problems. A crucial point here is that, by construction, neutrinos are massless in this model; as the neutrino writer would notice, this was indeed well accepted at the time.

\newpage

Moreover, it is clear that our knowledge has improved significantly after the neutrino and the weak interaction ideas came out. One of the triumphs of the Fermi model is explaining why stars shine \cite{Gamow:1938,Bethe:1939bt}. With an explanation beyond the imagination of any ancient civilization, we now understand the greatness of the Sun and its essential role in allowing for life on the Earth. In the solar model, nuclear reactions transform mainly hydrogen into helium and other elements, producing the energy which feeds most of the life. This solar description, the ingenious particle would certainly stress, was crucial for progress on the neutrino understanding. The first experiment, the Homestake experiment, was designed to detect solar neutrinos \cite{Cleveland:1998nv}. It consisted of a large tank full of tetrachloroethylene in which a neutrino could interact with a Chlorine nucleus, transforming it into Argon via charged-current interactions. Thus, by counting the number of Argon nuclei, it could be possible to measure the number of detected neutrinos. Neutrinos had there another surprise for us; the number of events was smaller than expected. Other experiments were performed to confirm or refute this result. All of them found smaller numbers than expected. Something was happening with solar neutrinos in their journey to the Earth.\\

Gribov and Pontecorvo \cite{Gribov:1968kq} suggested that if the neutrino masses were different from zero, and, if their mass eigenstates were a combination of the flavour eigenstates, neutrinos could undergo an {\it oscillation} process. Thus, if a neutrino is created in some definite flavour, there is a non zero probability for it to be detected in another flavour. This could explain the solar neutrino deficit since part of the neutrinos would not be detected by the experiment as they arrive in a distinct flavour than expected. This was actually proven by the Sudbury Neutrino Observatory (SNO) experiment \cite{Aharmim:2005gt}. This experiment first  measured neutrinos coming from the Sun using charged-current interactions, finding the same deficit encountered by previous experiments; however, it also was capable of measuring neutrinos through neutral-current interactions and elastic scattering. They found that the neutrino events were compatible with the number expected from neutrinos undergoing adiabatic conversion in the Sun. Further evidences came from neutrinos created in other independent sources, such as atmospheric neutrinos, detected by the SuperKamiokande experiment \cite{Fukuda:1998ah,Fukuda:1998ub}; reactor antineutrinos, detected for instance by the Kamioka Liquid scintillator AntiNeutrino Detector (KAMLAND) \cite{Abe:2008aa}. All of them showed that neutrinos do undergo oscillations. This was the final proof that neutrinos are massive, and it confirmed experimentally the existence of beyond SM physics. Still, the smallness of neutrino masses seems to be a difficulty. Nonetheless, this is a problem of the SM itself, and perhaps there is a unique solution for all fermions.\\

Thus, in principle, we could measure the values of the neutrino masses and mixing angles, which describe the mixing among mass and flavour eigenstates, and then we could obtain a final description of neutrinos. Nonetheless, and fortunately, the clever particle would assert, neutrinos are more complex than one may believe initially. Since such fermions are neutral particles, their nature is ambiguous to us. Let us remember that Dirac particles, evidently introduced by Dirac \cite{Dirac:1930ek}, are fermions which are different from their antiparticles. Meanwhile, E.\ Majorana \cite{Majorana:1937vz} described how a massive fermion can be identical to its antiparticle if it was neutral under any charge. As we see, neutrinos are the only elementary particles which can be Majorana or Dirac fermions. Unfortunately, there is no experimental evidence which allow us to corroborate the neutrino's true nature. Thus, we can speculate if there is a connection between the smallness of the neutrino mass and its nature. This will be one of the problems addressed in this thesis.\\

The other two problems that we will consider here are related to the confluence of Cosmology and Particle Physics, the Cosmic Neutrino Background detection and the Dark Matter identity. Our scientific cosmogony predicts the existence of a background composed by the archaic neutrinos which remained after the Big Bang \cite{Kolb:1990vq}. Such relic neutrinos are completely different from the neutrinos we are used to study since they are non-relativistic particles. Moreover, these neutrinos are fundamental to asseverate our understanding of the origin of the Universe. For the ingenious neutrino writer, such neutrinos would be compared to elderly wise ones which were witnesses to the beginning of the Universe. Nevertheless, they are enormously difficult to detect given their minuscule energy. There have been proposed many methods to observe these relic neutrinos, but most of them are beyond our current technology. The most promising method, however, uses a capture by a nucleus; a process closely related to beta decay. The main consequence of neutrinos being non-relativistic on the capture rate is that, when considering SM interactions, the rate for Dirac and Majorana neutrinos are different \cite{Long:2014zva}. Precisely, Majorana neutrinos expected rate is double the value for Dirac neutrinos. This nevertheless is a strong statement as one should take into account the possible existence of beyond SM physics and modifications on the cosmological model. Thus, we will analyse the consequences of both possibilities on the cosmic neutrino background detection.\\

On the other hand, Dark Matter (DM) composes approximately $25\%$ of our Universe, but we do not know its fundamental composition. We just know that DM has gravitational interactions, and it does not couple with photons \cite{Gelmini:2015zpa}. It is supposed however that DM has other interactions since it should have been created after the Big Bang. This is confirmed by studying the oldest light in the Universe, the Cosmic Microwave Background. The latest Planck results \cite{Ade:2015xua} confirm the existence of the unknown DM component. Among the many candidates to be DM, the Weakly Interacting Massive Particle (WIMP) emerges as one of the most studied and discussed. The main reason is that it can give the correct measured relic density and its cha\-rac\-te\-ris\-tics seem to agree with the expected for many beyond SM physics. Several experiments have been performed to test the WIMP hypothesis, but they have not found anything. Consequently, more precise and sensitive experiences are under planning, but they will suffer a difficulty. A special chapter of the novel would narrate how neutrinos became somewhat villains for the macroscopic creatures in their quest for knowledge. This is because neutrinos became an irreducible background in the WIMP detection experiments through a process called Coherent Neutrino Scattering off Nuclei. Thus, we need to analyse when neutrinos start to influence WIMP searches. Also, it seems that we are at a point in time in which neutrinos may become merely background to other breakthrough explorations. This couldn't be less true. If neutrinos are sensitive to some unknown physics which also affects WIMPs, experimental searches could constraint such interactions. This will be examined in detail here.\\ 

Regardless the specific topics we will discuss, neutrino physics is beyond any doubt one of the more active and compelling areas in Particle Physics. We certainly can imagine that the novel written by the clever neutrino would have an end. But before getting there, it could have chapters depicting the difficulties and wrong paths the inquisitive creatures found, and it may tell how those beings finally understood the neutrino. However, we, as main protagonists of such fictional history, do not know what awaits for us in the future, and what other surprises neutrinos have for us. 

\section*{About this Thesis}

This thesis intends to describe some phenomenological aspects of neutrino physics given the current status of the field. The main intention of the author is to give a friendly approach as complete as possible to the distinct issues and topics that he has addressed during his Doctoral studies in this fascinating and rich area. Keeping in mind this purpose, the document has been divided in two main parts. The first one contains the theoretical basis for a comprehension of the results obtained, and the second part includes the novel contributions that have resulted from the main research done in the past years.\\

The first part is composed by three chapters. The \hyperref[cha:nu-MP]{first} chapter encloses a brief description of the Standard Model; the details regarding the neutrino sources that will be used in subsequent chapters; and the basis of neutrino oscillations. Considering neutrinos as Majorana particles, in the \hyperref[cap:nuMaj]{second} chapter, we will give first a concise description of Majorana fermions, making explicit their peculiar properties. After that we will consider Majorana neutrinos in the SM framework from the point of view of the {\it see-saw} mechanism and one of its main consequences, leptogenesis. The other possibility for neutrinos, as being Dirac particles, is analysed in the \hyperref[nu-2HDM]{third} chapter. We will first describe the minimal SM extension, and, supposing that neutrino masses have a different origin from the other fermions, we will analyse the neutrinophilic two-Higgs-doublet models.\\

\newpage

The second part contains the original results of this thesis, as already mentioned. This part is also divided in three chapters.  First, in chapter \hyperref[nu-2HDM-feno]{four}, we will consider the phenomenological and theoretical limits on the neutrinophilic two-Higgs-doublet models coming from Electroweak precision measurements and flavour physics. Then, we will analyse the detection of the cosmic neutrino background in the \hyperref[cha:RelicNu]{fifth} chapter. Explicitly, we will describe the properties of such background and the detection by capture in tritium. Then, we will analyse the consequences of the possible existence of Non-Standard Interactions on the capture rate. After that, we will depart slightly from the main subject of the thesis in the \hyperref[cha:NeutrinoFloor]{sixth} chapter. We will study there the effect of the Coherent Neutrino Scattering off Nuclei on WIMP direct detection experiments. We will introduce the definition of the WIMP discovery limit considering only the SM interactions. Afterwards, we will study the impact of beyond SM physics, coupling with neutrinos and WIMP at the same experimental facilities. We will then give our \hyperref[cha:Conclu]{conclusions}. We also include an \hyperref[ap:Lgroup]{appendix} describing the fermion representation of the Lorentz Group and the construction of Weyl, Majorana and Dirac fields.\\ 

It is important to note that in each chapter of the second part we will use a different method to introduce new physics, namely, Ultraviolet complete models (neutrinophilic two-Higgs-doublet models); Effective Field Theory approach in the relic neutrino detection chapter; and simplified models, in the final chapter.\\

Throughout this Thesis, we will work with natural units in which the reduced Planck, the light speed and the Boltzmann constants are equal to the unity, $\hbar=c=k_B=1$. We also will make use of the Einstein notation, i.e. repeated indices indicate sum unless explicitly stated in the text. We will consider the Minkowski metric with trace $-2$ and the Dirac representation for the $\gamma^\mu$ matrices when necessary. We will also adopt the first letters of the Latin alphabet to indicate mass eigenstates, and the first Greek alphabet letters for flavour eigenstates. Greek letters starting from $\mu$ will indicate the space-time indices. Further definitions of the conventions used will be given in appendix \ref{ap:Conv}. 
	
		\part{Theoretical Basis} 
    	    
			%%%%%%%%%%%%%%%%%%%%%%%%%%%%%%%%%%%%%%%%%%%%%%%%%%%%%%%%%%%%%%%%%%%%%%%%%%%%%%%%%%%%%%%%%%%%%%%%%%
%%%%%%%%%%%%%%%%%%%%%%%%%%%%%%%%%%%%%%%%%%%%%%%%%%%%%%%%%%%%%%%%%%%%%%%%%%%%%%%%%%%%%%%%%%%%%%%%%%
%%%%%%%%%%%%%%%%%%%%%%%%%%%%%%%%%%%%%%%%%%%%%%%%%%%%%%%%%%%%%%%%%%%%%%%%%%%%%%%%%%%%%%%%%%%%%%%%%%

\chapter[Neutrinos in the Standard Model and Beyond]{Neutrinos in the Standard Model and Beyond}
\chaptermark{Neutrinos in the Standard Model and Beyond}\label{cha:nu-MP}

\lettrine{T}{he} two greatest milestones of the modern physics developed in the first
decades of the XX century, the Quantum Mechanics and the Relativity, have
become the keystones for any advancement in High Energy physics.
In other words, any quantum theory that attempts to describe consistently the physical phenomena
at high energies must be in accordance with the special relativity's principles.
The basic guidance to construct those theories is the lagrangian formalism, 
borrowed from the classical mechanics since it has the advantage of treating
equally space and time. In a relativistic compatible framework, the lagrangian, 
and therefore the action, must be invariant under the Lorentz transformations. 
On the other hand, it is firmly established that all matter fields, i.e.\ all 
quarks and leptons, are particles with spin one-half, i.e.\ they are {\it fermions}. 
Thus, it was necessary to build a invariant lagrangian for those fields, 
achievement accomplished by Dirac \cite{Dirac:1930ek} after the non-relativistic approach 
of Pauli. The current understanding of these fields, as belonging to a different representation 
of the Lorentz group from those describing scalar and vector fields, allows us to distinguish 
between two types of fermion representations, called, by historical reasons, {\it left}- and {\it right}-handed 
fermions \cite{Srednicki:2007qs}. These two distinct species of fermions emerge from the intrinsic 
properties of the Lorentz group, see appendix \ref{ap:Lgroup} for further 
details. Let us denote the type of representation as the \textit{\textbf{chirality}} of the field.
So, a fundamental question appears at this point: Is it strictly necessary
to have both chiralities for a complete description of an interaction? 
To answer this we need to notice that the parity operation converts one representation into the another. 
For this reason, Dirac indirectly included both species, making his theory 
parity-invariant. Nevertheless, as stated in the Introduction, the beta decay
does not conserve parity \cite{Wu:1957my}, and, consequently, we could have a unique fermion 
representation in a model for the Weak interactions \cite{Lee:1957qr}.

\newpage

Moreover, the smallness of the neutrino mass was established by direct measurements
in early studies of the weak interactions. Hence, physicists actually believed that 
the violation of the parity was a suggestion for a massless neutrino, 
represented by a left-handed chiral fermion. The justification for the last statement
is due to the work of Weyl \cite{Weyl:1927vd} where he proved that if a fermion was massless, 
one could describe such a particle either by a left- or a right-handed chiral field. This archetype
of a left-handed and massless neutrino was incorporated to the Standard Model (SM) 
in the 1960's \cite{Glashow:1961tr,Weinberg:1967tq,Salam:1968}. In this chapter we will 
present the neutrino as described in the SM. For that purpose, we will first consider briefly
the Weyl description of massless fermions and its most important properties. Then, we will 
introduce the SM and its basic characteristics, and, then, we will illustrate the relevant 
neutrino sources to be used in the development of the thesis. Finally, based on experimental
results, we will consider the current status of neutrino oscillations phenomena.

\section{Weyl Fermions}

Let us begin considering a massless left-handed two component spinor field $\psi_\alpha$, i.e.\ a 
Weyl fermion field whose lagrangian is given by
\begin{align}\label{eq:WeylLag}
	\mathscr{L}_{\rm Weyl}=\imath\psi^\dagger \bar{\sigma}^\mu \partial_\mu\psi
\end{align}
where $\bar{\sigma}^\mu=(I_{2\times 2}, -\vec{\sigma})$ is a set of Pauli 
matrices\footnote{One should be careful with the notation when stating that 
the set of matrices is a {\it four-vector}. Evidently, these matrices do not 
transform as a four-vector; they are independent of the inertial frame. Actually,
as shown in the appendix \ref{ap:Lgroup}, the current $\psi^\dagger \bar{\sigma}^\mu \psi$
do transform as a four-vector, and so we can write a invariant lagrangian as in 
\ref{eq:WeylLag}.}, see appendix \ref{ap:Conv}. This lagrangian is built considering 
the properties of spinors under Lorentz transformations, see appendix \ref{ap:Lgroup} 
for more details. The Weyl equation of motion,
\begin{align*}
	i\bar{\sigma}^\mu \partial_\mu\psi=0,
\end{align*}
has solutions that also solve the Klein-Gordon equation,
\begin{align*}
	\partial^\mu\partial_\mu\psi=0.
\end{align*}
Therefore, constructing the solutions is straightforward. A solution is
given by,
\begin{align*}
	\psi(x)=\psi_c\, e^{-ikx},\qquad \text{with}\quad k^2=0,
\end{align*}
with $\psi_c$ a constant two component spinor, depending on the direction of the 
propagation of the field.

\noindent \noindent The Weyl equation gives
\begin{align}
	\frac{\vec{\sigma}\cdot\vec{k}}{k^0}\psi&=-\psi,
\end{align}
showing that the solution is an eigenstate of the operator 
$(k^0)^{-1}[\vec{\sigma}\cdot\vec{k}]$. This operator, called {\it helicity}, is 
interpreted as the projection of the spin along the direction of motion. Let us note 
that this property is intrinsic to the Weyl fermion since it cannot be altered by a 
Lorentz transformation. In principle, if the particle was massive, one could boost
to another frame where the momentum is pointing in the opposite direction, changing the value of the 
projection. But this cannot be done for a massless particle. When the eigenvalue
of the helicity is negative, the particle is usually called left-handed. This is a
source of certain misunderstanding because it can be thought that helicity is 
equivalent to the fermion representation within the Lorentz group. {\it The type of 
representation, i.e., the chirality, will only coincide with the helicity in 
the case of a massless fermion.} Obviously, for a 
massive particle the helicity is frame-dependent while the chirality is not. 
Furthermore, a massless fermion with a definite helicity
violates the parity symmetry, as the parity reverts the linear momentum, keeping
at the same time the angular momentum invariant. Thus, to avoid confusion from now on, we will 
designate a particle with a negative (positive) helicity as {\it left-(right-)helical}.\\

All fundamental fermions are now known to have mass, but, in the 1960's, there was no unquestionable
evidences for that. The experiments showed that the neutrino mass was quite small, but there was 
no proof for it being different from zero. Invoking the Occam's razor, the models were 
built considering the neutrino as left-handed massless fermion \cite{Lee:1957qr,Landau:1957tp}, 
and the SM was assembled with this conjecture. Consequently, the SM is as a {\bf chiral theory} 
since the interactions affect differently the the two fermion chiral types. 
Hereafter, we will introduce the SM considering the basis for its construction, 
as the gauge principle and the Higgs mechanism.

\section{The Standard Model in a nutshell}

The modern theories are built considering the {\bf gauge principle}; this principle 
expresses that a theory must be invariant under local (gauge) phase transformations. 
Usually, when a free lagrangian possesses a global symmetry, in such a way that 
there exists a conserved charge due to the Noether's theorem, it is imposed that 
such symmetry has to be a local one. Under the new local symmetry the lagrangian is no 
longer invariant. It is necessary to introduce new fields, with specific transformation 
laws, that compensate for the extra terms. Afterwards, it is noticed that the new fields, 
called gauge fields, mediate the interactions among to particles present in the initial 
lagrangian. This also can be viewed as the substitution of the partial derivatives
for {\it covariant derivatives}, derivatives which contain the gauge fields in a specific manner.
The best known example of a gauge theory is the Quantum Electrodynamics (QED) \cite{Tomonaga:1946zz,Schwinger:1948iu,Feynman:1950ir},
the theory of the electromagnetic interactions among electrons and positrons, which is
mediated by a gauge field, the photon. The QED is a prototype to construct other gauge theories;
the most important of all is the SM.\\

The SM is a theory for the strong, electromagnetic and weak fundamental 
interactions \cite{Glashow:1961tr,Weinberg:1967tq,Salam:1968}. One of its most important 
results is the conspicuous unification 
between the electromagnetic and weak interactions into the so-called {\it Electroweak}
interaction. Given that our purpose is to study the different properties of 
neutrinos, we will concentrate ourselves on the electroweak part of the SM. \\

The accomplishments that the SM has presented since its formulation are beyond 
any doubt. Distinct tests, in both theoretical and experimental sides, have shown 
that this model gives an accurate description of nature. Undoubtedly, the SM
still has to resolve several issues concerning, for instance, the set of parameters 
contained in the model. The mathematical formulation of the SM has been the subject of 
innumerable books, papers and thesis, many of which are far more complete and
detailed than the description below. The purpose of this section will be to define 
the notation that will be used in this thesis and the relevant components necessary
for a complete subsequent comprehension.\\

Technically speaking, the SM is a gauge theory whose symmetry group, i.e.\ the group 
of the local transformations which leave the lagrangian invariant, is SU$(2)_L\times$U$(1)_Y$.
The subscripts denote that the weak interactions are left-handed ($L$), and there 
exists an additional abelian interaction, identified as {\it hypercharge} ($Y$). Regarding
the fields that compose the theory, we will classify them in three classes: matter
fields which are the fermion fields that constitute the matter of the Universe; gauge fields,
fields that carry the interactions, as stated before; and the symmetry breaking fields, 
which are responsible to give mass to the matter and gauge fields. In order to write
a consistent lagrangian, we need to define how our matter fields transform under the 
symmetry group, and define the gauge fields by an appropriate designation of the 
covariant derivatives. \\

The matter fields that compose the SM are divided in two categories, depending on whether
interact strongly or not: 6 quarks (up $u$, down $d$, charm $c$, strange $s$, top $t$ 
and bottom $b$) and 6 leptons (electron $e$, electron neutrino $\nu_e$, muon $\mu$, 
muon neutrino $\nu_\mu$, tau $\tau$, tau neutrino $\nu_\tau$). They are grouped in 
three {\bf generations}, each one composed by two quarks and two leptons. These groups 
are not arbitrary, instead they are arranged according to their increasing mass. From the point of 
view of the gauge symmetry, each chiral component of the matter fermions transforms in a 
different way. The left-handed matter fields will belong to the fundamental representation
of SU$(2)_L$\footnote{For convenience, we will consider the 4-component notation for the fermion
fields. The change among the notations is explicitly considered in the appendix \ref{ap:Lgroup}.},
\begin{align*}
	L_L^\alpha=\left\{
		\begin{pmatrix}
			\nu_{e L}\\
			e_L
		  \end{pmatrix},
		  \begin{pmatrix}
			\nu_{\mu L}\\
			\mu_L
		  \end{pmatrix},
		  \begin{pmatrix}
			\nu_{\tau L}\\
			\tau_L
		  \end{pmatrix}\right\},
	\qquad
	Q_L^\alpha=\left\{
		\begin{pmatrix}
			u_{L}\\
			d_{L}
		 \end{pmatrix},
		 \begin{pmatrix}
			c_{L}\\
			s_{L}
		 \end{pmatrix},
		 \begin{pmatrix}
			t_{L}\\
			b_{L}
		 \end{pmatrix}\right\},
\end{align*}
being $\alpha=1,2,3$ the generation (flavour) index, while the right-handed will be singlets of SU$(2)_L$,
\begin{align*}
	\ell_R^\alpha = \{e_R,\mu_R,\tau_R\},\qquad u^\alpha_R=\{u_R,c_R,t_R\},\qquad d^{\alpha}_R=\{d_R,s_R,b_R\}.
\end{align*}
Let us emphasize that the right-handed neutrinos are absent, in order to keep the neutrino 
massless as a result of our previous discussions. For the case of the hypercharge group,
the charges of the matter fields are given in table \ref{tab:mtfcharges}. The lagrangian for 
the matter fields is given by,
\begin{align}\label{eq:LagMF}
	\mathscr{L}_{\rm f}=\sum_{\alpha}&\left[\overline{L^\alpha_L} i\gamma^\mu\corc{\partial_\mu-igW^j_\mu\tau^j-ig^\prime Y B_\mu}L_L^\alpha
	                    +\overline{Q_L^\alpha} i\gamma^\mu\corc{\partial_\mu-igW^j_\mu\tau^j-ig^\prime Y B_\mu}Q_L^\alpha\right.\notag\\
					   &+\left.\overline{\ell^\alpha_R} i \gamma^\mu\corc{\partial_\mu-ig^\prime Y B_\mu}\ell^\alpha_R
						+\overline{u^{\alpha}_R} i \gamma^\mu\corc{\partial_\mu-ig^\prime Y B_\mu}u^\alpha_R
						+\overline{d^\alpha_R} i \gamma^\mu\corc{\partial_\mu-ig^\prime Y B_\mu} d^\alpha_R\right].
\end{align}
We introduced here the gauge fields, $W_\mu^j$, $j=1,2,3$, related to the SU$(2)_L$ symmetry, 
and $B_\mu$ to the U$(1)_Y$ one. These gauge fields are spin-1 bosons, and we will denominate
them simply by {\it gauge bosons}. The parameters $g,g^\prime$ are the coupling constants of 
the interactions, and $\tau^j$ are the generators of SU$(2)_L$, $\tau^j=\frac{1}{2}\sigma^j$. 
Let us note that there exists a gauge boson for each generator of the SM group. The gauge bosons 
also have a lagrangian that describes their kinetic terms,
\begin{align}\label{eq:LagGF}
	\mathscr{L}_{\rm g}=-\frac{1}{4}W^j_{\mu\nu}W^{j\,\mu\nu}-\frac{1}{4}B_{\mu\nu}B^{\mu\nu}
\end{align}
where the field strength tensors $W_{\mu\nu}^j$ and $B_{\mu\nu}$ are
\begin{subequations}
	\begin{align}
		W_{\mu\nu}^j&=\partial_\mu W_{\nu}^j-\partial_\nu W_\mu^j-g\varepsilon_{jkl}W_\mu^k W_\nu^l,\label{eq:KinTerWs}\\
		B_{\mu\nu}&=\partial_\mu B_{\nu}^j-\partial_\nu B_\mu^j.\label{eq:KinTerBs}
	\end{align}
\end{subequations}
%%%%%%%%%%%%%%%%%%%%%   table  %%%%%%%%%%%%%%%%%%%%%%%%%%%5
\begin{table}[t]
	\centering
	\caption{Charges of the matter and symmetry breaking fields.}
	\label{tab:mtfcharges}
	\begin{tabular}{l|cccccc}
		\toprule
    	          & $L_L^\alpha$        & $\ell^\alpha_R$ & $Q^\alpha_L$       & $u^\alpha_R$       & $d^\alpha_R$        & $\Phi$        \\ \midrule\midrule
		SU$(2)_L$ & $\dfrac{1}{2}$  & $0$      & $\dfrac{1}{2}$ & $0$           & $0$            & $\dfrac{1}{2}$ \\ \midrule
		U$(1)_Y$  & $-\dfrac{1}{2}$ & $-1$     & $\dfrac{1}{6}$ & $\dfrac{2}{3}$ & $-\dfrac{1}{3}$ & $\dfrac{1}{2}$ \\ \bottomrule
	\end{tabular}
\end{table}
%%%%%%%%%%%%%%%%%%%%%%%%%%%%%%%%%%%%%%%%%%%%%%%%%%%%%%%%%%%%
\negthickspace Let us mention here that the SU$(2)_L$ symmetry group in non-Abelian; 
therefore, we expect to have self-interactions among the gauge bosons related to 
this group. This is the reason why there is a term in \eqref{eq:KinTerWs} which 
is absent in \eqref{eq:KinTerBs}.\\

\newpage
The final interaction lagrangian will be simply the sum of the lagrangians for the
matter \eqref{eq:LagMF} and gauge \eqref{eq:LagGF} fields. All possible 
electroweak interactions among the fermion fields is contained there. This 
lagrangian is gauge invariant, by construction, and renormalizable \cite{Bollini:1972,tHooft:1972tcz}.
However, we encounter here three problems: the charged fermions should have masses, which 
has been well established by the experiments; second, it is not clear how the electromagnetic 
interaction emerge in this model; and, third, the gauge bosons 
that mediate the weak interaction should be also massive. Let us explore in more 
detail the last difficulty. Experiments show that the range of the weak interaction 
is finite. On the other hand, if one considers the temporal component of a massive spin-1 boson, 
one finds that the potential associated, denominated {\it Yukawa potential}, has a short range. 
Thus, to explain the weak interactions, we need the gauge bosons to be massive. The solution 
for this problem was found by Englert and Brout \cite{Englert:1964et}, Higgs \cite{Higgs:1964pj} 
and Guralnik, Hagen and Kibble \cite{Guralnik:1964eu}.

\subsection{Mass Generation in the SM}

The initial problem consisted in constructing a gauge invariant lagrangian that
possess mass terms for the gauge bosons and for the fermions. An explicit mass 
term for those fields is not gauge invariant since, for the case of the matter fields,
the left- and right-handed transform differently. The mechanism to give mass to the 
particles while maintaining the gauge invariance is known as the {\it Higgs mechanism}. 
A fundamental consequence of the application of this mechanism to the 
SM is that a symmetry will remain unbroken, corresponding to the electromagnetic 
interaction; or, in other words, the photon, will remain massless. To implement
the mechanism, we need to introduce a SU$(2)_L$ doublet, composed of complex
scalars,
\begin{align*}
	\Phi=\begin{pmatrix}
			\phi^+\\
			\phi^0
		\end{pmatrix},
\end{align*}
with a hypercharge given in table \ref{tab:mtfcharges}. Then, we will need a lagrangian
to describe the scalar doublet,
\begin{align}\label{eq:LagH}
	\mathscr{L} = \Bigg[\underbrace{\corc{\partial_\mu-igW^j_\mu\tau^j-ig^\prime Y B_\mu}}_{D_\mu}\Phi]^\dagger [\corc{\partial^\mu-igW^{j\,\mu}\tau^j-ig^\prime Y B^\mu}\Phi\Bigg]-\mathrm{V}(\Phi^\dagger \Phi);
\end{align}
the scalar potential $\mathrm{V}(\Phi^\dagger \Phi)$ is chosen to {\bf spontaneously break} the 
gauge symmetry. By this we mean that the potential has a minimum value, the vacuum state, 
which is not invariant under the gauge symmetry. Thus, the excited states over
the vacuum will not manifest explicitly the symmetry. Let us show this in some detail.
The potential is given by
\begin{align}
	\mathrm{V}(\Phi^\dagger \Phi)=-\mu^2 \Phi^\dagger \Phi + \lambda (\Phi^\dagger \Phi)^2,
\end{align}
with $\mu > 0$ and, also, $\lambda > 0$. We see that this potential has a minimum, 
$\frac{\partial V}{\partial \Phi} = 0$, when $\Phi^\dagger \Phi = \frac{\mu^2}{2\lambda}$. 
The crucial point here is that we can {\it choose} the vacuum state without loss of 
generality\footnote{Initially, such vacuum state can be taken in a general way, but after 
performing a gauge transformation one can obtain the case we are considering.}. We take
\begin{align*}
	\langle\Phi\rangle=\frac{1}{\sqrt{2}}\begin{pmatrix}
											0\\
											v\\
										\end{pmatrix},
\end{align*}
with $v$, the {\bf vacuum expectation value} (VEV) given by
\begin{align*}
	v\equiv \sqrt{\frac{\mu^2}{\lambda}}.
\end{align*}
The choice of the vacuum is done to break the gauge symmetry. For instance, applying
the hypercharge operator $Y$, we have,
\begin{align*}
	Y\langle\Phi\rangle = \frac{1}{2}\langle\Phi\rangle
\end{align*}
which is non-zero. This implies that the vacuum has an hypercharge! Now, let
us compute the case of the third SU$(2)_L$ operator, $\tau_3$,
\begin{align*}
	\tau_3\langle\Phi\rangle = -\frac{1}{2}\langle\Phi\rangle,
\end{align*}
that is also non-zero. However, the combination $\tau_3+Y$ gives us zero, 
so that the vacuum is invariant under that combination,
\begin{align*}
	e^{i\theta (\tau_3+Y)}\langle\Phi\rangle=\langle\Phi\rangle.
\end{align*}
We can now define the electric charge operator, called {\bf Gell-Mann--Nishijima} 
operator, as
\begin{align}
	\mathcal{Q}=\tau_3+Y,
\end{align}
so, the electromagnetic interaction will remain unbroken. We can now 
write explicitly the lagrangian in the broken phase. To do so, we write the
scalar doublet as
\begin{align}
	\Phi=\exp\left[\frac{i}{2v}\xi^a(x)\tau^a\right]
										\begin{pmatrix}
											0\\
											\frac{v+h(x)}{\sqrt{2}}\\
										\end{pmatrix},
\end{align}
being $\xi^a (x)$ and $h(x)$ scalar fields. The $\xi^a (x)$ field
are also known as {\bf Goldstone bosons} \cite{Goldstone:1962es}, and they will be massless.
Taking a gauge transformation, these Goldstone fields can be hidden
in the theory. In fact, they are {\it absorbed} by the gauge bosons,
becoming the longitudinal polarization which a massive spin-1
particle has, but a massless one does not. The kinetic part of the lagrangian \eqref{eq:LagH} contains the crucial
terms,
\begin{align}
	\left[D_\mu\Phi\right]^\dagger \left[D^\mu\Phi\right]
	&= \left|\corc{\partial_\mu-igW^j_\mu\tau^j-ig^\prime Y B_\mu}\begin{pmatrix}
											0\\
											\frac{v+h(x)}{\sqrt{2}}\\
										\end{pmatrix}\right|^2\notag\\
	&=\frac{1}{8}v^2\left[g^2 \left(W_\mu^1 W^{1\mu}+W_\mu^2 W^{2\mu}\right)+\left(gW_\mu^3-g^\prime B_\mu\right)\left(gW^{3\mu}-g^\prime B^\mu\right)\right]\notag\\
	&\quad+\frac{1}{2}(\partial_\mu h)(\partial^\mu h)+\frac{1}{2}\underbrace{2\lambda v^2}_{m_h^2} h^2+\mathrm{interactions},
\end{align}
so we can conclude here that, after the spontaneous symmetry breaking, we obtain
mass terms for the gauge bosons although there seems to appear a mixing between 
$W_\mu^3$ and $B_\mu$. This is solved when we define the combinations,
\begin{subequations}
	\begin{align}
		W^\pm_\mu&=\frac{1}{2}(W_\mu^1\pm iW^2_\mu),\\
		Z_\mu&=\cos\theta_W W_\mu^3 -\sin\theta_W B_\mu,\\
		A_\mu&=\sin\theta_W W_\mu^3 +\cos\theta_W B_\mu,
	\end{align}
\end{subequations}
where the {\bf weak angle} $\theta_W$ was introduced as
\begin{align}
	\tan\theta_W=\frac{g^\prime}{g}.
\end{align}
Substituting on the kinetic term, we have that
\begin{align}
	\left[D_\mu\Phi\right]^\dagger \left[D^\mu\Phi\right]
	&=\frac{1}{4}g^2v^2W_\mu^-W^{+\mu}+\frac{1}{8\cos^2\theta_W}g^2v^2 Z_\mu Z^\mu\notag\\
	&\quad+\frac{1}{2}(\partial_\mu h)(\partial^\mu h)+\frac{1}{2}m_h^2 h^2+\mathrm{interactions},
\end{align}
so now is completely clear that three weak gauge bosons have mass, $W^\pm, Z^0$,
$m_W=g v/2$, $m_Z=g v/(2 \cos\theta_W) $; while the fourth one, 
the photon $A_\mu$, is massless, as expected. Experimentally,
all these particles have been found which was one of the first triumphs of the 
SM. On the other hand, we did not comment about the scalar field $h$, the {\bf Higgs 
boson}. As we can see, this scalar have a mass $m_h=\sqrt{2\lambda v^2}$ which is 
not predicted by the model. Nonetheless, a particle close to what is expected of the 
Higgs boson behaviour was found in the LHC, with a mass of $\sim 125$ GeV \cite{Aad:2012tfa,Chatrchyan:2012xdj}.
Studies still need to be done to affirm without doubt that this particle is 
in fact the SM Higgs or other similar particle. The last scenario seems more
compelling, given that it opens a window to physics beyond the SM.\\

Now, we need to write masses for the fermions. For that purpose, we need to join 
the left- and right-handed chiral parts without explicitly breaking the symmetry.
A simple manner to do this is using the scalar doublet $\Phi$, for instance,
for the charged leptons
\begin{align}
	\mathscr{L}_{\rm Y}^\ell=- y_{\alpha\beta}^\ell\, \overline{L_L^\alpha} \Phi \ell^\beta_R +{\rm h.c.};
\end{align}
this term is gauge invariant. Note that this is a general term since the {\it Yukawa}
couplings matrix $y_{\alpha\beta}^\ell$ can be complex and non-diagonal. Thus, in principle, 
it will be necessary to rotate to the physical states with a defined mass. In the
case of the charged leptons, we have after the spontaneous symmetry breaking
\begin{align*}
	\mathscr{L}_{\rm Y}^\ell=-\frac{v+h}{\sqrt{2}} y_{\alpha\beta}^\ell \overline{\ell^\alpha_L} \ell^\beta_R+{\rm h.c.}.
\end{align*}
The rotation is achieved by defining the mass eigenstates $\ell^{a}_{L,R}$ as a linear combination
of the {\it flavour} eigenstates, $\ell^\alpha_{L,R}$, 
\begin{align*}
	\ell^\alpha_{L,R}= V_{L,R}^{\alpha a}\ell^{a}_{L,R},
\end{align*}
given that the matrices $V_{L,R}^{ab}$ diagonalize the Yukawa matrix,
\begin{align*}
	(V_L^{a \alpha})^*y_{\alpha\beta}^\ell V_R^{\beta b} = y_a^\ell \delta_{ab}.
\end{align*}
Therefore, we have that,
\begin{align}
	\mathscr{L}_{\rm Y}^\ell=&-\frac{v+h}{\sqrt{2}} y_{a}^\ell \overline{\ell^{a}_L} \ell^{a}_R+{\rm h.c.}\notag\\
		=&- \left(1+\frac{h}{v}\right) m_a^l\, \overline{\ell^a} \ell^a,
\end{align}
so, we find that the charged leptons have masses $m_a^l=\displaystyle{\frac{y_a^\ell v}{2}}$, $m_a^l=\{m_e,m_\mu,m_\tau\}$,
and couplings to the Higgs boson also proportional to their masses. Obviously,
the neutrinos are massless, as we wanted. But the SM do not predict the 
values of the charged lepton masses, as the Yukawas are free parameters.
Here, a simple question may be asked: are there any consequences of 
this mismatch between mass and flavour states? This question may appear simple, 
but it is the basis for the confirmation of the non-zero value of the neutrino masses. 
But before attacking the neutrino sector, let us complete the fermion discussion 
with the quarks. In this case, we have that in the Yukawa lagrangian we need to write two types of terms,
\begin{align}
	\mathscr{L}_{\rm Y}^q=- y_{\alpha\beta}^d\, \overline{Q_L^\alpha} \Phi d^\beta_R 
					+y_{\alpha\beta}^u\, \overline{Q_L^\alpha} \widetilde{\Phi} u^\beta_R +{\rm h.c.}.
\end{align}
This is because a term involving a quark doublet with the right-handed up-like
quarks $u^\alpha$ cannot be written with the scalar doublet since it would not be gauge 
invariant. Instead, it is necessary to consider the {\it conjugate} doublet, 
$\widetilde{\Phi}=i\sigma^2 \Phi^\dagger$, which also belongs to the fundamental 
representation but has the opposite hypercharge (notice the similarity with the two 
inequivalent representations of the Lorentz group, appendix \ref{ap:Lgroup}). 
On the other hand, the Yukawa matrices $y_{\alpha\beta}^d,y_{\alpha\beta}^u$ do not need to be 
diagonal, as in the charged lepton case; it is required to diagonalize those 
matrices by redefining the mass eigenstates, analogously to the charged leptons,
\begin{align*}
	u^\alpha_{L,R} = W_{L,R}^{\alpha a}u^{a}_{L,R}, \qquad d^\alpha_{L,R} = \widetilde{W}_{L,R}^{\alpha a}d^{a}_{L,R}
\end{align*}
where 
\begin{align*}
	(W_L^{a\alpha})^*y_{\alpha\beta}^u W_R^{\beta b} = y_a^u \delta_{ab}, \qquad (\widetilde{W}_L^{a \alpha})^*y_{\alpha\beta}^d \widetilde{W}_R^{\beta b} = y_a^d \delta_{ab}.
\end{align*}
Again, we find that diagonalized lagrangian is
\begin{align}
	\mathscr{L}_{\rm Y}^q=-\left(1+\frac{h}{v}\right) \left[m_a^u\, \overline{d^a} d^a 	+m_a^u\, \overline{u^a} u^a\right].
\end{align}
The quarks masses are not predicted by the SM, as in the case of the leptons.
However, there is a consequence of the discrepancy among eigenstates. Let us 
define the charged current for quarks as,
\begin{align}
	j_{Wq}^\mu = \overline{u_L^\alpha} \gamma^\mu d_L^\alpha.
\end{align}
After the diagonalization, we find that
\begin{align}
	\overline{u_L^\alpha} \gamma^\mu d_L^\alpha \tto  \overline{u_L^a} \gamma^\mu \underbrace{(W_L^{a\alpha})^*\widetilde{W}_{L}^{\alpha b}}_{U_{\rm CKM}^{ab}} d_L^b
\end{align}
where the {\it Cabibbo-Kobayashi-Maskawa} (CKM) matrix \cite{Cabibbo:1963yz, Kobayashi:1973fv}, 
$U_{\rm CKM}^{ab} = (W_L^{a\alpha})^*\widetilde{W}_{L}^{\alpha b}$,
was defined. This is a complex unitary matrix with 9 free parameters, in 
principle. Nonetheless, it is possible to eliminate five phases by re-shifting 
the quark fields, $q\tto e^{i\theta} q$, remaining only four parameters. 
This mixing matrix, understood as a rotation in the quark "three-dimensional" space, 
is parametrized by three Euler angles $\theta_{12}, \theta_{23}, \theta_{13}$,
and an additional phase, $\delta$. This last parameter is related with the CP
violation that appears in the quark sector. Therefore, we see manifestly the
importance of the mixing for the High Energy Physics. Experimentally, flavour-changing charged processes
have been found, proving of the non-diagonality of the CKM matrix.
On the other hand, Flavour-Changing Neutral Current (FNCN) processes in the SM are very 
suppressed. Actually, they only occur at loop level, given that at tree level the so-called
Glashow-Iliopoulos-Maiani (GIM) mechanism forbids these processes. In fact, there are 
experimental strong limits to these FNCN, and these will constraint any new physics beyond 
the SM. Now, we have completed our task. The fermions and the weak gauge bosons have masses, 
while the photon is massless.\\ 

Finally, and for future convenience, let us introduce the ladder operators,
\begin{align*}
	\tau^\pm=\tau^1\pm i \tau^2;
\end{align*} 
also, we define next the couplings of the left-handed ($L_L,Q_L$) and
right-handed ($\ell_R, u_R,$ $d_R$) matter fields with the $Z^0$ boson:
\begin{itemize}
	\item leptons,
			\begin{align*}
				g_Z^L=g\cos\theta_W\tau^3+\frac{1}{2}e\tan\theta_W, \qquad g_Z^\ell=-e\tan\theta_W;
			\end{align*}		
	\item quarks
			\begin{align*}
				g_Z^Q =g\cos\theta_W\tau^3+\frac{1}{6}e\tan\theta_W, \quad g_Z^u =\frac{2}{3}e\tan\theta_W, \quad g_Z^d = -\frac{1}{3}e\tan\theta_W,
			\end{align*}
\end{itemize}
where the electromagnetic coupling, $e=g\sin\theta_W$, appears explicitly. This is 
the last piece of our construction of the SM. Let us write down in its entire 
magnificence the SM lagrangian in the broken phase
\begin{mdframed}[backgroundcolor=gray!20]
	\begin{align}\label{eq:LagSM}
		\mathscr{L}_{\rm SM}&=\mathscr{L}_{\rm f}+\mathscr{L}_{\rm g}+\mathscr{L}_{\rm \Phi}+\mathscr{L}_{\rm y}\notag\\
		 &= \left.\overline{L^a_L} i\gamma^\mu\corc{\partial_\mu-ig(W_\mu^+\tau^++W_\mu^-\tau^-)-ig_Z^L Z_\mu-i e\mathcal{Q}A_\mu}L_L^a\right.\notag\\
		 &\qquad+\left.\overline{Q_L^a} i\gamma^\mu\corc{\partial_\mu-ig(W_\mu^+\tau^+U_{\rm CKM}+W_\mu^-\tau^-U_{\rm CKM}^\dagger)-ig_Z^Q Z_\mu-i e\mathcal{Q}A_\mu}Q_L^a\right.\notag\\
		 &\qquad+\left.\overline{\ell^a_R} i \gamma^\mu\corc{\partial_\mu-ig_Z^\ell Z_\mu-i e\mathcal{Q}A_\mu}\ell^a_R\right.\notag\\
	 	 &\qquad+\left.\overline{u^{a}_R} i \gamma^\mu\corc{\partial_\mu-ig_Z^u Z_\mu-i e\mathcal{Q}A_\mu}u^a_R
	 	  +\overline{d^a_R} i \gamma^\mu\corc{\partial_\mu-ig_Z^d Z_\mu-i e\mathcal{Q}A_\mu} d^a_R\right.\notag\\
	 	 &\quad-\frac{1}{4}W^j_{\mu\nu}W^{j\,\mu\nu}-\frac{1}{4}B_{\mu\nu}B^{\mu\nu}+m_W^2 W_\mu^-W^{+\mu}+\frac{1}{2}m_Z^2 Z_\mu Z^\mu\notag\\
     	 &\quad+\frac{1}{2}(\partial_\mu h)(\partial^\mu h)+\frac{1}{2}m_h^2 h^2-\lambda v h^3-\frac{\lambda}{4}h^4\notag\\
     	 &\quad+\frac{g^2}{2}\esp{W_\mu^\dagger W^\mu+\frac{1}{2\cos^2\theta_W}Z_\mu Z^\mu}h\corc{v+\frac{h}{2}}\notag\\
		 &\quad-\left(1+\frac{h}{v}\right) \left[m_a^l\, \overline{\ell^a} \ell^a + m_a^u\, \overline{d^a} d^a 	+m_a^u\, \overline{u^a} u^a\right].
	\end{align}
\end{mdframed}
\newpage 
\noindent We see that electromagnetic and weak interactions are two facets of a unique interaction, 
the electroweak interaction. The separation between the two 
forces is a result of the spontaneous breaking due to the non zero VEV of the scalar potential. 
Using our knowledge of Thermodynamics and Cosmology, we can imagine that the universe should 
have been in an unbroken phase where all the particles were massless and interacted 
with a unique electroweak force. Then, due to the expansion of the universe, a phase transition 
occurred, giving mass to the charged fermions and the weak gauge bosons, while 
maintaining the electromagnetic symmetry intact \cite{Kolb:1990vq}. We can think if there is a fundamental 
reason for the electromagnetic force be unaltered. However, any thoughts about this will 
belong to the speculative realm. In any case, we can now focus our study on the neutrino
sector relevant for our purposes, considering natural and artificial sources, and the experiments 
which have studied these particles.
%-
\begin{figure}[t]
	\begin{center}
    			\includegraphics[width=\textwidth]{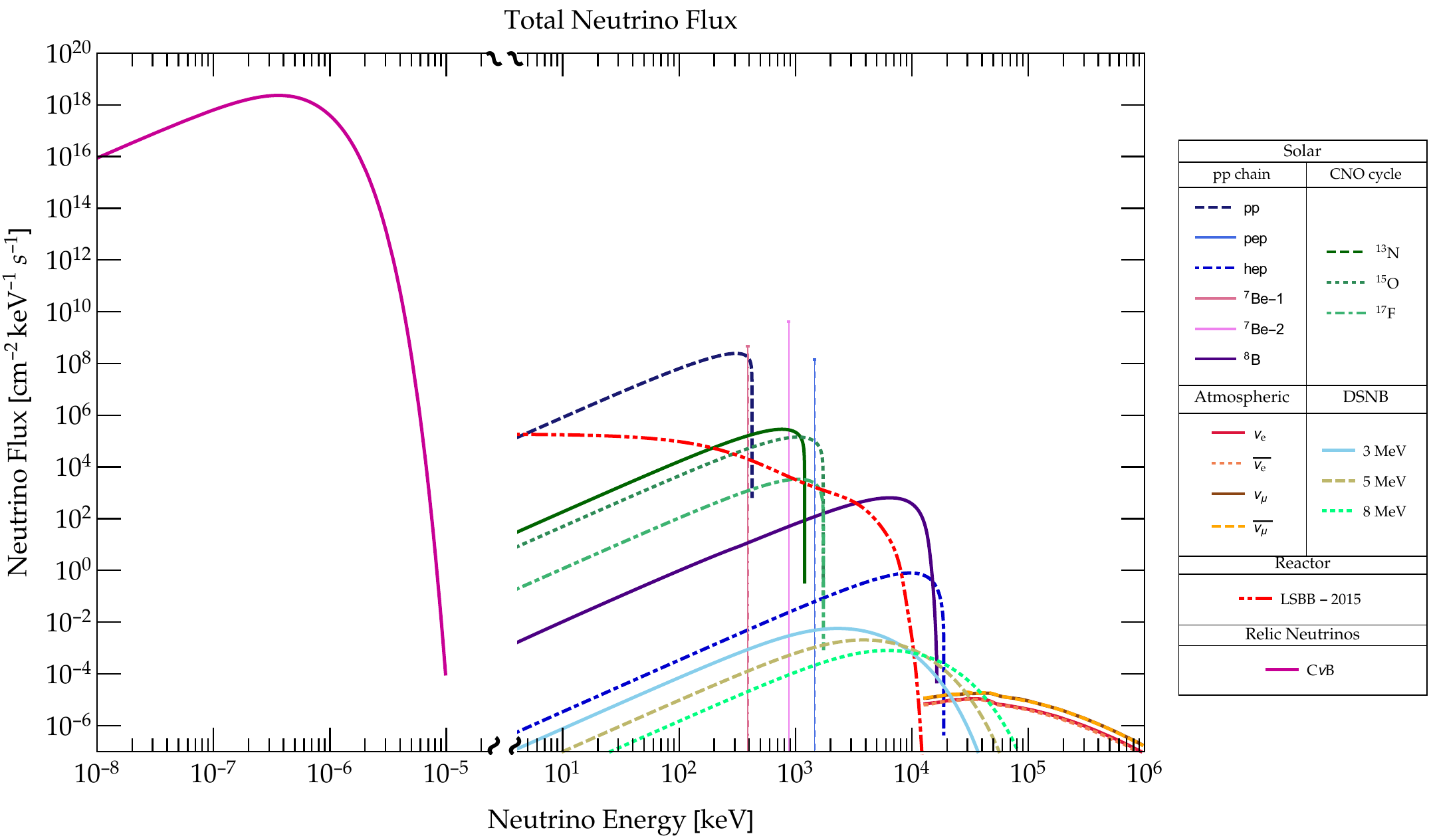}  
	\end{center}
  	\caption{Energy spectra of the neutrino fluxes considered in the present thesis.}
  	\label{fig:TotalNu}
\end{figure}
%-

\section{Neutrino Sources}\label{sec:NFl}

All possible neutrino interactions present in the SM lagrangian, equation \eqref{eq:LagSM},
allow us to understand several processes actually happening in nature. Moreover, the different 
neutrino sources are windows to comprehend the properties of the neutrino, and also 
analyse if there are deviations from what is expected from the SM. There are three basic types
of neutrino sources: with an astrophysical origin, such as the Solar, Supernova, galactic and 
cosmological neutrinos; terrestrial, as the atmospheric and geoneutrinos; and artificial, 
like the reactor and accelerator neutrinos. In the present thesis, we will concentrate ourselves
on the solar, atmospheric, reactor neutrinos, together with the diffuse supernova and cosmic 
neutrino background. The energy dependence of the flux of such neutrinos is in 
figure \ref{fig:TotalNu}. Let us now study the four cases separately; the fifth case, the cosmic
neutrino background, will be considered in chapter \ref{cha:RelicNu}.

\subsection{Solar Neutrinos}

\begin{wrapfigure}{R}{7.5cm}
	\begin{center}
    			\includegraphics[scale=0.5]{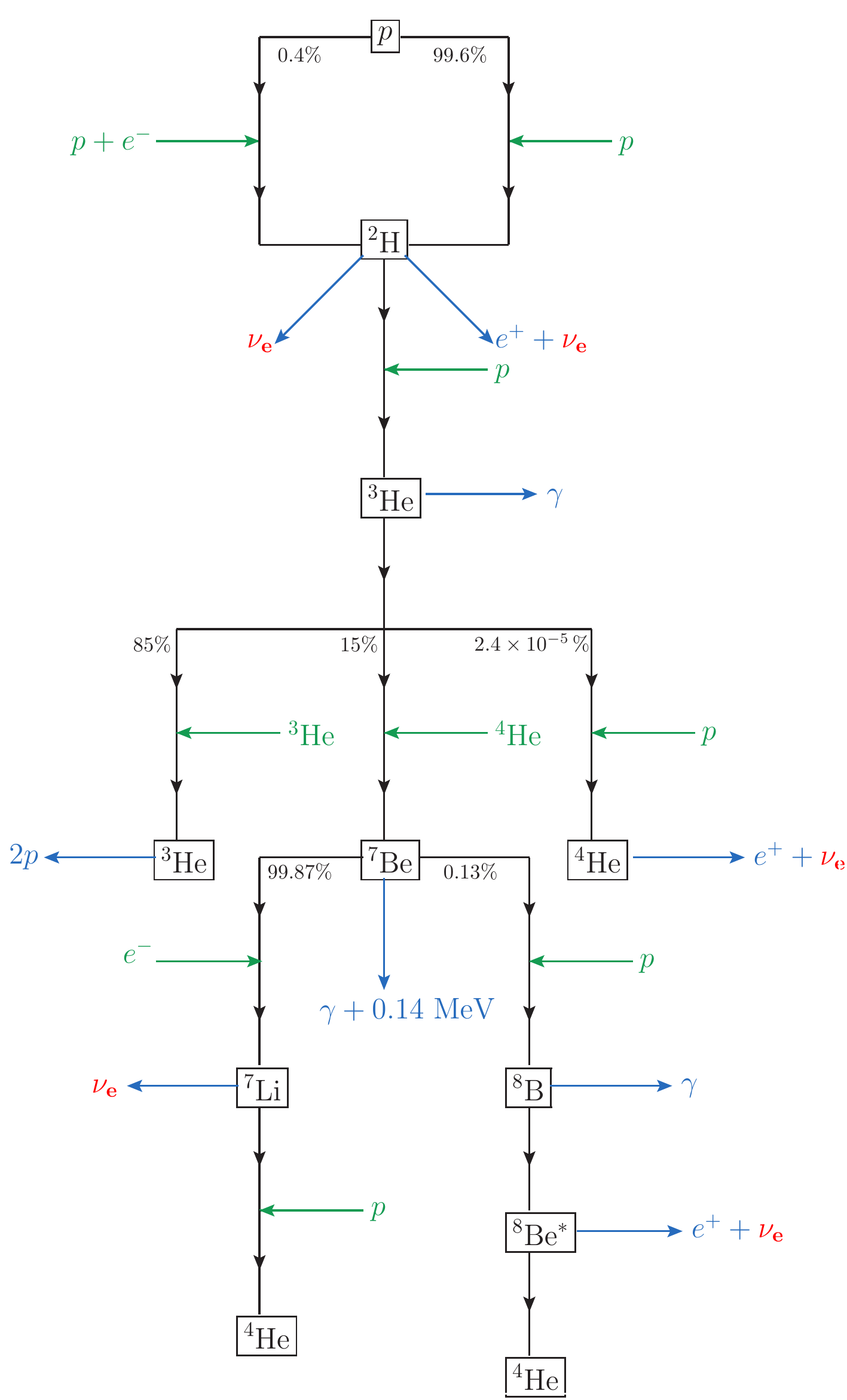}  
	\end{center}
  	\caption{Representation of the solar pp chain reactions.}
  	\label{fig:ppchain}
\end{wrapfigure}

Since the dawn of man, the Sun has been recognized by its immense significance for Earth, inspiring 
several myths about its origin and influence on the mankind. Nowadays, a complete 
picture about our star has been established, as a plasma sphere which is maintained 
due to the perfect balance between gravity and the radiation pressure, created by the 
thermonuclear fusion of several elements. A crucial consequence is that we now recognize
the Sun as a huge source of neutrinos~\cite{Giunti:2007ry}. Multiple experiments have detected neutrinos
coming from our star, allowing us to broaden our knowledge of these particles, and also about
the Sun itself, given that the neutrinos carry direct information concerning the solar interior.
The energy of the Sun is created by thermonuclear processes, as concluded by Gamow and Bethe 
in the late 30's~\cite{Gamow:1938,Bethe:1939bt}. There are two basic chains of these mechanisms in our star: the pp-chain 
and CNO-cycle. In the pp-chain, two protons are converted mainly in $^4$He through the fusion 
and/or decay of several isotopes. In figure \ref{fig:ppchain}, we present schematically this chain.
Let us explain this sequence in more detail.\\
%-
%-

\begin{figure}[t]
  		\begin{center}
    			\includegraphics[scale=0.45]{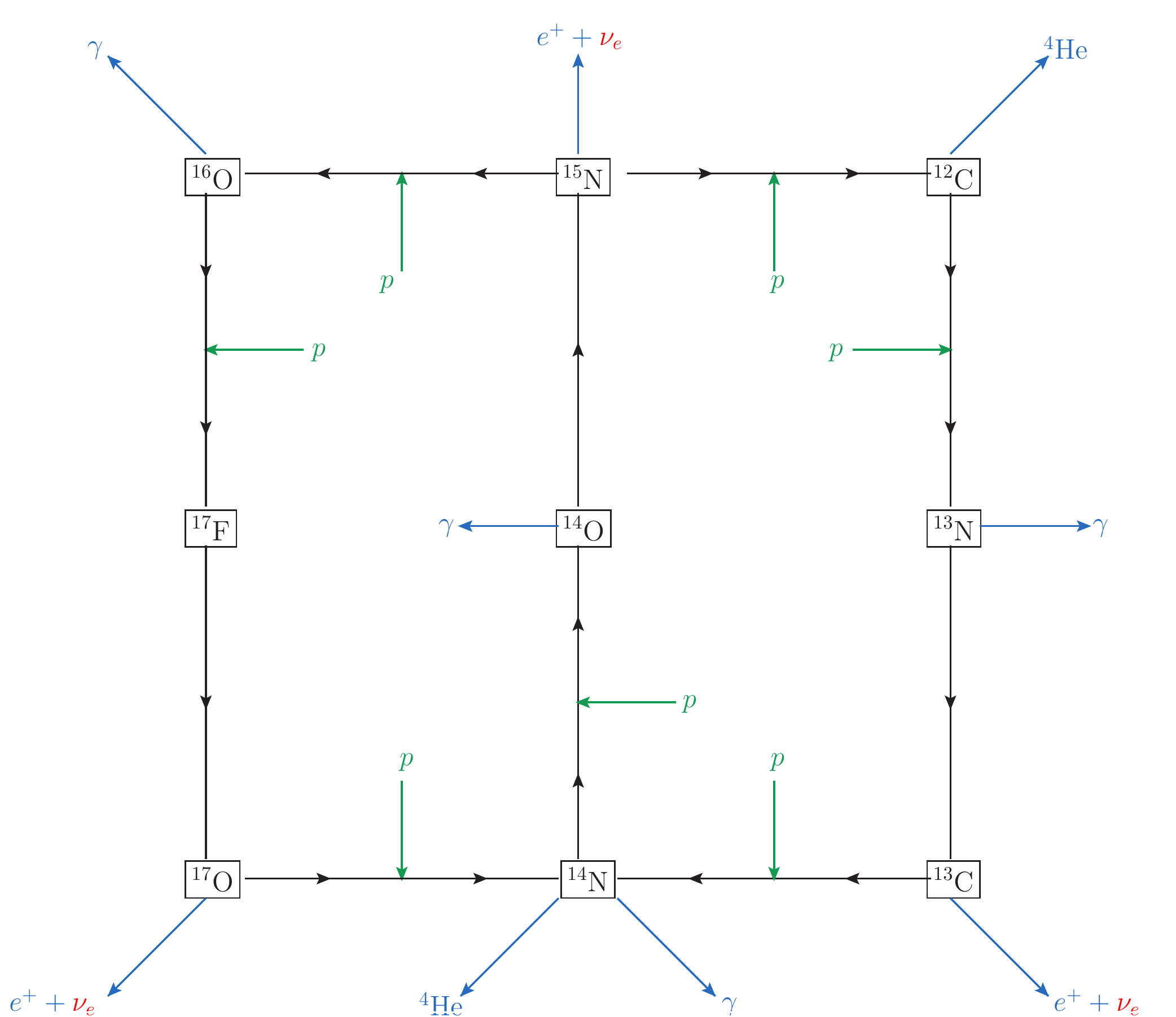}  
  		\end{center}
  	\caption{Representation of the solar CNO cycle reactions.}
  	\label{fig:CNO}
\end{figure}

Two hydrogen nuclei fuse together to form a deuterium nucleus in two different manners:
a direct $p+p$ fusion (99.6\%) and in the presence of an electron ($p+e^-+p$, 0.4\%). 
In both cases a neutrino is produced, which are denominated  $\bm{pp}$ and $\bm{pep}$ 
{\bf neutrinos}. The $pep$ neutrinos are mono-energetic since they are produced in a three
body collision. Next, the deuterium fuses with another proton to form helium-3. 
After this, the helium-3 has three possibilities to interact. In the first one, it fuses with another helium-3 to form 2 protons 
plus an $^{4}$He nuclei; this occurs 85\% of the times. The second possibility is to interact 
with an helium-4, to produce beryllium-7, and the third one is to fuse with a proton to 
create again an $^4$He isotope but with the production of a neutrino, called {\bf $\bm{hep}$ 
neutrino}. Later, the beryllium-7 isotope also interacts in two different fashions: 
with an electron, it produces lithium-7 with the emission of a neutrino,
labelled {\bf $^7$Be neutrino}; this lithium-7 fuses with a proton to form 2 helium-4 nuclei
together with the emission of energy. Besides, if the beryllium-7 interacts with a 
proton, this will create a boron-8 nuclei. The boron-8 nuclei decays to an excited state of beryllium-8 
with the emission of a neutrino, the {\bf $^8$B neutrino}. Finally, the excited state decays into two 
helium-4 nuclei, completing the chain. In addition to the pp-chain, the CNO cycle is also present in the Sun. This cycle, outlined in figure \ref{fig:CNO}, is composed by two different branches in which the $^{12, 13}$C, 
$^{13,14}$N, $^{15,17}$O isotopes interact with hydrogen nuclei to produce each other,
and the $^{15}$N, $^{16}$O, $^{17}$F nuclei. These last three isotopes decay producing neutrinos,
which are labelled according to the initial decaying isotope. It is important to note that in the case 
of the Sun, the CNO cycle is only responsible for $1.6\%$ of the energy it produces. 
However, for stars which higher temperatures, this cycle becomes dominant \cite{salaris2005evolution}.\\

These chains have been extensively stud\-ied to estimate the flux and the spectrum of the solar
neutrinos. The denominated {\it Standard Solar Model} \cite{Bahcall:2000nu} has been established from such studies.
The neutrino fluxes and spectra are computed considering the hydrodynamic evolution of our star
from some boundary conditions that reproduce the current values of the solar characteristics.
The complete solar neutrino flux is of the order of $\sim 10^{10}$ cm$^{-2}$ s$^{-1}$. For each 
type of the neutrino, the total flux with its corresponding uncertainty is given in table \ref{tab:SolarNuFlxu}.
In the present Thesis, we are considering the Bahcall-Serenelli-Basu (BSB05) Solar Standard Model~\cite{Bahcall:2005va}, 
with the input abundance of the heavy elements in the Sun given by Grevesse-Sauval work (GS98) \cite{Grevesse:1998bj}.\\

\begin{figure}[t]
	\begin{center}
    			\includegraphics[scale=0.495]{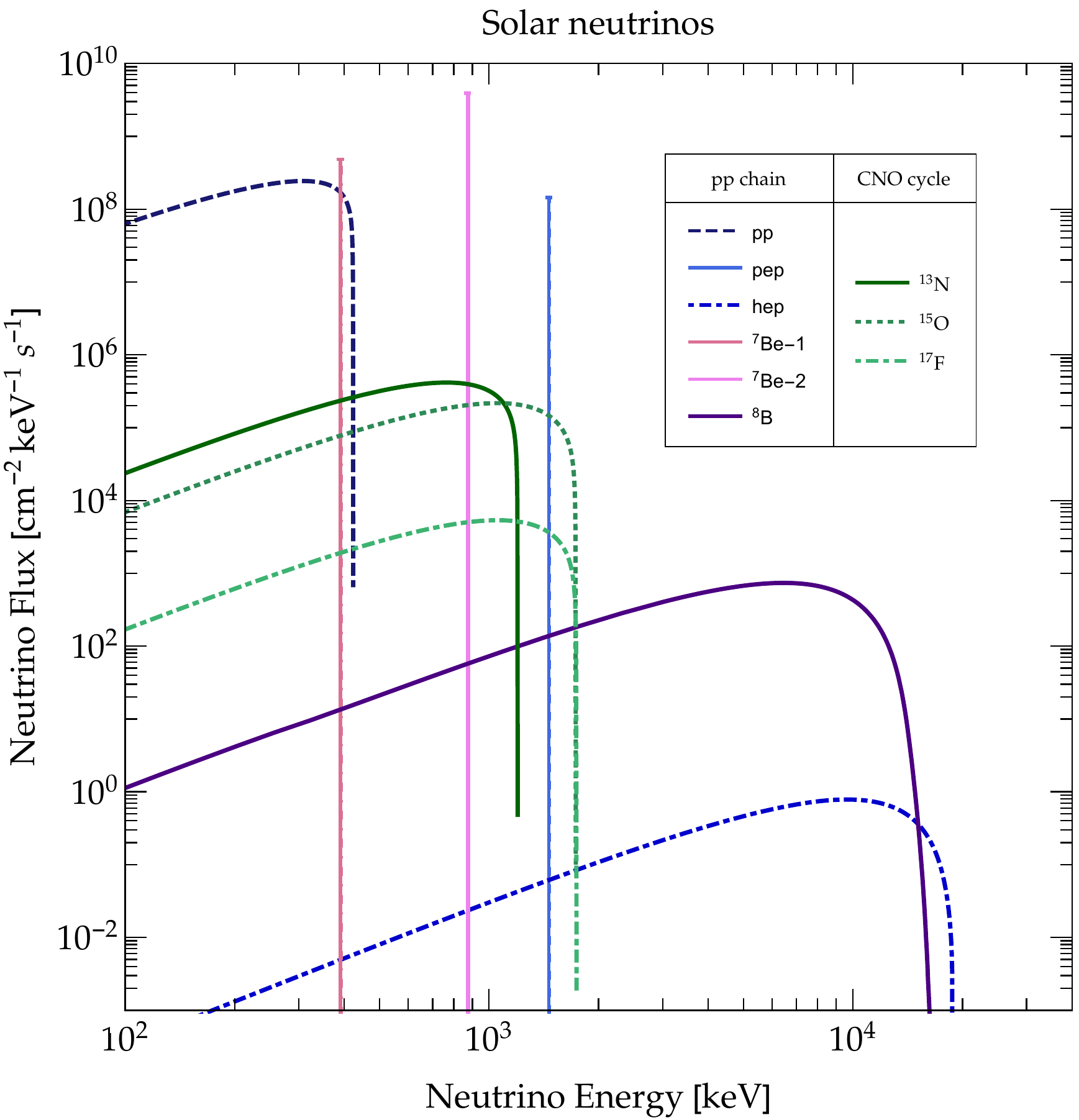}  
	\end{center}
  	\caption{Energy spectra of the solar neutrino fluxes, as predicted by the BSB05 solar model.}
  	\label{fig:SolarNu}
\end{figure}

The solar neutrino spectra obtained in the Solar Standard model can be fitted by a polynomial
of order nine \cite{Guetlein:2013},
\begin{align}
	\der{\Phi}{E_\nu}=\begin{cases}
					\displaystyle{\frac{1}{\mathcal{N}}} \displaystyle{\sum_{i=0}^8}\, a_i E_\nu^{i+1} & \text{for}\quad E_\nu < E_{\nu,\,\rm max},\\
					0 & \text{for}\quad E_\nu \ge E_{\nu,\,\rm max},
			   \end{cases}
\end{align}
with $\mathcal{N}$ a normalization factor
\begin{align}
	\mathcal{N} = \sum_{i=0}^8\,\frac{1}{i+2}\, a_i E_{\nu,\,\rm max}^{i+2},
\end{align}
$E_{\nu,\rm max}$ the maximum neutrino energy for each component, see table \ref{tab:SolarNuFlxu}, 
and $a_i$ are the fitting parameters, given in table \ref{tab:SolarNuFtab}. In figure \ref{fig:SolarNu} we show the 
dependence of the spectra on the energy. We see clearly that for small neutrino energies, the flux is dominated
by the $pp$ neutrinos, as expected, while for energies larger that $3$ MeV, the solar neutrino flux
is dominated by the $^8$ B and $hep$ neutrinos. This energy dependence will be important in the next 
chapters. \newline
%-
%-

%%%%%%%%%%%%%%%%%%%%%   table  %%%%%%%%%%%%%%%%%%%%%%%%%%%5
\begin{table}[t]
	\centering
	\caption{Flux and maximum energy of each solar neutrino component.
	For the case of the mono-energetic neutrinos, $pep$, $^7$Be, we quote the value of the energy.
	Note that for the beryllium-7 neutrino there are two different energies. This is due to the two 
	possible decays of this isotope. From the BSB05 Model \protect\cite{Bahcall:2005va}.}
	\label{tab:SolarNuFlxu}
	\begin{tabular}{c|c|c}
		\toprule
    	          & Flux [cm$^{-2}$ s$^{-1}$] & Maximum energy $E_{\nu,\,\rm max}$ [MeV] \\ \midrule\midrule
		$pp$ & $5.990(1\pm 0.009)\times 10^{10}$ & $0.423\pm 0.03$\\ \midrule
		$pep$ & $1.420(1\pm 0.015)\times 10^{8}$ & $1.445$ \\ \midrule
		$hep$ & $7.930(1\pm 0.155)\times 10^{3}$ &  $18.778$ \\ \midrule
		$^{7}$Be & $4.840(1\pm 0.105)\times 10^{9}$ & $0.3855$, $0.8631$ \\ \midrule
		$^8$B & $5.690(1^{+0.173}_{-0.147})\times 10^{6}$ & $14.88$ \\ \midrule
		$^{13}$N & $3.050(1^{+0.366}_{-0.268})\times 10^{8}$ & $1.1982\pm 0.0003$ \\ \midrule
		$^{15}$O & $2.310(1^{+0.374}_{-0.272})\times 10^{8}$ & $1.7317\pm 0.0053$ \\ \midrule
		$^{17}$F & $5.830(1^{+0.724}_{-0.420})\times 10^{6}$ & $1.7364\pm 0.0003$ \\ \bottomrule
	\end{tabular}
\end{table}
%%%%%%%%%%%%%%%%%%%%%%%%%%%%%%%%%%%%%%%%%%%%%%%%%%%%%%%%%%%%

Since the late 1960's, several experiments have detected the solar neutrino flux. The pioneering
Homestake Chlorine experiment \cite{Cleveland:1998nv} was capable of detecting specially the $^8$B neutrinos, 
given its threshold energy of $0.841$ MeV \cite{Cleveland:1998nv}. However, the measured flux was about one third of the expected one; this difference was denominated as the {\it solar neutrino problem}. Some other experiments,
such as the GALLium Experiment (GALLEX) \cite{Altmann:2005ix}, the Soviet-American Gallium Experiment (SAGE) 
\cite{Abdurashitov:2002nt}, had similar results: the flux of the solar neutrinos was lower than the models estimated.\\ 

Given that these three experiments were insensitive to the incoming direction, other types of experiments were proposed to 
confirm the solar origin of the detected neutrinos. These pioneering experiments used as physical principle of the detection the 
Cherenkov process. The Kamiokande \cite{Fukuda:1996sz}, and its successor, SuperKamiokande \cite{Hosaka:2005um}, 
and the Sudbury Neutrino Observatory (SNO) \cite{Aharmim:2005gt} experiments validated the solar provenance of the 
neutrinos and also their diminished flux. Nonetheless, the SNO experiment elucidated the situation given that they were capable of identifying solar neutrinos in three manners: through charged and neutral current interactions and electron scattering processes. 
For the charged interactions, sensitive to the neutrino flavour, they found that the flux was indeed 
smaller than predicted, but, in the neutral current and scattering cases, they discovered that 
the estimation from the Solar Standard Model was in agreement with their results \cite{Aharmim:2005gt}. This showed that
the "problem" was not related with the Sun's model, but with the neutrinos! In some way, the
electron neutrinos were metamorphosed to muon and tau neutrinos in its way to the Earth. The complete
explanation is that neutrinos suffer an adiabatic flavour conversion inside the solar medium \cite{Smirnov:2016xzf}. 
Yet, we will discuss this process in the final section of these chapter; hereafter, we will continue considering 
the neutrino sources. 
%%%%%%%%%%%%%%%%%%%%%   table  %%%%%%%%%%%%%%%%%%%%%%%%%%%5
\begin{table}[t]
	\centering
	\caption{Fit parameters for the each solar continuous spectra, taken from \protect\cite{Guetlein:2013}.}
	\label{tab:SolarNuFtab}
	\begin{tabular}{c|r||r||r}
		\toprule
    	          &  $pp$ & $hep$ & $^8$B \\ \midrule\midrule
		$a_0$ & $-2.87034\times 10^{-1}$ & $-2.04975\times 10^{-6}$ & $6.01473\times 10^{-4}$ \\ \midrule
		$a_1$ & $1.63559\times 10^2 $ & $4.22577\times 10^{-3}$ & $1.49699\times 10^{-2}$ \\ \midrule
		$a_2$ & $-1.22253\times 10^3$ & $-4.70817\times 10^{-4}$  & $-2.6481\times 10^{-3}$ \\ \midrule
		$a_3$ & $1.55085\times 10^4$ & $2.55332\times 10^{-5}$ & $-2.4141\times 10^{-5}$ \\ \midrule
		$a_4$ & $-1.465140\times 10^5$ & $-2.81714\times 10 ^{-6}$ & $6.22325\times 10^{-5}$ \\ \midrule
		$a_5$ & $7.843750\times 10^5$ & $3.06961\times 10^{-7}$ & $-9.84329\times 10^{-6}$ \\ \midrule
		$a_6$ & $-2.35727\times 10^6$ & $-1.87479\times 10^{-8}$ & $7.51311\times 10 ^{-7}$ \\ \midrule
		$a_7$ & $3.71082\times 10^6$ & $6.01396\times 10^{-10}$ & $-2.8606\times 10^{-8}$\\ \midrule
		$a_8$ & $-2.38308\times 10^6$ & $-7.88874\times 10^{-12}$ & $4.31572\times 10^{-10}$ \\ \midrule\midrule
		         & $^{13}$N & $^{15}$O & $^{17}$F \\ \midrule\midrule
		$a_0$ & $-2.59522\times 10^{-2}$ & $-1.36811\times 10^{-2}$ & $-6.74473\times 10^{-2}$ \\ \midrule
		$a_1$ & $1.05228\times 10^{1}$ & $3.94221$ & $4.63141$ \\ \midrule
		$a_2$ & $-3.2344\times 10^{1}$ & $-8.55431$ & $-1.1535\times 10^1$ \\ \midrule
		$a_3$ & $1.33211\times 10^2$ & $2.36712\times 10^1$ & $2.85067\times 10^1$ \\ \midrule
		$a_4$ & $-4.11848\times 10^2$ & $-5.00477\times 10^1$ & $-5.11664\times 10^1$ \\ \midrule
		$a_5$ & $7.28405\times 10^2$ & $6.07542\times 10^1$ & $5.51102\times 10^1$ \\ \midrule
		$a_6$ & $-7.26105\times 10^2$ & $-4.15278\times 10^1$ & $-3.45332\times 10^1$ \\ \midrule
		$a_7$ & $3.8137\times 10^2$ & $1.49462\times 10^1$ & $1.1655\times 10^1$ \\ \midrule
		$a_8$ & $-8.21149\times 10^1$ & $-2.20367$ & $-1.62814$ \\ \bottomrule
	\end{tabular}
\end{table}
%%%%%%%%%%%%%%%%%%%%%%%%%%%%%%%%%%%%%%%%%%%%%%%%%%%%%%%%%%%%

\subsection{Atmospheric Neutrinos}
%-
\begin{figure}[t]
	\begin{center}
    			\includegraphics[scale=0.5]{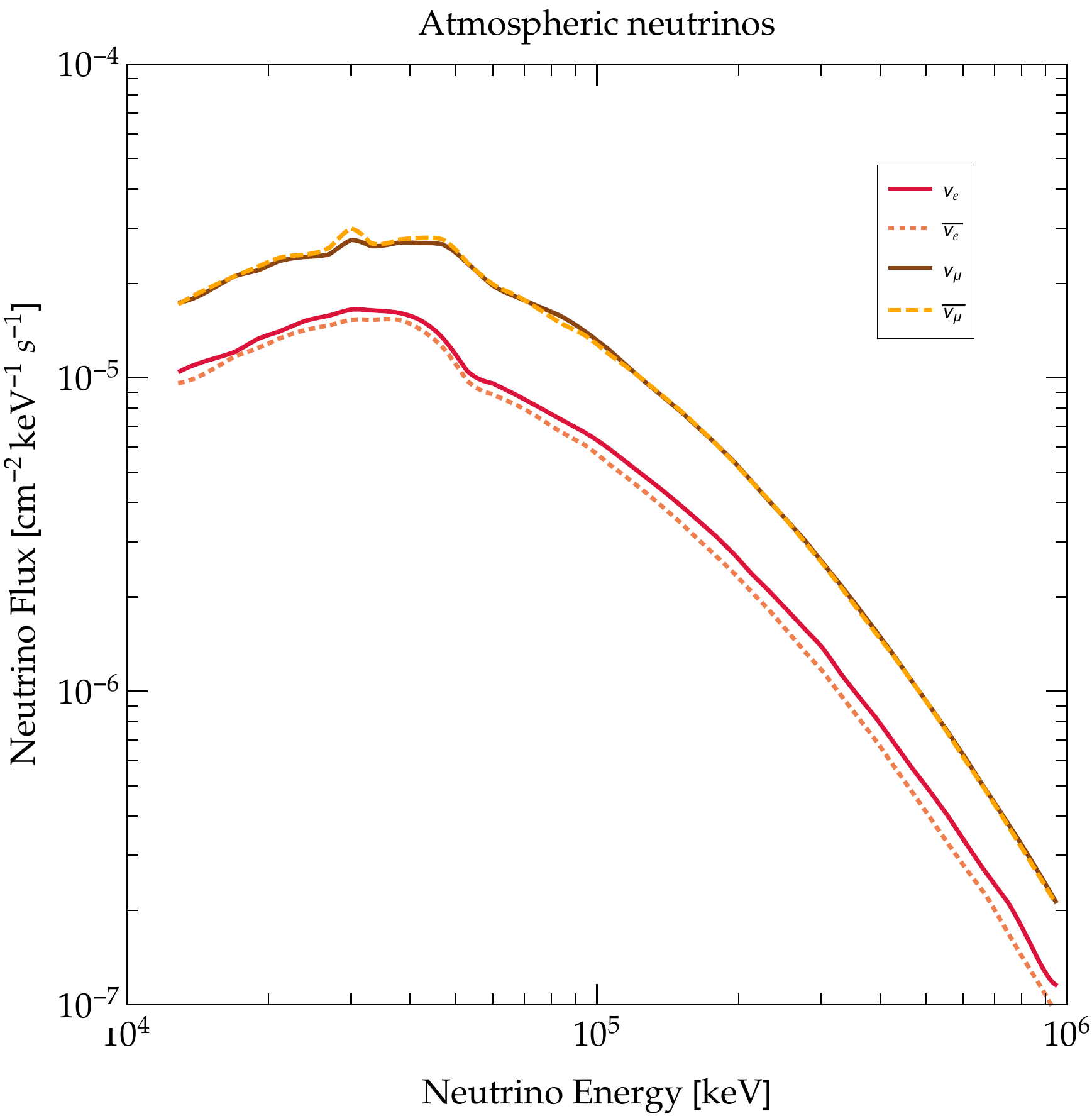}  
	\end{center}
  	\caption{Energy spectra of the solar averaged atmospheric neutrino fluxes, as predicted by the work of G. Battistoni et. al.~\protect\cite{Battistoni:2005pd}.}
  	\label{fig:AtmNu}
\end{figure}
%-
The Earth is constantly under a shower of cosmic rays, composed principally of protons. Their interactions
with the atmosphere produce a cascade of other particles, in special, pions and muons. The decay of these 
particles creates an additional source of neutrinos, called for clear reasons {\it atmospheric} neutrinos. 
For instance, the pions decay primarily to muons which in turn decay to electrons and positrons, with the 
emission of electron neutrinos and muon antineutrinos
\begin{align*}
	\pi^{\pm} \tto &\mu^\pm + \pbar{\nu}_\mu\\
				   &\downarrow\\
				   &\mu^\pm \tto e^\pm + \nu_e (\bar{\nu}_e) +\bar{\nu}_\mu(\nu_\mu).
\end{align*} 
The energy range of these neutrinos is quite broad, from $\sim 10$ MeV to $10$ TeV \cite{Giunti:2007ry}. For low energies,
$E_\nu\lesssim 1$ GeV, which corresponds to the case where almost all muons decay in the atmosphere, the
previous chain of decays shows that the following ratios between the fluxes should be satisfied in an experiment 
if the neutrinos do not mutate into other types,
\begin{align}
	\frac{\Phi_{\nu_\mu}+\Phi_{\bar\nu_\mu}}{\Phi_{\nu_e}+\Phi_{\bar\nu_e}}\approx 2, \quad \frac{\Phi_{\nu_\mu}}{\Phi_{\bar\nu_\mu}}\approx 1.
\end{align}
Given that in a unique pion decay a muon neutrino-antineutrino pair is produced, together with an
electron neutrino or antineutrino (depending on the charge of the initial pion), the ratio between the 
sum of the muon neutrino and antineutrino flux should be twice the sum of the electron neutrino and 
antineutrino flux. Let us note that if the experiment is sensitive to the direction of the incoming 
(anti)neutrino, we can even further determine the dependence of the fluxes with the zenith angle of the 
experiment. These angle distributions showed that in fact neutrinos suffer oscillations, a consequence of 
the existence of masses and mixing. This was demonstrated
by the SuperKamiokande \cite{Fukuda:1998ah,Fukuda:1998ub}, the MACRO \cite{Giacomelli:2002nn} and the more 
recent IceCube Neutrino Observatory \cite{Aartsen:2013jza}.\\

The complete computation of the atmospheric neutrino and antineutrino flux needs to take into account the 
full cosmic rays spectrum and all the possible interactions that can occur between the cosmic rays and the 
atmosphere. Also, it is needed to know the model for the atmosphere. For our future purposes, we will 
consider the results from the group of G. Battistoni et. al. \cite{Battistoni:2005pd}. Let us keep in mind that
these fluxes have an uncertainty of $\sim 20\%$. In figure \ref{fig:AtmNu}, we show 
the dependence of the atmospheric neutrinos fluxes with the energy. 

\subsection{Reactor Antineutrinos}\label{subsec:ReacFlux}

The first technological achievement related to the discoveries and theoretical advances in the Weak
interaction physics was the creation of a self-sustained nuclear chain reaction, and the subsequent elaboration of a 
nuclear reactor. In these electricity generators, a set of isotopes undergo fission due to the
absorption of a neutron. The fission products are usually unstable and rich in neutrons, generating approximately 
six antineutrinos after decaying weakly. In a\-ve\-ra\-ge, $6\times 10^{20}\ \bar{\nu}/{\rm s}$ are produced
in a 3 GW reactor~\cite{Baldoncini:2014vda}. Following closely~\cite{Baldoncini:2014vda}, we are going 
to introduce the main pieces to determine the reactor antineutrino flux for any place on the Earth. This
will be used in the chapter \ref{cha:NeutrinoFloor}.\\

Let us note that the determination of the antineutrino spectra has two separated contributions. The first one
is related to the specific properties of the reactor. A generic nuclear reactor is characterized by its thermal 
power ($P_{\rm th}$) and the Load Factor ($L_F$), corresponding to the percentage of energy that a reactor has 
produced over a time period compared to the energy it would have produced if it were operating continuously 
at the reference power in the same period. These two features are published each year by the International 
Atomic Energy Agency (IAEA). \\

\begin{figure}[t]
\centering
\includegraphics[width=0.6\linewidth]{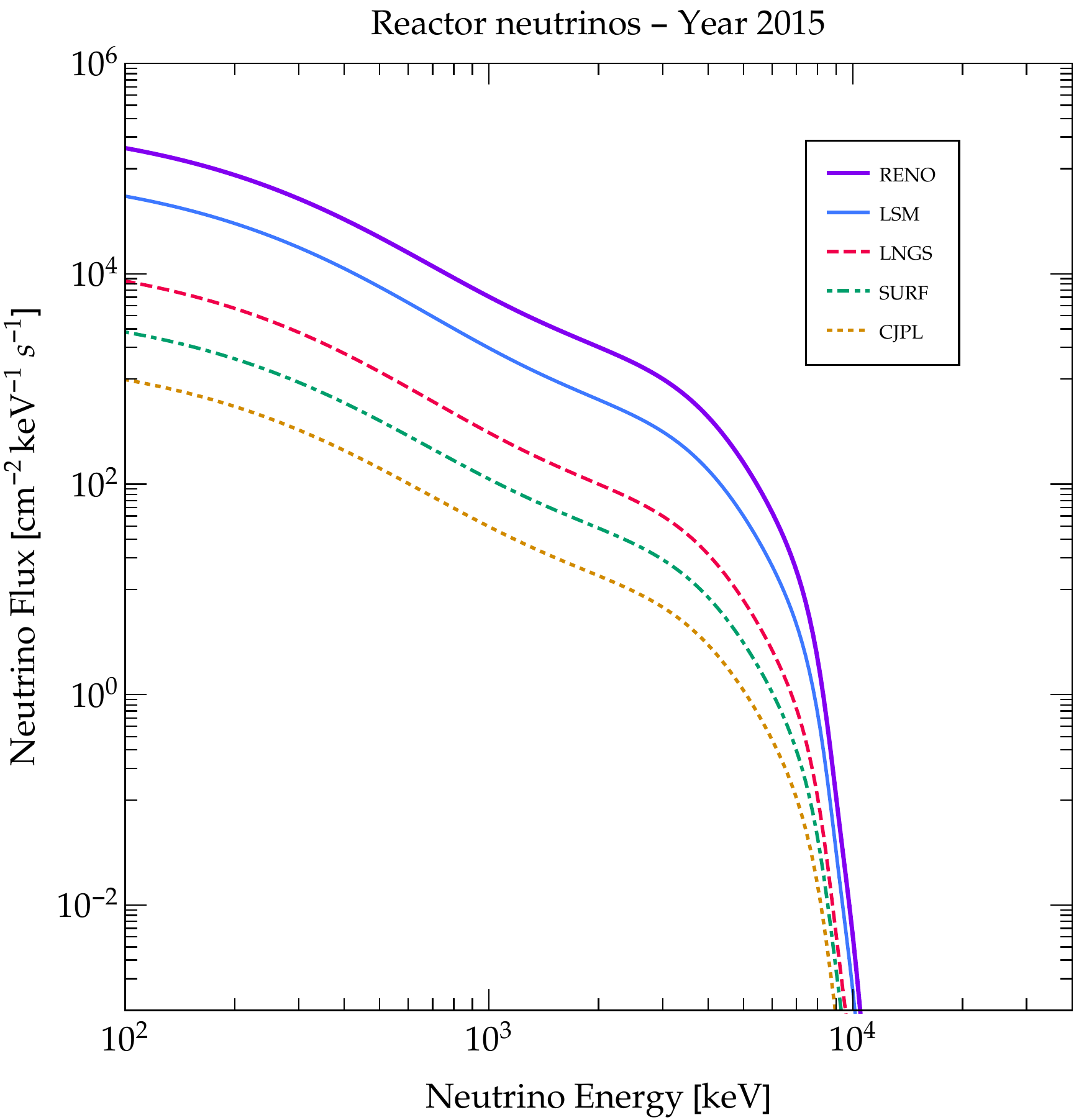}
\caption{Comparison among reactor antineutrino flux in five different laboratories.}
\label{fig:RNF}
\end{figure}

On the other hand, the antineutrino spectrum will depend on the details of the beta decays of the fission 
products. In a typical reactor, there are four isotopes, $^{235}{\rm U},\ ^{238}{\rm U}$, $^{239}{\rm Pu}, 
\ ^{241}{\rm Pu}$, that undergo fission. For a given reactor, the antineutrino
spectrum is~\cite{Baldoncini:2014vda}
\begin{align*}
	\mathcal{S}(E_{\bar\nu})= L_F\sum_{i=1}^4 \mathsf{N}_i \lambda_i (E_{\bar\nu})
\end{align*}
where the sum is over the four isotopes, $\mathsf{N}_i$ is the number of fissions per second for each isotope; 
$\lambda_i (E_{\bar\nu})$ is the antineutrino spectrum for one fission \cite{Baldoncini:2014vda,Huber:2004xh}. 
The thermal power produced by the reactor is given by
\begin{align*}
	P_{\rm th}=\sum_{i=1}^4 \mathsf{N}_i Q_i,
\end{align*}
with $Q_i$ the energy released by each isotope. The values of the $Q_i$ for the isotopes under consideration 
can be found in table \ref{tab:QEnergyTab}. Next, we introduce the {\it power fraction} $p_i$, corresponding to the fraction of the total 
thermal power created by the isotope $i$ \cite{Baldoncini:2014vda,Huber:2004xh}, as 
\begin{align}
	p_i=\frac{\mathsf{N}_i Q_i}{P_{\rm th}}.
\end{align}
%%%%%%%%%%%%%%%%%%%%%   table  %%%%%%%%%%%%%%%%%%%%%%%%%%%5
\begin{table}[t]
	\centering
	\caption{Energy released per fission for the four isotopes in consideration. Taken for \protect\cite{Ma:2012bm}.}
	\label{tab:QEnergyTab}
	\begin{tabular}{c|c}
		\toprule
    	    Isotope      &  $Q_ i$ [MeV] \\ \midrule\midrule
		$^{235}$U & $202.36\pm 0.26$  \\ \midrule
		$^{238}$U & $205.99\pm 0.52$  \\ \midrule
		$^{239}$Pu & $211.12\pm 0.34$  \\ \midrule
		$^{241}$Pu & $214.26\pm 0.33$   \\ \bottomrule
	\end{tabular}
\end{table}
%%%%%%%%%%%%%%%%%%%%%%%%%%%%%%%%%%%%%%%%%%%%%%%%%%%%%%%%%%%%
\negthickspace These power fractions depend on the type of reactor. We will consider basically five types of reactors:
Pressurized Water Reactors (PWR), Boiling Water Reactors (BWR), Pressurized Heavy Water Reactors (PHWR), Light
Water Graphite Reactors (LWGR) and Gas Cooled Reactors (GCR) \cite{Baldoncini:2014vda,Bellini:2013nah}. The power fractions for these reactors
are in table \ref{tab:PowerFracTab}. Also, if the reactor uses Mixed OXide fuel (MOX) as $30\%$ of the combustible, the power fractions
are slightly modified, see table \ref{tab:PowerFracTab} \cite{Baldoncini:2014vda,Bellini:2013nah}. So, we can write
\begin{align*}
	\mathcal{S}(E_{\bar\nu}) = P_{\rm th}\,L_F\,\sum_{i=1}^4 \frac{p_i} {Q_i} \lambda_i (E_{\bar\nu}).
\end{align*}
%%%%%%%%%%%%%%%%%%%%%   table  %%%%%%%%%%%%%%%%%%%%%%%%%%%5
\begin{table}[t]
	\centering
	\caption{Power fractions for the five types of reactors used in this work and for those which 
			 burn MOX as well. Taken for \protect\cite{Bellini:2013nah} and \protect\cite{Bellini:2010hy}.}
	\label{tab:PowerFracTab}
	\begin{tabular}{c|c|c|c|c}
		\toprule
    	    Reactor      & $^{235}$U & $^{238}$U & $^{239}$Pu & $^{241}$Pu \\ \midrule\midrule
		PWR & $0.560$ & $0.080$ & $0.300$ & $0.060$ \\ \midrule
		MOX &  $0.000$ & $0.081$ & $0.708$ & $0.212$ \\ \midrule
		PHWR & $0.543$ & $0.411$ & $0.022$ & $0.024$  \\ \bottomrule
	\end{tabular}
\end{table}
%%%%%%%%%%%%%%%%%%%%%%%%%%%%%%%%%%%%%%%%%%%%%%%%%%%%%%%%%%%%
\negthickspace Now, to obtain the antineutrino spectrum per fission, $\lambda_k(E_{\bar\nu})$, one has to analyze the 
chain of decays originated form the fission. But given that our purposes are not to study this computation,
we give next the result of M\"uller et.\ al.\ for the spectrum \cite{Mueller:2011nm}. In such a work, after the complete 
calculation, the spectrum is fitted for all those four contributing isotopes in terms of the exponential of a order 5 
polynomial,
\begin{align}
	\lambda_i(E_{\bar\nu})=\exp\left(\sum_{j=1}^6\,\tilde{a}_j^i E_{\bar\nu}^{j-1}\right);
\end{align}
note that this function has units of [energy]$^{-1}$. The values of the parameters $\tilde{a}_j^i$ are in table 
\ref{tab:ReacNuFtab}. The flux of reactor antineutrinos at any point on the Earth is then obtained supposing an 
isotropic emission,
\begin{align}
	\frac{d\Phi}{dE_{\bar\nu}}=\sum_{k} \frac{P_{\rm th}^k \langle L_{F}^k \rangle}{4\pi d_k^2} \sum_{i=1}^4\frac{p_i}{Q_i}\lambda_i(E_{\bar\nu}).
\end{align}
%%%%%%%%%%%%%%%%%%%%%   table  %%%%%%%%%%%%%%%%%%%%%%%%%%%5
\begin{table}[t]
	\centering
	\caption{Fit parameters for the each reactor spectra, taken from \protect\cite{Mueller:2011nm}.}
	\label{tab:ReacNuFtab}
	\begin{tabular}{c|r|r|r|r}
		\toprule
    	Isotope      & $^{235}$U & $^{238}$U & $^{239}$Pu & $^{241}$Pu \\ \midrule\midrule
		$\tilde{a}_1$ & $3.217$ & $4.833\times 10^{-1}$ & $6.413$ & $3.251$ \\ \midrule
		$\tilde{a}_2$ & $-3.111$ & $1.927\times 10^{-1}$  & $-7.432$ & $-3.204$ \\ \midrule
		$\tilde{a}_3$ & $1.395$ & $1.283\times 10^{-1}$ & $3.535$ & $1.428$ \\ \midrule
		$\tilde{a}_4$ & $-3.690\times 10 ^{-1}$ & $-6.762\times 10 ^{-3}$ & $-8.820\times 10^{-1}$ & $-3.675\times 10^{-1}$ \\ \midrule
		$\tilde{a}_5$ & $4.445\times 10 ^{-2}$ & $2.233\times 10^{-3}$ & $1.025\times 10^{-1}$ & $4.254\times 10^{-2}$ \\ \midrule
		$\tilde{a}_6$ & $-2.053\times 10 ^{-3}$ & $-1.536\times 10^{-4}$ & $-4.550\times 10 ^{-3}$ &  $-1.896\times 10^{-3}$ \\ \bottomrule
	\end{tabular}
\end{table}
%%%%%%%%%%%%%%%%%%%%%%%%%%%%%%%%%%%%%%%%%%%%%%%%%%%%%%%%%%%%
\negthickspace Here, the sum over $k$ is made over \textit{\textbf{all}} reactors on the Earth, $\langle L_F^k \rangle$ is the average of the 
Load Factor over a given time and $d_k$ is the distance between the reactor and the location on the Earth. 
In figure \ref{fig:RNF}, we considered the data corresponding to the year 2015, and the distance to the laboratory is 
computed considering an spherical Earth\footnote{The data to compute the flux has been taken from the source 
maintained by the same authors of~\cite{Baldoncini:2014vda}. Their website is:
 \url{http://www.fe.infn.it/antineutrino/}}. We determined the flux for five laboratories:\\

\begin{itemize}
	\item Laboratoire Souterrain de Modane (LSM), France.
	\item 원자로 중성미자 진동 실험 (Reactor Experiment for Neutrino Oscillation -- RENO), Korea.
	\item Laboratori Nazionali del Gran Sasso (LNGS), Italy.
	\item Sanford Underground Research Facility (SURF), USA.
	\item \begin{CJK}{UTF8}{gbsn} 中国锦屏地下实验室 \end{CJK} -- China Jinping Underground Laboratory (CJPL), China.
\end{itemize} 

\newpage

As expected, we see that in the places where there are several reactors near by the flux expected there is high, as for 
the RENO and LSM cases. For the LNGS, the flux is one order of magnitude less than in the LSM. For SURF and 
CJPL, the flux is even smaller. Finally, for the remainder of the work, we will consider a conservative $5\%$ uncertainty
in the neutrino fluxes \cite{Hayes:2013wra,Vogel:2015wua}.\\
 
The first detection of an antineutrino was done by Cowan and Reines using the reactor at the Savanna River 
Plant \cite{Cowan:1992xc}. Ever since, several experiments have been performed with the reactor antineutrinos. The principle 
of detection is quite simple, using the inverse beta decay, $\bar{\nu} + p \to n + e^+$. The reactor experiments
are divided according to the distance from the core. The {\it short-baseline} experiments have a distance of 
$L \sim (10-100)\ \text{m}$, while the {\it long-baseline} ones correspond to distances  of $L\sim 1$ km. The third 
category is for the case of $L \gtrsim 100$ km, corresponding to {\it very-long-baseline} experiments. 
Given the results from solar neutrinos, the reactor experiments
were searching the disappearance of antineutrinos and its dependence with the distance. The most recent 
experiments, the Daya-Bay Experiment \cite{An:2013uza}, RENO \cite{Ahn:2012nd}, and KamLAND \cite{Abe:2008aa} h
ave found results consistent with the disappearance of the antineutrinos, 
confirming the non zero value of the neutrino mass and the oscillation phenomena. 

\newpage

\subsection{Diffuse Supernova Neutrino Background}

\begin{figure}[t]
\centering
\includegraphics[width=0.6\linewidth]{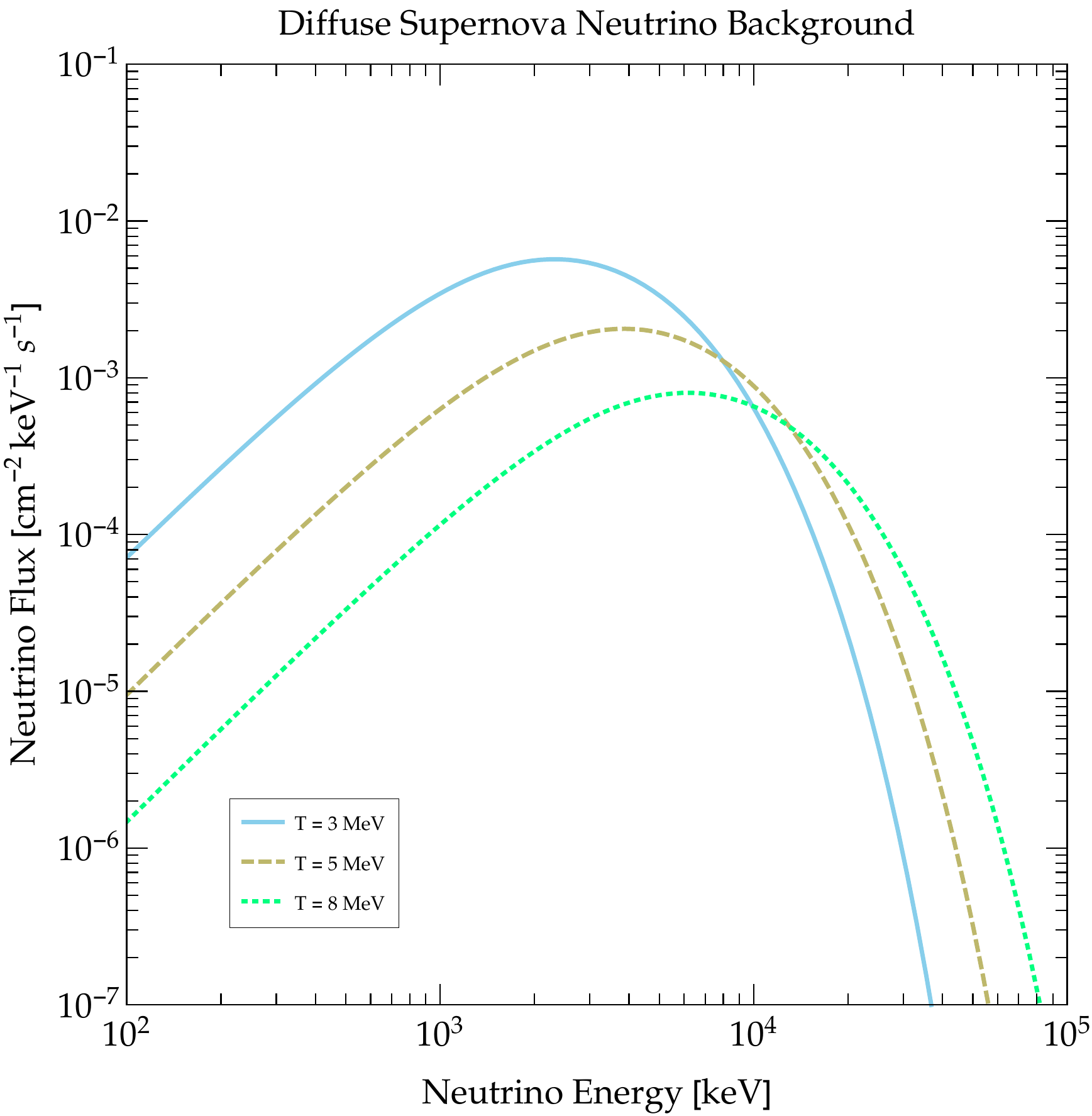}
\caption{Diffuse Supernova Neutrino Background flux for three different temperatures.}
\label{fig:DSNBF}
\end{figure}

From the time of the birth of the first star, supernovae explosions have been occurring in the Universe.
These gigantic events have created a flux of neutrinos and antineutrinos, denominated {\it diffuse supernova
neutrino background} (DSNB) \cite{Beacom:2005it}. Although this background has not been observed yet, it is expected to be seen
by future experiments \cite{Beacom:2010kk}. Let us note that this discovery would have a profound impact on our knowledge 
not only about neutrinos but also regarding supernovae, opening another window to understand our Universe. The DSNB depends on 
the rate in which supernovae happen, and on how neutrinos have been emitted in the explosion. As well stressed before, 
our purpose here will not be to give the details of the complete computation of these factors, but to make explicit
the parameters and definitions we will use later in the development of the thesis. We suggest the interested reader 
to see the J.\ Beacom review \cite{Beacom:2010kk} and some other papers \cite{Beacom:2005it,Horiuchi:2008jz}.\\

The first part, the rate in which supernovae occur in the Universe, can be related to the cosmic star formation 
history, which is obtained by direct measurements. We will adopt the continuous broken power law as a function
of the redshift \cite{Beacom:2010kk},
\begin{align*}
	\dot{\rho}_*(z)=\dot{\rho}_0\left[(1+z)^{\alpha\eta}+\left(\frac{1+z}{B}\right)^{\beta\eta}+\left(\frac{1+z}{C}\right)^{\gamma\eta}\right]^{\frac{1}{\eta}}
\end{align*}
where $\dot{\rho}_0$ is a normalization constant, $\eta\approx -10$, $\alpha$, $\beta$, $\gamma$ are constants 
related to the redshift regimes \cite{Horiuchi:2008jz}, and the $B,C$ parameters are related to the redshift breaks, 
given by \cite{Horiuchi:2008jz}
\begin{subequations}
	\begin{align}
		B&=(1+z_1)^{1-\frac{\alpha}{\beta}},\\
		C&=(1+z_1)^\frac{\beta-\alpha}{\gamma}(1+z_2)^{1-\frac{\beta}{\gamma}}.
	\end{align}
\end{subequations}
The values of the previous parameters are in table \ref{tab:DSNBFtab}. The rate of neutrino emitting supernovae, in terms
of the Solar mass $M_{\astrosun}$, is fitted to be \cite{Beacom:2010kk,Horiuchi:2008jz}
\begin{align*}
	R_{\rm CCSN}(z)\approx \frac{\dot{\rho}_*(z)}{143 M_{\astrosun}}.
\end{align*}
Now, the second component, the neutrino emission in the supernova, will depend on the fraction of the total 
energy that has been taken by these particles. Also, let us note that all three flavors of neutrinos and antineutrinos 
are created, and each species takes an equal part. Supernovae simulations have shown that the energy spectrum of 
the neutrinos is well approximated by a Fermi-Dirac thermal distribution \cite{Horiuchi:2008jz}
\begin{align}
	\frac{dN_{\bar{\nu}_e}}{d E^\prime_{\bar{\nu}_e}}=\frac{E_\nu^{\rm tot}}{6}\,\frac{120}{7\pi^4}\,\frac{E^{\prime\, 2}_{\bar{\nu}_e}}{T_{\bar{\nu}_e}^4}\,\frac{1}{\exp\left(\frac{E^{\prime}_{\bar{\nu}_e}}{T_{\bar{\nu}_e}}\right)+1}
\end{align}
where $E_\nu^{\rm tot} \approx 10^{59}$ MeV is the total energy, $T_{\bar{\nu}_e}$ is the effective antineutrino 
temperature outside the protoneutron star. Having the two main ingredients to construct the DSNB, we finally are
able to compute the flux \cite{Horiuchi:2008jz},
\begin{align}
	\frac{d\Phi}{dE_{\bar{\nu}}}=c\int_0^{z_{\rm max}}\, R_{\rm CCSN}(z)\,\frac{dN_{\bar{\nu}_e}}{d E^\prime_{\bar{\nu}_e}}\,(1+z)\,\left|\frac{dt}{dz}\right|dz
\end{align}
where $z_{\rm max}$ is the maximum redshift to compute the flux, $z_{\rm max}=5$; $E^\prime_{\bar{\nu}_e}=(1+z)E_{\bar{\nu}_e}$
is the relation between the energy at the creation time and the current energy, and
\begin{align}
	\left|\frac{dt}{dz}\right|^{-1}=H_0(1+z)[\Omega_m(1+z)^3+\Omega_\Lambda]^{\frac{1}{2}}.
\end{align}
In the literature \cite{Horiuchi:2008jz,Beacom:2010kk}, the DSNB is basically composed by the electron antineutrino flux with temperatures of
$T_{\bar{\nu}_e}=3,5,8$ MeV. In figure \ref{fig:DSNBF} we show the spectrum of the DSNB for these three 
scenarios. Let us stress that the systematic uncertainty of this flux is about $\sim 50\%$. 
Although the DSNB has not been found experimentally, there are good prospects to find them in the
Superkamiokande experiment \cite{Horiuchi:2008jz}. However, as we will see later, the existences of this cosmological flux has an
impact in some future experiments to be performed here on the Earth.\\
%%%%%%%%%%%%%%%%%%%%%   table  %%%%%%%%%%%%%%%%%%%%%%%%%%%5
\begin{table}[t]
	\centering
	\caption{Parameters for the cosmic star formation history, taken from \protect\cite{Horiuchi:2008jz}.}
	\label{tab:DSNBFtab}
	\begin{tabular}{c|c|c|c|c|c|c}
		\toprule
    	Fit Parameter      & $\dot{\rho}_0$ & $\alpha$ &  $\beta$ & $\gamma$ & $z_1$ & $z_2$ \\ \midrule\midrule
		Fiducial & $0.0178$ & $3.4$ & $-0.3$ &  $-3.5$ & $1$ & $4$ \\ \bottomrule
	\end{tabular}
\end{table}
%%%%%%%%%%%%%%%%%%%%%%%%%%%%%%%%%%%%%%%%%%%%%%%%%%%%%%%%%%%%

Nonetheless, before getting to that point, we have now seen that experiments from different sources has shown that 
neutrinos undergo flavour oscillations in their travel between the source and the detector. This is a fundamental
proof that the neutrinos are indeed massive particles. To understand in detail this, let us now introduce the neutrino
oscillations and its current status.

\section{Neutrino Oscillations}

As we have seen in the previous section, experimental evidences show that neutrinos metamorphose into another
flavour in its journey between a source and a detector. To describe in the standard manner this process, let us start
in an analogous way to what was done in the quark sector. Supposing that the neutrinos are massive particles, having masses equal to $m_a^\nu$, and the flavour states that do not have definite masses, we can write the flavour fields as a superposition of the mass fields \cite{Pontecorvo:1957cp,Maki:1962mu}
\begin{align}
	\nu_L^\alpha = \widetilde{V}_L^{\alpha a} \nu^{a}_L.
\end{align}
Then, the charged current for the leptons becomes,
\begin{align}\label{eq:PMNS}
	j^\mu_{Wl}= \overline{\ell_L^\alpha} \gamma^\mu \nu_L^\alpha \tto   \overline{\ell_L^a} \gamma^\mu \underbrace{(V_L^{a\alpha})^*\widetilde{V}_{L}^{\alpha b}}_{\widetilde{U}_{\rm PMNS}^{ab}} \nu_L^b,
\end{align}
\sloppy with the definition of the Pontecorvo-Maki-Nakagawa-Sakata (PMNS) matrix $\widetilde{U}_{\rm PMNS}^{ab}=(V_L^{a\alpha})^*\widetilde{V}_{L}^{\alpha b}$ \cite{Pontecorvo:1957cp,Maki:1962mu}, the a\-na\-lo\-gous to the CKM matrix in the lepton 
sector. Let us note that at this point we are not considering the origin of the masses and the mixing of 
neutrinos; this will be our task in the next chapters. Anyhow, if in a weak processes a neutrino is created
by a charged interaction, it will have a definite flavour, corresponding to the flavour of the associated charged
lepton created. Therefore, as we have seen, the flavour eigenstate will be a superposition of the mass eigenstate\footnote{For 
simplicity in the notation, from now on we will remove the PMNS and CKM indexes; so, to avoid confusion, the PMNS matrix 
will always have the symbol $\widetilde{U}$ and the CKM matrix $U$.}
\begin{align}
	\ket{\nu_\alpha}=\widetilde{U}_{\alpha a}^*\ket{\nu_a}.
\end{align}
Let us stress that in the previous relation between flavour and mass eigenstates appears the PMNS matrix. This is due to
the fact that a neutrino is created by a charged current process, which depends on such a matrix, eq. \ref{eq:PMNS}. If the
neutrino is created in a neutral current interaction, the neutrino will not have a definite flavour. However, given
that there are not flavour-changing neutral currents in the lepton sector, a neutrino with a definite flavour will not undergo
a flavour change by neutral interactions, i.e. the mixing is not affected by interactions with the Z boson.\\

If a neutrino is created in some point with a definite flavour at some initial time $\ket{\nu_\alpha(t_0)}$, $t_0=0$,
we will be interested in determine the probability of a detection of the neutrino with flavour $\beta$ in some
different point, given by \cite{Giunti:2007ry}
\begin{align}
	P(\nu_\alpha\to \nu_\beta)=\abs{\bra{\nu_\beta}|\mathrm{U}(T,L)\ket{\nu_\alpha(0)}}^2,
\end{align}
with $\mathrm{U}(T,L)$ the evolution operator in space and time. Since the mass of the neutrinos is usually smaller than the
energy in which they are produced, we can safely approximate that they travel at the light speed $c$, so the distance
between the production and the detection is equal to the time spent in the process, $L\approx T$. Furthermore, if the neutrino 
propagates in vacuum, we can take it as a free particle; the evolution operator will be related to the free hamiltonian
in a simple manner,
\begin{align}
	\mathrm{U}(T,L) = \mathrm{U}(L) = e^{-i H L}.
\end{align}
The mass eigenstates are also energy eigenstates as the hamiltonian is a free one. Therefore
\begin{align}
	e^{-i H L}\ket{\nu_\alpha} &= e^{-i H L}\, \widetilde{U}_{\alpha a}^*\ket{\nu_a}\notag\\
	& =\widetilde{U}_{\alpha a}^*\,e^{-i E_a L}\ket{\nu_a},
\end{align}
with the energy $E_a$ well aproximated to \cite{Giunti:2007ry}
\begin{align*}
	E_a&=\sqrt{p_a^2+m_a^{\nu\,2}}\notag\\
	   &\approx p_a+\frac{m_a^{\nu\,2}}{2p_a}=E+\frac{m_a^{\nu\,2}}{2E}.
\end{align*}
Here we used that the momentum of the neutrino is approximately equal to the neutrino energy $p_a\approx E$, 
which is practically equal for all flavours in this ultrarelativistic regime. Using these approximations,
we have that the probability $P(\nu_\alpha\to \nu_\beta)$ is then
\begin{align}\label{eq:Palphabeta}
	P(\nu_\alpha\to \nu_\beta)= \sum_{a,b}\widetilde{U}_{\alpha a}^* \widetilde{U}_{\beta a} \widetilde{U}_{\alpha b}  \widetilde{U}_{\beta b}^*\, e^{i\Delta_{ab} L}
\end{align}
where $\Delta_{ab} = \frac{\Delta m_{ab}^{2}}{2 E}$, and $\Delta m_{ab}^{2} = m_a^{\nu\,2}-m_b^{\nu\,2}$. In \ref{eq:Palphabeta}
we wrote explicitly the sums over $a,b$ to avoid confusion, given that there is no sum over $\alpha$ and $\beta$.\\ 

Let us comment here that a neutrino oscillation experiment is only sensitive to the squared mass difference, $\Delta m_{ab}^2$; the 
real scale of the neutrino masses needs to be studied by other phenomena. Experimentally, there are two types of processes 
corresponding to the possible values of $\beta$. If $\beta=\alpha$, the experiment will be a disappearance one since the experiment 
will seek a difference between the expected number of neutrinos in the detector compared to the observed number. This is the case of 
the solar and reactor experiments. While if the final flavour is different from the initial one, $\beta\neq \alpha$, we will 
consider an appearance experiment inasmuch as the detector will observe neutrinos in a flavour that is not expected in the
production process. The most known appearance experiments are those that study neutrinos from accelerators \cite{Giunti:2007ry}. In 
this case, the flux of neutrinos is build to have some specific flavour, and the detectors look for the other flavours.\\ 

Keeping in mind the analogy between the PMNS matrix with a rotation in the flavour space as in the quark sector, it 
is costumary to write this mixing matrix in terms of three angles $\theta_{12},\theta_{23},\theta_{13}$ and a CP-violating
phase $\delta$ as
\begin{align}
	\widetilde{U} = \begin{pmatrix}
						c_{12} c_{13} & s_{12} c_{13} & s_{13}\, e^{-i\delta} \\
						-s_{12} c_{23} - c_{12}s_{23}s_{13}\, e^{i\delta} & c_{12} c_{23} - s_{12}s_{23}s_{13}\, e^{i\delta} & s_{23} c_{13}\\
						s_{12} s_{23} - c_{12}c_{23}s_{13}\, e^{i\delta}  & - c_{12} s_{23} - s_{12}c_{23}s_{13}\, e^{i\delta} & c_{23} c_{13} 
					\end{pmatrix}
\end{align}
where we used a simplified notation with $s_{ij}=\sin\theta_{ij}$ and $c_{ij}=\cos\theta_{ij}$. In a given experiment, the 
measurements will constrain both the mixing angle and the squared mass difference of relevance for the channel in which the 
searching of oscillations has been performed. A global fit for all the neutrino oscillation experiments has been done by 
Esteban et.\ al.\ \cite{Esteban:2016qun}\footnote{These authors maintain a website with the updated global fit for the neutrino 
oscillations, \url{http://www.nu-fit.org/}.}. The last results of the oscillation parameters global fit, from November - 2017, are 
given in table \ref{tab:GlobalFitNeuOsc}. From these results, it is clear that the CP-violation phase is unknown. However, several 
experiments are being proposed to measure this very important phase \cite{Acciarri:2015uup, C.:2014ika}. The relevance of this phase 
will be clear in the next chapter.\\ 

%%%%%%%%%%%%%%%%%%%%%   table  %%%%%%%%%%%%%%%%%%%%%%%%%%%5
\begin{table}[t]
	\centering
	\caption{Global fit of the neutrino oscillation parameters, taken from \protect\cite{Esteban:2016qun}.}
	\label{tab:GlobalFitNeuOsc}
	\begin{tabular}{c|c|c}
		\toprule
               & Normal Ordering & Inverted Ordering  \\ \midrule\midrule
		$\sin^2\theta_{12}$ & $0.307^{+0.013}_{-0.012}$ & $0.307^{+0.013}_{-0.012}$ \\ \midrule
		$\sin^2\theta_{23}$ & $0.565^{+0.027}_{-0.120}$ & $0.572^{+0.021}_{-0.028}$  \\ \midrule
		$\sin^2\theta_{13}$ & $0.02195^{+0.00075}_{-0.00074}$ & $0.02212^{+0.00074}_{-0.00073}$  \\ \midrule
		$\delta$ & $228^{+51}_{-33}$ & $281^{+30}_{-33}$  \\ \midrule
		$\Delta m_{21}^2$ [eV] & $7.40^{+0.21}_{-0.20}\times 10 ^{-5}$ & $7.40^{+0.21}_{-0.20}\times 10 ^{-5}$  \\ \midrule
		$\Delta m_{3\mathpzc{a}}^2$ [eV] & $+2.515^{+0.035}_{-0.035}\times 10 ^{-5}$ & $-2.483^{+0.034}_{-0.035}\times 10 ^{-5}$ \\ \bottomrule
	\end{tabular}
\end{table}
%%%%%%%%%%%%%%%%%%%%%%%%%%%%%%%%%%%%%%%%%%%%%%%%%%%%%%%%%%%%

\begin{figure}[t]
\centering
\includegraphics[width=0.5\linewidth]{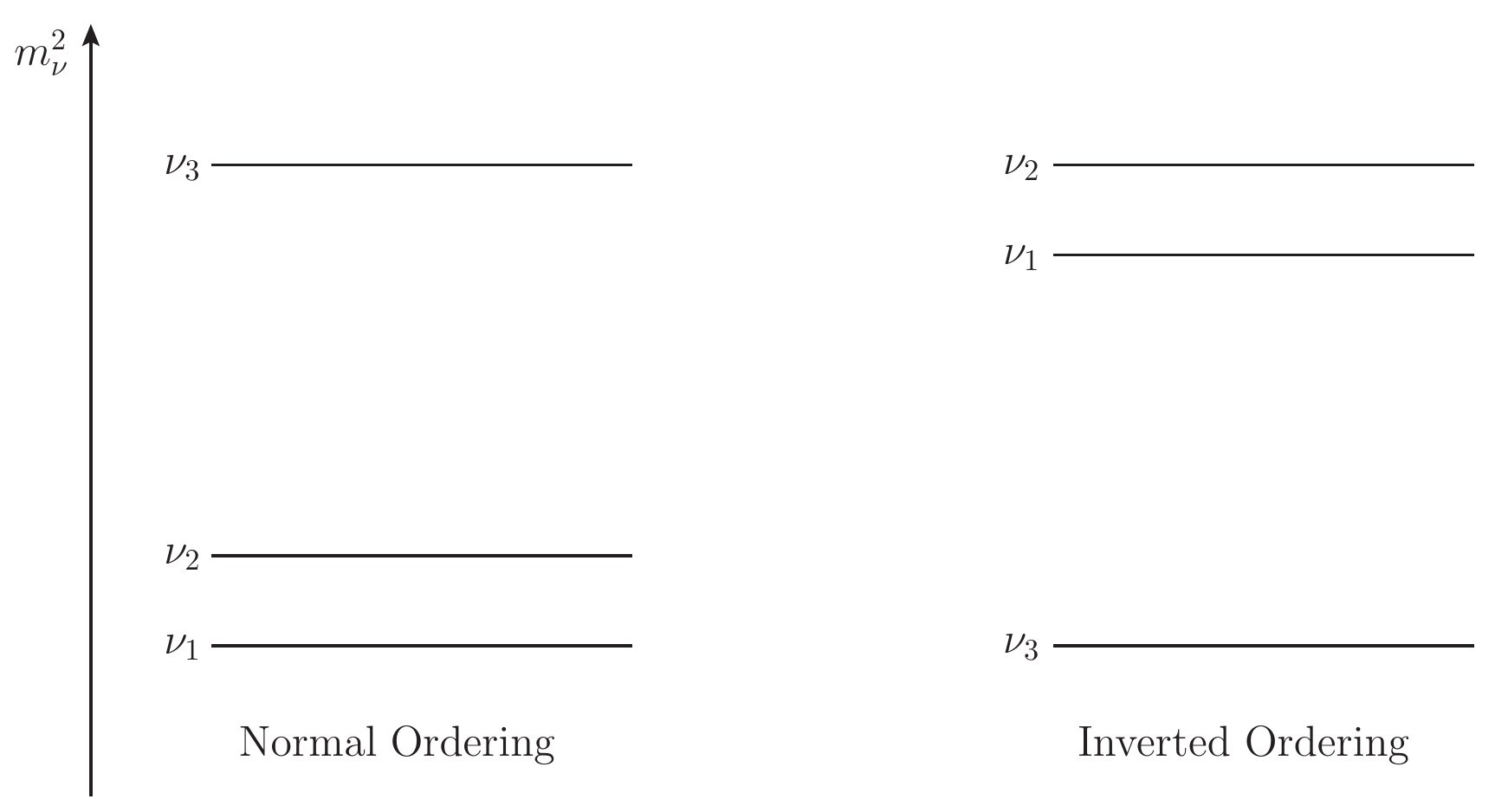}
\caption{Schematic representation of the two possible mass orderings allowed by neutrino oscillations data.}
\label{fig:MO}
\end{figure}

On the other hand, since the real value of the neutrino masses are not known, there are two possible structures 
for these masses that are in agreement with the data. These arrangements, called {\it Normal} and {\it Inverted 
Orderings}, are possible given that the lightest neutrino is unknown. The lightest eigenstate can be the $\nu_1$ (normal 
ordering), which corresponds to the neutrino mass eigenstate mostly composed by electron flavour state, or the $\nu_3$ (inverted ordering) mass eigenstate, the eigenstate which has the least electron flavour composition \cite{Giunti:2007ry}. In figure \ref{fig:MO} we show schematically 
these possibilities.\\

In order to obtain the previous relations, we considered that the neutrino was propagating in vacuum. However, there exist
the possibility of propagating in a dense medium such as the case of solar neutrinos. In that case, it will be necessary
to solve the complete Sch\"odinger equation in the presence of the medium,
\begin{align}
	i\partial_t\ket{\nu_\alpha(t)} = H \ket{\nu_\alpha(t)}
\end{align}
where the hamiltonian will be 
\begin{align*}
	H = H_0 + V,
\end{align*}
being $H_0$ the free hamiltonian and $V$ the effective potential. This potential can be obtained considering the
interactions that a neutrino will have in its journey in the medium. In the SM, both charged and neutral current
interactions can affect the neutrino, creating the effective matter potential for a neutral homogeneous medium
\begin{align}
	V = \sqrt{2} G_F \left(N_e \delta_{\alpha e} -\frac{1}{2} N_n\right)
\end{align}
where $N_e$ is the electron density and $N_n$ is the neutron one. It is possible to show that the neutral current 
contribution will be irrelevant for neutrino oscillations since it is common for the three flavours. In practice,
the matter potential can be interpreted as a modification of the mixing angle and the squared mass difference. 
In a medium, the neutrino has an effective mixing and mass difference dependent on the effective potential \cite{Giunti:2007ry}. 
This has an important consequence \cite{Mikheev:1986gs,Mikheev:1986wj}: it is possible to have a resonance of the 
probability when the effective mixing angle is $\pi/4$. This effect, called the Mikheev-Smirnov-Wolfstein (MSW) 
effect, is the final explanation of the solar neutrino problem \cite{Mikheev:1986gs,Mikheev:1986wj}. It shows that
the deficit in the solar neutrino experiments is due to the adiabatic flavour conversion in the solar medium 
\cite{Hosaka:2005um,Aharmim:2005gt,Smirnov:2016xzf}.\\

In summary, in the present chapter we have introduced the basis for the development of this thesis. We have presented 
the basics of the SM, giving emphasis to the mass generation in the model since the Higgs sector is under experimental
test in the LHC. We also introduced the important definitions of mixing, showing that the SM does not predict the 
precise values of the fermion masses, but it predicts the absence of flavour-changing neutral current processes. The SM also 
predicts the masses of the $W^\pm$ and $Z^0$ bosons and the massless photon. Nonetheless, the mass of the residual 
scalar, the Higgs boson, is not predicted.\\

Afterwards, we focused on the relevant neutrino sources to be used in the present manuscript, and showed the main components for 
studying them. Solar, atmospheric neutrinos and reactor antineutrinos have been observed experimentally while the diffuse supernova 
neutrino background lacks a experimental proof. The experiments have shown that the neutrinos metamorphose into each other after 
being created in a specific processes. This conversion is a proof of the non zero value of the neutrino masses, see the discussion from A.\ Smirnov \cite{Smirnov:2016xzf}. The neutrino oscillations phenomena gives a simple explanation: the flavour states, created in charged current interactions, do not have a definite mass which creates a non zero probability of detecting a different flavour to the one created. Now we have at first glance a picture of neutrino oscillations and the basic phenomena related to it. However, as we mentioned before, the origin of the neutrino mass was not taken into account. Furthermore, we can ask ourselves if there is a fundamental reason for the PMNS matrix to 
have the form \cite{Esteban:2016qun}
\begin{align}
	|\widetilde{U}|_{3\sigma} = 
		\begin{pmatrix}
			0.799 \to 0.844 & 0.516 \to 0.582 & 0.140 \to 0.156 \\
			0.234 \to 0.502 & 0.452 \to 0.688 & 0.626 \to 0.784 \\
			0.273 \to 0.527 & 0.476 \to 0.705 & 0.604 \to 0.765 
		\end{pmatrix},
\end{align}
which is completely different from the CKM matrix, corresponding approximately to an identity matrix.  There are still some other 
open questions about the neutrino masses regarding the ordering, which eigenstates is the lightest neutrino, and also the 
value of the CP-violation phase. Future experiments are being planned to shed some light on these issues \cite{Acciarri:2015uup, C.:2014ika}. But, from the point of view of the author, the basic question concerning the neutrinos is related to the fundamental 
nature of these particles. As we have seen before, neutrinos are the only fundamental fermions that can be Dirac or Majorana 
particles.  So, we can ask ourselves if there is a connexion between the true nature of the neutrino and the smallness of their 
masses. Let us note that there is no definitive answer to these questions. Several models have been proposed to understand the 
true nature of the neutrino. Such discussion will be the task for the next chapters. Concretely, we will consider the neutrinos as 
Majorana fermions in the next chapter, and see the different consequences of this case.
%\newpage
			%%%%%%%%%%%%%%%%%%%%%%%%%%%%%%%%%%%%%%%%%%%%%%%%%%%%%%%%%%%%%%%%%%%%%%%%%%%%%%%%%%%%%%%%%%%%%%%5
%%%%%%%%%%%%%%%%%%%%%%%%%%%%%%%%%%%%%%%%%%%%%%%%%%%%%%%%%%%%%%%%%%%%%%%%%%%%%%%%%%%%%%%%%%%%%%%
\chapter{Majorana Neutrinos}\label{cap:nuMaj}
\chaptermark{Majorana Neutrinos}
%%%%%%%%%%%%%%%%%%%%%%%%%%%%%%%%%%%%%%%%%%%%%%%%%%%%%%%%%%%%%%%%%%%%%%%%%%%%%%%%%%%%%%%%%%%%%%%
%%%%%%%%%%%%%%%%%%%%%%%%%%%%%%%%%%%%%%%%%%%%%%%%%%%%%%%%%%%%%%%%%%%%%%%%%%%%%%%%%%%%%%%%%%%%%%%

% A ¨navalha de Ockham¨\footnote{Entia non sunt multiplicanda praeter necessitatem, as entidades não devem ser multiplicadas além da necessidade.} 

\lettrine{N}{eutrinos} stand out undoubtedly from the other elementary particles due to their peculiarities.
Notably, the absence of electric charge make their nature unclear; this is an additional
problems which charged leptons and quarks do not possess. Majorana particles appeared
as a solution to the negative energy problem, intrinsic to the Dirac formalism, by 
introducing an improved quantization method \cite{Majorana:1937vz}. Majorana obtained that it is
not necessary to presume the existence of antiparticles when the particle is neutral. In other words,
the Majorana fermion is its own antiparticle, analogous to a real scalar field. Evidently, we 
see that this is a quite simpler description for a neutral fermion. Nonetheless, Physics is an experimental
science, and experiences will be the ultimate source of elucidation regarding the true neutrino nature. On the other
hand, the smallness of neutrino masses constitutes an enigma to elementary particle physics.
Certainly, the origin of all fermion masses is by itself an unsolved problem since the Yukawa couplings
are not predicted by the SM as previously shown. However, the large difference between neutrinos 
and charged leptons masses may indicate a different origin for neutrino masses. The {\it see-saw}
mechanism gives an explanation to this difference; the left-handed neutrinos masses are suppressed
by the mass of heavy states. The main inconvenient of this scenario is the lack of a direct proof
of its validity since the involved energy scales go beyond the current reaches. Yet, 
Fukugita and Yanagida \cite{Fukugita:1986hr} found a con\-ne\-xion between the {\it see-saw} mechanism
and the absence of antimatter in our Universe through the process named {\bf leptogenesis}.
In this chapter, we will consider the essential properties of neutrinos supposing them as Majorana fermions.
For that purpose the fundamental properties of such fermions will be analysed as basis to introduce 
the {\it see-saw} mechanism in its three forms. Thenceforth, we will investigate the leptogenesis
scenario in its standard form, addressing its main characteristics.

\newpage

%%%%%%%%%%%%%%%%%%%%%%%%%%%%%%%%%%%%%%%%%%%%%%%%%%%%%%%%%%%%%%%%%%%%%%%%
\section{Majorana Fermions}\label{sec:FerMaj}
%%%%%%%%%%%%%%%%%%%%%%%%%%%%%%%%%%%%%%%%%%%%%%%%%%%%%%%%%%%%%%%%%%%%%%%%

The pioneering work of Majorana \cite{Majorana:1937vz} proposed a symmetric theory of the electron
and the positron through a generalization of the variational principle to Grassmann (anticommuting) 
variables. Majorana asked what are the conditions for a massive fermion to be described by a Weyl spinor;
he found that the basic prerequisite is that the particle has to be its own antiparticle. Evidently, this means
that Majorana fermions are completely neutral. Let us describe below the principal characteristics of this
class of fermions. In this section, we will consider the notation of two-component fermions; after that, we 
will return to the usual four-component notation. The lagrangian describing a left-handed Majorana particle
in this case will be\footnote{The details of the derivation of this lagrangian are given in appendix 
\ref{ap:Lgroup}.}~\cite{Case:1957zza}
\begin{align}\label{eq:DLM}
	\mathscr{L}_M=\phi^{\dagger}i\bar\sigma^{\mu}\partial_{\mu}\phi-\frac{m}{2}\corc{\phi^{T}i\sigma_2\phi-\phi^{\dagger}i\sigma_2\phi^*}.
\end{align}
This lagrangian is built considering the anticommuting character of $\phi$. If this condition were not true, 
the mass term would be trivially zero (see appendix \ref{ap:Lgroup}). We are also writing the terms with dimension 
equal to four, without introducing new particles, i.e., supposing a free particle. The equation of motion for the 
particle, which we will denote as Majorana equation, is given by
\begin{align}\label{eq:EMDC}
	i\bar{\sigma}^\mu\partial_\mu\phi+im\sigma_2\phi^*=0.
\end{align}
An important consequence appears at this point; the Majorana fermion $\phi$ cannot be treated as a particle
in the usual sense, but it needs to be treated as a field from the beginning. In the rest frame of the 
fermion, we have that there are two independent solutions \cite{Mannheim:1980eb}
\begin{align*}
	\phi_1=\begin{pmatrix}
				\theta e^{-imt}\\
				\theta^* e^{+imt}
			\end{pmatrix},
	\quad
	\phi_2=\begin{pmatrix}
				-\theta^* e^{+imt}\\
				\theta e^{-imt}
			\end{pmatrix},
\end{align*}
with $\theta$ and $\theta^*$ Grassmann variables; it is evident that
\begin{align*}
	i\partial_0\phi_i\neq\pm m \phi_i,
\end{align*}
so the solutions are not eigenstates of the $i\partial_0$ operator. In other words, there are not
travelling wave solutions that could be interpreted as particles \cite{Mannheim:1980eb}. Therefore,
we need to start by considering the quantization of the Majorana field.\\

On the other hand, regarding the number of degrees of freedom, it is possible to show that a Majorana fermion
possesses half the degrees of freedom of a Dirac one. This can be proved considering its properties under
the CPT operations and Lorentz transformations \cite{Giunti:2007ry}. A Majorana field will have two possible states for a given momentum
\begin{align*}
	\phi(\vec{p},h=1/2),\quad \phi(\vec{p},h=-1/2)
\end{align*}
where $h$ stands here for the helicity. Let us stress that the field $\phi$ has a definite chirality, left-handed in
the specific case we are treating, but it also has the two distinct helicities, showing definitely the unequivalence between 
chirality and helicity. In addition to this, a crucial property of Majorana fermions is explicit in this two-component notation. 
After quantizing the field in a canonical manner by introducing appropriate anticommutators between the fields and their canonical
conjugate momenta, it is possible to determine the two-point function associated to the Majorana fermion. Actually, there
are two possible manners to define the Feynman propagator in this case,
\begin{align*}
	\bra{0}|\mathscr{T}[\phi(x)\phi(y)^\dagger]\ket{0},\quad \bra{0}|\mathscr{T}[\phi(x)\phi(y)]\ket{0},
\end{align*}
with $\mathscr{T}$ the time-ordering operator. This is a consequence of the same field $\phi$ being able to create and annihilate
the fermion. Therefore, there are additional Wick contractions when compared to the Dirac case \cite{Akhmedov:2014kxa}. Explicitly,
we have that
\begin{subequations}
	\begin{align}
		\bra{0}|\mathscr{T}[\phi(x)\phi(y)^\dagger]\ket{0} &= \int\,\frac{d^4 p}{(2\pi)^4}\frac{\sigma^\mu p_\mu}{p^2-m^2+i\epsilon}\,e^{-ip(x-y)},\\ 
		\bra{0}|\mathscr{T}[\phi(x)\phi(y)^T]\ket{0} &= \int\,\frac{d^4 p}{(2\pi)^4}\frac{im\sigma_2}{p^2-m^2+i\epsilon}\,e^{-ip(x-y)},
	\end{align}
\end{subequations}
with $\sigma^\mu=(I_{2\times 2}, \vec{\sigma})$. To comprehend these two different propagators, let us make the comparison
with the Dirac case. There, the existence of two different particles would make the second type of propagator vanishing; this
is due to the conservation of the charge that a Dirac fermion carries, denominated fermion number. The Feynman propagators for Dirac
particles indicate this conservation graphically using an arrow. However, for a Majorana fermion, there is no conservation of the 
fermion number, which is described symbolically by a line with two arrows pointing in opposite directions \cite{Akhmedov:2014kxa}. 
This is the information contained in the second type of propagator. Let us stress that this additional Feynman propagator is 
proportional to the mass of the field. Thus, we expect that the processes where there is a violation of the fermion number are 
proportional to the mass of the particle in the propagator. This has an important consequence in the {\it neutrinoless double beta 
decay}, decay which may occur if neutrinos are Majorana particles. There are many works considering this neutrinoless decay; for the 
interested reader we indicate the works from Schechter et.\ al.\ \citep{Schechter:1981bd} and Duerr et.\ al.\ \citep{Duerr:2011zd}, 
which show what are the minimal requirements for this decay to happen and for it to be a definite proof of the Majorana 
nature of the neutrino.

%%%%%%%%%%%%%%%%%%%%%%%%%%%%%%%%%%%%%%%%%%%%%%%%%%%%%%%%%%%%%%%%%%%%%%%%
\section{Weinberg Operator}\label{sec:WeinOp}
%%%%%%%%%%%%%%%%%%%%%%%%%%%%%%%%%%%%%%%%%%%%%%%%%%%%%%%%%%%%%%%%%%%%%%%%

The previous discussion was done for a generic Majorana fermion; let us now consider the case we are interested on, the neutrino
case. We now return to the four-component notation by making the substitutions
\begin{align}
	\phi\tto \nu_L,\quad i\sigma^2\tto \MCC,
\end{align}
where $\nu_L$ is a four-component field and $\MCC$ is the charge conjugation matrix, see appendix \ref{ap:Conv}. 
The Majorana lagrangian is given by
\begin{align}\label{eq:LagMAj}
  \mathscr{L}_M=\overline{\nu_L}i\slashed\partial\nu_L+\frac{m_a^\nu}{2}\overline{\nu_L^c}\nu_L+{\rm h.c.},
\end{align}
with the usual definition of the charged conjugated field,
\begin{align*}
	\nu_L^c\equiv \MCC (\overline{\nu_L})^T.
\end{align*}
Our task next will be to consider how to obtain a Majorana mass term, as the one appearing in \ref{eq:LagMAj},
in terms of SU$(2)_L\times$U$(1)_Y$ invariant operators. Let us note that the mass term is built with 
the left-handed component of the neutrino. However, such term has an hypercharge of $Y=-2$; we would 
need a triplet to construct a SU($2$)$_L$ invariant term. Such triplet does not exists in the SM. 
Nonetheless, let us remember that the SM is not considered a complete and final description of the nature 
given that there are many aspects it cannot explain in a satisfactory way. Thus, it is an essential work to consider
non-renormalizable terms. This is not new at all; the SM was built from a non-renormalizable lagrangian, the Fermi
description of the weak interaction. Therefore, the SM is itself a great example of a theory built as a Ultraviolet
completion of an approximate model. Weinberg \cite{Weinberg:1979sa} considered the lowest non-renormalizable
terms built with the SM matter fields which are gauge invariant. He wrote the following dimension-5 operator 
\begin{align}\label{eq:LMM5D}
\mathscr{L}_5^W=\frac{g_{\alpha\beta}^M}{\Lambda}\overline{\corc{L_L^{\alpha}}^c}\widetilde{\Phi}^*\widetilde{\Phi}^\dagger L_L^\beta+{\rm h.c.},
\end{align}
with $g_{\alpha\beta}^M$ some coupling constants and $\Lambda$ is a parameter with dimension of mass. After the 
electroweak symmetry breaking, we get a term
\begin{align}
\mathscr{L}^{W}_5=\frac{1}{2}g_{\alpha\beta}^M\frac{v^2}{\Lambda}\overline{(\nu_L^{\alpha})^c}\nu_L^\beta+{\rm h.c.}
\end{align}
which corresponds to a Majorana mass term. We see that the lowest non-re\-nor\-ma\-li\-za\-ble o\-pe\-ra\-tor, which respects 
the SM gauge group, gives mass to the neutrino. Another important point should be noted at this point. Given that
this term should be suppressed at electroweak scales, the mass scale is larger than the VEV of the Higgs field, $\Lambda\gg v$,
making the neutrino mass also suppressed when compared to the rest of the fermions. This is a significant 
result, we have proven that the dimension-5 operator, which we will call as {\it Weinberg operator}
~\cite{Weinberg:1979sa,Ma:1998dn,Giunti:2007ry}, explains the smallness of mass and the nature of the neutrino.

%%%%%%%%%%%%%%%%%%%%%%%%%%%%%%%%%%%%%%%%%%%%%%%%%%%%%%%%%%%%%%%%%%%%%%%% 
\section{See-saw Mechanism}\label{sec:MassEfecSS}
%%%%%%%%%%%%%%%%%%%%%%%%%%%%%%%%%%%%%%%%%%%%%%%%%%%%%%%%%%%%%%%%%%%%%%%%

It has been shown that the Weinberg operator can be obtained from a UV theory in three different minimal 
manners \cite{Ma:1998dn}. These three cases are denominated as See-saw mechanisms of type I, II and III. 
The reason for such a name will be clear below. Evidently, there exists a large variety of non-minimal models
that can give rise to the dimension-five operator. Nonetheless, we will consider here the basic three cases.

%%%%%%%%%%%%%%%%%%%%%%%%%%%%%%%%%%%%%%%%%%%%%%%%%%%%%%%%%%%%%%%%%%%%%%%%%
\subsection{Type I}\label{subsec:MSSTI}
%%%%%%%%%%%%%%%%%%%%%%%%%%%%%%%%%%%%%%%%%%%%%%%%%%%%%%%%%%%%%%%%%%%%%%%%%

In the simplest case we will introduce three right-handed singlet neutrinos $N_R^\alpha$ ($\alpha=1,2,3$) to the
particle content of the SM \citep{Minkowski:1977sc,Mohapatra:1979ia,Schechter:1980gr}. These neutrinos can have a Majorana mass term given that they have no SM charge
\begin{align*}
	\mathscr{L}^{MR}_N&=\frac{1}{2}\overline{N_R^{\alpha c}}M_R^{\alpha\beta}N_R^\beta+{\rm h.c.}
\end{align*}
On the other hand, we can write also terms coupling left- and right-handed neutrinos as the quarks' case,
\begin{align*}
	\mathscr{L}^{D}_N=- y_{\alpha\beta}^\nu\, \overline{L_L^\alpha} \widetilde{\Phi} N_R^\beta +{\rm h.c.};
\end{align*}
notice the equivalence with the term for the up-type quarks. Therefore, in general, we will 
have a mass term composed of these two terms,
\begin{align}\label{eq:TerMasDM}
	\mathscr{L}^{M+D}_N&=\frac{1}{2}\overline{N_R^{\alpha c}}M_R^{\alpha\beta}N_R^\beta
	- y_{\alpha\beta}^\nu\, \overline{L_L^\alpha} \widetilde{\Phi} N_R^\beta +{\rm h.c.}
\end{align}
Such lagrangian is usually called Dirac-Majorana mass term. Let us stress here that, differently from the
other fermions, neutrinos are the only SM matter fields which can have Majorana and Dirac terms simultaneously. 
However, the true nature of neutrinos is hidden here. If we rewrite the Dirac-Majorana mass term after the
electroweak symmetry breaking as
\begin{align}\label{eq:TerMasDMArr}
	\mathscr{L}^{M+D}_N=\frac{1}{2}\overline{\tilde{N}_L^{\alpha c}} M_\nu^{\alpha\beta}\tilde{N}_L^\beta+h.c.,
\end{align}
with
\begin{align}
\tilde{N}_L =\begin{pmatrix}
		\nu_L\\
		N_R^c
	 \end{pmatrix},
\qquad
M_\nu=\begin{pmatrix}
		0 & \frac{y^\nu}{\sqrt{2}} v\\
		\frac{y^\nu}{\sqrt{2}} v & M_R
	 \end{pmatrix},
\end{align}
it is clear here that left- and right-handed neutrinos do not possess definite mass due to the existence of the Dirac term. 
However, it is possible to diagonalize the general term by considering a rotation in the fields
\begin{align}
	\tilde{N}_L^\alpha=\widetilde{V}^{\alpha a}N_L^a
\end{align}
where
\begin{align}
	\widetilde{V}^{a\alpha}M_\nu^{\alpha\beta}\widetilde{V}^{\beta b}=m^\nu_{a}\delta_{ab}.
\end{align}
Let us note that, differently from the charged fermions, the previous diagonalization is achieved by an 
orthogonal transformation. This is originated in the structure of the Majorana mass term since it can be
written as
\begin{align*}
	\mathscr{L}^{M+D}_N=\frac{1}{2}\tilde{N}_L^\alpha\MCC^\dagger M_\nu^{\alpha\beta} \tilde{N}_L^\beta.
\end{align*}
After the diagonalization, it is clear that the mass eigenstates are Majorana fermions: their mass term is
\begin{align}
	\mathscr{L}^{M+D}_N=\frac{1}{2}m^\nu_{a}\overline{N_L^{a c}} N_L^a+{\rm h.c.}
\end{align}
which correspond to a Majorana mass term. The previous discussion was general in the sense 
that we have not considered any explanation for the smallness of neutrino masses. However, it is
important to note that the right-handed masses are not constrained by the electroweak scale; they 
can be originated from new physics at higher scales. For instance, in Grand Unified Theories (GUT), theories
that consider the unification of the Electroweak and Strong forces, it is usually needed to include right-handed
neutrinos, in such a way that their masses can be of the order of the GUT symmetry breaking scale, approximately $10^{12}$ GeV.\\

In any case, we can suppose that the right-handed neutrinos masses are larger compared to the electroweak scale.
This allow us to diagonalize the mass matrix explicitly in the case in which the {\it eigenvalues} of the right-handed mass
matrix are larger than those of the Dirac term, i.e
\begin{align*}
	M_R^a\gg \frac{v}{\sqrt{2}}y^{\nu\, a}.
\end{align*}
Thus, it is possible to show that the active neutrino mass matrix is then
\begin{align}
	M_L\approx -\left(\frac{v}{\sqrt{2}}y^\nu\right)^T M_R^{-1} \left(\frac{v}{\sqrt{2}}y^\nu\right),
\end{align}
therefore, we see that the mass of active neutrinos is proportional to the inverse of the mass
of the right-handed neutrinos. Furthermore, considering that $\mathcal{O}(M_R)\approx 10^{12}$ GeV, 
and choosing a yukawa of order one, we get that active neutrino masses can be of order $\mathcal{O}(M_L)\approx 1$ eV,
which is of the same order of the limit from PLANCK recent result on the sum of neutrino masses \cite{Ade:2015xua}. 
This is the {\it see-saw} mechanism, given that increasing the right-handed mass values, the active neutrino 
masses decreases. However, a direct detection of a right-handed neutrino with a mass as large as the one 
considered previously is beyond any future experiment. Thus, there are several models in which the 
$\mathcal{O}(M_R)\sim$ TeV, making it possible to be searched at the LHC. Nonetheless, up to this moment 
the {\it see-saw} mechanism has not yet been tested.

%%%%%%%%%%%%%%%%%%%%%%%%%%%%%%%%%%%%%%%%%%%%%%%%%%%%%%%%%%%%%%%%%%%%%%%%%
\subsection{Type II}\label{subsec:MSSTII}
%%%%%%%%%%%%%%%%%%%%%%%%%%%%%%%%%%%%%%%%%%%%%%%%%%%%%%%%%%%%%%%%%%%%%%%%%

The second possibility to obtain the Weinberg operator consists in introducing a 
scalar triplet $\Delta$ with an hypercharge $Y=1$ \cite{Mohapatra:1980yp}. This 
triplet will transform in the adjoint representation of the SU$(2)_L$ group. Our purpose 
here does not consist in studying in a great detail this scenario, so we will briefly 
give the main results. In this case, the lagrangian is given by
\begin{align}
  \mathscr{L}_\Delta=(D_\mu\Phi)^\dagger(D^\mu\Phi)+\Tr\esp{(D_\mu\Delta)^\dagger(D^\mu\Delta)}-V(\Phi,\Delta)+\mathscr{L}_Y,
\end{align}
with the scalar potential $V(\Phi,\Delta)$ being  
\begin{align}
  V(\Phi,\Delta)=&-m_\Phi^2\Phi^\dagger\Phi+\frac{\lambda}{4}(\Phi^\dagger\Phi)^2+M_{\Delta}^2\Tr\esp{\Delta^\dagger\Delta}+[\mu\Phi^Ti\sigma_2\Delta^\dagger\Phi+\text{h.c.}]\notag\\
  &+\lambda_1(\Phi^\dagger\Phi)\Tr\esp{\Delta^\dagger\Delta}+\lambda_2(\Tr\esp{\Delta^\dagger\Delta})^2+\lambda_3\Tr\esp{\Delta^\dagger\Delta}^2+\lambda_4\Phi^\dagger\Delta^\dagger\Delta\Phi.
\end{align}
On the other hand, the Yukawa lagrangian, responsible to give mass to neutrinos, is
\begin{align}
  \mathscr{L}_Y=-Y_\nu \overline{L_{L}^c}\,i\sigma_2\,\Delta\, L_{L}+\text{h.c.},
\end{align}
in such a way that, when the neutral component of the triplet gets a vev, the neutrino eigenstate $a$
gets a mass of
\begin{align}
	m^\nu_a=\sqrt{2}Y_\nu^a v_\Delta=\frac{\mu v^2}{M_\Delta^2}Y_\nu^a.
\end{align}
\newpage

\noindent The relationship for the triplet vev $v_\Delta$ was obtained considering the properties 
of the scalar potential \cite{Perez:2008ha}. We can see that the $\mu M_\Delta^{-2}$ factor
controls the smallness of neutrino masses, in a similar fashion as in the type I case. In general,
the $\mu$ parameter is of the same order as the mass of the triplet $M_\Delta$, 
$\mathcal{O}(M_\Delta)\sim 10^{14-15}$ GeV. Contrary to the previous {\it see-saw} scenario,
the type II case is rich in phenomenology since the scalar sector is constituted of seven bosons 
after the symmetry breaking. However, this scenario suffers the same issue as the 
type I {\it see-saw}, is nearly impossible to test it in its simpler form with the current 
and proposed accelerators. There are forms to circumnavigate this problem. In the Perez et.\ al.\ 
work \cite{Perez:2008ha}, for instance, it is considered a case in which the $M_\Delta$
parameter is of order TeV, making it reachable by the LHC.
 
%%%%%%%%%%%%%%%%%%%%%%%%%%%%%%%%%%%%%%%%%%%%%%%%%%%%%%%%%%%%%%%%%%%%%%%%%
\subsection{Type III}\label{subsec:MSSTIII}
%%%%%%%%%%%%%%%%%%%%%%%%%%%%%%%%%%%%%%%%%%%%%%%%%%%%%%%%%%%%%%%%%%%%%%%%%

The last possibility to obtain a Majorana mass term at tree level consist in introducing the 
a right-handed fermion triplet $\vec\Sigma_R=(\Sigma^1_R,\Sigma^2_R,\Sigma^3_R)$ with
zero hypercharge. This scenario is known as type III {\it see-saw} \cite{Foot:1988aq}.
It is described by
\begin{align}
	\mathscr{L}_\Sigma=i\overline{\vec{\Sigma}_R}\gamma^\mu D_\mu\vec{\Sigma}_R-\esp{\frac{1}{2}\overline{\vec{\Sigma}_R}M_T\vec{\Sigma}_R^c+\overline{\vec{\Sigma}_R}y_T\widetilde{\Phi}^{\dag}\vec\tau L_L+\text{h.c.}}
\end{align}
where $\vec\tau=\frac{1}{2}\vec{\sigma}$ are the SU$(2)_L$ generators. After the 
electroweak symmetry breaking, the sector responsible for the neutrino mass is
\begin{align}\label{eq:DenYukTrip'}
	\mathscr{L}_Y&=-\frac{vy_T}{\sqrt 2}\overline{\Sigma_R^0}\nu_L-\frac{M_T}{2}\overline{\Sigma_R^0}\Sigma_R^{0c}+{\rm h.c.}\notag\\
	&=-\frac{1}{2}\overline{N_L^c} M_\nu N_L,
\end{align}
with
\begin{align}\label{eq:ArrNeut}
N_L=\begin{pmatrix}
		\nu_L\\
		\MCC\overline{\Sigma_R^0}^T
	 \end{pmatrix}
 =\begin{pmatrix}
		\nu_L\\
		\Sigma_R^{0c}
	 \end{pmatrix},
\qquad
M_\nu=\begin{pmatrix}
		0 & \frac{v}{\sqrt 2}y_{\scriptscriptstyle T}\\
		\frac{v}{\sqrt 2}y_{\scriptscriptstyle T} & M_T
	 \end{pmatrix}.
\end{align}
Let us note here that the structure of the mass matrix is similar to the type
I scenario. Therefore, in the case which $vy_T\ll M_T$ the active neutrinos will 
have a small mass. An important difference between type I and type III scenarios 
is the existence of the charged fermions $\Sigma_R^\pm$ having gauge interactions,
enriching the phenomenology of this case.\\ 

Finally, it is possible to show that in the three types of {\it see-saw} mechanisms the Weinberg operator is obtained 
when integrating out the heavy states \cite{Giunti:2007ry}. It is also worth to mention
here that a Majorana mass term can be obtained by radiative corrections, such in
GUT and Supersymmetric (SUSY) models \cite{Fukuyama:2008sz}.

%\newpage

%%%%%%%%%%%%%%%%%%%%%%%%%%%%%%%%%%%%%%%%%%%%%%%%%%%%%%%%%%%%%%%%%%%%%%%%
\section{Leptogenesis}\label{sec:leptopad}
%%%%%%%%%%%%%%%%%%%%%%%%%%%%%%%%%%%%%%%%%%%%%%%%%%%%%%%%%%%%%%%%%%%%%%%%

Among the possible consequences of neutrinos being Majorana particles 
is the generation of lepton asymmetry in the early Universe. This process, denominated
{\it Leptogenesis}, has attracted a lot of attention given that it connects two open problems
in High Energy Physics, the smallness of the neutrino masses and the matter-antimatter asymmetry.
We will concentrate ourselves in the remaining of this chapter on the discussion of such process,
considering as basis the type I {\it see-saw} mechanism.

%%%%%%%%%%%%%%%%%%%%%%%%%%%%%%%%%%%%%%%%%%%%%%%%%%%%%%%%%%%%%%%%%%%%%%%%%%%%%%%%%%%%%%%%%%%%%%%%%%
\subsection{Evidences of a Baryonic Asymmetry}\label{sec:EvAss}
%%%%%%%%%%%%%%%%%%%%%%%%%%%%%%%%%%%%%%%%%%%%%%%%%%%%%%%%%%%%%%%%%%%%%%%%%%%%%%%%%%%%%%%%%%%%%%%%%%

Nowadays, diverse evidences indicate that in our Universe exists a baryonic asymmetry, i.e.,
the number density of antibaryons (antiprotons and antineutrons) is smaller than the baryon density 
(protons and neutrons) \cite{Kolb:1990vq}. Such hypothesis was built from several astrophysical
measurements. In the first place, it is possible is assess that our Solar System is made only from
matter. For instance, it was observed that cosmic rays are made of antiprotons in a quantity 
$10^{-4}$ times smaller than the number of protons. This number can be obtained considering the 
quantity of antiprotons created in the atmosphere as collision sub-products. Therefore, we see that
the cosmic rays are a good evidence that exists a difference between the baryon and antibaryon numbers 
in our surroundings \cite{Kolb:1990vq}.\\

On the other hand, in the scales of galaxy clusters, the asymmetry evidence is weaker. For instance,
if there were galaxies made completely of antimatter in the same cluster, we would expect a large emission 
of gamma rays due to the large quantities of matter and antimatter in an annihilation process. However,
the lack of those emissions shows that, in general, the closest clusters are made entirely of baryons
or antibaryons \cite{Kolb:1990vq}.\\ 

Now, to have large quantities of antimatter in a baryonically symmetric Universe, it would be necessary
that baryons and antibaryons were separated in scales bigger than  $10^{12}M_{\astrosun}$ \cite{Kolb:1990vq}. 
Nonetheless, the current description of the evolution of our Universe given by the Standard Cosmological Model ($\Lambda$CDM)
shows that, in order to have a Universe without baryonic asymmetry, some unknown process should have
acted to separate baryons and antibaryons. Otherwise, an ``annihilation catastrophe'' would have occurred
due to the absence of an asymmetry, and the Universe would only be composed by radiation after the annihilation.\\

A simple question can be asked at this point; is it possible that the asymmetry was an initial condition 
in the Universe? Such a solution has an inconvenient. If the Inflation mechanism is actually correct,
any initial condition should have been erased by the accelerated expansion. Therefore, we can imagine
then that the Universe started in a symmetric state and a dynamical mechanism acted to create the
baryon asymmetry  \cite{Kolb:1990vq}.

\newpage

The baryon asymmetry is characterized by the ratio \cite{Davidson:2008bu}
\begin{align*}
  \eta&\defm\frac{n_B-n_{\bar B}}{n_\gamma}=(6.21\pm 0.16)\times 10^{-10}
\end{align*}
where $n_{B}$ in the baryon number density, $n_{\bar B}$ in the antibaryon number density, $n_\gamma$ 
is the photon number density. The baryon asymmetry $\eta$ has been obtained from the observations of the Wilkinson Microwave
Anisotropy Probe (WMAP) of the Cosmic Microwave Background (CMB). This implies that for each
$10^{10}$ antiquarks there were $10^{10}+1$ quarks that created the structures we see.\\

Another evidence in favour of the asymmetry is obtained considering the primordial abundance of the
light elements D, $^3$He,$^4$He and $^7$Li in the $\Lambda$CDM and comparing with the observational data \cite{Kolb:1990vq}.
These results also indicate the existence of a baryon asymmetry. So, we can assert that the asymmetry
hypothesis has strong experimental bases. Therefore, from a theoretical point of view, the explanation of 
this asymmetry is one of the principal open problems in elementary particle physics and cosmology.

%%%%%%%%%%%%%%%%%%%%%%%%%%%%%%%%%%%%%%%%%%%%%%%%%%%%%%%%%%%%%%%%%%%%%%%%%%%%%%%%%%%%%%%%%%%%%%%%%%
\subsection{Sakharov's Conditions}\label{sec:ConSak}
%%%%%%%%%%%%%%%%%%%%%%%%%%%%%%%%%%%%%%%%%%%%%%%%%%%%%%%%%%%%%%%%%%%%%%%%%%%%%%%%%%%%%%%%%%%%%%%%%%

Sakharov \cite{Sakharov:1967dj} established the three basic ingredients that a model needs to contain
to generate dynamically a baryonic asymmetry. Let us revise them briefly:
\begin{description}
\item [1. Baryonic Number Violation] It is necessary to have a source of baryon number violation to generate
	the asymmetry. This violation is present in several GUT models given that leptons and quarks
	usually belong to a irreducible representation of the gauge group in the model. However, in the SM
	it is possible to have a baryon number violation; processes known as instantons, created by non-perturbative
	phenomena, violate the total baryon number \cite{Davidson:2008bu}.
\item [2. C and CP Violation] When a baryon number violation exists, it is also needed that the model
	contains some source of C and CP violation since the baryon number non-conserving processes create
	baryons and antibaryons in the same quantities. 
\item [3. Out-of-equilibrium conditions] Since the mass of particles and antiparticles are equal in a 
	CPT invariant model, which will respect also Lorentz invariance, the baryon and antibaryon numbers 
	need to be the same in a equilibrium situation. Let us write the operator of a simultaneously
	charge conjugation, parity and temporal inversion  operations as $\Theta=CPT$, and $\rho(t)=e^{-\beta(t)H(t)}$
	a density matrix in a thermal equilibrium state, with $H$ the hamiltonian of the system. Then, the expectation
	value of the baryon number is $B$
 	 \begin{align*}
    		\langle B\rangle&=\tr\llav{e^{-\beta H}B}=\tr\llav{\Theta^{-1}\Theta e^{-\beta H}B}\\
    		&=\tr\llav{\Theta e^{-\beta H}\Theta^{-1}\Theta B\Theta^{-1}}=\tr\llav{ e^{-\beta H}(- B)}\\
    		\langle B\rangle&=-\langle B\rangle
  	\end{align*}  
  	where we used that $\Theta B\Theta^{-1}=-B$ as, by definition, the antibaryon number is the opposite
  	of the baryon number. So, we see that it is not possible to create a baryon asymmetry when a system 
  	is in thermal equilibrium.
\end{description}

Although all these ingredients are present in the SM, the CP violation produced by the
complex phase of the CKM matrix is not enough to create an asymmetry of the order $10^{-10}$.
Furthermore, the discovery of a SM-like Higgs boson, mentioned in the previous chapter, implies
that the Electroweak phase transition, which would be the responsible for the 
deviation from thermal equilibrium, is not of the first order \cite{Kuzmin:1985mm,Kajantie:1996mn,Chung:2012vg}. 
This proofs that {\it baryogenesis} needs beyond SM physics to be understood.\\

There are many models which allow to generate dynamically a baryon asymmetry. Usually, those
models contain new sources of CP violation and additional deviations from thermal equilibrium.
Some of these mechanisms are GUT baryogenesis, Affleck-Dine mechanism and leptogenesis \cite{Davidson:2008bu}.
The last one, leptogenesis, or generation of lepton asymmetry is a cosmological consequence of 
neutrinos being Majorana particles. We will consider next this scenario.

%%%%%%%%%%%%%%%%%%%%%%%%%%%%%%%%%%%%%%%%%%%%%%%%%%%%%%%%%%%%%%%%%%%%%%%%%%%%%%%%%%%%%%%%%%%%%%%%%%
\subsection{Standard Leptogenesis}\label{sec:Lepto}
%%%%%%%%%%%%%%%%%%%%%%%%%%%%%%%%%%%%%%%%%%%%%%%%%%%%%%%%%%%%%%%%%%%%%%%%%%%%%%%%%%%%%%%%%%%%%%%%%%

The baryon asymmetry generation through a lepton asymmetry is a scenario, which has attracted a lot of
attention in the last years, since it allows to connect two seemingly uncorrelated problems:
the smallness of neutrino masses and the baryogenesis. This is possible due to the {\it see-saw}
mechanism, as we will see. Let us note that lepton number is not conserved by a Majorana mass term. This
is due basically to the fact that a Majorana fermion cannot have a conserved charge. Moreover, new CP 
violation phases can exist both in active and sterile neutrino sectors, in the mixing 
matrices $\widetilde{V}^{a\alpha}$. Let us notice that this matrix has more CP violation phases 
than in the Dirac case as the Majorana mass term in \eqref{eq:LagMAj} is not invariant under 
phase transformations. In the case of 3 right-handed heavy neutrinos, there are two 
additional phases to the usual phase appearing in the Dirac case. These phases are known as Majorana phases. 
One important consequence is that these phases do not modify the neutrino oscillation
probability, making the differentiation between Majorana and Dirac natures impossible in an oscillation
experiment.\\ 

On the other hand, the deviation from thermal equilibrium can be present in the decay of the
heavy neutrinos. Thus, in principle, leptogenesis obeys all Sakharov conditions. However, it is
necessary to determine if this scenario can give rise an asymmetry compatible with the observed one. In the
type II {\it see-saw}, described previously, is it possible to generate the lepton asymmetry in different manners.
The most well known, and the one we will study here, is thermal leptogenesis. In that case, the right-handed
neutrinos are created by scatterings in the thermal bath, and, after their decay, a lepton asymmetry
is generated.\\ 

In the present section we will describe the standard formalism for thermal leptogenesis.
Let us recall the Majorana lagrangian \eqref{eq:TerMasDM} for $n=3$ right-handed neutrinos as
\begin{align}
	\mathscr{L}^{M+D}_N&=\frac{1}{2}\overline{N_R^{\alpha c}}M_R^{\alpha\beta} N_R^\beta
	-y_{\alpha\beta}^\nu\, \overline{L_L^\alpha} \widetilde{\Phi} N_R^\beta +{\rm h.c.}, \tag{\ref{eq:TerMasDM}}
\end{align}
where we will consider the basis in which the heavy right-handed neutrinos mass matrix is diagonal
$M_R^{\alpha\beta}=M_R^\alpha\delta_{\alpha\beta}$. Lepton asymmetry is generated in the following scheme. 
Scatterings in the thermal bath create an important population of $N_a$ at temperatures of the same order 
of the masses $T\sim M_a$. Then, these neutrinos decay through the channels 
\begin{align*}
  N_a&\tto l_b\phi,\\
  N_a&\tto \bar l_b\phi^*.
\end{align*}
If there is CP violation, asymmetries will be created in all channels; 
meanwhile, if the interactions are out-of-equilibrium, those asymmetries will remain.
Finally, the processes which occur in the Electroweak transition phase will 
transform part of the lepton asymmetry into a baryonic one. An important
fact that is noted in \cite{Davidson:2008bu} is that the creation and annihilation is
controlled by the same set of coupling constants, $y_{\alpha\beta}^\nu$. Therefore,
we should notice that the same CP asymmetry is involved in the creation and the 
decays of the $N_a$. Thus, as a consequence of this, the total asymmetry can be
zero since the initial asymmetry is ``washed out'' by the decays, inverse decays and
scatterings \cite{Davidson:2008bu}. Such washout will be critical for thermal leptogenesis
Furthermore, flavour effects will be very important, since the washout depend on how the 
leptons are distinguished in the processes. Anyhow, we will not address the washout here.
Let us discuss in more detail the Sakharov conditions in this scenario.

%%%%%%%%%%%%%%%%%%%%%%%%%%%%%%%%%%%%%%%%%%%%%%%%%%%%%%%%%%%%%%%%%%%%%%%%%%%
\subsubsection{CP Violation}\label{subsec:ViolCP}
%%%%%%%%%%%%%%%%%%%%%%%%%%%%%%%%%%%%%%%%%%%%%%%%%%%%%%%%%%%%%%%%%%%%%%%%%%%

\begin{figure}[!h]
  \centering
  \includegraphics[scale=0.6]{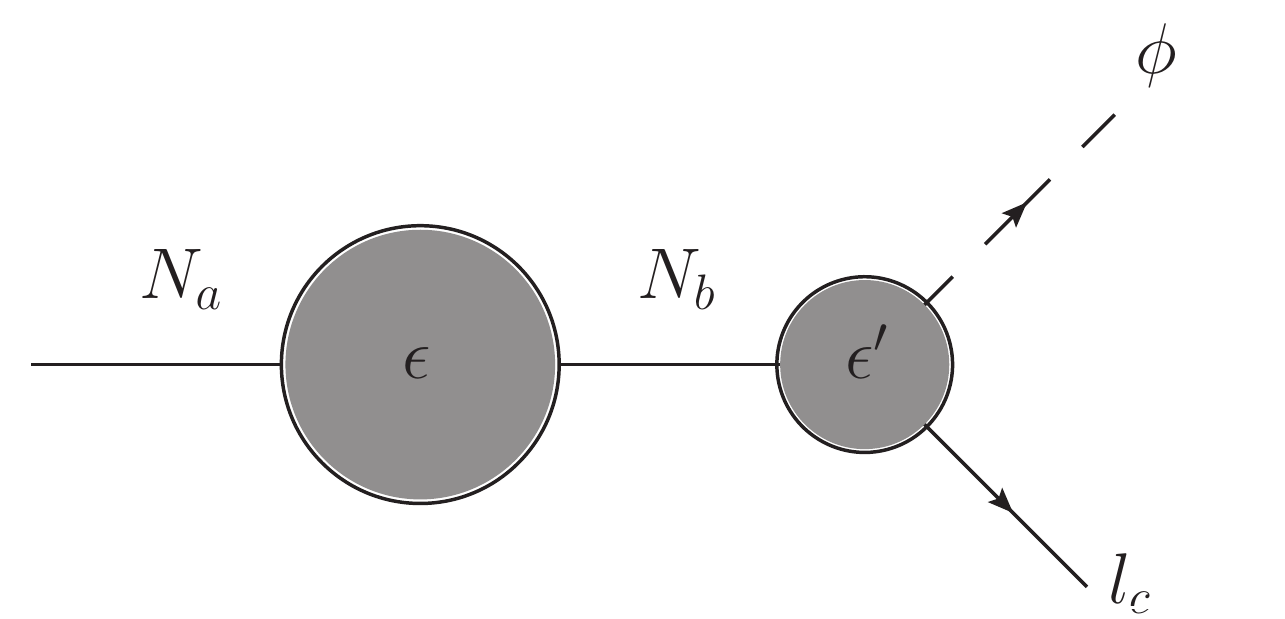}
  \caption{Types of CP violation $\epsilon$ e $\epsilon'$}
  \label{fig:ViolCP}
\end{figure}

\begin{figure}[!htb]
  \centering
  \includegraphics[scale=0.45]{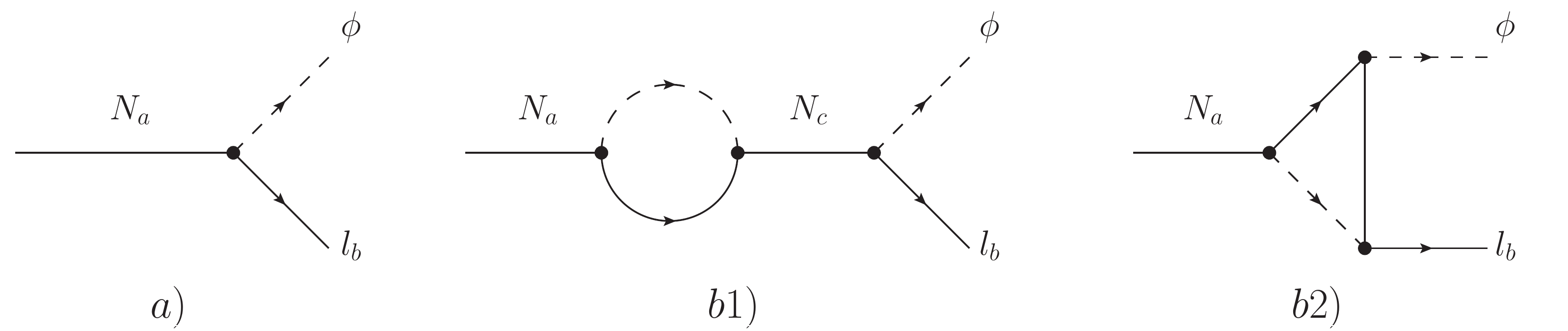}
  \caption{Contributing diagrams to CP asymmetry. $a)$ tree level, $b1)$ self-energy at 1-loop, $b2)$ vertex correction.}
  \label{fig:ViolCP-Decay}
\end{figure}

To consider the CP violation we need to keep in mind that there are two different contributions:
the asymmetry coming from the decays of the right-handed neutrinos
\begin{align*}
  \epsilon_{ab}\defm\frac{\Gamma(N_a\tto l_b\phi)-\Gamma(N_a\tto \bar l_b\phi^*)}{\Gamma(N_a\tto l_b\phi)+\Gamma(N_a\tto \bar l_b\phi^*)}
\end{align*}
and the asymmetry from the scatterings
\begin{align*}
  \epsilon_{ab}\defm\frac{\sigma(l_a\phi\tto \bar l_b\phi^*)-\sigma(\bar l_a\phi^*\tto l_b\phi)}{\sigma(l_a\phi\tto \bar l_b\phi^*)+\sigma(\bar l_a\phi^*\tto l_b\phi)}.
\end{align*}
Let us study the asymmetry from the decays as the scattering case is analogous to the following
discussion. The CP asymmetry is obtained considering the quantum interference between
the diagrams at tree level with the 1-loop corrections; such asymmetry is zero when
considering only tree level processes. The self-energies and vertex contributions are denominated as
$\epsilon$ and $\epsilon'$ respectively, see figure \eqref{fig:ViolCP}. For the $1$-loop contributions we will
have the diagrams in figure \eqref{fig:ViolCP-Decay}.\\

For the \emph{hierarchical} case, in which one of the right-handed neutrino masses is smaller
that the other masses, $M_1\ll M_2,M_3$, we have that the contribution $\epsilon$ is not
relevant since the transitions among neutrinos is suppressed by the mass difference. On the other 
hand, the $\epsilon'$ contribution can be computed, and is given by \cite{Fukugita:1986hr,Davidson:2008bu}
\begin{align}
  \epsilon_{1b}'=\frac{1}{8\pi\,(YY^\dag)_{11}}\sum_{c=2,3}\esp{\mathfrak{Im}[Y^*_{b1}(YY^\dag)_{1c}Y_{bc}]f\corc{\frac{M_c^2}{M_1^2}}+\mathfrak{Im}[Y^*_{b1}(YY^\dag)_{c1}Y_{bc}]\corc{1-\frac{M_c^2}{M_1^2}}^{-1}}
\end{align}
where the Fukugita-Yanagida 1-loop function is \cite{Fukugita:1986hr,Pilaftsis:2003gt}
\begin{align*}
  f(x)&=\sqrt{x}\esp{\frac{1}{1-x}+1-(1+x)\ln\corc{\frac{1+x}{x}}}.
\end{align*}\\
\sloppy \negthickspace Before considering the leptonic asymmetry, it is necessary to mention a problem which appears
in the computation of the CP violation. There exists a double counting in the previous CP
asymmetries. In the scattering $|\mathcal{M}(l_a\phi\to l_b\phi)|^2$,  we are considering
processes where neutrinos are created on-shell and then decaying $\abs{\mathcal{M}(l_a\phi\to N_c,N_c\to l_b\phi)}^2$.
This process was already included in the decay asymmetry.
\newpage 
\noindent Therefore, we need to subtract the ``real intermediate
state'' by hand,
\begin{align}
  \abs{\mathcal{M}'(l_a\phi\to l_b\phi)}^2&\defm\abs{\mathcal{M}(l_a\phi\to l_b\phi)}^2-\abs{\mathcal{M}^{os}(l_a\phi\to l_b\phi)}^2,\\
  \abs{\mathcal{M}'(l_a\phi\to l_b\phi)}^2&\defm\abs{\mathcal{M}(l_a\phi\to \bar l_b\phi^*)}^2-\abs{\mathcal{M}^{os}(l_a\phi\to \bar l_b\phi^*)}^2,
\end{align}
\sloppy where $\abs{\mathcal{M}^{os}(l_a\phi\to \bar l_b\phi^*)}^2$ corresponds to the matrix element in which the neutrino is on-shell. We can then rewrite the previous term using the branching ratio $\abs{\mathcal{M}^{os}(l_a\phi\to \bar l_b\phi^*)}^2$  for $N_c\tto l_b\phi$,
\begin{align}
  \abs{\mathcal{M}'(l_a\phi\to l_b\phi)}^2&\defm\abs{\mathcal{M}(l_a\phi\to l_b\phi)}^2-\sum_c\abs{\mathcal{M}(l_a\phi\to N_c)}^2B^{N_c}_{l_b\phi}.
\end{align}
When considering the Boltzmann equations, it will be necessary to take into account this subtraction of the 
real intermediate states in order to obtain the correct result.

%%%%%%%%%%%%%%%%%%%%%%%%%%%%%%%%%%%%%%%%%%%%%%%%%%%%%%%%%%%%%%%%%%%%%%%%%%%
\subsubsection{Boltzmann Equations}\label{subsec:EqBoltz}
%%%%%%%%%%%%%%%%%%%%%%%%%%%%%%%%%%%%%%%%%%%%%%%%%%%%%%%%%%%%%%%%%%%%%%%%%%%

The standard procedure to determine the generation and evolution of a lepton asymmetry
uses the Boltzmann equations as a fundamental equation in the description of
the right-handed neutrinos out-of-equilibrium. In the case in which the Universe 
is described by the Friedmann-Lemaître-Roberston-Walker (FLRW) metric, the equation for
the number density for the $k$ species, defined as
\begin{align*}
  n_k(t)=g_k\int\,d^3\vec{p}\,f_{k}(p,t),
\end{align*}
with $f_{k}(p,t)$ the distribution function, $g_k$ the internal degrees of freedom
of the species, is given by\footnote{We are considering here the notation $\int_{\vec{p}_i}\defm\int\frac{d^3\vec{p}_i}{(2\pi)^32p^0_i}$}
\begin{align}
  \der{n_k}{t}+3Hn_k=\sum_{int}\int_{\vec{p}_k}\llav{C^{1\leftrightarrow 2}[f_k]+C^{2\leftrightarrow 2}[f_k]}.
\end{align}
Here $H$ corresponds to the Hubble expansion parameter; $C^{1\leftrightarrow 2,2\leftrightarrow 2}[f_k]$ are
the collision terms for decays and scatterings ($f_{k}\defm f_{k}(p_k,t)$)
\begin{subequations}
  \begin{align}
    C^{1\leftrightarrow 2}[f_k]&=\frac{1}{2}\int_{\vec{p}_m,\vec{p}_n}[f_mf_n(1\pm f_k)W_{mn|k}-(1\pm f_m)(1\pm f_n)f_kW_{k|mn}],\\
    C^{2\leftrightarrow 2}[f_k]&=\frac{1}{2}\int_{\vec{p}_l,\vec{p}_m,\vec{p}_n}[f_mf_n(1\pm f_k)(1\pm f_l)W_{mn|kl}-(1\pm f_m)(1\pm f_n)f_kf_lW_{kl|mn}],
  \end{align}
\end{subequations}
being the signal $\pm$ an indicator of the fermionic or bosonic nature, respectively, and we have defined
the relativistic invariant probability densities of interaction as
\begin{align*}
  W_{mn|k}&\defm\frac{1}{2p^0_k}(2\pi)^4\delta^4(p_k-p_m-p_n)\overline{\abs{\mathcal{M}(mn\to k)}}^2,\\
  W_{mn|kl}&\defm\frac{1}{2p^0_k}(2\pi)^4\delta^4(p_k+p_l-p_m-p_n)\overline{\abs{\mathcal{M}(mn\to kl)}}^2.
\end{align*}
$\overline{\abs{\mathcal{M}(mn\to k(l))}}^2$ are the averaged in the internal degrees of freedom
probabilities of occurring an interaction $mn\tto k(l)$.
The non-linear system of Boltzmann equations for all the species present in the 
thermal bath is difficult to solve since for each species there will be an equation 
which, in general, will depend on the other species. Nonetheless, to solve such system, 
we need to understand first how the out-of-equilibrium conditions appear in the early Universe.
The expansion of the Universe will determine when a species is in equilibrium or not \cite{Kolb:1990vq}.
If the interaction rate $\Gamma$ is of the same order or less than the Hubble expansion $H$  
\begin{align*}
\Gamma\lesssim H,
\end{align*}
the species will not be able to balance the reactions, and will be out-of-equilibrium;
meanwhile. if the interaction rates are larger than $H$, the species will be in thermal equilibrium
with the primordial plasma \cite{Kolb:1990vq}. Such criterion regulates which species can be considered
in equilibrium, simplifying the whole system of Boltzmann equations.\\

For the initial conditions, we will assume that the Universe, composed by SM particles
interacting through gauge interactions, remains in equilibrium. A thermal density of right-handed
neutrinos can be produced if the time scale of production $1/\Gamma_{pr}$, $\Gamma_{pr}$ the decay width 
of the right-handed neutrinos, is smaller
than the age of the Universe in that epoch $H^{-1}$ where \cite{Kolb:1990vq}
\begin{align*}
  H(T=M_a)=1.66g_*^{\frac{1}{2}}\left.\frac{T^2}{M_{pl}}\right|_{T=M_a},
\end{align*}
with $g_*=106.75$ the effective number of relativistic degrees of freedom of the thermal bath and $M_{pl}$
the Planck mass.\\

At this point we need to obtain the Boltzmann equations which govern the generation and evolution of 
the asymmetry. However, since our purpose is to give a general vision of the leptogenesis, we will
give some results taken from the literature. The procedure used consists in solving the Boltzmann
equation for the right-handed neutrino number density and then solve the equation of motion
for the lepton asymmetry. Such equations are given by \cite{Kolb:1990vq,Luty:1992un,Pilaftsis:1997jf}
\begin{subequations}
  \begin{align}
    \der{n_{N_a}}{t}+3Hn_{N_a}&=-\corc{\frac{n_{N_a}}{n_{N_a}^{\text{eq}}}-1}\gamma_{N_a},\\
    \der{n_L}{t}+3Hn_L&=\sum_{a=1}^3\esp{\epsilon_{aa}\corc{\frac{n_{N_a}}{n_{N_a}^{\text{eq}}}-1}-\frac{n_L}{2n_{l}^{\text{eq}}}}\gamma_{N_a}-\frac{n_L}{2n_{l}^{\text{eq}}}\gamma_{\sigma},
  \end{align}
\end{subequations}
with $n_{N_a}$, $n_L\defm n_l-n_{\bar l}$ the number densities of the right-handed neutrinos
and the lepton asymmetry, respectively; $n_{N_a}^{\text{eq}},n_{l}^{\text{eq}}$ are their values
at thermal equilibrium and $\gamma_{N_a},\gamma_{\sigma}$ are the reduced collision terms,
\begin{subequations}
  \begin{align}
    \gamma_{N_a}&=n_{N_a}^{\text{eq}}\frac{K_1(M_{a}^2\,T^{-1})}{K_2(M_{a}^2\,T^{-1})}\Gamma_a,\\
    \gamma_{\sigma}&=\frac{T}{8\pi^4}\int_0^\infty ds\,s^{\frac{3}{2}}K_1(\sqrt{s}\,T^{-1})\sigma'(s),
  \end{align}
\end{subequations}
being $K_1(z)$ and $K_2(z)$ the modified Bessel functions; $\Gamma_a$, 
$\sigma'(s)$ are the expressions at zero temperature ($T=0$) of the total decay width
and cross section of the scatterings $l\phi\tto\bar l \phi^*$, respectively.
We should notice here that the contributions of the real intermediate states
have been subtracted from the cross sections \cite{Pilaftsis:1997jf}. Some suppositions have been made
to obtain these equations \cite{Davidson:2008bu, Pilaftsis:1997jf}. First, it was supposed
that particles obey the Maxwell-Boltzmann distribution functions, which is reasonable 
in the absence of effects related to the statistics, such as Bose-Einstein condensates
or degeneracies due to fermionic degrees of freedom. Second, it was assumed that the leptons
and Higgs bosons are in thermal equilibrium. Third, interactions which create
a washout are weak, and we can neglect them in the first discussion.\\

\begin{figure}[!tb]
  \centering
  \includegraphics[scale=0.4]{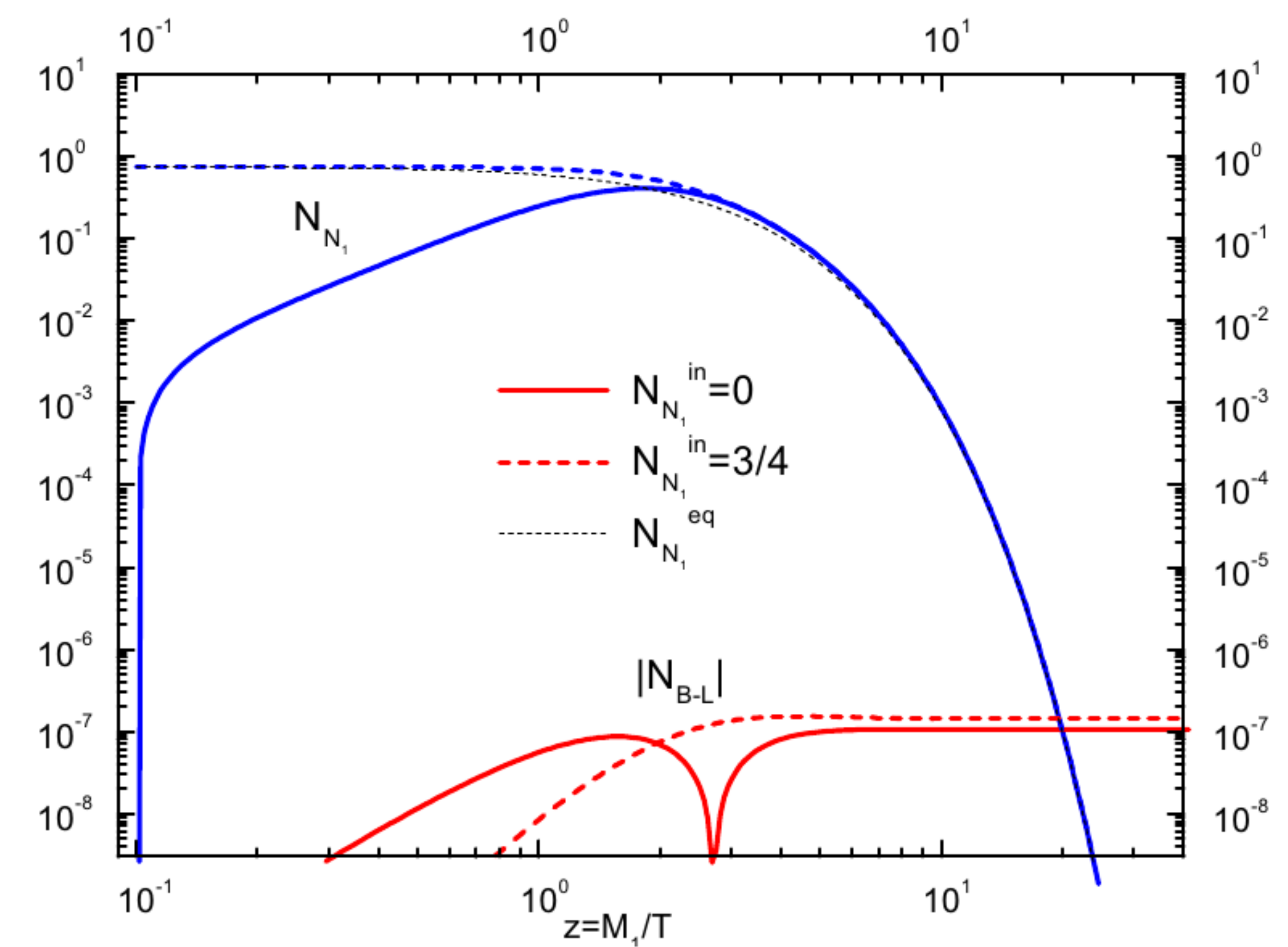}
  \caption{Number neutrino densities and lepton asymmetry for the hierarchical case with vacuum
  (full line) and thermal (dashed line) initial conditions. Taken from \protect\cite{Buchmuller:2002rq}.}
  \label{fig:ResulHierar}
\end{figure}

The figure \ref{fig:ResulHierar}, taken from \cite{Buchmuller:2002rq}, provides the results
of the numerical study of the Boltzmann equations for two different types of initial conditions.
Moreover, it presents the evolution of the abundance of the right-handed neutrinos and the
lepton asymmetry. For the case in which the initial heavy neutrino abundance is zero (full blue line), 
we see that such abundance grows until it exceeds the equilibrium abundance (dotted blue line) 
creating an asymmetry  at $T\approx 0.5M_1$; then, it decreases and goes to the equilibrium
value. For that case, an asymmetry appear even before the neutrino density goes to its equilibrium value
(full red line) and, starting in $T\sim 0.3M_1$ the asymmetry becomes a constant due to the freeze out
process. Notice that in the point in which the neutrino abundance gets to the thermal value, the lepton
asymmetry goes to zero, as expected. Now, for the second case, the initial abundance corresponds to a 
thermal value (dashed blue line),
choosing $n_{N_1}(0)=\frac{3}{4}$\footnote{This choice is not arbitrary as it corresponds
to the number of neutrinos per comoving volume, which, in general, is equal to $\frac{3}{8}g$,
$g$ being the internal degrees of freedom of the field. For Majorana neutrinos we have that $g=2$.}.
The time evolution of the neutrino abundance is such that it becomes out-of-equilibrium, and in that
point a lepton asymmetry is created (dashed red line). An important feature of this is that in both cases the 
lepton asymmetry is of the same order.\\

In conclusion, in this chapter we have considered a general formalism for the case in which neutrinos are Majorana
particles. We have seen how Majorana fermions have peculiar properties, such as 
having half of the degrees of freedom of a Dirac particle; they are their own antiparticles,
making them completely neutral; and the quantization procedure shows that there are two different types
of Feynman propagators. In the set of the known fundamental particles, the only ones that can be Majorana are neutrinos. 
However, since a Majorana term is forbidden by the SM symmetry group at tree level, we considered the 
lowest non-renormalizable term which is invariant under the SU$(2)_L\times$U$(1)_Y$ group. That term
is a five dimensional operator, known as Weinberg operator, and, after the electroweak symmetry breaking,
it gives mass to the neutrinos. Such term is suppressed by a energy scale larger than the Electroweak scale, 
justifying the smallness of the neutrino mass. We also considered the simplest Ultraviolet complete theories
which can give rise to the Weinberg operator when integrating out the heavy states. These 
theories are known as the {\it see-saw} mechanisms in its three different types.\\ 

The first type corresponds to the introduction of heavy right-handed neutrinos while the second and third
are for the cases in which a scalar triplet or a fermion triplet is included to the SM set of fields,
respectively. Briefly, in this scenarios the mass of the active (left-handed) neutrinos is inversely
proportional to the new particles, so, increasing the mass of the additional fields, the neutrino mass
becomes smaller; hence, the name of {\it see-saw} mechanism. A problem common to the minimal three {\it see-saw} 
scenarios is that the new states are quite heavy, making their production beyond the reach of current and 
future experiments. Nonetheless, we showed a crucial consequence of the {\it see-saw} mechanism, the explanation
of the matter-antimatter asymmetry through the denominated {\it leptogenesis}. In this case, the baryon
asymmetry is created by a lepton asymmetry which, in turn, is originated by the out-of-equilibrium decay
of the heavy states. We have seen in some detail how the asymmetry is produced and maintained in the Universe
evolution. However, a definite proof of neutrinos being Majorana is still to be found. The search of neutrinoless
double beta decay is the foremost candidate to establish the neutrino nature, although there are some caveats
in the interpretation of this decay. In the next chapter, we will concentrate ourselves in the case of neutrinos 
being Dirac fermions, focusing in two simple models for that case.%\newpage
			%%%%%%%%%%%%%%%%%%%%%%%%%%%%%%%%%%%%%%%%%%%%%%%%%%%%%%%%%%%%%%%%%%%%%%%%%%%%%%%%%%%%%%%%%%%%%%%%%%
%%%%%%%%%%%%%%%%%%%%%%%%%%%%%%%%%%%%%%%%%%%%%%%%%%%%%%%%%%%%%%%%%%%%%%%%%%%%%%%%%%%%%%%%%%%%%%%%%%
\chapter[Dirac Neutrinos]{Dirac Neutrinos}\label{nu-2HDM}
\chaptermark{Dirac Neutrinos}
%%%%%%%%%%%%%%%%%%%%%%%%%%%%%%%%%%%%%%%%%%%%%%%%%%%%%%%%%%%%%%%%%%%%%%%%%%%%%%%%%%%%%%%%%%%%%%%%%%
%%%%%%%%%%%%%%%%%%%%%%%%%%%%%%%%%%%%%%%%%%%%%%%%%%%%%%%%%%%%%%%%%%%%%%%%%%%%%%%%%%%%%%%%%%%%%%%%%%

\lettrine{P}{lanck} scale is usually considered as a final frontier since it is thought that all fundamental 
forces will have the same strength in that scale. Furthermore, all the interactions could become one 
breaking the laws of physics as we know them. Between the Planck and Electroweak scales there is 
a large set of possible models that can exist, such as GUT theories, SUSY, technicolor, extradimensions, etc, as solutions 
of the hierarchy problem. It is also possible the existence of other physics that are currently unknown. 
Experimentally, we have just started to explore the TeV scale.
Anyhow, neutrinos can be charged under new interactions at higher energy scales, but the current sensitivity of 
our experiments can not detect any evidences of it. Thus, in principle, neutrinos can be Dirac 
particles in those possible scenarios since Majorana terms would be forbidden. Nonetheless, how to explain 
the smallness of Dirac neutrino masses with our current knowledge? We can consider and constrain possible
SM extensions that could give rise neutrino masses; the simplest possibility is to extend minimally the SM
and give mass to neutrinos in a analogous manner to the other particles. In that case, there
is not an explanation of the smallness of neutrino masses as there is not for the exact values of the 
other particles' masses.\\ 

However, inspired by the {\it see-saw} mechanism, we can speculate that the neutrino mass has a different origin from the charged fermions. We can then ask ourselves what is the simplest scenario in which masses are generated keeping $\mathcal{O}(1)$ couplings.
As in the SM, we can suppose that the mass has an origin in a spontaneous symmetry breaking 
created by a new doublet whose VEV is of the same order of the neutrino masses.
The scenario in which we have a second doublet responsible for the neutrino mass will be the 
object of our study in the present chapter. We will consider first briefly the problems regarding the
minimal SM extension; after that, we will introduce the second Higgs doublet Model that will be named 
{\it neutrinophilic}, as the second scalar doublet manly couples with those fermions while the first doublet
will couple basically with charged fermions. We will analyse two different cases, 
related to the symmetries imposed to the model.

%\newpage

%%%%%%%%%%%%%%%%%%%%%%%%%%%%%%%%%%%%%%%%%%%%%%%%%%%%%%%%%%%%%%%%%%%%%%%%%%%%%%%%%%%%%%%%%%%%%%%%%%
\section{Minimal SM extension}\label{sec:nuMP}
%%%%%%%%%%%%%%%%%%%%%%%%%%%%%%%%%%%%%%%%%%%%%%%%%%%%%%%%%%%%%%%%%%%%%%%%%%%%%%%%%%%%%%%%%%%%%%%%%%

The simplest extension that can be done in the SM to address neutrino masses corresponds to introduce three
right-handed neutrinos which will be singlets of the SM gauge group
\begin{align*}
	\nu_R^\alpha=\{\nu_{eR},\nu_{\mu R},\nu_{\tau R}\},
\end{align*}
in such a way that we will be able to write down Yukawa terms
\begin{align*}
	\mathscr{L}_\nu=- y_{\alpha\beta}^\nu\, \overline{L_L^\alpha} \widetilde{\Phi} \nu_R^\beta +{\rm h.c.}
\end{align*}
We need to prevent to have a Majorana mass term since such term will imply that neutrinos are Majorana fermions as we saw before. 
This can be done by keeping in mind that the total lepton number is conserved, so, if neutrinos
are charged with such number, Majorana terms are forbidden\footnote{It has been shown that global symmetries are
violated by quantum gravity effects \cite{Banks:2006mm}. This shows that symmetries as lepton number conservation should be gauged
at some point before the Planck scale, avoiding this issue. This is also true for the symmetries that will be considered
in the next section.}. Leptons will have then a Yukawa lagrangian given by
\begin{align}
	\mathscr{L}_{\rm Y}^\ell=-y_{\alpha\beta}^\ell\, \overline{L_L^\alpha} \Phi \ell^\beta_R
					+y_{\alpha\beta}^\nu\, \overline{L_L^\alpha} \widetilde{\Phi} \nu_R^\beta +{\rm h.c.},
\end{align}
then, after the Electroweak symmetry breaking, the charged leptons and neutrinos will have definite masses
after the diagonalization of the mass terms. That diagonalization is performed similarly to the quarks, by 
defining 
\begin{align*}
	\ell^\alpha_{L,R} = V_{L,R}^{\alpha a}\ell^{a}_{L,R}, \qquad \nu^\alpha_{L,R} = \widetilde{V}_{L,R}^{\alpha a}\nu^{a}_{L,R},
\end{align*}
where
\begin{align*}
	(V_L^{a\alpha})^*y_{\alpha\beta}^\ell V_R^{\beta b} = y_a^\ell \delta_{ab}, \qquad (\widetilde{V}_L^{a \alpha})^*y_{\alpha\beta}^\nu \widetilde{V}_R^{\beta b} = y_a^\nu \delta_{ab},
\end{align*}
to obtain the mass eigenstate lagrangian
\begin{align}
	\mathscr{L}_{\rm Y}^\ell=-\left(1+\frac{h}{v}\right) \left[m_a^\ell\, \overline{\ell^a} \ell^a 	+m_a^\nu\, \overline{\nu^a} \nu^a\right].
\end{align}
The neutrino masses explicitly are given by
\begin{align}
	m_a^\nu = \frac{v}{\sqrt{2}}y_a^\nu,
\end{align}
so, to have neutrino masses of order $m_a^\nu\sim\mathcal{O}({\rm eV})$, we need that $y_a^\nu\sim\mathcal{O}(10^{-11})$
since $v\sim 10^2$ GeV. This is an extremely small number even if compared to the electron Yukawa, which is of the order
$y_e^\ell\sim\mathcal{O}(10^{-6})$. However, as we have seen, the SM does not address the origin of the Yukawa
couplings. Indeed, the neutrino Yukawa is small, but, analysing the difference between the top quark ($y_t^q\sim\mathcal{O}(1)$) 
and the electron Yukawa couplings, we see that there is also a large hierarchy. Thus, what is different in the 
neutrino case? We can consider the following argument. Since the doublet components behave as one single field 
from the Electroweak interaction's point of view, we see that the difference among the masses of 
the particles belonging to the same doublet is not so large in the quark sector's first and second families. Explicitly, 
the up and down quarks have masses of $\sim 2$ MeV and $\sim 4$ MeV, respectively, while the strange and charm 
quarks have masses of $\sim 0.1$ GeV and $\sim 1$ GeV. Nonetheless, in the lepton sector, we see that the mass difference 
between the doublet constituents is indeed large. This may suggest that neutrino masses have a different origin and 
that is the reason of such large hierarchy. Of course, in the third quark family there is an important difference
between the bottom ($\sim 4$ GeV) and top ($\sim 170$ GeV) quark masses, yet such difference is not as large as 
in the lepton case. Anyhow, we can simply suppose that the Yukawa neutrino has such small value without
trying to justify it, as done for the other particles. Still, it is important to consider and constrain 
other possible models for neutrino masses as we will analyse in the next section.

%%%%%%%%%%%%%%%%%%%%%%%%%%%%%%%%%%%%%%%%%%%%%%%%%%%%%%%%%%%%%%%%%%%%%%%%%%%%%%%%%%%%%%%%%%%%%%%%%%
\section{Neutrinophilic two-Higgs-doublets Models}\label{sec:nu2HDM}
%%%%%%%%%%%%%%%%%%%%%%%%%%%%%%%%%%%%%%%%%%%%%%%%%%%%%%%%%%%%%%%%%%%%%%%%%%%%%%%%%%%%%%%%%%%%%%%%%%

Although there is not a fundamental reason for the neutrino masses in the Dirac case to have a 
different origin from the rest of fermions, we can study possible models that can give rise to  
small Dirac neutrino masses. We can think about other possible extensions that can be done to the
SM. Our guide for such scrutiny is that we prefer neutrino masses generated without 
any extremely small parameter, or, in other words, the neutrino mass will be protected by some symmetry. 
Anyhow, to obtain neutrino masses of order $\mathcal{O}({\rm eV})$ in a similar fashion to the other 
SM particle, we would need a small VEV to evade small couplings. On the other hand, the next to minimal 
extension to the SM would be to include a second Higgs doublet, see \citep{Branco:2011iw} and references therein. 
Let us stress that this two-Higgs-doublets models ($2$HDM) are interesting by themselves since, for 
instance, supersymmetry requires the presence of two scalar doublets.\\ 

For neutrino masses we can adapt $2$HDM to induce small neutrino masses. Thus, we will suppose 
that the two doublet neutral components of the doublets 
get a VEV. The first doublet will be responsible to give masses to the charged leptons while the second 
one will give mass to neutrinos. We also require that the first doublet does not couple with neutrinos; 
otherwise, the mass contribution from such doublet to the neutrino will dominate. Consequently, it is necessary to introduce a new 
symmetry in which right-handed neutrinos and the second doublet are charged while the rest of particles 
are not. The two simplest cases which have been considered are a $\Z_2$ symmetry, proposed by Gabriel and 
Nandi \cite{Gabriel:2006ns}, and a U$(1)$ scenario in the Davidson Logan work \cite{Davidson:2009ha}. 
In general, the lagrangian associated to a $2$HDM is given by
\begin{align}
	\mathscr{L}=(D_\mu\Phi_1)^\dagger(D^\mu\Phi_1)+(D_\mu\Phi_2)^\dagger(D^\mu\Phi_2)+V(\Phi_1,\Phi_2),
\end{align}
with the scalar potential
\begin{align}
\label{v2hdm}
V(\Phi_1,\Phi_2)&=m_{11}^2\Phi_1^\dagger\Phi_1 +m_{22}^2\Phi_2^\dagger\Phi_2-(m_{12}^2\Phi_1^\dagger\Phi_2+\text{h.c.})\nonumber\\
&+\dfrac{\lambda_1}{2}(\Phi_1^\dagger \Phi_1)^2+\dfrac{\lambda_2}{2}(\Phi_2^\dagger \Phi_2)^2
+\lambda_3 \Phi_1^\dagger\Phi_1 \Phi_2^\dagger\Phi_2+\lambda_4 \Phi_1^\dagger\Phi_2 \Phi_2^\dagger\Phi_1\notag \\
&+\left[\dfrac{\lambda_5}{2}(\Phi_1^\dagger\Phi_2)^2 +\left(\lambda_6 \Phi_1^\dagger\Phi_1 + \lambda_7 \Phi_2^\dagger\Phi_2\right)\Phi_1^\dagger\Phi_2 + \text{h.c.}\right],
\end{align}
being $\Phi_{1,2}$ two scalar doublets with hypercharge $Y=+1$. The neutral components
of these doubles will have VEVs given by $\langle \Phi_i\rangle=\frac{v_i}{\sqrt{2}}$,
$i=1,2$. Recalling our purposes, we will impose that the second doublet generates neutrino masses,
so $\mathcal{O}(v_2)\sim$ eV. Besides, $\Phi_1$ will break the Electroweak symmetry, and we have 
that $v_1\sim$ $246$ GeV. Clearly, $v_2\ll v_1= v$, with $v^2\defm v_1^2+v_2^2$. 
Analysing the scalar potential, we have that the $m_{12}^2, \lambda_{5,6,7}$ parameters
can be complex.\\ 

However, the new symmetry will forbid in general the terms proportional
to $\lambda_6$ and $\lambda_7$ while the $m_{12}^2$ and $\lambda_5$ parameters may
be different from zero according to the symmetry. In the Gabriel-Nandi scenario,
we have that $m_{12}^2$ must be zero since the term proportional to such coupling is
not invariant under the symmetry. For the U$(1)$ case both $m_{12}^2$ and $\lambda_5$
need to be zero. Nonetheless, as we will see shortly, to avoid a Nambu-Goldstone boson
in the spectrum, $m_{12}^2$ cannot be zero. We will consider in this section 
$\lambda_6=\lambda_7=0$ only to be as general as possible.\\

The stationary conditions $\partial V/\partial \Phi_i=0$ for our potential are
\begin{subequations}\label{eq:stationary}
\begin{align}
	\dfrac{\lambda_1}{2}v_1^3+\dfrac{\lambda_{345}}{2}v_1v_2^2+
m_{11}^2v_1-m_{12}^2v_2&=0,\\
	\dfrac{\lambda_2}{2}v_2^3+\dfrac{\lambda_{345}}{2}v_2v_1^2+
m_{22}^2v_2-m_{12}^2v_1&=0.
\end{align}
\end{subequations}
Thus, we have that the parameters $m_{ii}^2$ can be written as functions of the
VEVs and the $m_{12}^2$ mass term. For the case in which such mass term is zero, 
$m_{12}^2=0$, we will have two stable solutions $(v,0)$ or $(0,v)$, so one VEV will
correspond to the SM one while the second one will be inert. In general, 
for $m_{12}^2 \neq 0$ the equations \eqref{eq:stationary} do not have analytical 
solutions; nevertheless, supposing that  $m_{12}^2\ll v^2$ we can obtain
\begin{equation}
   v_1\approx v, \qquad v_2\approx\dfrac{m_{12}^2}{\frac{\lambda_{345}}{2}\, v^2+m_{22}^2}v,
\end{equation}
and a symmetric solution interchanging the indices $1\leftrightarrow
2$. Therefore, we can conclude that a small VEV requires a correspondingly small
$m_{12}^2$ parameter. Finally, let us notice that there can be more than
one solutions to the stationary conditions \eqref{eq:stationary}, creating
the possibility of having simultaneously different minima $(v_1,v_2)$ and
$(v_1',v_2')$. However, it can be checked analytically if the vacuum is a
global one, see \cite{1303.5098,Machado:2015sha}. We have included such 
conditions to our analysis.\\

Given that there is a large hierarchy between the scales of the model, we can wonder if the second VEV $v_2$
is stable under radiative corrections. It has been shown that the radiative corrections for both 
$\Z_2$ model \cite{Haba:2011fn} and the U($1$) case \cite{Morozumi:2011zu} do not spoil the smallness of the second VEV. 
In the case of the
$\Z_2$ including a non-zero value $m_{12}^2$, it was shown that loop corrections induces
the terms which breaks the symmetry, similar to the terms proportional to $\lambda_6$ and 
$\lambda_7$ \cite{Haba:2011fn}. Such additional contributions modify the stability conditions
\eqref{eq:stationary}, and it can be shown that the ratio between the induced and 
the $m_{12}^2v_1$ terms are given by \cite{Haba:2011fn}
\begin{align}
	\abs{\frac{\lambda_6 v_1^3}{2m_{12}^2v_1}}\sim \frac{3}{4\pi^2}\log\abs{\frac{v_1}{v_2}}\sim 2,
\end{align}
where
\begin{align}
	\lambda_6 \sim -\frac{3\lambda_1\lambda_5}{4\pi^2}\frac{m_{12}^2}{(m_{22}^2-m_{11}^2)^2}\corc{m_{22}^2-m_{11}^2+m_2\log\frac{\abs{m_{11}^2}}{\abs{m_{22}^2}}}.
\end{align}
Therefore, the order of $v_2$ will not be modified by radiative corrections in this case. On the other hand,
in the Davidson-Logan model, it has been shown that the radiative corrections are also proportional to
the $m_{12}^2$ term, and they are dependent on the scalar spectrum \cite{Morozumi:2011zu}. 
In the case in which the charged and pseudoscalar scalars are mass degenerated, the corrections
are small. As we will see afterwards and in the next chapter, the theoretical and experimental
constraints actually imply that those particles need to be degenerated.\\

After the Electroweak symmetry breaking, we will parametrize the two doublets as
\begin{align}\label{Higgs}
\Phi_a 	=
	\begin{pmatrix}
		\phi_a^+\\ (v_a+\rho_a+i\eta_a)/\sqrt{2} 
	\end{pmatrix},\qquad a=1,2.
\end{align}
The scalar spectrum is composed by four charged particles $\phi_1^\pm,\phi_2^\pm$,
two CP-even neutral bosons $\rho_1$ e $\rho_2$, and two CP-odd neutral scalars, $\eta_{1,2}$.
However, due to the mixing terms, we have that these particles do not possess definite masses.
Introducing the rotations 
\begin{subequations}
\begin{align}
    \begin{pmatrix}
      \rho_1\\
      \rho_2
    \end{pmatrix}&=-
    \begin{pmatrix}
      c_{\hat{\alpha}} & -s_{\hat{\alpha}}\\
      s_{\hat{\alpha}} & c_{\hat{\alpha}}
    \end{pmatrix}
    \begin{pmatrix}
      h\\
      H
    \end{pmatrix},\\
    \begin{pmatrix}
      \phi_1^-\\
      \phi_2^-
    \end{pmatrix}&=-
    \begin{pmatrix}
      c_{\hat{\beta}} & -s_{\hat{\beta}}\\
      s_{\hat{\beta}} & c_{\hat{\beta}}
    \end{pmatrix}
    \begin{pmatrix}
      G^-\\
      H^-
    \end{pmatrix},\\
    \begin{pmatrix}
	\eta_1\\
	\eta_2
	\end{pmatrix}&=-
	\begin{pmatrix}
	c_{\hat{\beta}}&-s_{\hat{\beta}}\\
	s_{\hat{\beta}}&c_{\hat{\beta}}
	\end{pmatrix}
	\begin{pmatrix}
	G^0\\
	A
	\end{pmatrix},
\end{align}
\end{subequations}
being $c_{\hat{\alpha}(\hat{\beta})}=\cos\hat{\alpha}({\hat{\beta}})$ and $s_{{\hat{\alpha}}({\hat{\beta}})}=\sin\hat{\alpha}({\hat{\beta}})$, we can diagonalize 
the mass terms of the potential. The angles $\hat{\alpha}$ and $\hat{\beta}$ must be
\begin{align}
	\tan(2\hat{\alpha})&=\dfrac{2(-m_{12}^2+\lambda_{345}\; v_1 v_2)}{m_{12}^2(v_2/v_1-v_1/v_2)+\lambda_1 v_1^2-\lambda_2 v_2^2},\\
	\tan{\hat{\beta}} &= \dfrac{v_2}{v_1},
\end{align}
where $\lambda_{345}\defm\lambda_3+\lambda_4+\lambda_5$. The physical states will
be then a charged scalar $H^\pm$, two neutral scalars $h$ and $H$, and a pseudoscalar $A$
\begin{subequations}
	\begin{align}
		h &=-\rho_1\cos\hat{\alpha} -\rho_2\sin{\hat{\alpha}}, \\ 
	H &=\rho_1\sin\hat{\alpha} -\rho_2\cos\hat{\alpha},\\
		H^+ &=\phi_1^+\sin{\hat{\beta}} -\phi_2^+\cos{\hat{\beta}}, \\ 
		A &=			\eta_1\sin{\hat{\beta}}-\eta_2\cos{\hat{\beta}},
	\end{align}
\end{subequations}
whose masses are given by
\begin{subequations}
\begin{align}
	m_h^2 &= M^2 \cos^2({\hat{\alpha}}-{\hat{\beta}})\nonumber\\ 
	&\quad+\left(\lambda_1 \sin^2 {\hat{\alpha}} \cos^2{\hat{\beta}}+\lambda_2 \cos^2 {\hat{\alpha}} \sin^2{\hat{\beta}}-\frac{\lambda_{345}}{2}\sin 2{\hat{\alpha}} \sin 2{\hat{\beta}}\right)v^2, \label{eq:mh}\\
	 m_H^2 &= M^2 \sin^2({\hat{\alpha}}-{\hat{\beta}})\nonumber\\ 
	&\quad +\left(\lambda_1 \cos^2 {\hat{\alpha}} \cos^2{\hat{\beta}}+\lambda_2 \sin^2 {\hat{\alpha}} \sin^2{\hat{\beta}}+\frac{\lambda_{345}}{2}\sin 2{\hat{\alpha}} \sin 2{\hat{\beta}}\right)v^2, \label{eq:mH}\\
	m_{H^\pm}^2 &= M^2 - \frac{\lambda_{45}}{2} v^2,\label{eq:mHpm}\\ 
	m_A^2 &= M^2 - \lambda_5 v^2,\label{eq:mA}
\end{align}
\end{subequations}
being $M^2\equiv m_{12}^2(\sin{\hat{\beta}} \cos{\hat{\beta}})^{-1}$. The inverse relations are straightforwardly obtained \cite{Kanemura:2004mg}
%% 
%\begin{subequations}\label{eq:lambdas}
	\begin{align}
		\label{eq:lambda1}
		\lambda_1  &= \frac{1}{v^2}\left(-\tan^2{\hat{\beta}} M^2+\frac{\sin^2{\hat{\alpha}}}{\cos^2{\hat{\beta}}} m_H^2+\frac{\cos^2{\hat{\alpha}}}{\cos^2{\hat{\beta}}}m_h^2\right), \\
		\label{eq:lambda2}
		\lambda_2  &=\frac{1}{v^2}\left(-\cot^2{\hat{\beta}} M^2+\frac{\cos^2{\hat{\alpha}}}{\sin^2{\hat{\beta}}} m_H^2+\frac{\sin^2{\hat{\alpha}}}{\sin^2{\hat{\beta}}}m_h^2\right),\\
			\label{eq:lambda3}
		\lambda_3 &=\frac{1}{v^2}\left(-M^2+2 m_{H^\pm}^2+\dfrac{\sin(2{\hat{\alpha}})}{\sin(2{\hat{\beta}})}(m_h^2-m_H^2)\right),\\
		\label{eq:lambda4}
		\lambda_4 &= \frac{1}{v^2} \left(M^2+m_A^2-2 m_{H^\pm}^2\right), \\
	\label{eq:lambda5}
		\lambda_5 &= \frac{1}{v^2} \left(M^2- m_A^2\right).
	\end{align}
These relations will be useful for studying the physical parameter space of the model.
On the other hand, the $G^\pm$ and $G^0$ fields are massless corresponding to the Nambu-Goldstone
bosons appearing due to the symmetry breaking. These bosons will be absorbed by the Gauge bosons
$W^\pm$, $Z$. Let us see this in more detail. Rewriting the doublets in terms of the physical 
states
\begin{align*}
  \Phi_1=
  \begin{pmatrix}
    c_{\hat{\beta}} G^+-s_{\hat{\beta}} H^+\\
    \frac{1}{\sqrt{2}}\llav{v_1+c_{\hat{\alpha}} h-s_{\hat{\alpha}} H-ic_{\hat{\beta}} G^0+is_{\hat{\beta}} A}
  \end{pmatrix},\\ 
  \Phi_2=
  \begin{pmatrix}
    s_{\hat{\beta}} G^++c_{\hat{\beta}} H^+\\
    \frac{1}{\sqrt{2}}\llav{v_2+s_{\hat{\alpha}} h+c_{\hat{\alpha}} H-is_{\hat{\beta}} G^0-ic_{\hat{\beta}} A}
  \end{pmatrix},
\end{align*}
we can show that it is equivalent to
\begin{subequations}
  \begin{align}
    \Phi_1&=
  \exp\llav{\frac{i}{v}\corc{\zeta^a(x)\tau^a+\frac{\xi(x)}{2}}}
  \begin{pmatrix}
    -s_{\hat{\beta}} H^+\\
    \frac{1}{\sqrt{2}}\llav{v_1+c_{\hat{\alpha}} h-s_{\hat{\alpha}} H+is_{\hat{\beta}} A}
    \end{pmatrix},\\
    \Phi_2&=
  \exp\llav{\frac{i}{v}\corc{\zeta^a(x)\tau^a+\frac{\xi(x)}{2}}}
  \begin{pmatrix}
    c_{\hat{\beta}} H^+\\
    \frac{1}{\sqrt{2}}\llav{v_2+s_{\hat{\alpha}} h+c_{\hat{\alpha}} H-ic_{\hat{\beta}} A}
  \end{pmatrix},
  \end{align}
\end{subequations}
where the $\zeta^a,\xi$ functions are related to the Nambu-Goldstone fields
in the following manner
\begin{align*}
	G^+&=\frac{1}{2\sqrt{2}}(\zeta^2+i\zeta^1),\\
	G^0&=\frac{1}{2\sqrt{2}}(\zeta^3-\xi).\\
\end{align*}
\newpage
We can now do a gauge transformation to eliminate the non-physical degrees of freedom, going to
the so-called {\it Unitary gauge}
\begin{align*}
  \Phi_1&\to {\cal U}(x)\exp\llav{-\frac{i}{2v}\xi(x)}\Phi_1
  =\begin{pmatrix}
    -s_{\hat{\beta}} H^+\\
    \frac{1}{\sqrt{2}}\llav{v_1+c_{\hat{\alpha}} h-s_{\hat{\alpha}} H+is_{\hat{\beta}} A}
    \end{pmatrix},\\
  \Phi_2&\to {\cal U}(x)\exp\llav{-\frac{i}{2v}\xi(x)}\Phi_2
  =\begin{pmatrix}
    c_{\hat{\beta}} H^+\\
    \frac{1}{\sqrt{2}}\llav{v_1+s_{\hat{\alpha}} h+c_{\hat{\alpha}} H-ic_{\hat{\beta}} A}
  \end{pmatrix},\\
  \vec{W}_\mu&\to {\cal U}(x)\vec{W}_\mu {\cal U}(x)^\dagger+\frac{i}{g}(\partial_\mu {\cal U}(x)){\cal U}(x)^\dagger, \qquad
  B_\mu\to B_\mu-\frac{i}{2g'v}\partial_\mu\xi(x),
\end{align*}
where ${\cal U}(x)=\exp\llav{-\frac{i}{v}\zeta^a(x)\tau^a}$. After performing this gauge transformation, it 
is necessary to study the gauge boson sector.; nonetheless, it will not be affected by the new particles, 
remaining the same as studied previously in the SM.\\

Now, we can turn our attention to the neutrino sector. In this case, the Yukawa lagrangian is given by
\begin{align}
	-\mathscr{L}_{\rm Y}^\ell &=y_{\alpha\beta}^\ell\, \overline{L_L^\alpha} \Phi_1 \ell^\beta_R
					+y_{\alpha\beta}^\nu\, \overline{L_L^\alpha} \widetilde{\Phi}_2 \nu_R^\beta +{\rm h.c.},\notag\\
	&=m_a^\ell\, \overline{\ell^a}\ell^a+\frac{m_a^\ell}{v_1}(c_{\hat{\alpha}} h-s_{\hat{\alpha}} H)\,\overline{\ell^a}\ell^a+i\frac{m_a^\ell}{v_1}s_{\hat{\beta}} A\,\overline{\ell^a}\gamma^5\ell^a-s_{\hat{\beta}}(\sqrt{2}\, y_{\alpha\beta}^\ell \overline{\nu^\alpha_L} H^+ \ell^\beta_R+\mathrm{h.c.})\notag\\
	&\quad+m_a^\nu\overline{\nu^a} \nu^a+\frac{m_a^\nu}{v_2}(s_{\hat{\alpha}} h+c_{\hat{\alpha}} H)\overline{\nu^a} \nu^a-i\frac{m_{\nu_i}}{v_2}c_{\hat{\beta}} A\,\overline{\nu^a}\gamma^5 \nu^a-c_{\hat{\beta}}(\sqrt{2}\, y_{\alpha\beta}^\nu \overline{\ell^\alpha_L}H^-\nu^\beta_R+\mathrm{h.c.}).
\end{align}
Let us stress that we have chosen a basis in which the charged leptons and neutrinos mass matrices are diagonal,
which implies that the interaction term with the charged scalar is non-diagonal in the flavour space. Moreover,
as we will see later the angles $\hat\alpha$ and $\hat\beta$ are small due to the large hierarchy between the 
scales $v_1$ and $v_2$. Subsequently, the Yukawa lagrangian can be written as
\begin{align}
	-\mathscr{L}_{\rm Y}^\ell&=m_a^\ell\, \overline{\ell^a}\ell^a+\frac{m_a^\ell}{v_1}h\overline{\ell^a}\ell^a\notag\\
	&\quad+m_a^\nu\overline{\nu^a} \nu^a+\frac{m_a^\nu}{v_2}H\overline{\nu^a} \nu^a-i\frac{m_{\nu_i}}{v_2} A\,\overline{\nu^a}\gamma^5 \nu^a
	-\frac{\sqrt{2}m_a^\nu}{v_2} [\widetilde{U}_{\beta a}^\ast H^+ \overline{\nu^a} P_L \ell^\beta+ \mathrm{h.c.}],
\end{align}
where we see that the charged lepton $\ell_a$ masses basically come from the doublet $\Phi_1$,
while the neutrino masses depend on the VEV of the second doublet $\Phi_2$. Let us also 
note that in the previous lagrangian appears explicitly the PMNS matrix. This can 
induce FNCN processes; such processes will be considered in the following chapter.
Now, let us consider each scenario for the symmetry introduced to protect the neutrino
masses.

\newpage

%%%%%%%%%%%%%%%%%%%%%%%%%%%%%%%%%%%%%%%%%%%%%%%%%%%%%%%%%%%%%%%%%%%%%%%%%%%%%%%%%%%%%%%%%%%%%%%%%%
\subsection{$\mathbb{Z}_2$ Symmetry Model}\label{sec:Z2}
%%%%%%%%%%%%%%%%%%%%%%%%%%%%%%%%%%%%%%%%%%%%%%%%%%%%%%%%%%%%%%%%%%%%%%%%%%%%%%%%%%%%%%%%%%%%%%%%%%

For the Gabriel-Nandi scenario \cite{Gabriel:2006ns} a discrete $\Z_2$ symmetry is
introduced in such a way that only the second doublet $\Phi_2$ and right-handed neutrinos
are odd under that symmetry while the SM particles are even. Besides, lepton-number conservation is imposed to 
make neutrinos Dirac particles. However, if such symmetry is not imposed, neutrinos can
have Majorana mass terms, given that the symmetry is a discrete one, the Majorana condition 
for neutralness is still satisfied. Anyhow, we will consider the conservation of lepton number,
maintaining the Dirac nature\footnote{In the work published by the author with collaborators \cite{Machado:2015sha},
it was not considered the lepton number conservation; thus, it was possible to write down 
Majorana mass terms. However, we found that such mass term do not affect the results.}.
Also, as previously mentioned, the $m_{12}^2$ parameter will be zero due to the symmetry,
therefore, when  $v_2/v_1\to 0$ is imposed in the angles $\hat{\alpha}$
and $\hat{\beta}$ definitions, it is clear that they must be small. This implies that the mixture
between the doublets is also small. Phenomenologically speaking, this has many consequences. The
first one is that the $\Phi_1$ doublet behaves as the SM Higgs doublet so that the neutral
scalar $h$ can be identified as the $\sim 125$ GeV scalar boson found at the LHC. The second
consequence is that the neutrinophilic scalar $H$ has a quite small mass, of order of the 
second doublet's VEV, i.e. $m_H\sim\mathcal{O}(\mathrm{eV})$. Besides, the masses of the charged
scalar and the pseudoscalar will be in the TeV scale, for values of the quartic couplings not 
very large. Such a quite specific spectrum will have problems with the electroweak precision
measurements, as we will describe in detail in the next chapter.

%%%%%%%%%%%%%%%%%%%%%%%%%%%%%%%%%%%%%%%%%%%%%%%%%%%%%%%%%%%%%%%%%%%%%%%%%%%%%%%%%%%%%%%%%%%%%%%%%%
\subsection{$U(1)$ Symmetry Model}\label{sec:U(1)}
%%%%%%%%%%%%%%%%%%%%%%%%%%%%%%%%%%%%%%%%%%%%%%%%%%%%%%%%%%%%%%%%%%%%%%%%%%%%%%%%%%%%%%%%%%%%%%%%%%

In the case of the model with U$(1)$ symmetry, described in the work of Davidson-Logan 
\cite{Davidson:2009ha}, the spontaneous symmetry breaking of that symmetry by the 
VEV $v_2$ induces the appearance of a Nambu-Goldstone boson. This is quite problematic
due to the many existing limits regarding a massless scalar coming from electroweak precision
test and cosmology. This is the reason why it is necessary to have a term which breaks explicitly the symmetry. Therefore, we will impose that $\lambda_5$ is zero, while $m_{12}^2$ is small.
From equation  \eqref{eq:mA} we can write
\begin{equation}
  m_{12}^2 = \sin\beta\cos\beta \; m_A^2
\end{equation}
so that to have $v_2\sim$ eV and a mass of the pseudoscalar of $m_A\sim\mathcal{O}$($100$ GeV)
it is necessary that $m_{12}^2\sim$ ($200$ keV)$^2$. The spectrum will be different from the $\Z_2$
model due to the small value of $m_{12}^2$. The main characteristic will be that the 
neutrinophilic scalar $H$ mass depends on $m_{12}^2$ through the $M$ parameter, see 
\eqref{eq:mH}, which is not be necessary small. On the other hand, it is possible to show that
the quartic coupling $\lambda_2$, given by the expression \eqref{eq:lambda2} in the case
of $v_2\ll v_1$, is
\begin{align}
\label{eq:lambda2v2}
\lambda_2  &=\frac{1}{v^2}\left(-\cot^2\hat{\beta} M^2+\frac{\cos^2\hat{\alpha}}{\sin^2\hat{\beta}} m_H^2+\frac{\sin^2\hat{\alpha}}{\sin^2\hat{\beta}}m_h^2\right) \simeq \frac{1}{v_2^2}\left(m_H^2-m_{12}^2 \frac{v}{v_2}\right)+\frac{\sin^2\hat{\alpha}}{\sin^2\hat{\beta}}\frac{m_h^2}{v^2}.
\end{align}
This means that
\begin{equation}
  |m_H^2-m_{12}^2 v/v_2|\lesssim \O(v_2^2).
\end{equation}
Therefore, if we assume that $m_H^2=m_{12}^2 v/v_2$, we see that the $H$ scalar
and the $A$ pseudoscalar are degenerated in mass, see equations \eqref{eq:mH} and \eqref{eq:mA}.
This is a consequence of the large hierarchy between the VEV's and the symmetry imposed to the 
potential. Therefore, we find that the scalar spectra of the two models are quite different which
will make our theoretical and phenomenological limits also distinct. Regarding the
neutrino nature, let us note that the U$(1)$ symmetry is a continuous one, preventing 
the development of a Majorana mass term, since the right-handed neutrinos need to be charged 
under this symmetry. Neutrinos will be Dirac particles.\\

We find then that it is possible to construct models for Dirac neutrinos having naturally small masses
although there is not a reason for avoiding small Yukawas in the minimal SM extension.
Accordingly, we have introduced the neutrinophilic $2$HDMs by introducing a second doublet to the 
set of SM fields. Let us stress here that this $2$HDM scenario is different from those considered
in the literature. Usually it is considered that the second VEV is of the same order, larger
than the Electroweak one, or even sometimes is zero \citep{Branco:2011iw}. In the present case, it is supposed
by hypothesis that the VEV of the new doublet is of the same order of the neutrino masses. Furthermore,
a new symmetry is needed in order to forbid the coupling of the first doublet, having a VEV similar to the
SM one, with the neutrino. There are two simple cases, a $\Z_2$ and U$(1)$ scenarios. We have discused in detail
the properties of the scalar potential, stressing the differences in the two models we are interested.
Now, having established the models, we will consider in the next chapter the theoretical and experimental
constraints of the neutrinophilic $2$HDM; such constraints are imposed from Electroweak precision 
measurements, $Z^0$ and Higgs decays, and flavour physics.

%\newpage
    	    
		\part{Phenomenology}
    
			%%%%%%%%%%%%%%%%%%%%%%%%%%%%%%%%%%%%%%%%%%%%%%%%%%%%%%%%%%%%%%%%%%%%%%%%%%%%%%%%%%%%%%%%%%%%%%%%%
%%%%%%%%%%%%%%%%%%%%%%%%%%%%%%%%%%%%%%%%%%%%%%%%%%%%%%%%%%%%%%%%%%%%%%%%%%%%%%%%%%%%%%%%%%%%%%%%%%
%%%%%%%%%%%%%%%%%%%%%%%%%%%%%%%%%%%%%%%%%%%%%%%%%%%%%%%%%%%%%%%%%%%%%%%%%%%%%%%%%%%%%%%%%%%%%%%%%%
\chapter[Constraints on Neutrinophilic two-Higgs-doublets Models]{Constraints on Neutrinophilic two-Higgs-doublets Models}
\label{nu-2HDM-feno}
\chaptermark{Constraints on Neutrinophilic $2$HDM}

\lettrine{S}{ince} the experimental discovery of a scalar boson compatible with the SM Higgs by the ATLAS 
and CMS collaborations, many theories have been under scrutiny given the updated information. The SM itself
has been analysed under the light of the new boson discovery. At first sight, such discovery would not have
a crucial impact on our purpose of understanding neutrino's nature and its relationship with neutrino masses. 
However, as we have seen, in the Dirac scenario
the structure of the Higgs potential can teach us about the existence of further scalar doublets. Other ways
have been proposed some time ago to constrain the existence of beyond SM physics through direct measurements
or global fits of the model. We can divide them in two basic categories, theoretical and phenomenological bounds. 
Theoretical limits are related to the properties that any model needs to fulfil to avoid contradictions in some
computations, while the phenomenological bounds come from experimental data that the model has to agree with.
We will consider in this chapter the constraints on neutrinophilic $2$HDM that are imposed on two different sectors,
the scalar potential and the flavour leptonic sector; in the scalar sector theoretical bounds are related with
the properties of the potential while phenomenological limits are related with unobserved decays and precision
measurements in both scalar and flavour sectors. The results presented in this chapter constitute an original 
contribution, and have been published in \cite{Machado:2015sha} and \cite{Bertuzzo:2015ada}.

\newpage

%%%%%%%%%%%%%%%%%%%%%%%%%%%%%%%%%%%%%%%%%%%%%%%%%%%%%%%%%%%%%%%%%%%%%%%%%%%%%%%%%%%%%%%%%%%%%%%%%%
\section{Theoretical and Phenomenological Constraints on the Scalar Potential}\label{sec:TEC-SE}
%%%%%%%%%%%%%%%%%%%%%%%%%%%%%%%%%%%%%%%%%%%%%%%%%%%%%%%%%%%%%%%%%%%%%%%%%%%%%%%%%%%%%%%%%%%%%%%%%%

We will first consider theoretical and phenomenological constraints imposed on the scalar sector of the models.
Theoretical limits will allow to reduced the models parameter space, in such a way that phenomenological limits
coming from LEP and LHC experiments will be only applied to the theoretically allowed parameter space. We will 
present first the limits in a general way applying them to our models of interest in a further section.
\subsection{Theoretical Constraints}
Our scalar potential needs to obey certain constrains, such as it should be stable and unitarity should be respected. This
will impose some specific inequalities that should be respected by the $\lambda_i$ couplings, appearing in the potential.
Let us examine these constraints. First, it is necessary that the potential is stable at tree level, i.e. the scalar potential
can not take large negative values for large values of the couplings. To do this, we must consider only the quartic terms in the 
potential. Proceeding as in \cite{Gunion:2002zf} and defining $a\defm \Phi_1^\dagger\Phi_1$, $b\defm \Phi_2^\dagger\Phi_2$, $c\defm\mathfrak{Re}\,\Phi_1^\dagger \Phi_2$ e $d=\mathfrak{Im}\,\Phi_1^\dagger \Phi_2$, we have that 
\begin{align*}
	V_4&=\frac{1}{2}\corc{\sqrt{\lambda_1}\,a-\sqrt{\lambda_2}\,b}^2+\corc{\lambda_3+\sqrt{\lambda_1\lambda_2}}(ab-c^2-d^2)+2\corc{\lambda_3+\lambda_4+\sqrt{\lambda_1\lambda_2}}c^2\notag\\
	&\quad+\corc{\mathfrak{Re}\,\lambda_5-\lambda_3-\lambda_4-\sqrt{\lambda_1\lambda_2}}(c^2-d^2)-2cd\,\mathfrak{Im}\,\lambda_5.
\end{align*}
To determine the conditions that the quartic couplings should obey, we impose that there should not be directions in
the parameter space in which $V_4\to-\infty$. For the direction in which $c=d=0$, we obtain
\begin{align}\label{eq:estab1}
\lambda_{1,2}>0,\qquad{}\lambda_3>-(\lambda_1\lambda_2)^{1/2}
\end{align}
while that considering the direction in which$\sqrt{\lambda_1}\,a=\sqrt{\lambda_2}\,b$ and $ab=c^2+d^2$ and putting $c=\xi d$,
we obtain a quadratic inequality for the parameter $\xi$. This implies an inequality for the couplings, given by \cite{Gunion:2002zf}
\begin{align}\label{eq:estab2}
\lambda_3+\lambda_4-|\lambda_5|>-(\lambda_1 \lambda_2)^{1/2}.
\end{align}
The conditions \eqref{eq:estab1} and \eqref{eq:estab2} are the constraints that we will impose on the parameter space. On the other 
hand, perturbative unitarity must be respected at tree level; in other words, at higher energies, scalar-scalar
scatterings should not violate unitarity \cite{Kanemura:1993hm,Arhrib:2000is}. 
We will see how these constraints are obtained \cite{Arhrib:2000is}.
The total cross-section for the process $S_1\ S_2 \to S_3\ S_4$ is given by \cite{Arhrib:2000is}
\begin{align}
	\sigma=\frac{16\pi}{s}\sum_{l=0}^\infty (2l+1)\abs{a_l}^2.
\end{align}
where $s$ is the Mandelstam variables, $a_l(s)$ is a partial $l$ spin wave. Using the optical theorem, 
we obtain the condition for unitarity
\begin{align}
	\abs{a_l}^2=\Re\, (a_l)^2+\Im\,(a_l)^2=\Im\,(a_l),\qquad\quad\text{for all}\qquad l.
\end{align}
Graphically, the previous equation represents a circle in the plane $(\Re\ a_l, \Im\  a_l)$ with radius $\frac{1}{2}$
and centred in $(0,\frac{1}{2})$. Therefore, it is necessary that \cite{Arhrib:2000is}
\begin{align}
	\abs{\Re(a_l)}<\frac{1}{2},\qquad\qquad\text{for all}\quad l.
\end{align}
Inverting the partial wave $a_l(s)$ in terms of the amplitude, and considering only the $s$ wave for $J=0$, one finds \cite{Arhrib:2000is}
\begin{align}
	a_0(s)=\frac{1}{16\pi}\llav{Q+T_h^{12}T_h^{34}\frac{1}{s-m_h^2}-\frac{1}{s}(c_t\,T_h^{13}T_h^{24}+c_u\,T_h^{14}T_h^{23})\ln\corc{1+\frac{s}{m_h^2}}},
\end{align}
where $Q=S_1S_2S_3S_4$  is the four-point vertex and $T_h^{ij}$ is the trilinear vertex of the interaction $hS_iS_j$,
$c_{t(u)}=1,0$ for processes with and without the t-channel, respectively. As we see here, the first $s$ wave contribution
comes from the quartic coupling and the other contributions come from diagrams in which there is an exchange of a third 
particle $h$ in the $s,t,u$ channels (second and third terms) \cite{Arhrib:2000is}. For high energy collisions the 
dominant term is the quartic interaction term since the other terms are suppressed by energy; thus, 
the unitarity limit should be applied to such term. Specifically, if $\abs{a_0}\le 1/2$, the quartic term should obey 
$\abs{Q(S_1S_2S_3S_4)}\le 8\pi$ \cite{Arhrib:2000is}.\\

To apply the previous results to a specific model, we should consider all possible scalar-scalar, gauge boson-gauge boson, 
scalar-gauge boson scattering process provided that they should satisfy $\abs{a_0}\le 1/2$ in the high energy limit.
Nonetheless, in such limit, we can use the {\it equivalence theorem} \cite{Cornwall:1973tb}. Such theorem establishes that
a scattering amplitude containing longitudinal gauge bosons is well approximated by the amplitude obtained substituting the gauge
bosons with their corresponding Nambu-Goldstone bosons. Therefore, we have that the unitarity limit is 
enforced studying only scalar scatterings \cite{Arhrib:2000is}. For the specific case of a $2$HDM, and taking into account all 
possible scattering processes, we have to deal with a $22\times 22$ amplitude matrix composed by four sub-matrices $\M_1(6\times 
6)$, $\M_2(6\times 6)$, $\M_3(6\times 6)$ and $\M_4(8\times 8)$, whose elements are the quartic couplings of the processes 
\cite{Arhrib:2000is}.
\newpage
\noindent Determining the eigenvalues of such matrix, one obtains the unitarity constraints
\begin{equation}\label{eq:unitarity}
	|a_{\pm}|,|b_{\pm}|,|c_{\pm}|,|f_{\pm}|,|e_{1,2}|,|f_1|,|p_1|\le  8 \pi,
\end{equation}
where
\begin{subequations}
\begin{align}
a_\pm &= \frac{3}{2}(\lambda_1+\lambda_2)\pm \sqrt{\frac{9}{4}(\lambda_1-\lambda_2)^2+(2\lambda_3+\lambda_4)^2}, \\
b_\pm &= \frac{1}{2}(\lambda_1+\lambda_2)\pm \frac{1}{2}\sqrt{(\lambda_1-\lambda_2)^2+4\lambda_4^2}, \\
c_\pm &= \frac{1}{2}(\lambda_1+\lambda_2)\pm \frac{1}{2}\sqrt{(\lambda_1-\lambda_2)^2+4\lambda_5^2}, \\
f_+ &= \lambda_3 + 2\lambda_4 + 3 \lambda_5, \\
f_- &= \lambda_3 + \lambda_5, \\
e_1 &= \lambda_3 + 2\lambda_4 - 3 \lambda_5, \\
e_2 &= \lambda_3- \lambda_5, \\
f_1 &= \lambda_3 + \lambda_4, \\
p_1 &= \lambda_3 - \lambda_4. 
\end{align}
\end{subequations}
Subsequently, the parameter space must obey the stability conditions, equations \eqref{eq:estab1} and \eqref{eq:estab2}
as well as the unitarity conditions, equations \eqref{eq:unitarity}. Therefore, as already mentioned, we will scan
the parameter space and select the points which obey these theoretical conditions; then, we will apply the phenomenological
constraints.

\subsection{Electroweak Oblique Parameters}

The oblique parameters codify the impact of beyond SM physics in the electroweak precision measurements realized by LEP I and II
experiments. These parameters were introduced by Peskin and Takeuchi \cite{Peskin:1990zt,Peskin:1991sw}, and they are based in three
suppositions related to the new physics: i) the new physics does not change the gauge group SU$(2)_L\times$U$(1)_Y$; therefore, 
there are no new gauge bosons beyond $W^\pm$, $Z$ and $\gamma$; ii) the couplings of the new physics with the lightest fermions
are suppressed compared to the gauge boson couplings; iii) the intrinsic scale of the new interaction is larger than the Electroweak
scale, i.e., compared to $m_W$ and $m_Z$. Thus, if the previous hypothesis are obeyed, the impact of the new physics will appear 
through the contributions to the gauge bosons self-energies.
\begin{align*}
	\Pi_{VW}^{\mu\nu}(q)=\Pi_{VW}(q^2)g^{\mu\nu}+(\text{terms}\ q^\mu q^\nu ),
\end{align*}
with $V,W=\{W^\pm,Z,\gamma\}$. A new contribution is added to the SM term as
\begin{align*}
	\Pi_{VW}(q^2)=\Pi_{VW}^{\rm SM}(q^2)+\delta\Pi_{VW}(q^2). 
\end{align*}
Peskin and Takeuchi supposed that it is possible to do a Taylor expansion in the new contributions when the new physics exists
at a higher energy scale
\begin{align}
	\delta\Pi_{VW}(q^2)=\delta\Pi_{VW}(0)+\underbrace{\left.\der{\delta\Pi_{VW}}{q^2}\right|_{q^2=0}}_{\delta\Pi_{VW}'(0)}q^2 ,
\end{align}
being $\delta\Pi_{VW}'(0)$ the derivative with $q^2$ evaluated at $q^2=0$. With this assumption, the number of independent 
parameters is simple to obtain. In principle, there are eight quantities which describe the new physics, since $(V,W)={(\gamma,
\gamma),(\gamma,Z),(Z,Z),(W,W)}$. However, $\delta\Pi_{\gamma\gamma}(0)$ and $\delta\Pi_{\gamma Z}(0)$ are zero due to the gauge 
invariance. On the other hand,three combinations of the six remaining quantities can be eliminated when the parameters $\alpha_{\rm 
EM}, m_Z, G_F$ are renormalized. Therefore, all the effect of the new physics are codified in three combinations of $\delta 
\Pi_{VW}$\footnote{We should take into account that, if the new physics scale assumptions is disregarded, there will be six 
independent parameters, denominated $S,T,U,V,W,X$. See, for instance, \cite{Maksymyk:1993zm}}. These combinations are denominated 
$S,T,U$ parameters, given by \cite{Haber:2010bw}
\begin{subequations}
\begin{align}
	\frac{\alpha_{\rm EM}\, S}{4 s_W^2 c_W^2}&\defm\delta\Pi_{ZZ}'(0)-\frac{c_W^2-s_W^2}{s_Wc_W}\delta\Pi_{Z\gamma}'(0)-\delta\Pi_{\gamma\gamma}'(0),\\
	\alpha_{\rm EM}\, T&\defm \frac{\delta\Pi_{WW}(0)}{m_W^2}-\frac{\delta\Pi_{ZZ}(0)}{m_Z^2},\\
	\frac{\alpha_{\rm EM}\, U}{4 s_W^2}&\defm \delta\Pi_{WW}'(0)-c_W^2\delta\Pi_{ZZ}'(0)-2s_Wc_W\delta\Pi_{Z\gamma}'(0)-s_W^2\delta\Pi_{\gamma\gamma}'(0),
\end{align}
\end{subequations}
where $s_W=\sin\theta_W$ and $c_W=\cos\theta_W$. The physical interpretation of theses parameters is given next. 
The $S$ parameter codifies the running of the two-point functions of the neutral gauge bosons between zero momentum and 
the $Z$ pole. Therefore, this parameter will be very sensitive to the presence of new physics in scales lower than the $Z$
mass. The $T$ parameter is sensitive to the custodial symmetry breaking at zero momentum, i.e., to the difference between the two-
point functions of the $W^\pm$ and $Z$ bosons; thus, this parameter measures the isospin violation. The $U$ parameter, or the $S+U$
combination, determines the presence of new light charged scalars in the radiative corrections.\\

For the specific case of the neutrinophilic $2$HDM, we will use the expressions for a general $2$HDM obtained in 
\cite{Haber:2010bw}. Let us stress here that the assumption for the new physics scale is not valid for the case that
we are considering since the scale of the second VEV is smaller than the electroweak one. Thus, we should use the full set
$S,T,U,V,W,X$. Nevertheless, for simplicity, we will consider only the $S,T,U$ parameters, which will give us  
strong limits in both models. We can expect a general behaviour for these parameters. The $S$ parameter will constrain strongly the 
models, due to the presence of the light scalar in the case of the $\Z_2$ symmetry model. Meanwhile, the difference between the 
masses of the charged scalar and the pseudoscalar will be limited by the $T$ parameter given that those particles belong to the same 
doublet; thus, any large mass difference will imply an isospin breaking. The $U$ parameter will not be very important since direct 
measurements put already a limit of $\sim100$ GeV on the charged scalars mass \cite{Achard:2001qw}. Therefore, the $U$ parameter 
will not be affected.\\

Experimentally, the oblique parameters are obtained fitting the electroweak precision measurements, specially the LEP data regarding
measurements at the $Z$ pole and the Higgs mass, determined by the LHC. Computing the relative values compared to the SM 
predictions, the GFITTER group obtained the best fit results, uncertainties and the covariance matrix given by  \cite{Baak:2014ora}
\begin{equation}
  \begin{aligned}
    & \Delta S^{SM} = 0.05\pm 0.11,\\
    & \Delta T^{SM} = 0.09\pm 0.13,\\
    & \Delta U^{SM} = 0.01\pm 0.11,\\
  \end{aligned}
\qquad\qquad
V = \left(\begin{array}{ccc}
1 & 0.90 & -0.59\\
0.90 & 1 & -0.83\\
-0.59 & -0.83 & 1
\end{array}\right). 
\end{equation}
Hence, to determine the impact of electroweak precision measurements in the models, we construct the $\chi^2$ function
\begin{equation}
  \chi^2= \sum_{i,j}(X_i - X_i^{\rm SM})(\sigma^2)_{ij}^{-1}(X_j - X_j^{\rm SM}),
\end{equation}
with $X_i=\{\Delta S, \Delta T, \Delta U\}$ and the covariance matrix $\sigma^2_{ij}\equiv\sigma_iV_{ij}\sigma_j$ where $(\sigma_1,\sigma_2,\sigma_3)=(0.11,0.13,0.11)$. We compute then the goodness of the fit for each model codified through the allowed regions at
$1\sigma$, $2\sigma$ e $3\sigma$ levels, which correspond to $\chi^2 = 3.5,8.0,14.2$ for two degrees of freedom, respectively. Thus, concerning the oblique parameters, we will study the behaviour of the parameter space's regions allowed theoretically, and we will establish the viability of each models.
\subsection{Higgs Invisible Width}
When we analise in detail the scalar potential after the mass matrices diagonalization, it is clear that
triple couplings among the scalars are generated. Specifically, in the case in which the neutral scalar
masses are smaller than the mass of the scalar behaving as the SM Higgs, that we will denominate hereafter
simply as Higgs boson, $2m_\mathcal{S}<m_h$, ($\mathcal{S}=H,A$), we can have decays $h\to \mathcal{S}\mathcal{S}$
with $\S$ decaying later as $\mathcal{S}\to \bar{\nu}\nu$. Now, keeping in mind that the Higgs couplings are practically unaltered
by the smallness of $\tan\hat{\beta}$, the only important modification to the Higgs branching ratios studied at the LHC will be 
the addition of an invisible channel. The combination of ATLAS+CMS we considered gives BR$(h\to\mathrm{invisible})<0.13$ at $95\%$ 
CL of the invisible branching ratio \cite{Ellis:2013lra}. The decay width in our case is given by \cite{Bernon:2014nxa}
\begin{align}
&\Gamma(h\to \S \S) = \frac{g_{h\S\S}^2}{32 \pi m_h} \sqrt{1-\frac{4 m_\S^2}{m_h^2}}, 
\qquad
\end{align}
with couplings
\begin{align}
&g_{hAA}=\frac{1}{2v}\Bigg{[}(2 m_A^2-m_h^2)\frac{\sin(\hat{\alpha}-3\hat{\beta})}{\sin 2\hat{\beta}}
\nonumber\\
&\qquad\qquad\qquad +(8m_{12}^2-\sin 2\hat{\beta}(2m_A^2+3 m_h^2)) 
  \frac{\sin(\hat{\beta}+\hat{\alpha})}{\sin^2 2\hat{\beta}}\Bigg{]},\label{eq:HAA}\\
&g_{hHH}=-\frac{1}{v} \cos(\hat{\beta}-\hat{\alpha})\Bigg[\frac{2 m_{12}^2}{\sin 2\hat{\beta}} 
  +\left(2 m_H^2+m_h^2-\frac{6 m_{12}^2}{\sin 2\hat{\beta}}\right)
  \frac{\sin 2\hat{\alpha}}{\sin 2\hat{\beta}}\Bigg].\label{eq:Hhh}
\end{align}
We should address here that these trilinear couplings can be large, $g_{h\S\S}\sim m_h^2/v\sim$ 60 GeV.
Moreover, they can have a phenomenological important impact since the SM total width is small, approximately
$4.07$ MeV \cite{Denner:2011mq}.
\subsection{$Z^0$ Invisible Width}
Other important contribution to the precision measurements in this model is related to the $Z$ boson properties,
specially to the invisible width. The neutrinophilic $2$HDM contribute with such width through the decays
$Z\to\S\bar{\nu}\nu$, $\mathcal{S}=H,A$. Let us notice that the two scenarios contribute differently to the
width given that in the U$(1)$ softy broken symmetry both $H$ and $A$ can have masses smaller than the $Z$ mass.
In that case, the width will be the superposition $\Gamma(Z\to \S \nu\bar{\nu})=\Gamma(Z\to H \nu\bar{\nu})+\Gamma(Z\to A \nu\bar{\nu})$. For the $\Z_2$ model, only $H$ will contribute to the decay. Computing the width for a scalar $\S$ with
mass $m_\S$, we obtain
\begin{align}
\label{eq:zinv}
  \Gamma(Z\to \S \nu\bar{\nu}) &= 
       \dfrac{1}{384 \pi^3 m_Z^5} \left(\frac{g}{2 \cos\theta_W}\right)^2 \frac{m^2_{\nu,\mathrm{tot}}}{v_2^2} \int_0^{(m_Z-m_\S)^2}\mathrm{d}q^2\, \frac{\lambda^{1/2}(q^2,m_Z^2,m_\S^2)}{(q^2-m_\S^2)^2+m_\S^2 \Gamma_\S^2}\notag\\
       &\times\Bigg{[}g_\S(q^2)+\frac{f_\S(q^2)}{\lambda^{1/2}(q^2,m_\S^2,m_Z^2)}\coth^{-1}\left(\frac{m_Z^2+m_\S^2-q^2}{\lambda^{1/2}(q^2,m_Z^2,m_\S^2)}\right) \Bigg{]},
\end{align}
 where the total scalar decay width $\Gamma_\S$ is simply -- $m_{\nu,\mathrm{tot}}^2=\sum m_{\nu_i}^2$ is the neutrino mass 
squared sum --
\begin{equation}
\Gamma_\S = \frac{m_\S}{8\pi}\frac{m_{\nu,\mathrm{tot}}^2}{v_2^2}.
\end{equation}
\newpage
\noindent In the width \eqref{eq:zinv} it was defined the phase space function $\lambda(a^2,b^2,c^2)=(a^2-(b-c)^2)(a^2-(b+c)^2)$,
and the functions $f_\S(q^2)$, $g_\S(q^2)$ are
\begin{align}
	f_\S(q^2) &= 4 m_Z^2 \left[ (m_\S^2-q^2)(m_\S^4 - m_Z^2 q^2+ q^4+m_\S^2(m_Z^2-4 q^2))+\Gamma_\S^2 m_\S^2 (m_\S^2+m_Z^2-q^2)\right],\\
	g_\S(q^2) &= 4 m_\S^4(q^2-m_Z^2)+m_\S^2[4 m_Z^2(2q^2-\Gamma_\S^2)+q^2(\Gamma_\S^2-8 q^2)]+q^2(m_Z^4-8m_Z^2 q^2+4q^4).
\end{align}
As experimental constraint we will use the LEP results, $\Gamma^{\mathrm{exp}}(Z\to\mathrm{invisible})=499.0(15)$ MeV \cite{Agashe:2014kda},
and the SM prediction $\Gamma^{\mathrm{SM}}(Z\to\mathrm{invisible})=501.69(6)$ MeV, \cite{Agashe:2014kda}. Then, the limit
in the invisible width coming from new physics is $\Gamma^{NP}(Z\to\mathrm{invisible})<1.8$ MeV at $3\sigma$ level. For the case
of the $\Z_2$ model, the width expression contains an infrared divergence, given that $m_H\ll m_Z$; such divergence is
cured by considering the $1$-loop contributions coming from the $Z\to\bar{\nu} \nu$ decay.

%%%%%%%%%%%%%%%%%%%%%%%%%%%%%%%%%%%%%%%%%%%%%%%%%%%%%%%%%%%%%%%%%%%%%%%%%%%%%%%%%%%%%%%%%%%%%%%%%%
\subsection{Results}
%%%%%%%%%%%%%%%%%%%%%%%%%%%%%%%%%%%%%%%%%%%%%%%%%%%%%%%%%%%%%%%%%%%%%%%%%%%%%%%%%%%%%%%%%%%%%%%%%%

Using the theoretical and phenomenological constraints from last section, we will study the parameter space
of the $\Z_2$ e $U(1)$ models. Such parameter space is the physical parameter space, i.e the parameter
space of scalar masses and mixing angles. The quartic couplings can be expressed in terms of these physical
parameters, equations \eqref{eq:lambda1} - \eqref{eq:lambda5}. We performed an scan with approximately $10^7$ points. 
For each point we applied the theoretical constraints first to determine if such point is viable; then, if it is viable,
we considered the phenomenological limits. In the plots of this section we will present the phenomenological
constraints to the viable points in different planes of the parameter space.
\subsubsection{$\Z_2$ Symmetry Model}

\begin{figure}[t]
  \begin{center}
    \includegraphics[scale=0.9]{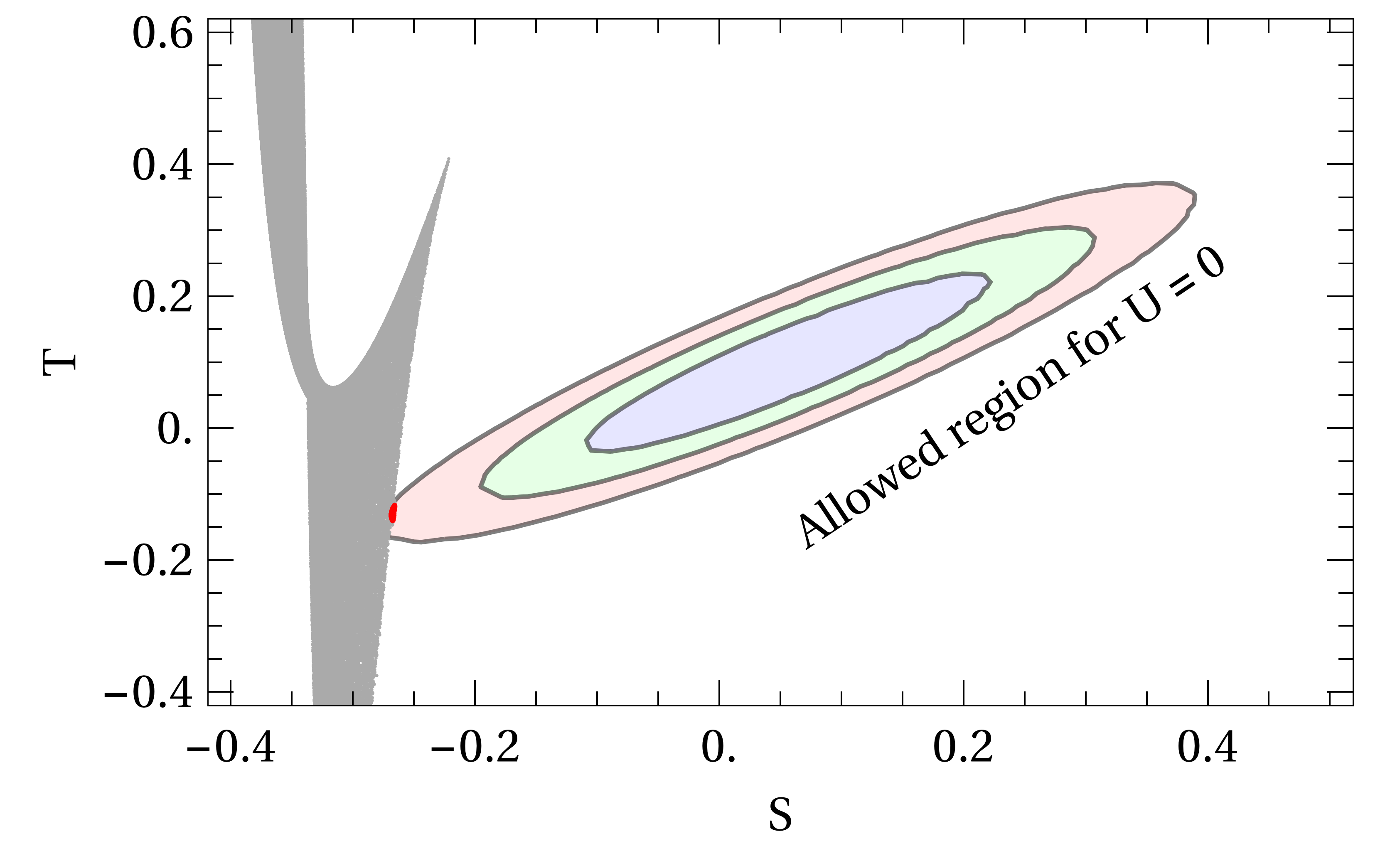}  
  \end{center}
  \caption{Constraints to the $\Z_2$ model coming from Electroweak precision measurements, expressed in the $S\times T$ plane.
  Red points are allowed by Peskin-Takeuchi parameters at $3\sigma$ level, while the gray points are excluded at $3\sigma$ level
  or more.}
  \label{fig:gabriel-nandi-TH}
\end{figure}
For the $\Z_2$ symmetry model we performed the scan over the parameter space imposing the following relations,
\begin{align*}
  0.01~{\rm eV} < &m_H < 1~\GeV, \\
  124.85~\GeV < &m_h < 125.33~\GeV, \\
  70~\GeV < &m_{H^\pm}< 1~\TeV,\\
  1~\GeV < &m_A < 1~\TeV,\\
  0.01~{\rm eV}<&v_2<1~\MeV.
\end{align*}
where the value of the Higgs mass $m_h$ was taken from the combined results of ATLAS+CMS experiments \cite{Aad:2015zhl} 
and $-\pi/2<\alpha<\pi/2$.

\newpage

The scan results, presented in figure \ref{fig:gabriel-nandi-TH}, show the large tension that this model possesses.
In the same figure, we see the theoretically allowed points in gray, in the $S\times T$ plane, and the 
allowed regions at 1$\sigma$ (blue region), 2$\sigma$ (green region), 3$\sigma$ (red region) levels. Basically,
we find that Electroweak precision measurements exclude this model since practically all gray points are out of the
experimentally allowed region. This is due to the presence of a quite light scalar $H$, $m_H\ll\mathcal{O}(\mathrm{GeV})$,
which contributes negatively to the $S$ parameter. The $T$ parameter imposes certain degeneracy between $m_A$ and $m_{H^\pm}$, $m_A\approx m_{H^\pm}$, as explained before. This fact creates a displacement to negative values of $S$. It was found an
allowed region at $3\sigma$. However, when we consider a stronger perturbativity limit of $4\pi$, this region
disappears. On the other hand, keeping in mind that these results were obtained considering the spectrum at
tree level, we can ask ourselves if loop effects can modify significantly the mass of the scalar $H$, in such
a way that constraints are avoided. In a general fashion, we have that the charged scalar and pseudoscalar masses are not
affected by loop corrections while, for the CP even scalars, we have that the mass matrix can receive corrections
of the form \cite{Lee:2012jn},
\begin{align}
M_\rho = 
\begin{pmatrix}
\lambda_1 v_1^2 & \lambda_{345} v_1 v_2\\ 
\lambda_{345} v_1 v_2 &  \lambda_2 v_2^2 
\end{pmatrix}+\frac{1}{64\pi^2}
\begin{pmatrix}
\Delta m_{11}^2 v_1^2 & \Delta m_{12}^2 v_1 v_2 \\
\Delta m_{12}^2 v_1 v_2  & \Delta m_{22}^2 v_2^2 
\end{pmatrix}.
\end{align}
The second term comes from the effective potential, being $\Delta m_{ij}$ functions on masses and quartic couplings.
Therefore, we see that the mass matrix structure is preserved by loop corrections; this implies that if $v_2$ is small,
$m_H$ will always be light. We checked numerically this, finding that the parameter $S$ still presents problem, despite the 
mass of the scalar $H$ being increased by a factor of $100$.\\

\begin{figure}[t]
  \begin{center}
    \includegraphics[scale=0.7]{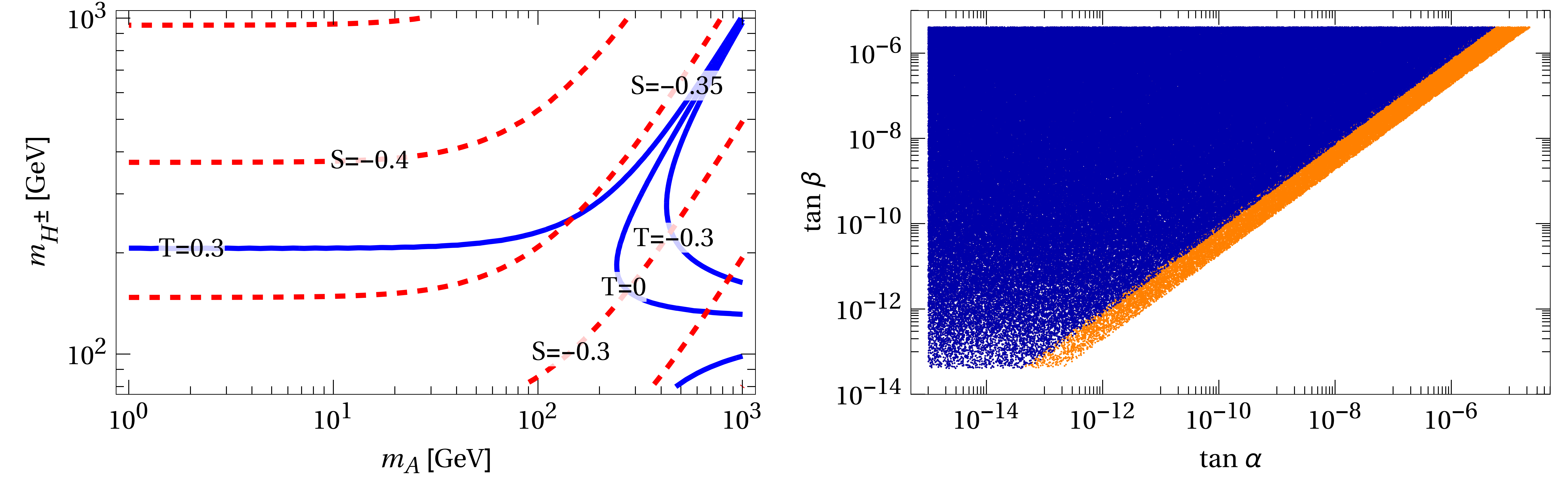}
  \end{center}
  \caption{$\Z_2$ symmetry model. Left: Values of the parameters $S$ and $T$ as a function of $m_A$ , $m_{H^\pm}$.
  Right: Exclusions in the plane $\tan\hat{\alpha}\times\tan\hat{\beta}$ coming from the invisible Higgs width. Orange points are 
  excluded, while blue ones are allowed.}
  \label{fig:gabriel-nandi-TH-2}
\end{figure}

In figure \ref{fig:gabriel-nandi-TH-2} we present the allowed region by the Higgs invisible decay (right panel) in the plane 
$\tan\hat{\alpha}\times\tan\hat{\beta}$. Given that $A$ is, in general, heavier than the Higgs, we have that only the $h\to HH$ decay will be allowed. For $\hat{\alpha}$ and $\hat{\beta}$ small, the $g_{hHH}$ coupling is approximately
\begin{equation}
  g_{h H H} \approx -\frac{m_h^2}{v} \dfrac{\sin(2\hat{\alpha})}{\sin(2\hat{\beta})}.
\end{equation}
This shows that $g_{hHH}$ will have a value compatible with the experimental limit when $\alpha \gtrsim \beta$.
This fact explains the region excluded in figure \ref{fig:gabriel-nandi-TH-2}. Anyhow, as already mentioned, 
the electroweak precision measurements exclude this model, and any other phenomenological bound will be weaker is 
comparison. For this, we do not present the results for the $Z$ invisible decay width.
\subsubsection{$U(1)$ Symmetry Model}
In the $U(1)$ global symmetry scenario, the scan needs to take into account that the $m_{12}^2$ parameter is 
non-zero. Thus, the mass of the neutrinophilic scalar $H$ can have a larger value. Besides, as showed in the
previous chapter, the masses of the pseudoscalar $A$ and the neutral scalar $H$ are degenerated at first order
in $v_2$; this was considered in the scan. We scanned over the region
\begin{align*}
  10~\GeV < &\,m_H < 1~\TeV, \\   
  124.85~\GeV <&\, m_h < 125.33~\GeV, \\
  70~\GeV < &\,m_{H^\pm}< 1~\TeV,\\
   -\pi/2<&\,\alpha<\pi/2\\
  0.01~{\rm eV}<&\,v_2<1~\MeV,
\end{align*}
\begin{figure}[t]
  \begin{center}
    \includegraphics[scale=0.6]{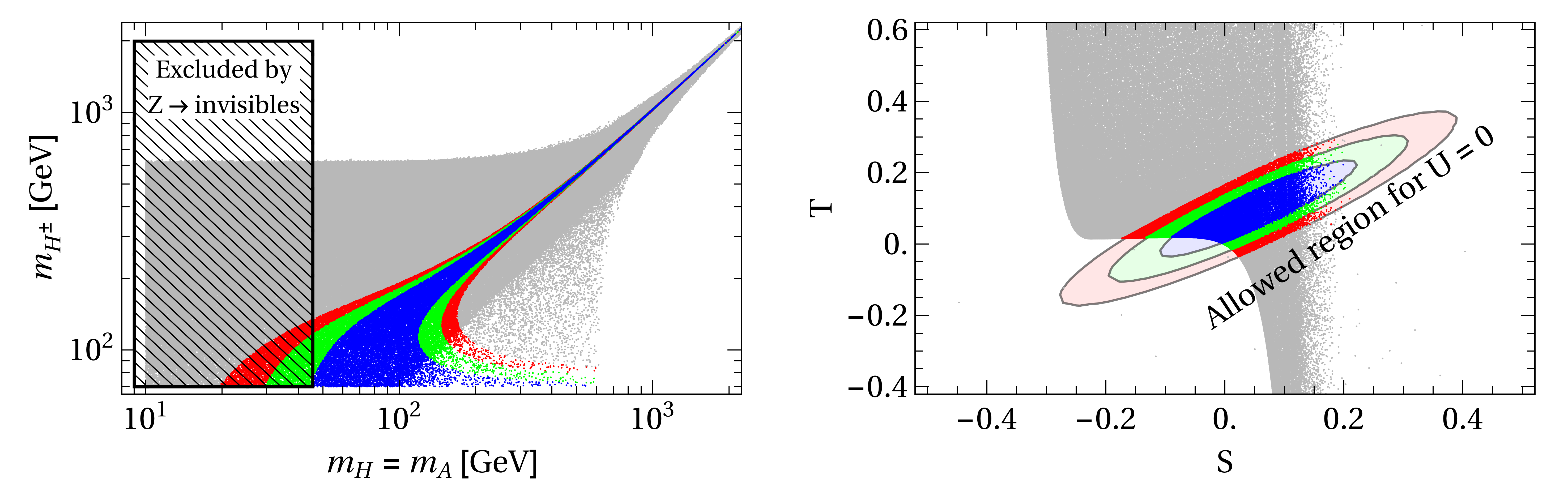}
    \includegraphics[scale=0.65]{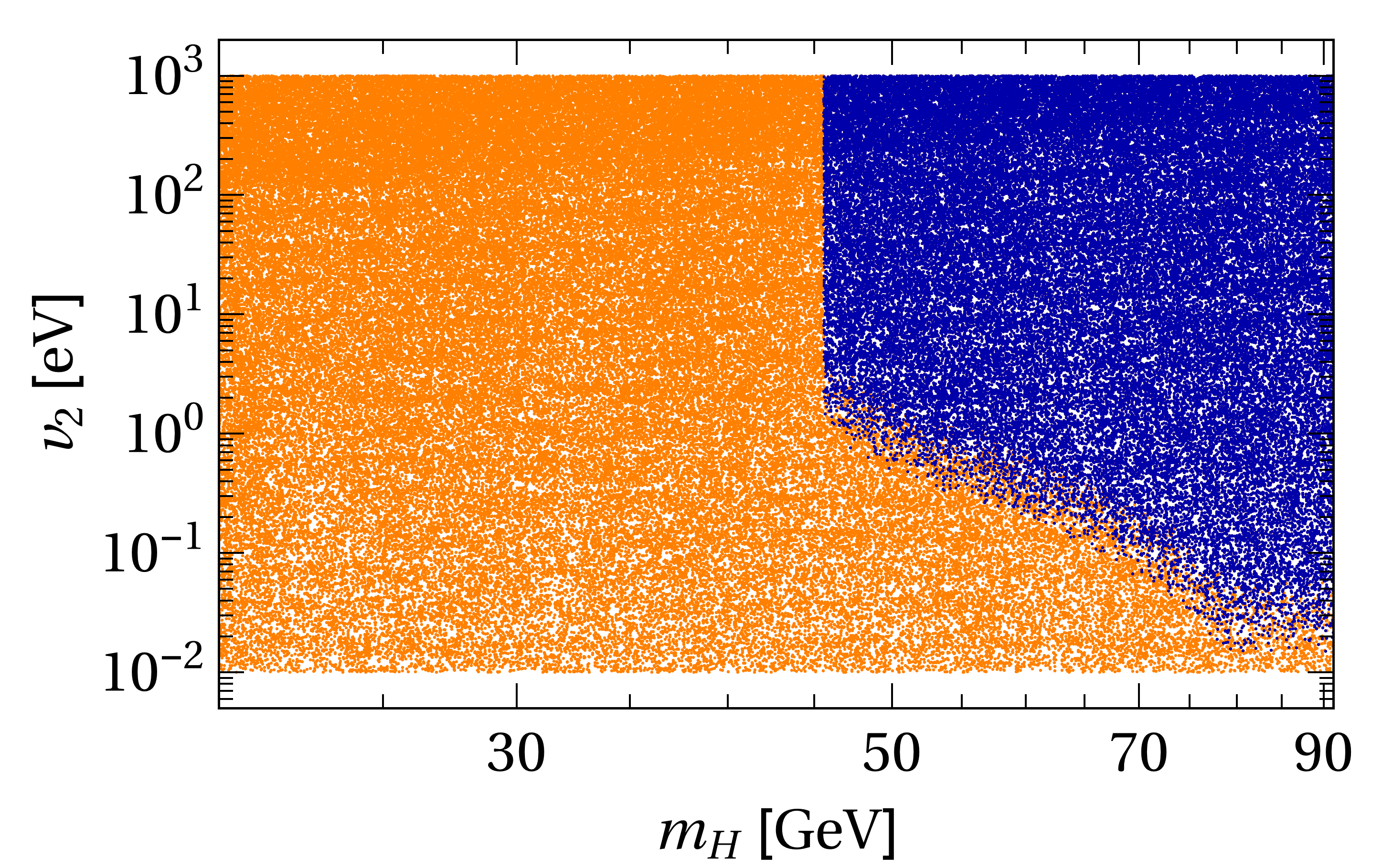}
  \end{center}
  \caption{$U(1)$ global symmetry softly broken model. The blue, green and red regions are allowed at $1\sigma$, $2\sigma$, e 	
  $3\sigma$ level, respectively, while gray points are excluded at $3\sigma$ level or more. Top left: $m_H \times m_{H^\pm}$
  plane with the points satisfying electroweak bounds, besides theoretical constraints. Top right: Projection of those points
  in the $S\times T$ plane. Bottom: Region in the $m_H\times v_2$ plane excluded by the $Z$ invisible width (orange region).}
  \label{fig:davidson-logan-TH}
\end{figure}
\negthickspace We present the results in figure \ref{fig:davidson-logan-TH}. We see here that, in the $S\times T$ plane, there is a
region which passes the electroweak precision tests and theoretical constraints simultaneously. An important
characteristic which appears in the analysis is that the difference between the pseudoscalar $A$ mass and the charged
scalar is approximately $\sim$ $80$ GeV. Also, if the charged scalar has a mass of $100$ GeV, the pseudoscalar and neutrinophilic
scalar should be bigger than $m_H=m_A>$ 150 GeV. Now, if the masses of the scalars are of $\mathcal{O}(\mathrm{TeV})$,
we find that the precision measurements impose that the mass of the charged scalars is also degenerated with the neutral
scalars. For the $\hat{\alpha}$ and $\hat{\beta}$ angles, we find that $\tan\hat{\beta}\lesssim 10^{-6}$ and $\hat{\alpha}\lesssim 
5\hat{\beta}$. \\

Furthermore, in the case in which the neutral scalars are light and satisfy $m_\S<m_h/2$, $\S=H,A$, the bound from the
Higgs invisible width results similar to the $\Z_2$ scenario, and also do not imposes strong limits. However, if the channel
$Z\to\S \bar{\nu} \nu$ is kinematically allowed, the $Z$ invisible width imposes strong constraints to the parameter space.
We can see this in figure \ref{fig:davidson-logan-TH}, bottom. To obtain such result, we also performed an scan on the
oscillation parameters, given in table \ref{tab:GlobalFitNeuOsc}. This is necessary since the width depends on the
sum of the neutrino masses, see equation \eqref{eq:zinv}. Imposing perturbativity in the scalar $\S$ decay, $\Gamma_\S<m_\S/2$,
we find that the region $m_\S<m_Z/2$ is completely excluded due to the on-shell contribution of the scalar particles, 
$Z\to H (A^\ast\to \nu\bar{\nu})$ and $Z\to A (H^\ast\to \nu\bar{\nu})$. This increases the width by several orders
of magnitude.\\

Briefly, we see that in this case the Electroweak precision bounds constrain this model in such a way that 
the spectrum gets very limited. This results as the symmetry imposes that the neutral scalar masses to be the same,
and the $T$ parameter constraints the difference between the pseudoscalar and charged scalar masses. Therefore, all
particles end up with very similar masses. Beyond this, the $Z$ invisible width excludes the region $m_A=m_H<m_Z/2$.
Thus, the spectrum of this scenario becomes a very specific one. This may create additional problems 
when the oblique parameters be measured with a better precision.

%%%%%%%%%%%%%%%%%%%%%%%%%%%%%%%%%%%%%%%%%%%%%%%%%%%%%%%%%%%%%%%%%%%%%%%%%%%%%%%%%%%%%%%%%%%%%%%%%%
\section{Flavour Constraints on Charged Scalar Sector}\label{sec:TEC-HPM}
%%%%%%%%%%%%%%%%%%%%%%%%%%%%%%%%%%%%%%%%%%%%%%%%%%%%%%%%%%%%%%%%%%%%%%%%%%%%%%%%%%%%%%%%%%%%%%%%%%

Other bounds which should be considered in neutrinophilic $2HDM$ come from flavour physics. Flavour bounds will 
only constrain the charged scalar and leptons interactions since, as we have seen, couplings with quarks are very 
suppressed; quarks will couple with the second doublet proportionally to $\tan\hat{\beta}$. Therefore, limits coming from 
hadron observables as leptonic and semi-leptonic B meson decays, for instance, will not be relevant for the model.
Thus, we will concentrate ourselves on leptonic observables. An important fact appears as consequence of the 
specific form we are considering here: flavour limits will not depend on the specific form of the symmetry imposed
on the scalar potential. Hence, the constraints we will obtain here will be applicable to both $\Z_2$ and $U(1)$
symmetry models. Let us write here the part of the lagrangian which presents flavour changing interactions, in 
terms of neutrino flavour eigenstates
\begin{align}
-\mathscr{L}_{\rm Y}^{\rm charged}&=\frac{\sqrt{2}m_a^\nu}{v_2} [\widetilde{U}_{\beta a}^\ast H^+ \overline{\nu^a} P_L \ell^\beta+ \mathrm{h.c.}],\notag\\
 &=-\sqrt{2}\frac{\widetilde{U}_{\alpha a} m_a^\nu \widetilde{U}_{\beta a}^\ast}{v_2} [H^+ \overline{\nu^\alpha} P_L \ell^\beta+ \mathrm{h.c.}].
\end{align}
We can integrate out the charged scalar since we know that its mass needs to be larger than $100$ GeV. Consequently, for 
the energy scales we will consider here ($\sim$ MeV), we can work in an effective theory framework. We find
\begin{align}\label{eq:eff_L}
  {\displaystyle -{\cal L}_{\mathrm{eff}} = \frac{1}{m^2_{H^\pm}}\frac{\langle m_{\alpha\beta} \rangle}{v_2} \frac{\langle m_{\rho\sigma} \rangle}{v_2} \left( \overline{\nu}_\alpha \gamma^\mu P_R \nu_\sigma \right) \left(\overline{\ell}_\rho \gamma_\mu P_L \ell_\beta \right) + \dots \, ,}
\end{align}
where we defined
\begin{align}
	\langle m_{\alpha\beta} \rangle = \widetilde{U}_{\alpha a} m_a^\nu \widetilde{U}^*_{\beta a},
\end{align} 
and we presented only the relevant interaction term for our purposes. Let us notice that the previous effective lagrangian depends 
only in the right-handed neutrino state; thus, this new interaction will not modify the propagation of neutrinos created in nuclear 
processes since in such processes neutrino are left-handed, as we have seen before. We will divide our study in two categories: 
processes with or without flavour conservation. Our results are summarized in \ref{fig:limits}.
\subsection{Flavour-conserving Decays}
\begin{figure}[tb]
 \begin{center}
   \includegraphics[width=\textwidth]{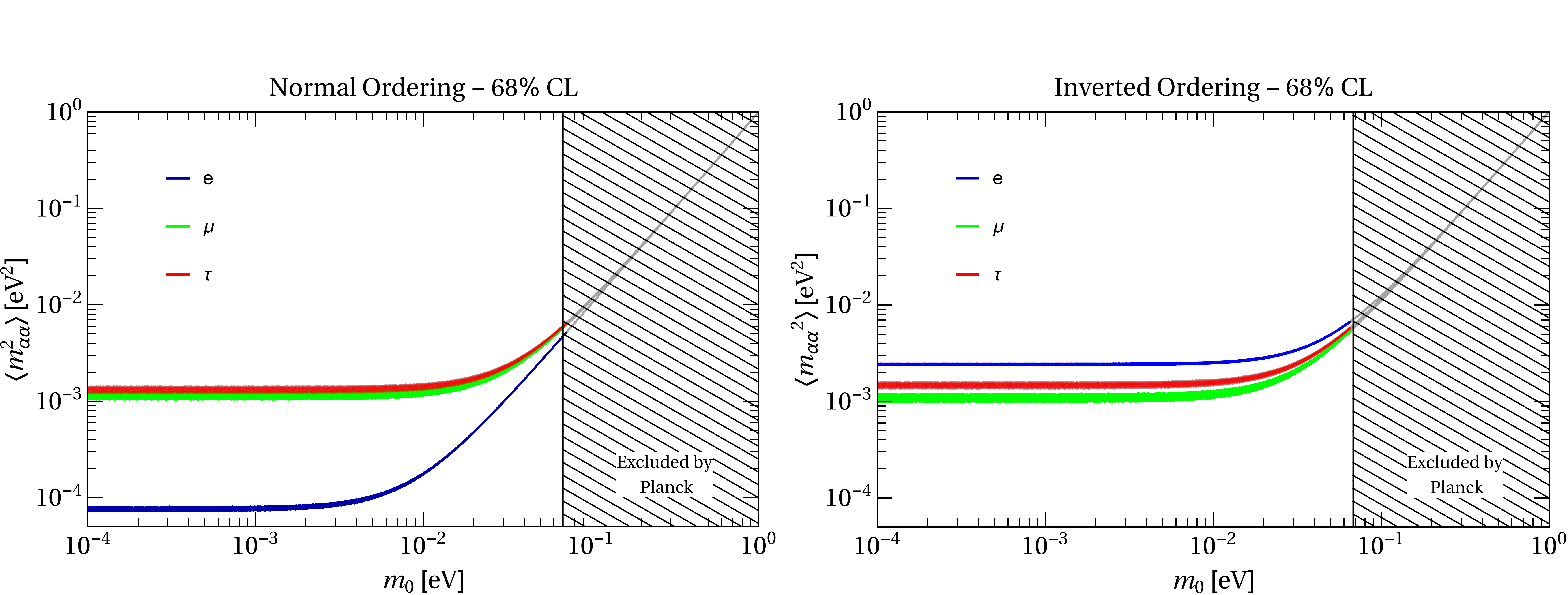}
 \end{center}
 \caption{\label{fig:maa} $\langle m^2_{\alpha\alpha}\rangle$  values as function of the lightest neutrino mass eigenstate. These values were obtained performing an scan over the oscillation parameters, table \protect\ref{tab:GlobalFitNeuOsc}, at $1\sigma$ level. Blue region: $\alpha=e$, green region: $\alpha=\mu$, red region: $\alpha=\tau$. In the left (right) panel we consider a normal (inverted) ordering of the neutrino masses. Gray shaded region is excluded by the Planck limit on the sum of the neutrino masses.}
\end{figure}

For the case in which the interactions are flavour-conserving, we will consider the tree level decays
$\ell_\alpha\to \ell_\beta+\bar{\nu}_\beta+\nu_\alpha$. The contribution of the scalar can be codified through
the {\it flavour} gauge couplings $g_\alpha$ \cite{Roney:2007zza}. The total width in the presence of a charged
scalar is \cite{Davidson:2009ha,Fukuyama:2008sz}
\begin{align*}
	\Gamma(\ell_\alpha \to \ell_\beta \overline{\nu}\nu) = \Gamma^{\rm SM}(\ell_\alpha \to \ell_\beta \overline{\nu}\nu) \left(1+ \langle m^2_{\alpha\alpha} \rangle \langle m^2_{\beta\beta} \rangle \frac{\rho^2}{8}\right),
\end{align*}
which allows to obtain the relation for the {\it flavour} couplings as
\begin{align}\label{eq:ratioga}
  \left(\dfrac{g_\mu}{g_e}\right)^2 &\simeq  1+ \dfrac{\langle m_{\tau\tau}^2 \rangle (\langle m_{\mu\mu}^2 \rangle-\langle m_{ee}^2 \rangle)}{8 } \rho^2, 
\end{align}
\begin{align}\label{eq:ratiogb}
\left(\dfrac{g_\mu}{g_\tau}\right)^2 &\simeq1+ \dfrac{\langle m_{ee}^2 \rangle (\langle m_{\mu\mu}^2 \rangle-\langle m_{\tau\tau}^2 \rangle)}{8} \rho^2,\,
\end{align}
where we defined $\langle m_{\alpha \beta}^2 \rangle = U_{\alpha i} m_{\nu_i}^2 U^*_{\beta i}$ and 
$\rho =(G_F m_{H^\pm}^2 v_2^2)^{-1}$. Notice that the new physics is codified through the $\rho$ parameter. From expressions
\eqref{eq:ratioga}, \eqref{eq:ratiogb}, can be obtained bounds on the $\rho$ parameter from the $\mu$ and $\tau$ leptons 
half-lives. However, experimentally we know that such half-lives are compatible with the lepton universality at $1\sigma$ level. A possible explanation for these results appears when we study the dependence of $\langle m^2_{\alpha\alpha}\rangle$ with 
the mass of the lightest neutrino, $m_0$, figure \ref{fig:maa}. We see the specific values for 
$\langle m^2_{ee}\rangle$ and $\langle m^2_{\tau\tau}\rangle$ which can be small for $m_0^2\ll \Delta m_{ij}^2$.
Thus, the {\it flavour} couplings will be also small, and we will not be able to extract information from these
observables.
\subsection{Flavour-violating Processes}
Next, we will analyse processes in which there is flavour violation. These processes are, or will be, constrained
by experimental data. For the neutrinophilic case, such processes come from 1-loop corrections mediated
by the charged scalar, and they will constrain the model.
%--
%--
\subsubsection{$\bullet\ l_\alpha \to l_\beta \gamma$}
\begin{figure}[tb]
  \begin{center}
    \includegraphics[width=0.475\textwidth]{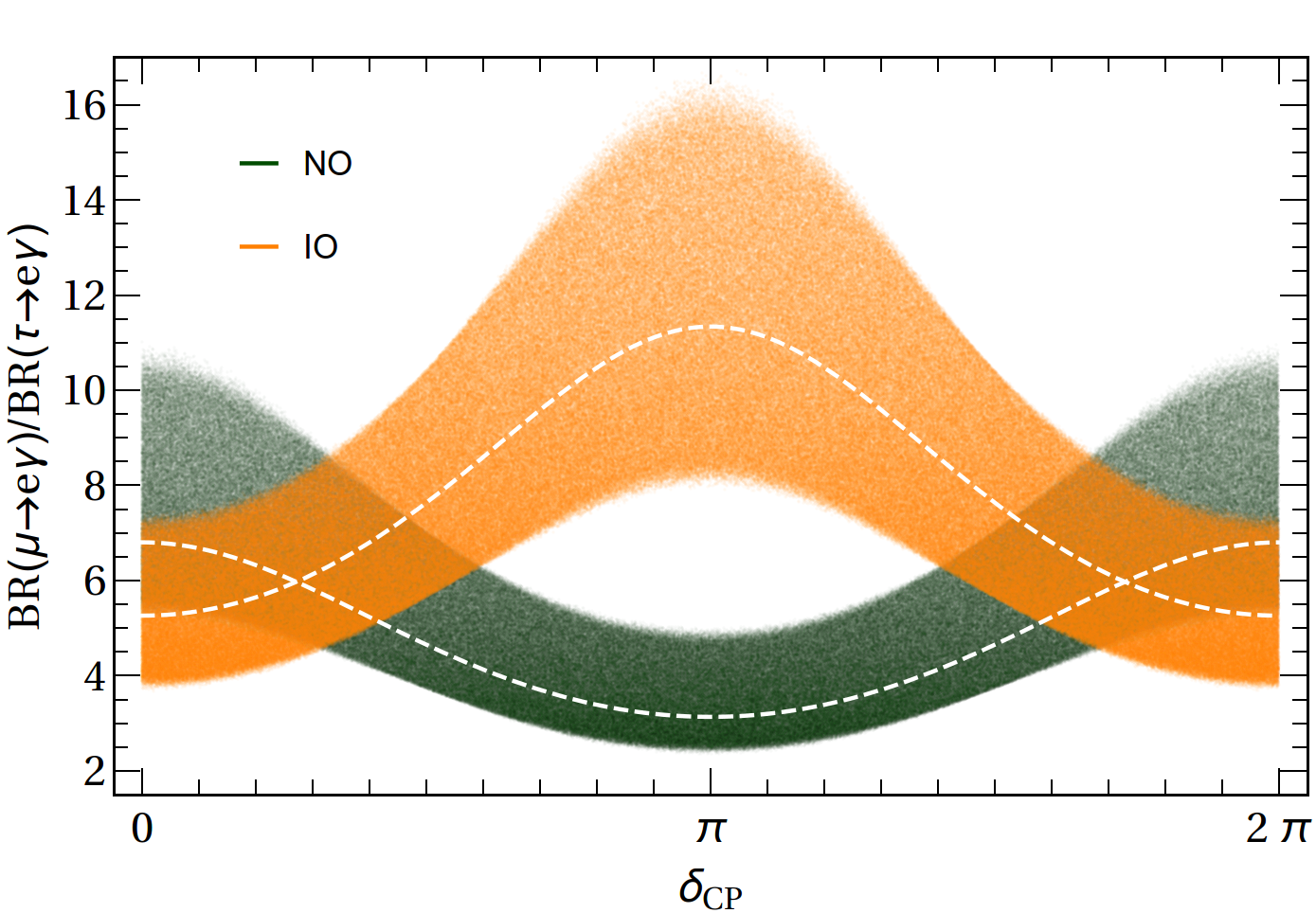} 
    \includegraphics[width=0.475\textwidth]{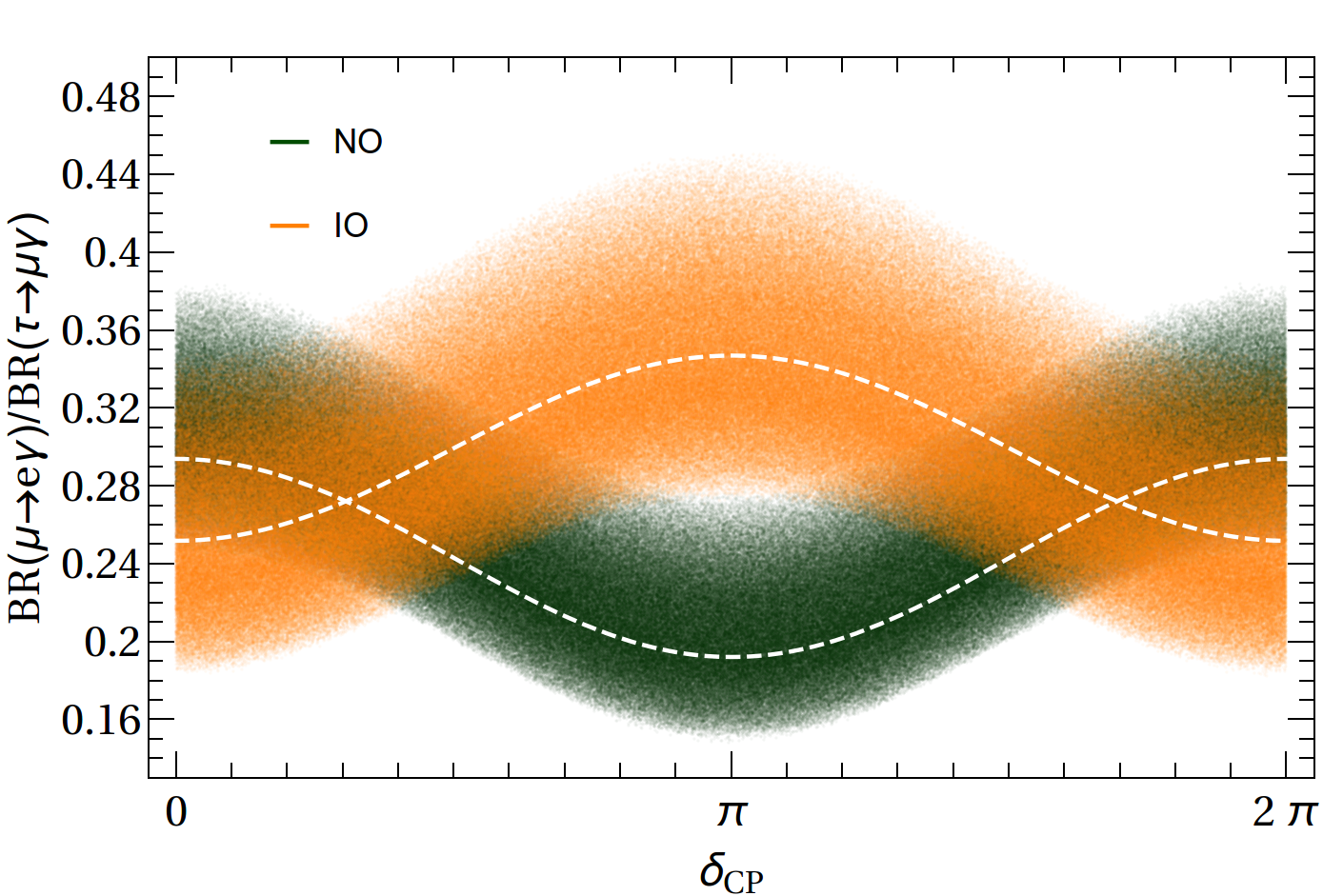} 
  \end{center}
  \vspace{-0.5cm} 
  \caption{\label{fig:cp} Ratios $\mathrm{BR}(\mu \to e \gamma)/\mathrm{BR}(\tau \to e \gamma)$ (right) and $\mathrm{BR}(\mu \to e \gamma)/\mathrm{BR}(\tau \to \mu \gamma)$ (left) as function of the CP phase $\delta$. These regions
  where obtained performing an scan on the neutrino oscillation parameters in the $2\sigma$ range. The white line represents
  the ratios using the neutrino oscillation best fit.}
 \end{figure}

%--
%--
The processes currently possessing the strongest experimental bounds is $\mu\to e \gamma$. Considering in general
the decay $l_\alpha \to l_\beta \gamma$, it is possible to find the branching ratio of this process as a function
of the parameter $\rho$ \cite{Fukuyama:2008sz}
\begin{align}
\mathrm{BR}(\ell_\alpha\to \ell_\beta\gamma) = \mathrm{BR}(\ell_\alpha\to e\bar{\nu}\nu)\frac{\alpha_\text{EM}}{192 \pi} |\langle m_{\alpha \beta}^2 \rangle|^2 \rho^2 \, .
\end{align}
The strongest experimental bound comes form Mu to E Gamma-II (MEG-II) experiment, $\mathrm{BR}(\mu\to e \gamma)<5.7 \times 10^{-13}$
\cite{Adam:2013mnn} while, for the other channels, there are weaker limits, $\mathrm{BR}(\tau \to e \gamma) < 3.3 \times 10^{-8}$ and $\mathrm{BR}(\tau \to \mu \gamma)< 4.4 \times 10^{-8}$, obtained by the BaBar experiment \cite{Aubert:2009ag}.
We can then set limits to our model as
\begin{align*}
 \begin{array}{ll}
  \rho \lesssim 1.2 \, \mathrm{eV}^{-2} & ~~~~~[\mu \to e \gamma]\, , \\
  \rho \lesssim 730 \, \mathrm{eV}^{-2} & ~~~~~[\tau \to e \gamma]\, , \\
  \rho \lesssim 793 \, \mathrm{eV}^{-2} & ~~~~~[\tau \to \mu \gamma]\, . \\
 \end{array}
\end{align*}
The bound $\rho \lesssim 1.2 \, \mathrm{eV}^{-2}$ is the best current limit obtained for the $v_2$ and $m_{H^\pm}$
parameters since it implies that, supposing $v_2\lesssim 1$ eV, the charged scalar mass will be $m_{H^\pm}\gtrsim$ $250$ GeV. In the future, if the predicted sensitivity of MEG will be achieved, $\mathrm{BR}(\mu \to e \gamma) \sim 5 \times 10^{-14}$ \cite{Baldini:2013ke,Iwamoto:2014tfa}, the limit on the $\rho$ parameter can be improved by an order of magnitude, $\rho \lesssim 0.4 \; \mathrm{eV}^{-2}$. The current bound and future sensitivity 
are presented in figure \ref{fig:limits}, as the blue line. Furthermore, we should emphasize that if the $\mu\to e\gamma$ decay is observed by some experiment, the neutrinophilic $2$HDM will predict a specific relation between BR$(\mu\to e\gamma)$ and BR$(\tau\to e\gamma, \mu\gamma)$ branching ratios, depending on the CP violation phase $\delta$ in the leptonic sector. We present the ratio between these branching ratios as a function of the CP phase in figure \ref{fig:cp}. There we can see that a limit on BR$(\mu\to e\gamma)$ puts a stronger limit on BR$(\tau\to e\gamma, \mu\gamma)$, independently on the phase. If the CP phase is determined by oscillation experiments, for instance, we will obtain a correlation among the branching ratios.

%--
%--
\subsubsection{$\bullet\ l_\alpha \to 3 l_\beta$}
%--
%--
An analogous process that appears in these scenarios consists in the decay into three charged leptons. Such
process can be described by the effective lagrangian
\begin{align}
\mathcal{L}_{\mathrm{eff}}=  \frac{e \, m_{\ell_\alpha}}{2}A_D\bar{\ell}_\beta \sigma_{\mu\nu}\ell_\alpha F^{\mu\nu}+ e A_{ND} \bar{\ell}_\beta \gamma_\mu P_L \ell_\alpha A^\mu+e^2 B (\bar{\ell}_\alpha \gamma_\mu P_L \ell_\beta) (\bar{\ell}_\beta \gamma_\mu P_L \ell_\beta)+\mathrm{h.c.},
\end{align}
where $A_{(N)B}$ and $B$ are the Wilson coefficients associated to the penguin diagrams of the photon and the box diagrams
of the Higgs, respectively. Neglecting the neutrino masses and taking $m_\beta\ll m_\alpha$, we find
\begin{align}
\label{eq:AD}A_D &= \frac{1}{6(4\pi)^2}\frac{1}{m_{H^\pm}^2}\frac{\langle m_{\alpha \beta}^2\rangle}{v_2^2}\, ,\\
\label{eq:AND}A_{ND} &= \frac{1}{9(4\pi)^2}\frac{q^2}{m_{H^\pm}^2}\frac{\langle m_{\alpha \beta}^2\rangle}{v_2^2}\, , \\
\label{eq:B}e^2 B &= -\frac{2}{(4\pi)^2}\frac{\langle m_{\alpha \beta}^2\rangle\langle m_{\beta \beta}^2\rangle}{m_{H^\pm}^2 v_2^4},
\end{align}
where $q^2$ is the squared momentum of the photon. We should mention here that we are not considering the penguin diagrams
of the $Z$ boson since those diagrams are suppressed by the $Z$ and $m_\beta$ masses. Computing the branching ratio in terms
of the previous coefficients, we have
\begin{align}
\mathrm{BR}(\ell_\alpha\to \ell_\beta\ell_\beta\ell_\beta) = \mathrm{BR}(\ell_\alpha\to \ell_\beta\bar{\nu}\nu) &\frac{3 (4\pi)^2\alpha_{\mathrm{EM}}^2}{8 \, G_F^2} \Bigg{[} \frac{|A_{ND}|^2}{q^4}+|A_D|^2\left(\frac{16}{3}\log\left(\frac{m_\alpha}{m_\beta}\right)-\frac{22}{3} \right) \nonumber \\
&\hspace{-1.5cm}+\frac{1}{6}|B|^2+2\,\mathrm{Re}\left(-2 \frac{ A_{ND}}{q^2} A_D^\ast+\frac{1}{3}\frac{A_{ND}}{q^2}B^\ast-\frac{2}{3}A_D B^\ast \right)\Bigg{]}.
\end{align}
Let us note that the Wilson coefficients associated to the box diagrams have a different dependence on $v_2$. This is the
reason why we can not write this ratio in terms of the parameter $\rho$.\\

\begin{figure}[t!]
 \begin{center}
   \includegraphics[width=0.49\textwidth]{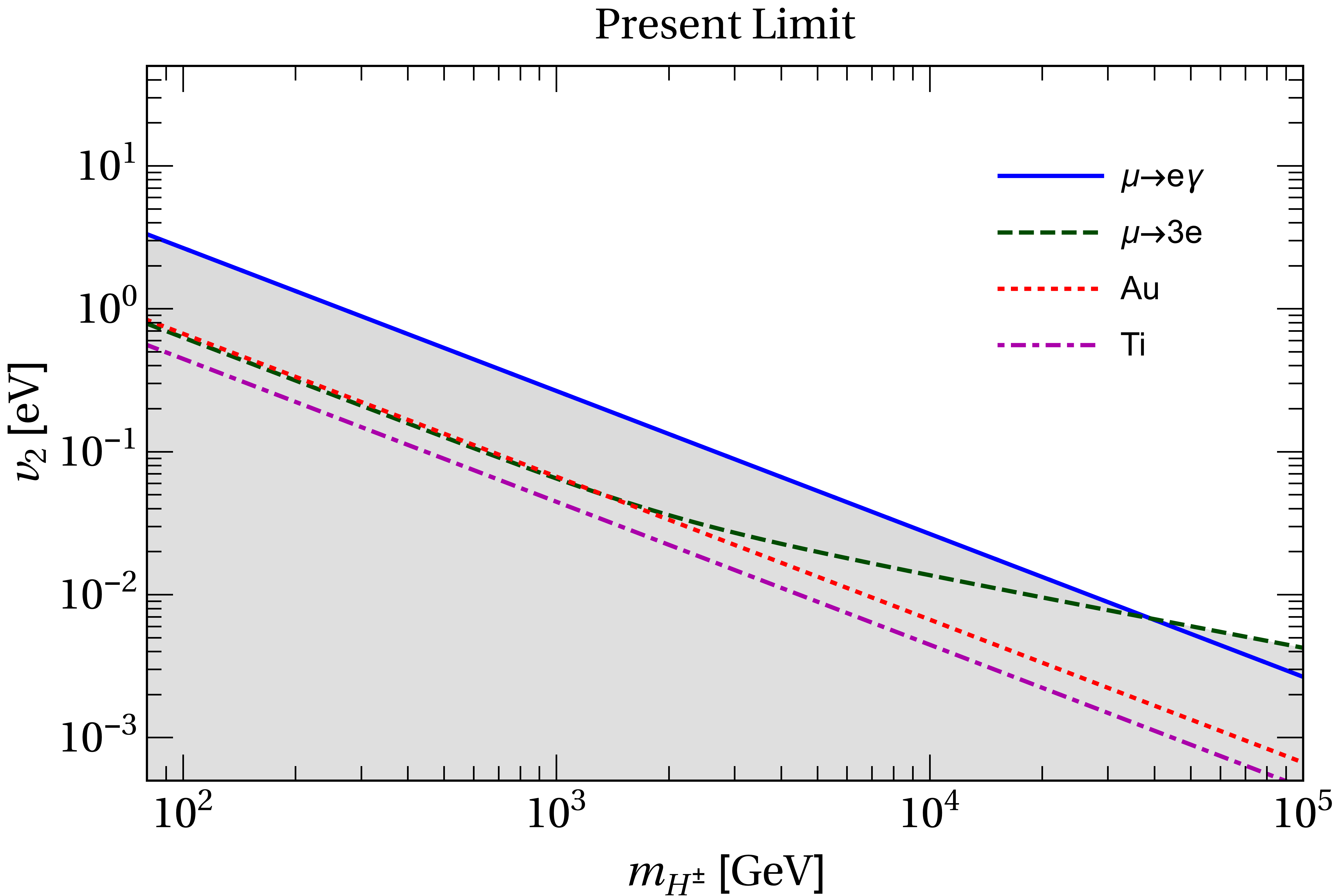}
   \includegraphics[width=0.49\textwidth]{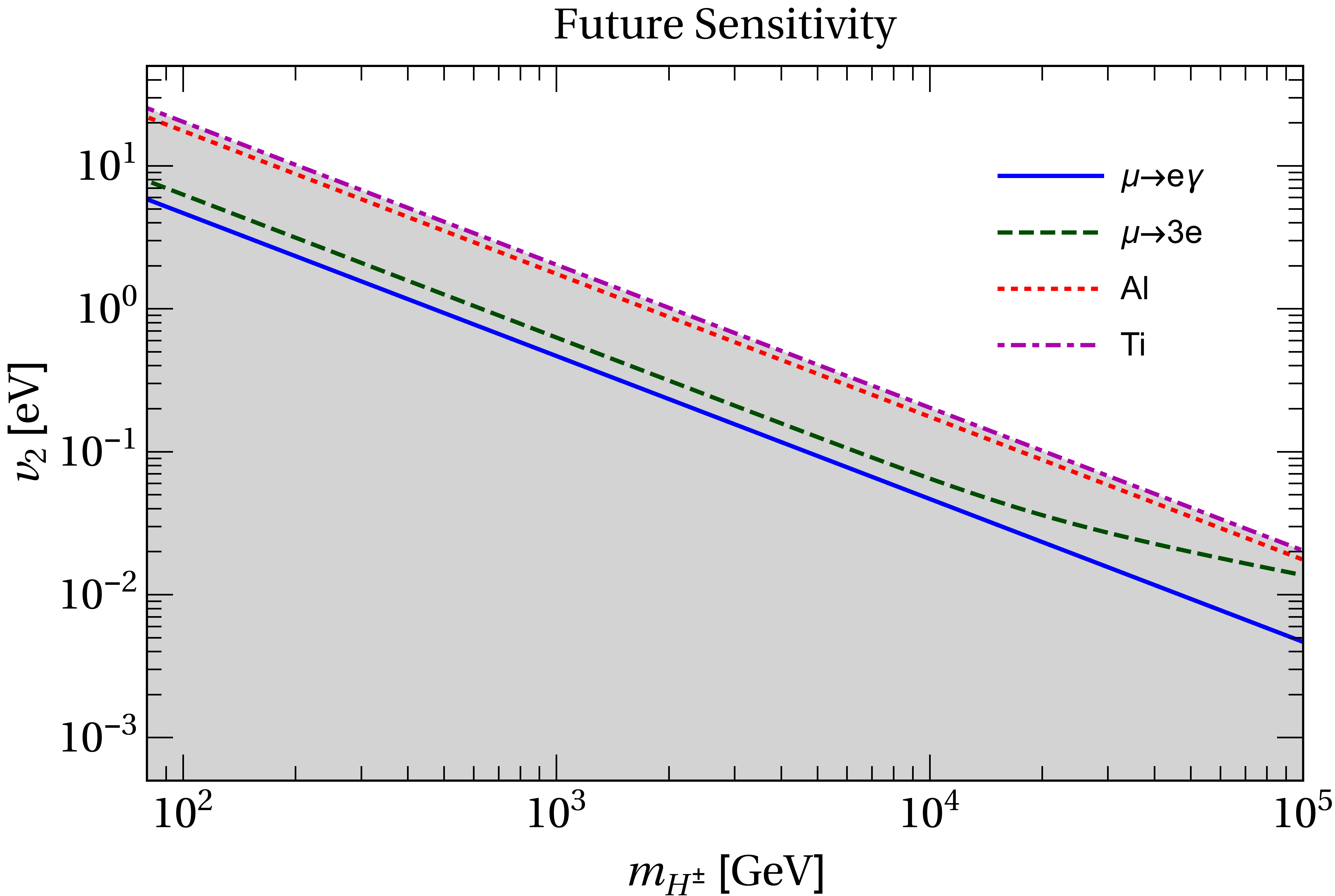}
 \end{center}
 \caption{\label{fig:limits}
 Flavour physics bounds on the charged part of the neutrinophilic $2$HDM. In the left panel we present the current limits on the
 plane $m_{H^\pm}\times v_2$ coming from $\mu \to e \gamma$, $\mu \to eee$ and $\mu \to e$ conversion in nuclei.
 The region in gray is excluded at 99\% CL. In the right panel we show the sensitivities of future experiments
 regarding the same processes. The region below the lines can be explored by the respective experiment.}
\end{figure}

Taking into account the previous results, we see that the box diagrams dominate when $v_2$ is small, while
the penguin ones dominate for larger values. Using the experimental bound $\rm{BR}(\mu \to e^{-} e^{-} e^{+})< 1 \times 10^{-12}$ 
\cite{Bellgardt:1987du}, and supposing that $v_2$ is in the region where penguins dominate, we find that
$\rho \lesssim 22$ eV$^{-2}$. The result independent on the choice of $v_2$ is in figure \ref{fig:limits},
green dashed line. We should notice that the current value for $l_\alpha \to 3 l_\beta$ is stronger than
the bound obtained in the previous subsection when $v_2 \lesssim 0.01$ eV, region where box diagrams dominate. 
Furthermore, using the sensitivity expected
for the future experiment Mu$3$e, $\rm{BR}(\mu \to e^{-} e^{-}e^{+}) \sim 1 \times 10^{-16}$ \cite{Blondel:2013ia},
we find that $\mu\to 3 e$ future sensitivity is bigger than $\mu\to e\gamma$, see figure \ref{fig:limits},
right panel.\\
%--
%--
\subsubsection{$\bullet\ \mu \to e$ conversion in nuclei}
%--
%--
%
\begin{table}[tb]
	\centering
	\begin{tabular}{|c|c|c|c|}\hline
		$^A_Z$ Nuclei  & $Z_{\mathrm{eff}}$ & $F_p$ & $\Gamma_{\rm capt}$ (GeV)\\ \hline\hline
		$^{27}_{13}\text{Al}$ & 11.5 & 0.64 & $4.64079 \,\times\,10^{-19}$\\ 
		$^{48}_{22}\text{Ti}$ & 17.6 & 0.54 & $1.70422\,\times\, 10^{-18}$\\ 
		$^{197}_{\phantom{1}79}\text{Au}$ & 33.5 & 0.16 & $8.59868\,\times\,10^{-18}$\\ \hline
	\end{tabular}
	\caption{Nuclear parameters used in our study.\label{tab:TabNuclear}}
\end{table}
The $\mu \to e$ conversion is a process which may become an important bound for different neutrino mass models. For the neutrinophilic $2$HDM, the dominant contributions come only from penguin photon diagrams given that other diagrams are suppressed by the number of protons $Z$, the electron mass or the tiny couplings between quarks and the neutrinophilic charged scalar. The conversion rate in nuclei is given by \cite{Kuno:1999jp,Arganda:2007jw,Toma:2013zsa}
\begin{align}
  \text{CR}(\mu-e,\text{nucleus})=\frac{p_e\, E_e\, m_\mu^3\, G_F^2\, \alpha_\text{EM}^3\,Z_{\text{eff}}^4\,F_p^2}{8\pi^2\,Z\,\Gamma_{\text{capt}}}&\ \left|(Z+N)g^{(0)}_{LV}+(Z-N)g^{(1)}_{LV}\right|^2,%\right.%\notag\\
 % &\left.+\left|(Z+N)g^{(0)}_{RV}+(Z-N)g^{(1)}_{RV}\right|^2\right\},
\end{align}
being $p_e,E_e\approx m_\mu$ the electron momentum and energy, and they are approximately equal to the muon mass;
$Z,N$ are the proton and neutron numbers of the nucleus, respectively; $Z_{\rm eff}$ is the effective
atomic charge and $F_p$ is the nuclear matrix element given in table \ref{tab:TabNuclear} for the nuclei
we will consider \cite{Arganda:2007jw}. Let us note that the conversion rate is normalized to the muon
capture rate $\Gamma_{\rm capt}$. The coefficients $g^{(0,1)}_{LV}$ are given by \cite{Kuno:1999jp,Arganda:2007jw}
\begin{subequations}
  \begin{align}
    g^{(0)}_{LV}&=\frac{1}{2}\sum_{q=u,d}\left(G_V^{(q,p)}g_{LVq}+G_V^{(q,n)}g_{LVq}\right),\\
    g^{(1)}_{LV}&=\frac{1}{2}\sum_{q=u,d}\left(G_V^{(q,p)}g_{LVq}-G_V^{(q,n)}g_{LVq}\right).
  \end{align}
\end{subequations}
We have to emphasize here that only the vector couplings are relevant given that only the photon penguins diagrams contribute.
Therefore, valence quarks ($u,d$) will be relevant because {\it see} quarks, as the strange quarks, interact effectively through the 
scalar part. The couplings $g_{LVq}$ are
\begin{align}
  g_{LVq}&=\frac{\sqrt{2}}{G_F}e^2Q_q\left(\frac{A_{ND}}{q^2}-A_D\right). 
\end{align}
\begin{table}[tb]
\centering
\begin{tabular}{|c|c|c|}\hline
Nucleus  & Present Bound & Future Sensitivity\\ \hline\hline
Al & $-$ & $10^{-15}-10^{-18}$ \cite{Kuno:2013mha}\\ 
Ti & 4.3$\,\times\,10^{-12}$ \cite{Dohmen:1993mp}& $\sim 10^{-18}$ \cite{Alekou:2013eta}\\ 
Au & 7$\,\times\,10^{-13}$ \cite{Bertl:2006up}& $-$\\ \hline
\end{tabular}
\caption{Current limits and future sensitivities on $\mu\to e$ nuclear conversion $\mu\to e$.
\label{tab:TabBounds}}
\end{table}
For completeness, we quote the values of the coefficients $G_V^{(q,p)}$ \cite{Arganda:2007jw}
\begin{align}
   G_V^{(u,p)}=G_V^{(d,n)}=2, ~~~~~ G_V^{(u,n)}=G_V^{(d,p)}=1.
\end{align}
From the current limits on $\nu\to e$ conversion, given in table \ref{tab:TabBounds} and taken from \cite{Abada:2014cca},
we obtain the following bounds
\begin{align*}
 \begin{array}{ll}
  \rho \lesssim 30 \, \mathrm{eV}^{-2} & ~~~~~{\rm Titanium\ (Ti)}\, , \\
  \rho \lesssim 13.5 \, \mathrm{eV}^{-2} & ~~~~~{\rm Gold\ (Au)}\, ,
 \end{array}
\end{align*}
while, using the future sensitivities proposed by future experiments, we find the region which can be constrained in the  
neutrinophilic $2$HDM
\begin{align*}
 \begin{array}{ll}
  \rho \lesssim 0.015 \, \mathrm{eV}^{-2} & ~~~~~{\rm Titanium\ (Ti)}\, , \\
  \rho \lesssim 0.020 \, \mathrm{eV}^{-2} & ~~~~~\text{Aluminium\ (Al)}\, .
 \end{array}
\end{align*}
We see then that the $\mu\to e$ conversion in nuclei will be the most sensitive processes, if the collaborations
achieve their proposed sensitivities. In figure \ref{fig:limits}, we see the current limits and
future sensitivities obtained from this process.
%--
%--
\subsubsection{$\bullet\ Z \to l_\alpha l_\beta$}
\begin{figure}[tb]
	\includegraphics[width=\textwidth]{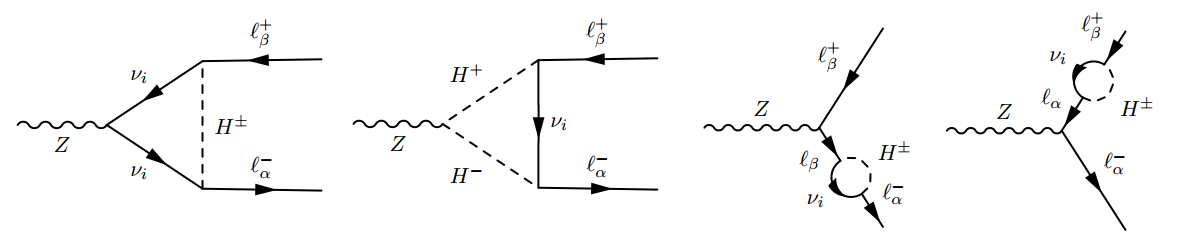} 
	\vspace{-0.5cm} 
	\caption{\label{fig:lfvz} 
	Diagrams contributing to $Z$ flavour-violating decays.}
\end{figure}

%--
%--
Another process which can violate flavour due to the possible existence of charged scalars is 
the $Z$ boson decay, $Z \to l_\alpha l_\beta$, $\alpha\neq\beta$. In the models we are interested in, 
we have loop processes that generate this decay, see figure \ref{fig:lfvz}. The effective hamiltonian in 
this case is
\begin{align}
\mathcal{H}_{\mathrm{eff}} =   C_V \bar{\ell}_\alpha \gamma^\mu P_L \ell_\beta Z_\mu \, +\mathrm{h.c.}
\end{align}
where the Wilson coefficient $C_V$ is given by
\begin{align}
C_V =\frac{1}{64\pi^2} \frac{ g\cos (2\theta_W)}{\cos \theta_W}\frac{\langle m_{\alpha\beta}^2\rangle}{v_2^2} \Bigg{\lbrace} &4 \left( \frac{2}{x_Z}-1\right) \left(\frac{4}{x_Z}-1\right)^{1/2}\arctan\left[\left(\frac{4}{x_Z}-1\right)^{-1/2} \right]\notag\\
&-\frac{16}{x_Z^2}\arctan^2\left[\left(\frac{4}{x_Z}-1\right)^{-1/2}\right]+\left(5-\frac{4}{x_Z}\right)\Bigg{\rbrace},
\end{align}
being $x_Z=m_Z^2/m^2_{H^\pm}<1$. In the case in $x_Z \ll 1$, the previous expression can be approximated to
\begin{equation}
C_V = \frac{x_{Z}}{288 \pi^2}\frac{ g\cos (2\theta_W)}{\cos \theta_W}\frac{\langle m_{\alpha\beta}^2\rangle}{v_2^2} +\mathcal{O}(x_Z^2).
\end{equation} 
The $Z$ boson decay width is
\begin{equation}
\Gamma(Z\to \ell^\pm_\alpha \ell^\mp_\beta) = \dfrac{m_Z^2}{12 \pi \Gamma_Z}  |C_V|^2
\end{equation}
where we see that the Wilson coefficient appears explicitly in the expression. Here $\Gamma(Z\to \ell^\pm_\alpha \ 
\ell^\mp_\beta)=\Gamma(Z\to \ell^-_\alpha \ell^+_\beta)+\Gamma(Z\to \ell^+_\alpha \ell^-_\beta)$.

\newpage
The strongest experimental bound corresponds to the ATLAS superior limit, $\mathrm{BR}(Z\to e^\pm\mu^\mp)<7.5 \times 10^{-7}$ 
\cite{Aad:2014bca}. The channels where there is a tau in the final state were studied by LEP, and possess
a weaker bound $\mathrm{BR}(Z\to e^\pm\tau^\mp)<9.8 \times 10^{-6} $ and $\mathrm{BR}(Z\to \mu^\pm\tau^\mp)<1.2 \times 10^{-5} $
\cite{Abreu:1996mj}. Thus, taking into account the previous results, we find that $\rho\lesssim 3.5 \times 10^{3}$ eV$^{-2}$,
which is weaker than the results from previous processes. Now, considering the future sensitivity of a electron-positron
collider, we have found that the bound will not be improved significantly \cite{Bertuzzo:2015ada}.
%--
%--
\subsubsection{$\bullet\ h \to l_\alpha l_\beta$}
%--
%--
Finally, we will consider the flavour-violating Higgs decay. In an analogous manner as previously done for the $Z$ boson decay, 
we have that the effective hamiltonian is given by
\begin{equation}
\mathcal{H_\mathrm{eff}} = C_L \bar{\ell}_\alpha P_L \ell_\beta h \,+ \mathrm{h.c.}
\end{equation}
with the Wilson coefficient $C_L$
\begin{align}
C_L = -\frac{1}{8\pi^2}\frac{\langle m_{\alpha\beta}^2\rangle}{v_2^2} \frac{m_\alpha g_{hH^+ H^-}}{m_{H^\pm}^2} \frac{1}{y_H} \Bigg\lbrace & 1-2\left( \frac{4}{y_{H}}-1\right)^{1/2}\arctan\left[\left(\frac{4}{y_H}-1\right)^{-1/2}\right]\notag\\
&+\frac{4}{y_H}\arctan^2\left[\left(\frac{4}{y_H}-1\right)^{-1/2}\right]\Bigg\rbrace \, .
\end{align}
Here, we assumed that $m_\beta/m_\alpha\ll 1$ and defined $y_H=m_h^2/m_{H^\pm}^2$.
The coupling $g_{hH^+H^-}$ is the trilinear coupling among $hH^+H^-$, and will depend on the specific
case of the symmetry considered. The coefficient in the asymptotic limit is given by
\begin{align}
C_L= -\frac{1}{32\pi^2}\frac{\langle m_{\alpha\beta}^2\rangle}{v_2^2} \frac{m_\alpha\, g_{hH^+ H^-}}{m_{H^\pm}^2} \left[1+\frac{y_H}{9} +\mathcal{O}(y_H^2)\right].
\end{align}
\begin{figure}[tb]
  \begin{center}
    \includegraphics[width=0.6\textwidth]{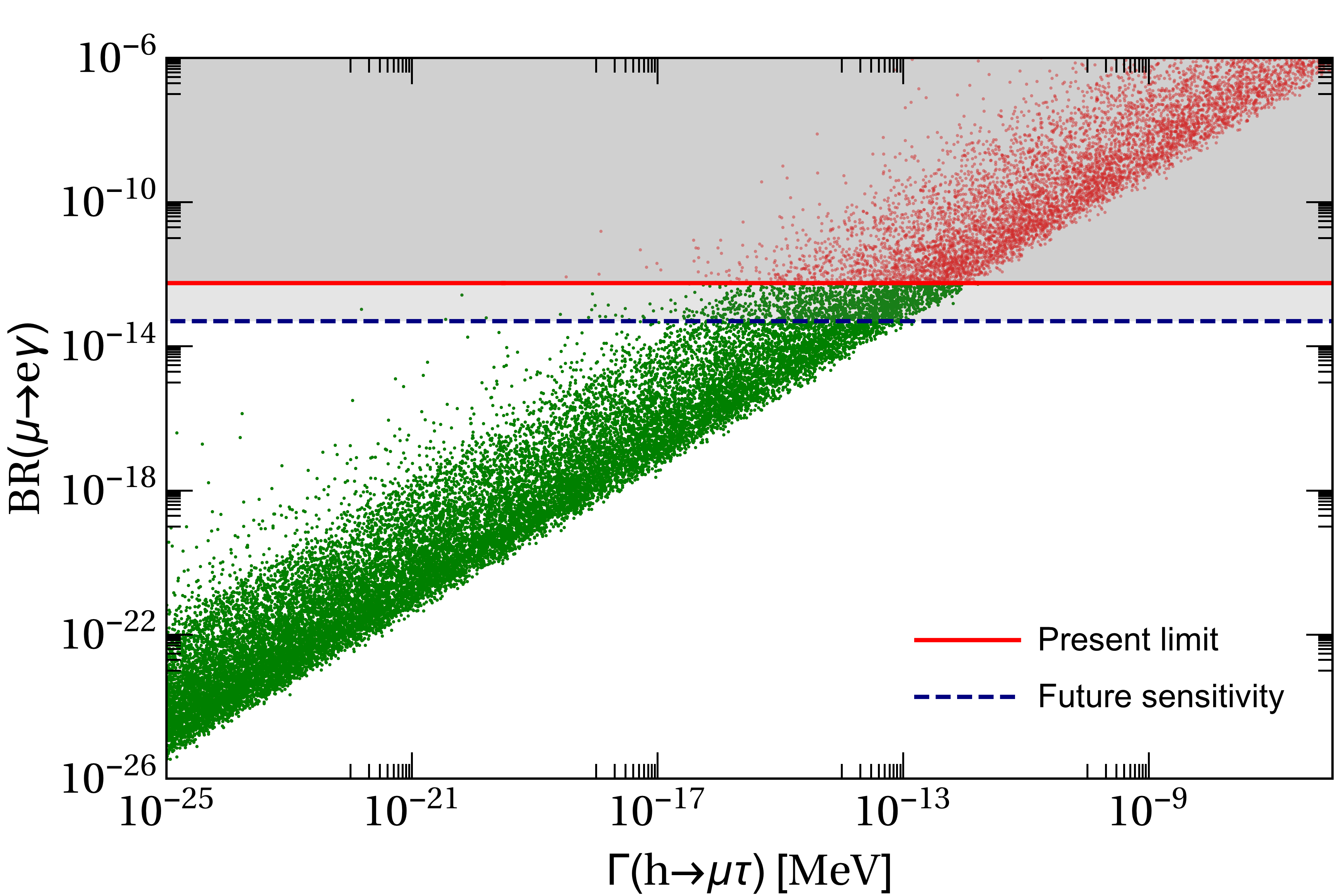}
  \end{center}
  \caption{\label{fig:higgs} $\Gamma(h \to \mu \tau)$ as function of $\mathrm{BR}(\mu \to e \gamma)$ for the
  allowed parameter space of the U$(1)$ global symmetry model. The gray regions correspond to the excluded region
  by MEG-II (red line) and the future sensitivity (dashed blue line).}
\end{figure}

\noindent The decay width $\Gamma(h\to \ell_\alpha^\pm \ell_\beta^\mp)=\Gamma(h\to \ell_\alpha^+ \ell_\beta^-)+\Gamma(h\to \ell_\alpha^- \ell_\beta^+)$ is
\begin{align}
\Gamma(h\to \ell_\alpha^\pm \ell_\beta^\mp)=\frac{(m_h^2-m_\alpha^2)^{2
}}{8\, \pi \, m_h^3}|C_L|^2.
\end{align}
To obtain the bound from this case, we will consider only the $U(1)$ scenario since we have seen that
the other model is strongly constrained by Electroweak precision tests. Using the same scan performed in the 
previous section, we obtained a value for the decay width $\Gamma(h \to \mu \tau)$, corresponding to 
the largest flavour violation, in terms of the branching ratio $\mathrm{BR}(\mu \to e \gamma)$, 
as shown in figure \ref{fig:higgs}. We see that $\Gamma(h \to \mu \tau)$ can not be larger than $10^{-9}$ MeV,
due to the strong limit on $\mu\to e\gamma$ of MEG-II. As consequence, we see that such small value of the branching
ratio is beyond the reach any future Higgs precision experiment.\\

We have seen in this chapter the limits coming from theoretical and phenomenological characteristics that 
the neutrinophilic model must fulfil. From the Electroweak precision tests, we see that the simplest cases of the symmetries
applied to the scalar potential have problems to accommodate the data. In special, the $\Z_2$ symmetry model is practically excluded 
after the Higgs discovery at the LHC. The second model does not possess problems with the precision data, but 
it has a very limited spectrum. It is possible that these problems can be solved by choosing other symmetries, or modifying
the value of the $m_{12}^2$ parameters. Let us notice here that such parameter is the crucial one for the phenomenology. On the
other hand, we have seen how lower energy bounds on flavour violating decays constrain the mass of the charged scalar and
the VEV of the second doublet. Basically, the $\mu\to e\gamma$ decay imposes that the mass of the charged scalar
has to be bigger than $250(2500)$ GeV  when the VEV is smaller than $v_2\lesssim 1(0.1)$ eV. A crucial characteristic
of this limit is that it is independent on the additional symmetry chosen for the model; thus, in principle, it can be applied
to other possible scenarios. In the next chapter we will consider a possible manner to differentiate between Dirac and Majorana 
neutrinos by the detection of the cosmic neutrino background.%\newpage
			%%%%%%%%%%%%%%%%%%%%%%%%%%%%%%%%%%%%%%%%%%%%%%%%%%%%%%%%%%%%%%%%%%%%%%%%%%%%%%%%%%%%%%%%%%%%%%%%%
%%%%%%%%%%%%%%%%%%%%%%%%%%%%%%%%%%%%%%%%%%%%%%%%%%%%%%%%%%%%%%%%%%%%%%%%%%%%%%%%%%%%%%%%%%%%%%%%%%
%%%%%%%%%%%%%%%%%%%%%%%%%%%%%%%%%%%%%%%%%%%%%%%%%%%%%%%%%%%%%%%%%%%%%%%%%%%%%%%%%%%%%%%%%%%%%%%%%%
\chapter[Non-Standard Interactions and the Detection of the Cosmic Neutrino Background]{Non-Standard Interactions and the Detection of the Cosmic Neutrino Background}\label{cha:RelicNu}
\chaptermark{NSI and the Cosmic Neutrino Background}

\lettrine{H}{ubble's} discovery about the expansion of the Universe led to the development of Cosmology. Works from several great physicists showed that the Universe had a beginning, and its evolution can be tracked to times larger than $10^{-42}$ s. After the Big Bang, all known particles were initially at thermal equilibrium provided by the interactions. Nonetheless, the expansion of the Universe acted in such a way that some particles decoupled from the rest of the particles as their interaction rate became weaker than the expansion rate. More and more particles decoupled from the bath until only neutrinos, electrons, positrons and photons remained in thermal equilibrium at temperatures of ${\cal O}($MeV$)$. Then, neutrinos decoupled $\sim 1$ s after the Big Bang, creating the {\it cosmic neutrino background} ($\CNB$). Afterwards, electrons and positrons annihilated each other, creating the cosmic microwave background (CMB); such radiation background, created $\sim 3.8\times 10^5$ years after the Big Bang, have been already detected and is still being analysed \cite{Ade:2015xua}. Even though the $\CNB$ contains information on a very early epoch of the Universe, it has not been observed yet; thus, the detection of these relic neutrinos is fundamental. However, such discovery does not seem to be an easy one. The main reason is the small interaction between very low energy neutrinos and the rest of matter.\\ 

Distinct methods have been proposed, but many of them are beyond the current tech\-no\-lo\-gy. The most promising one consists in the capture of the neutrinos belonging to the $\CNB$ by a nucleus, creating an electron. The energy of such electron will depend directly on the neutrino mass. Supposing only SM interactions, previous works \cite{Long:2014zva} have shown that the capture rate is different according to the neutrino's nature (Dirac or Majorana). Then, one can ask what could be the impact on the capture rate if there were Non-Standard Interactions (NSI) couplings with neutrinos? We will present hereafter the repercussion of those NSI on the $\CNB$ capture rate. Accordingly, we will first analyse the properties of the neutrino background and its detection. Subsequently we will determine the consequence of the possible existence of NSI. This chapter contains novel results which have been published in \cite{Arteaga:2017zxg}.

%\newpage

%%%%%%%%%%%%%%%%%%%%%%%%%%%%%%%%%%%%%%%%%%%%%%%%%%%%%%%%%%%%%%%%%%%%%%%%%%%%%%%%%%%%%%%%%%%%%%%%%%%%%%%%%%%%%%%%%%%%%%%%%%%%%%
\section{Cosmic Neutrino Background}

Let us start considering the generation of the $\CNB$. To do so, we will analyse the primordial plasma at a temperature of $\sim 10$ MeV \cite{Weinberg:2008zzc}, in which left-handed neutrinos were in thermal equilibrium with electrons, positrons and photons. All the other particles are supposed to be already decoupled. Neutrinos were maintained in equilibrium due to charged- and neutral-current interactions $\nu e\longleftrightarrow \nu e $ and $\nu \bar{\nu}\longleftrightarrow e^+ e^-$.  As noted before, weak interactions are the responsible for the preservation of the thermal equilibrium. In such equilibrium situation, we can compute the number density of one neutrino species $\nu_a$, having a mass $m_a^\nu$, at a temperature $T$
\begin{align}
	n_{\nu_a}(T)&=\frac{g_a}{8\pi^3}\int d^3p\, \frac{1}{\exp\left[(E_a-\mu_a)/T\right]+1},
\end{align}
with $E_a^2=\abs{\vec{p}}^2+m_a^{\nu\,2}$, $g_a = 2$ the number of internal degrees of freedom and $\mu_a$ the chemical potential. Since we are considering neutrinos at temperatures larger than the masses $T\gg m_a$ and we suppose that there is a negligible lepton asymmetry $\mu_a\approx 0$, we can approximate the integrals to obtain \cite{Kolb:1990vq}
\begin{align}
	n_{\nu_a}&=\frac{3\zeta(3)}{4\pi^2}g_a T^3.\label{eq:nunumber}
\end{align}
However, at some point, the expansion rate of the Universe became stronger than the weak interaction rate, and neutrinos froze out from the primordial plasma. To determine the temperature $T_\nu$ in which the freeze out happened, let us remember the criterion established in chapter \ref{cap:nuMaj}. When the interaction rate is smaller than the Hubble expansion rate, the species becomes out-of-equilibrium. For the weak interaction we have \cite{Kolb:1990vq,Weinberg:2008zzc}
\begin{align}
	\frac{\Gamma_W}{H}\approx \frac{G_F^2 T_\nu^5}{T_\nu^2/M_{Pl}}=\left(\frac{T_\nu}{1\ {\rm MeV}}\right)^3,
\end{align}
so, near a temperature of $1$ MeV, neutrinos decouple from the primordial plasma. After the neutrino decoupling, when the plasma becomes colder than the electron mass, electrons start to annihilate with positrons, increasing the photon temperature. Considering the entropy conservation, one can show that the neutrino and photon temperatures are related by \cite{Kolb:1990vq}
\begin{align}
	T_\nu=\left(\frac{4}{11}\right)^{\frac{1}{3}}T_\gamma;
\end{align}
thus, as the photon temperature measured at the present time is $T_\gamma =2.725$ K ($0.235$ meV), neutrinos should have a temperature of $T_\nu=1.945$ K ($0.168$ meV) today. To comprehend the properties of the $\CNB$, let us compute the root mean square momentum per neutrino species at the present time
\begin{align}
	\bar{p}_a\equiv\sqrt{\langle p_a^2 \rangle}=\sqrt{\frac{p_{\nu_a}^2}{n_{\nu_a}^0}}
\end{align}
where the number density today per species is
\begin{align}
	n_{\nu_a}^0&=\frac{3\zeta(3)}{4\pi^2}g_a T_\nu^3
\end{align}
and
\begin{align*}
	p_{\nu_a}^2&=\frac{g_a}{8\pi^3}\int d^3p\, p^2 \frac{1}{\exp\left[(E_a-\mu_a)/T_\nu\right]+1},\\
		   &=\frac{45 \zeta(5)}{4\pi^2}g_i T^5_\nu.
\end{align*}
Therefore,
\begin{align}
	\bar{p}_a=\sqrt{15\frac{\zeta(5)}{\zeta(3)}} T_\nu = 0.604\ {\rm meV}.
\end{align}
We see that neutrinos have a very small momentum today, so the $\CNB$ is actually composed by non-relativistic particles unless one of them is massless. This will have crucial consequences for the detection of these particles. In the first place, neutrino flavour eigenstates have suffered decoherence into their mass eigenstates due to the separation of the mass eigenstates wave packets; the $\CNB$ is then basically composed by these independent mass eigenstates. Second, at the present time, the neutrino background is composed not by chiral but by helicity eigenstates. Let us analyse this in more detail.\\ 

\sloppy When neutrinos decouple from the primordial plasma, they have momenta of $\mathcal{O}({\rm MeV})$; thus, they are ultrarelativistic particles. We can then approximate helical and chiral states. Due to the expansion of the Universe, neutrinos evolved into a non-relativistic state, and chirality became different from helicity. At first sight, neutrinos are free particles, but what are the states that are conserved in the evolution? Let us remember that the free Dirac hamiltonian conserves helicity and violates chirality due to the mass term. Therefore, neutrinos were created with a definite chirality, but the free evolution conserved only the helical states\footnote{If the neutrino underwent a clustering process, we have that the helicity is not conserved either. However, the final $\CNB$ will be composed by equal abundances of the helicity states \cite{Long:2014zva}}. Consequently, an essential differentiation between Majorana and Dirac neutrinos arises in the $\CNB$. If neutrinos are Dirac fermions, the left- and right-handed components $\nu_{L,\, R}^a$ behave differently under SM gauge interactions. Only left-handed neutrinos are in thermal equilibrium with the plasma as a right-handed neutrino cannot be in equilibrium through Yukawa interactions only \cite{Antonelli:1981eg,Chen:2015dka}. Thus, neutrinos $\nu_{L(R)}^a$ and antineutrinos $(\nu_{R(L)}^a)^c$\footnote{$\psi^c$ corresponds to the charge conjugate of $\psi$.} before the freeze out have the following abundances per species \cite{Duda:2001hd,Long:2014zva}
\begin{subequations}
	\begin{align}
		n_{\nu_L^a} &= n_{(\nu_R^a)^c} = \frac{3\zeta(3)}{4\pi^2}T^3, \notag \\
		n_{\nu_R^a} &= n_{(\nu_L^a)^c} \approx 0,
	\end{align}
\end{subequations}
being $T$ the temperature before the decoupling. Notice the absence of the $g_a$ factor; this is due to the fact that we are explicitly writing the densities for each internal degree of freedom. After neutrinos froze out, the previous abundances become abundances of helical states $\nu_{\pm}^{a}$, with $+(-)$ being the left-(right-) helicity at the present time per species
\begin{subequations}
	\begin{align}
		n_{\nu_-^a} &= n_{(\nu^a_+)^c} = n_0, \\
		n_{\nu_+^a} &= n_{(\nu^a_-)^c} \approx 0,
	\end{align}
\end{subequations}
with 
\begin{align*}
	n_0=\frac{3\zeta(3)}{4\pi^2} T_\nu^3\approx 56\ {\rm cm^{-3}}.
\end{align*}
On the other hand, if neutrinos are Majorana fermions, we will need two different particles per generation, a left-handed neutrino $\nu_L^i$ and a right-handed neutrino $N_R^i$. Only the left-handed neutrino interacts weakly while, if we consider the {\it seesaw} scenario, the right-handed neutrinos are heavy, making their abundances zero at the present time. This due to the fact that they decayed into leptons and Higgs way before the neutrino decoupling as we saw in chapter \ref{cap:nuMaj} when we considered leptogenesis. Therefore, their abundances per species {\bf before} the freeze out are
\begin{subequations}
	\begin{align}
		n_{\nu_L^a} &= \frac{3\zeta(3)}{4\pi^2}T^3, \\
		n_{N_R^a} &= 0,
	\end{align}
\end{subequations}
This means that today the abundances of the helicity states in the Majorana case are
\begin{subequations}
	\begin{align}
		n_{\nu_-^a} &= n_{\nu_+^a} = n_0, \\
		n_{N_+^a} &= n_{N_-^a} = 0,
	\end{align}
\end{subequations}
\newpage
\noindent as a left(right)-handed Majorana neutrino has two helical states $\nu_{\pm}^a(N_\pm^a)$. We can see then that the abundances of left- and right-helical states is different in Dirac and Majorana cases. This discrepancy is due to the distinct behaviour of chiralities and helicities for Dirac and Majorana neutrinos since only left-handed chiral states interact weakly. If neutrinos are Dirac particles, just left-handed neutrinos and right-handed antineutrinos are decoupled at $T\sim 1$ MeV, and their free streaming makes the left-handed chiral states to be actually left-helical particle states, and right-handed chiral states become right-helical antiparticle states. In the Majorana scenario, the left-handed neutrino freezes out alone; nonetheless, it is composed by the two helical states. Therefore, the Dirac $\CNB$ is composed by left-helical particles and right-helical antiparticles while, in the Majorana case, it will be composed by left- and right-helical particle states. This difference will be decisive when we consider the possible detection of the $\CNB$.

%\newpage

%%%%%%%%%%%%%%%%%%%%%%%%%%%%%%%%%%%%%%%%%%%%%%%%%%%%%%%%%%%%%%%%%%%%%%%%%%%%%%%%%%%%%%%%%%%%%%%%%%%%%%%%%%%%%%%%%%%%%%%%%%%%%%
\section{Detection of the $\CNB$}

The detection of the $\CNB$ in Particle Physics and Cosmology is comparable to the quest of the Holy Grail. The $\CNB$ can confirm several properties predicted by the $\Lambda$CDM cosmological model since it contains information from an era before the CMB creation. Thus, during the past decades, several methods have been proposed to detect this background. We will present next briefly some of them. Let us stress here that we will not consider the possible clustering of neutrinos around massive objects, for simplicity. For more information, see the work from A.\ Ringwald and Y.\ Y.\ Y. Wong \cite{Ringwald:2004te} and the more
recent work from P.\ F.\ de Salas et.\ al. \cite{deSalas:2017wtt}.\\

\noindent{\bf Stodolsky Effect.} This effect, proposed by Stodolsky \cite{Stodolsky:1974aq}, considers the generation of an energy difference between the two electron helicity states in a ferromagnet due to the presence of the $\CNB$. Such energy shift will depend on the neutrino asymmetry. It has been shown that it is different for Dirac and Majorana neutrinos \cite{Duda:2001hd}. For the Dirac case, the energy difference created per species is proportional to 
\begin{align}
	\Delta E_a^D=\begin{cases}
					2\sqrt{2}G_F g_A \abs{\vec{\beta}_{\oplus}} (n_{\nu_L^a}-n_{(\nu_R^a)^c}), & {\rm for\ R},\\
					1.7\sqrt{2}G_F g_A\abs{\vec{\beta}_{\oplus}}\,\sqrt{\frac{m_a^\nu}{\xi\, 1.7\times 10^{-4}\ {\rm eV}}}  (n_{\nu_-^a}+n_{(\nu^a_-)^c}-n_{\nu_+^a}-n_{(\nu^a_+)^c}), & {\rm for\ NR},\\
				\end{cases}
\end{align}
where $\vec{\beta}_{\oplus}$ is the Earth velocity with respect to the neutrino background, $g_A$ is the axial SM coupling, $\xi_a$ is the chemical potential $\xi_a=\mu_a/T_\nu$, and R(NR) are for relativistic (non-relativistic) relic neutrinos. Therefore, any given neutrino can be relativistic or not since we are considering any possible value for the masses. For Majorana neutrinos \cite{Duda:2001hd},
\begin{align}
	\Delta E_a^M=\begin{cases}
					2\sqrt{2}G_F g_A \abs{\vec{\beta}_{\oplus}} (n_{\nu_L^a}-n_{\nu_R^a}), & {\rm for\ R},\\
					3.4\sqrt{2}G_F g_A\abs{\vec{\beta}_{\oplus}}\,\sqrt{\frac{m_a^\nu}{\xi_a\, 1.7\times 10^{-4}\ {\rm eV}}}  (n_{\nu_-^a}-n_{\nu_+^a}), & {\rm for\ NR}.\\
				\end{cases}
\end{align}
Considering the energy difference previously shown, we can conceptualize an experimental set up for the detection. Supposing that we have a magnetized spherical material, a ferromagnet, we find that the $\CNB$ will create a torque on the sample, given by \cite{Duda:2001hd,Strumia:2006db,Domcke:2017aqj}
\begin{align*}
	\tau=\frac{N \Delta E}{\pi},
\end{align*}
with $N$ the number of polarized electrons. Supposing Dirac and relativistic neutrinos, the $\CNB$ will produce an acceleration on the material \cite{Duda:2001hd,Domcke:2017aqj}
\begin{align}
	a_S&=\frac{N_A}{A}\frac{\Delta E}{\pi}\frac{\gamma}{R}\notag\\
	   &=10^{-27} \eta_\nu \left(\frac{\gamma}{10}\right)\left(\frac{100}{A}\right)\left(\frac{{\rm cm}}{R}\right)\left(\frac{\beta_{\oplus}}{10^{-3}}\right)\frac{{\rm cm}}{{\rm s^2}},
\end{align}
being $\gamma$ a geometrical factor related to the detector's moment of inertia, $A$ the atomic number, $R$ the radius of the material, and $\eta_\nu=\sum_a(n_{\nu_L^a}-n_{\nu_R^a})$ is the neutrino-antineutrino asymmetry number density. For non-relativistic Dirac and Majorana neutrinos, the acceleration can be one order of magnitude larger, at most \cite{Duda:2001hd}. Anyhow, this is an extremely small acceleration, faraway from the current sensitivity.\\

\noindent{\bf CNSN and a Cavendish-like torsion balance.} Relic neutrinos can be scattered coherently by a nucleus. This specific interaction, the Coherent Neutrino Scattering off Nuclei, will be studied deeply in the chapter \ref{cha:NeutrinoFloor}. Integrating the differential cross section for such process and considering small neutrino energies, one finds \cite{Duda:2001hd,Strumia:2006db,Domcke:2017aqj}
\begin{align}
	\sigma^{\nu_a} = \frac{G_F^2}{4\pi}[Q_V^{\rm SM}]^2 E_a^2,
\end{align}
where the neutrino energy is
\begin{align*}
	E_a=\begin{cases}
			3.15\, T_\nu & {\rm for\ R},\\
			m_a^\nu  & {\rm for\ NR},\\
		  \end{cases}
\end{align*}
with the factor $3.15$ coming from the thermal average over the Fermi-Dirac distribution function. This is a quite small cross section; however, relic neutrinos have macroscopic de Broglie wavelengths $\lambda$. Therefore, the cross section is also enhanced by the scattering on nuclei distributed in a volume with size $R\lesssim \lambda/2\pi$. In an experiment like a Cavendish torsion balance, we can detect the force exerted on a pendulum by the cosmic neutrino background. This will result in a acceleration \cite{Duda:2001hd}
\noindent
\begin{align}
	a_{\rm CNSN}	&=2\pi^2G_F^2 \sum_a n_{\nu_a}^0 [Q_V^{\rm SM}]^2{\cal N}^2\rho\begin{cases}
										\beta_{\oplus}^{\rm CMB}  & {\rm for\ R},\\
										\frac{m_a^\nu \beta_{\oplus}^{\rm CMB}}{3.15 T_\nu}  & {\rm for\ NR},\\
									\end{cases}
\end{align}
where ${\cal N}$ is the number of nuclei, $\rho$ is the mass density of the target and $\beta_{\oplus}^{\rm CMB}$ is the Earth velocity in the CMB frame. For non-relativistic neutrinos, we have \cite{Duda:2001hd,Strumia:2006db,Domcke:2017aqj}
\begin{align}
	a_{\rm CNSN}	\sim 10^{-28}\left(\frac{\sum_a n_{\nu_a}^0}{10^3 {\rm \ cm^{-3}}}\right)\left(\frac{10^{-3}}{\beta_{\oplus}^{\rm CMB}}\right)\left(\frac{\rho}{\rm g\ cm^{-3}}\right)\left(\frac{R}{1/(\sum_a m_a^\nu \beta_{\oplus}^{\rm CMB})}\right)^3\frac{{\rm cm}}{{\rm s^2}},
\end{align}
which is also a tiny acceleration. Let us also stress here that the previous value corresponds to Dirac neutrinos; for the Majorana case, we have an additional suppression since these particles only have axial couplings.\\

\noindent{\bf Scattering from Ultra-high-energy Cosmic Rays.} Another possibility to detect the relic neutrino background is by its interaction with Ultra-High-Energy (UHE) neutrinos, neutrinos with energy of ${\cal O}(10^{21})$ GeV \cite{Strumia:2006db},
\begin{align*}
	\nu_{\rm UHE} + \bar{\nu}_{\CNB} \tto Z^0  \tto {\rm hadrons};
\end{align*}
then, a detector could see either a $Z^0$ burst, as an excess of cosmic rays above the Greisen-Zatsepin-Kuzmin (GZK) cutoff, or a diminution of the UHE cosmic neutrino flux as consequence of the interaction with the $\CNB$. Nevertheless, the rates for both cases are small and beyond the current tecnology \cite{Strumia:2006db}.\\

\noindent{\bf Neutrino Capture.} A more promising manner to detect the $\CNB$ is through neutrino capture by a nuclei \cite{Weinberg:1962zza}. This capture will create a peak in the value of the neutrino masses. We will concentrate ourselves in the remaining of the chapter on this method, following the description given by the Long et.\ al.\ work \cite{Long:2014zva}. This approach is based on the {\it threshold-less} interaction
\begin{align*}
	\nu_a+\!\, ^3{\rm H}\tto\, ^3{\rm He}+e^-,
\end{align*}
to capture a relic neutrino. We have chosen tritium as target material since it will be used by the Princeton Tritium Observatory for Light, Early-Universe, Massive-Neutrino Yield (PTOLEMY) experiment, which is being developed \cite{Betts:2013uya}. The capture rate of the $\CNB$ will be computed in detail, given its peculiarities compared to rates and cross sections computed with ultrarelativistic neutrinos. Let us first compute the neutrino capture by a neutron,
\begin{align*}
	\nu_a + n \tto p + e^-
\end{align*}
where $\nu_a$ is the $a$-th neutrino mass eigenstate field. The effective lagrangian for this process will be the Fermi lagrangian,
\begin{align}
	\mathscr{L}_{\rm eff}=-\frac{G_F}{\sqrt{2}} U_{ud} \widetilde{U}_{ea}\, [\bar{e}\gamma^\mu(1-\gamma^5)\nu_a][\bar{p}\gamma_\mu(g_V-g_A\gamma^5)n]+{\rm h.c.}.
\end{align}
Here we have $U_{ud}$ and $\widetilde{U}_{ea}$, the CKM and PMNS matrix elements relevant for the process; $g_V, g_A$ are the nuclear form factors, related to the vector and axial nucleon structures. Computing the amplitude squared for this process and summing over the spin indexes, we can obtain the total rate. However, we need to take into account that neutrinos are prepared in an specific helicity state while the neutron, proton and electron are not. Therefore, we will just sum over the neutron, proton and electron spins. We use the usual completeness relation for $i=n,p,e^-$
\begin{align*}
	\sum_{s_i} u_{i}(p_i,s_i) \overline{u_{i}}(p_i,s_i)=\slashed{p}_i+m_i
\end{align*}
while for the relic neutrino we use the relation for a fixed helicity ($h_\nu=\pm 1/2$)
\begin{align*}
	u_{a}(p_{a},s_{a}) \overline{u_{a}}(p_{a},s_{a}) = \frac{1}{2} (\slashed{p}_a+m_a^\nu)(1+2h_\nu\gamma^5\slashed{s}_a),
\end{align*}
being the spin four-vector $s_a^\mu$ given by
\begin{align*}
	s_a^\mu=\left\{\frac{\abs{\vec{p}_a}}{m_a^\nu},\frac{E_a}{m_a^\nu}\frac{\vec{p}_a}{\abs{\vec{p}_a}}\right\}.
\end{align*}
Now, we can determine the differential cross section for the neutrino capture in a standard way. We get in the rest frame of the neutron
\begin{align}\label{eq:SMdiffCrossSec}
	\frac{d\sigma_a(h_\nu)}{d\cos\theta}=\frac{G_F^2}{4\pi} \abs{U_{ud}}^2|\widetilde{U}_{ea}|^2 F_Z(E_e) \frac{m_p}{m_n}\frac{E_e p_e}{v_a}\esp{{\cal A}(h_\nu)(g_V^2+3g_A^2)+{\cal B}(h_\nu)(g_V^2-g_A^2)v_e\cos\theta}
\end{align}
where $v_a,v_e$ are the neutrino and electron velocities, respectively; $E_e,p_e$ is the energy and momentum of the electron, $\theta$ the angle between the electron and the neutrino. $F_Z(E_e)$ is the Fermi function which takes into account the enhancement of the cross section due to the electromagnetic attraction between the proton and electron,
\begin{align*}
	F_Z(E_e)=\frac{2\pi\eta}{1-e^{-2\pi\eta}},
\end{align*}
with $\eta=Z \alpha_{\rm EM} E_e/p_e$, $\alpha_{\rm EM}$ the fine structure constant.
The functions ${\cal A}(h_\nu)$ and ${\cal B}(h_\nu)$ appearing in \ref{eq:SMdiffCrossSec} are given by
\begin{align}
	{\cal A}(h_\nu)&=1-2h_\nu v_a,\qquad	{\cal B}(h_\nu)=v_a-2h_\nu.
\end{align}
We see here that if neutrinos are ultrarelativistic, $v_a\sim 1$, ${\cal A},{\cal B}$ are zero for right-helical neutrinos, making their capture impossible, while for left-helical ${\cal A}={\cal B}=2$. This is actually an expected result; let us remember that, in the massless limit, chirality and helicity coincide and only left-chiral neutrinos interact weakly. On the other hand, if neutrinos are non-relativistic we have that ${\cal A}(\pm 1/2) = \mp {\cal B}(\pm 1/2) =1$, showing the possibility of detecting both kinds of helical states. The integration in the angle is trivial, and we can obtain the capture cross section multiplied by the neutrino velocity, which is the relevant quantity for the $\CNB$ capture rate,
\begin{align}
	\sigma_a(h_\nu)v_a=\frac{G_F^2}{2\pi} \abs{U_{ud}}^2|\widetilde{U}_{ea}|^2 F_Z(E_e) \frac{m_p}{m_n}E_e p_e\, {\cal A}(h_\nu)(g_A^2+3g_V^2).
\end{align}
In the case of interest, the neutrino capture of a tritium nuclei, we simply need to make the substitutions of $m_n\tto m_{^3 {\rm H}}$ and $m_p\tto m_{^3{\rm He}}$. Notice that for simplicity we are not considering the modification of the nuclear form factors, as done in \cite{Ludl:2016ane}\footnote{We will not consider the vector and axial form factor for tritium as we do not have defined values for scalar and tensor ones. So, to avoid having two different sets of parameters, we will use only neutron form factors, as done in \cite{Long:2014zva}}. This will introduce a difference of $\sim 6\%$ with the values presented in \cite{Ludl:2016ane}. The total capture rate expected in a sample of tritium is the sum over the cross section for each of the three mass eigenstates ($j=1,2,3$) weighted by the appropriate flux
\begin{align}\label{eq:caprate}
	\Gamma_{\CNB} = N_T\sum_{a=1}^3\,\esp{\sigma_{a} (+1/2)v_a n_{\nu_+^a}+\sigma_a(-1/2)v_a n_{\nu_-^a}}
\end{align}
where $N_T$ is the number of nuclei present in the sample. In the case of non-relativistic neutrinos, we get a simpler relation,
\begin{align}
	\Gamma_{\CNB} = N_T \bar\sigma\sum_{a=1}^3|\widetilde{U}_{ea}|^2[n_{\nu_+^a} + n_{\nu_-^a}],
\end{align}
being
\begin{align*}
	\bar\sigma = \frac{G_F^2}{2\pi} \abs{U_{ud}}^2 F_Z(E_e) \frac{m_{^3{\rm He}}}{m_{^3 {\rm H}}}\,E_e p_e\, (g_A^2+3g_V^2)\approx 4.05\times 10^{-45}\ {\rm cm}^2.
\end{align*}
\newpage
\noindent Applying these results to the Dirac and Majorana cases we have that
\begin{align}
	\Gamma_{\CNB}^{\rm D} = N_T \bar\sigma n_0,
\end{align}
for Dirac neutrinos, while
\begin{align}
	\Gamma_{\CNB}^{\rm M} = 2 N_T \bar\sigma n_0,
\end{align}
for the Majorana case. Evidently,
\begin{align}
	\boxed{\Gamma_{\CNB}^{\rm M} = 2 \Gamma_{\CNB}^{\rm D},}
\end{align}
which shows that the capture rate of Majorana neutrinos is twice as the Dirac case. This can be explained keeping in mind that the $\CNB$ in the Dirac case consists of left-helical neutrinos and right-helical antineutrinos. However, the right-helical antiparticle states cannot be captured because the process involving an antineutrino $\bar{\nu} + p \tto n + e^+$ is kinematically forbidden. Thus, Dirac neutrinos have only half of the $\CNB$ abundance available to be captured. In the Majorana case, we will have left- and right-helical neutrinos, which can be captured. This is a crucial result since, in principle, the neutrino capture experimental technique not only can be used to detect the neutrino background, but also to shed some light on the neutrino nature. Numerically, we have that
\begin{align}
	\Gamma_{\CNB}^{\rm M} = 85.73\ [{\rm kg\ yr}]^{-1}, \qquad \Gamma_{\CNB}^{\rm D} = 42.87\ [{\rm kg\ yr}]^{-1},
\end{align}
which corresponds approximately to the values given in \cite{Long:2014zva} when the sample is composed by $100$ g of material, as in the case of PTOLEMY \cite{Betts:2013uya}.\\ 

\begin{figure}[t!]
    \centering 
        \subfloat[$\Delta = 10^{-2}$ eV]{\includegraphics[width=\columnwidth]{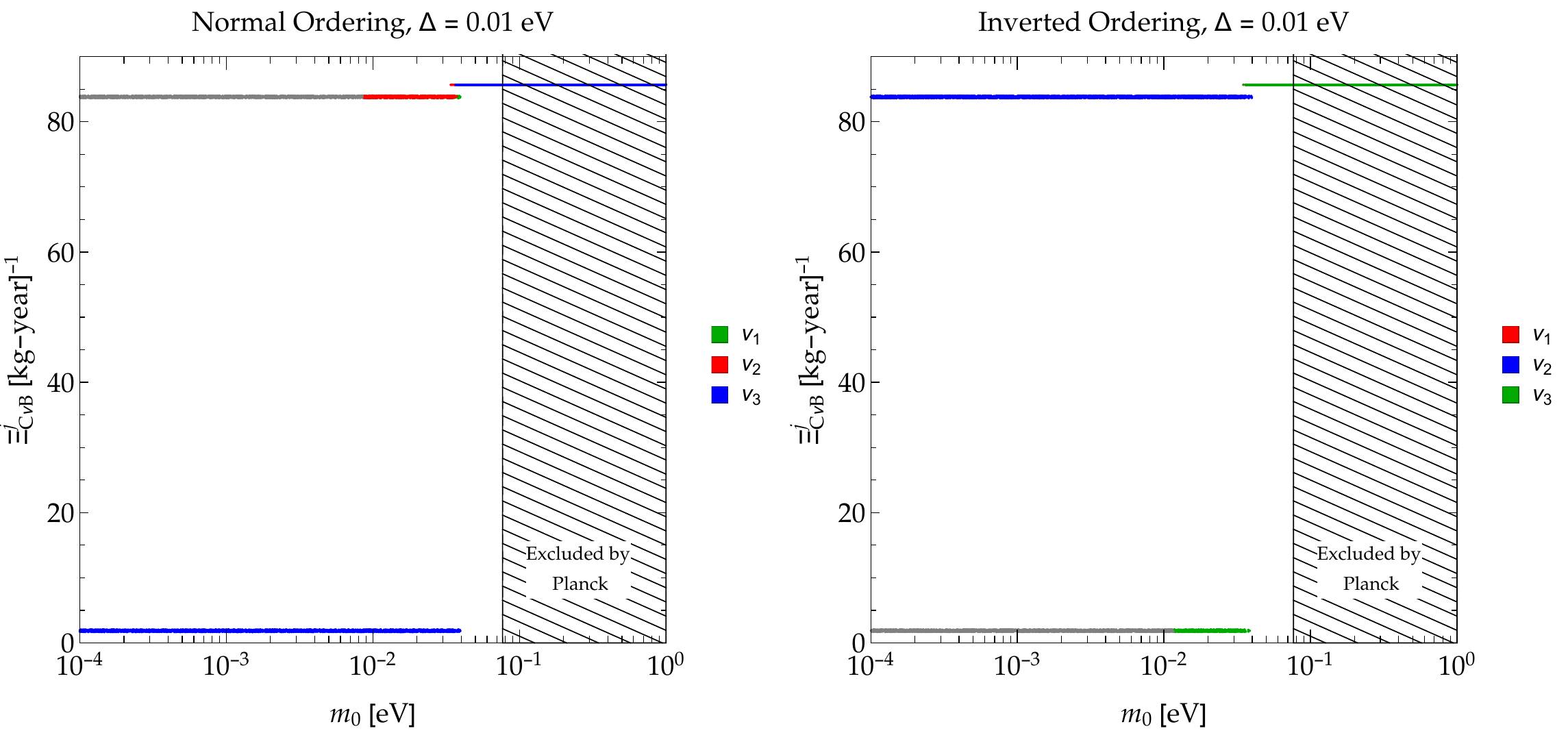}\label{fig:Xivsm0a}}\\
        \subfloat[$\Delta = 10^{-3}$ eV]{\includegraphics[width=\columnwidth]{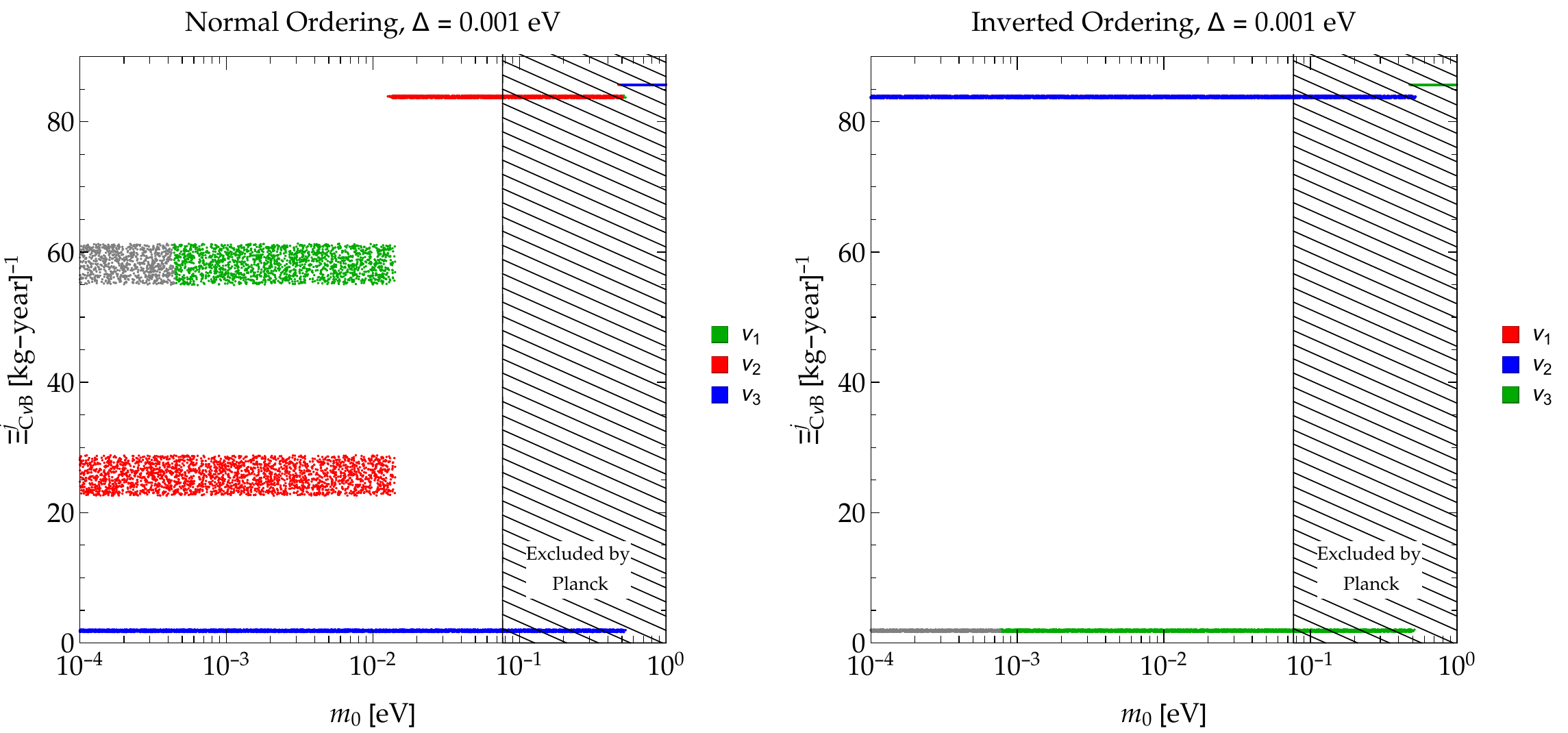}\label{fig:Xivsm0b}}
        \caption{\label{fig:Xivsm0} Dependence of the discrimination function on the lightest neutrino mass for two different values of the FWHM and for each mass eigenstate, $\nu_1$ (green), $\nu_2$ (red), $\nu_3$ (blue). The gray points correspond to the regions that are hidden in the beta decay background. The shaded region is excluded by the Planck limit on the sum of neutrino masses \protect\cite{Ade:2015xua}.}
\end{figure}

We should ask ourselves at this point how would be the signal in the neutrino capture rate, and what are the main difficulties of this method. PTOLEMY intends to detect relic neutrinos by the measurement of the electron created in the process. Nonetheless, tritium atoms undergo beta decay, in which the electrons emitted have a wide energy spectrum. Thus, one should be able to discriminate electrons from neutrino capture from the beta decay electrons. From a kinematic point of view, electrons produced by relic neutrinos will have a definite energy \cite{Long:2014zva}
\begin{align}
	E_e^{\CNB}\cong m_e+K_{\rm end}^0+2m_a^\nu
\end{align}
where $K_{\rm end}^0$ corresponds to the beta decay endpoint energy. This shows that relic neutrinos can produce a peak in an energy bigger than the beta endpoint one, making it possible to discriminate them. Nevertheless, this is a questionable affirmation since it does not take into account the finite resolution that a real detector has. In order to do a more realistic study, we will convolute the $\CNB$ capture rate with a Gaussian resolution function \cite{Long:2014zva}
\begin{align}
	\frac{d\Gamma_{\CNB}}{dE_e}=\frac{1}{\sqrt{2\pi\sigma^2}}\int_{-\infty}^\infty\, dE_e^\prime \,\Gamma_{\CNB} \,\exp\left[-\frac{(E_e^\prime-E_e)^2}{2\sigma^2}\right]\,\delta(E_e^\prime-m_e+K_{\rm end}^0+2m_a^\nu).
\end{align}
Furthermore, we will need the beta decay rate for tritium. We will use the more recent result from \citep{Ludl:2016ane},
\begin{align}
	\frac{d\Gamma_{\beta}}{dE_e^\prime}=\sum_{a=1}^3\,\left.\frac{d\Gamma_{\beta}}{dE_e^\prime}\right|_{\overline{\nu}_a}\Theta(E_e^{\rm max}-E_e^\prime)\Theta(m_{^3{\rm H}}-m_{^3{\rm He}}-m_e-m_a^\nu),
\end{align}
where the rate for one neutrino mass eigenstate is
\begin{align}
	\left.\frac{d\Gamma_{\beta}}{dE_e^\prime}\right|_{\overline{\nu}_a}=\frac{G_F^2 \abs{U_{ud}}^2 |\widetilde{U}_{ea}|^2}{4\pi^2}\frac{Q(E_e^\prime)}{m_{^3{\rm H}}}\left[A+BE_e^\prime+CP(E_e^\prime)+DE_e^{\prime\, 2}+F\left(P^2(E_e^\prime)+\frac{1}{3}Q^2(E_e^\prime)\right)\right],
\end{align}
with the definitions \citep{Ludl:2016ane}
\begin{subequations}
	\begin{align}
		P(E_e)&=-\frac{(m_{^3{\rm H}}-E_e)(E_e m_{^3{\rm H}}-\kappa)}{m_{^3{\rm H}}^2-2E_e m_{^3{\rm H}}+m_e^2},\\
		Q(E_e)&=\frac{\abs{\vec{p}_e}\sqrt{(E_e m_{^3{\rm H}}-\kappa-m_a^{\nu\, 2})^2-m_{^3{\rm He}}m_a^{\nu\, 2}}}{m_{^3{\rm H}}^2-2E_e m_{^3{\rm H}}+m_e^2},\\
		A&= m_{^3{\rm H}} m_{^3{\rm He}}(g_V^2-g_A^2)(m_{^3{\rm H}}^2-m_{^3{\rm He}}^2+m_e^2+m_a^{\nu\, 2}),\\
		B&= m_{^3{\rm H}}\llav{(g_V-g_A)^2(m_{^3{\rm H}}^2-m_{^3{\rm He}}^2+m_e^2-m_a^{\nu\, 2})-2m_{^3{\rm H}} m_{^3{\rm He}}(g_V^2-g_A^2)},\\
		C&= m_{^3{\rm H}}\llav{(g_V+g_A)^2(m_{^3{\rm H}}^2-m_{^3{\rm He}}^2-m_e^2+m_a^{\nu\, 2})-2m_{^3{\rm H}} m_{^3{\rm He}}(g_V^2-g_A^2)},\\
		D&=-2 m_{^3{\rm H}}^2(g_V-g_A)^2,\\
		F&=-2 m_{^3{\rm H}}^2(g_V+g_A)^2,
	\end{align}
\end{subequations}
and
\begin{align*}
	\kappa=\frac{1}{2}(m_{^3{\rm H}}^2-m_{^3{\rm He}}^2+m_e^2+m_a^{\nu\, 2}).
\end{align*}
The beta decay rate will be also convoluted with a Gaussian resolution function,
\begin{align}
	\frac{d\Gamma_{\beta}}{dE_e}=\frac{1}{\sqrt{2\pi\sigma^2}}\int_{-\infty}^\infty\, dE_e^\prime \, \frac{d\Gamma_{\beta}}{dE_e^\prime}\,\exp\left[-\frac{(E_e^\prime-E_e)^2}{2\sigma^2}\right].
\end{align}
To estimate the region that can be distinguished from the beta decay background, we will compute the number of events around the energy created by the relic neutrino capture as \cite{Long:2014zva}
\begin{subequations}
	\begin{align}
		{\cal N}_{\CNB}(\Delta)&=\int_{E_e^{\CNB}-\Delta/2}^{E_e^{\CNB}+\Delta/2}\, dE_e \frac{d\Gamma_{\CNB}}{dE_e},\\
		{\cal N}_{\beta}(\Delta)&=\int_{E_e^{\CNB}-\Delta/2}^{E_e^{\CNB}+\Delta/2}\, dE_e \frac{d\Gamma_{\beta}}{dE_e},
	\end{align}
\end{subequations}
where $\Delta$ corresponds the full width at half maximum (FWHM) of the Gaussian function, related to the standard deviation $\sigma$ as
\begin{align*}
	\Delta = \sqrt{8\ln 2}\, \sigma.
\end{align*}
Therefore, to distinguish the $\CNB$ signal from the beta decay background, we will consider the ratio,
\begin{align}
	r_{\CNB}=\frac{{\cal N}_{\CNB}(\Delta)}{\sqrt{{\cal N}_{\beta}(\Delta)}},
\end{align}
in such a way that the signal can be discriminated when $r_{\CNB}\geq 5$. However, we are laying aside the fact that both signal and background depend on the values of the neutrino masses. Thus, we should understand the regions where the mass eigenstates are degenerated or not, and see that if there is a possible differentiation among each contribution. This can, in principle, be possible as each mass eigenstate is captured independently from the others. Even though we computed the total neutrino capture as the sum over the three mass eigenstates, we can consider each contribution in \eqref{eq:caprate} as
\begin{align}\label{eq:capratej}
	\Gamma_{\CNB}^a = N_T\esp{\sigma_{a} (+1/2)v_a n_{\nu_+^a}+\sigma_a(-1/2)v_a n_{\nu_-^a}};
\end{align}
thus, we also can compute the number of events after convoluting with a proper Gaussian
\begin{align*}
		{\cal N}_{\CNB}^a(\Delta)&=\int_{E_e^{\CNB\, a}-\Delta/2}^{E_e^{\CNB\, a}+\Delta/2}\, dE_e \frac{d\Gamma_{\CNB}^a}{dE_e}.
\end{align*}
with $E_e^{\CNB\, a}$ the energy of the electrons produced by each mass eigenstate. Then, we have two tasks, first see if it possible to discriminate the $\CNB$ signal over the background and, second, comprehend if a mass eigenstate can be distinguished from the others. The first discrimination can be achieved by considering the ratio $r_{\CNB}^j$ for each eigenstate; for the second one, we should ask ourselves how to differentiate two Gaussians according to their mean values and standard deviations. We will use the Bhattacharya distance between two Gaussians \cite{MR0010358} as a basic condition for discrimination, given by
\begin{align}
	D_B(p,q)=\frac{1}{4}\ln\llav{\frac{1}{4}\corc{\frac{\sigma_p^2}{\sigma_q^2}+\frac{\sigma_q^2}{\sigma_p^2}+2}}+\frac{1}{4}\frac{(\mu_p-\mu_q)^2}{\sigma_p^2+\sigma_q^2},
\end{align}
with $p,q$ two different Gaussians.
We will define a discrimination function to determinate the separation of each neutrino capture rate contribution
\begin{align}\label{eq:DefXi}
	\Xi^a_{\CNB}=\sum_{b=1}^3\llav{1-\Theta\left(D_B\left(\frac{d\Gamma_{\CNB}^a}{dE_e},\frac{d\Gamma_{\CNB}^b}{dE_e}\right)-4.5\right)}\Gamma_{\CNB}^b.
\end{align}
This function has been constructed to fulfil the following purpose. When the mass eigenstates are degenerated, the function will give the value of the total neutrino capture. This is clear as the Bhattacharya distance in such case goes to zero, making the Heaviside theta to be null. Then, if the third mass eigenstate\footnote{The third neutrino can be separated first from the other two as the quadratic mass difference $|\Delta m_{3l}^2|$ ($l=1,2$) is larger than $\Delta m_{21}^2$.} has a mass different enough to be distinguished from the other two masses, the $\Xi^3_{\CNB}$ will correspond to the value of neutrino capture for $\nu_3$, while $\Xi^1_{\CNB}$ and $\Xi^2_{\CNB}$ will be equal to the sum of the rates of $\nu_1$ and $\nu_2$. The last possibility consists in the separation of the three eigenstates; thus, for each case, $\Xi^a_{\CNB}$ will give the value of the individual neutrino capture rate. In the definition of $\Xi^a_{\CNB}$, equation \eqref{eq:DefXi}, we chose a value of $4.5$ as a parameter to distinguish the two Gaussians. This number has been adopted noting that the Bhattacharya distance for two functions with equal standard deviation is
\begin{align*}
	\left.D_B(p,q)\right|_{\sigma_p=\sigma_q}=\frac{(\mu_p-\mu_q)^2}{8\sigma_p^2},
\end{align*}
so, to completely separate two Gaussians, we require that the difference of their mean values to be at least equal to $6\sigma$, 
 \begin{align*}
	\mu_p-\mu_q=6\sigma^2.
\end{align*}
Therefore,
\begin{align}
	\left.D_B(p,q)\right|_{\sigma_p=\sigma_q}=\frac{36}{8}=4.5. 
\end{align}
Let us note that this definition does not take into account the cases in which the Gaussians are superimposed but distinguishable since it depends on the number of events and the analysis performed on the data. Therefore, we will consider only when the Gaussian functions are completely separated. To understand the behaviour the $\Xi^a_{\CNB}$ function, we performed a scan over the neutrino oscillations parameters at $3\sigma$ level. In figure \ref{fig:Xivsm0}, we show the discrimination function dependence on the lightest neutrino mass $m_0$, for the mass eigenstates $\nu_1$ (green), $\nu_2$ (red), $\nu_3$ (blue), for both types of mass orderings and two values of $\Delta$. The gray points can not be distinguished from the beta decay background.\\

We see that for both orderings and $\Delta=10^{-2}$ eV and for $m_0\gtrsim 0.03$ eV that the relic neutrinos will appear as a single peak while for $m_0\lesssim 0.03$ eV it could be possible to differentiate two peaks. For a Normal Ordering and the smaller example of $\Delta=10^{-3}$ eV there are two possibilities: for $m_0\gtrsim 0.015$ eV, it would be possible to distinguish two peaks whilst, for $m_0\lesssim 0.015$ eV the three peaks could be discriminated. For the Inverted Ordering case, there will be still two peaks. This is related to the mass difference in this ordering, as the neutrinos $\nu_1$ and $\nu_2$ become almost degenerated when $m_3\tto 0$.\\

\begin{figure}[t!]
    \centering
        \subfloat[$m_0 = 0.075$ eV]{\includegraphics[width=0.9\columnwidth]{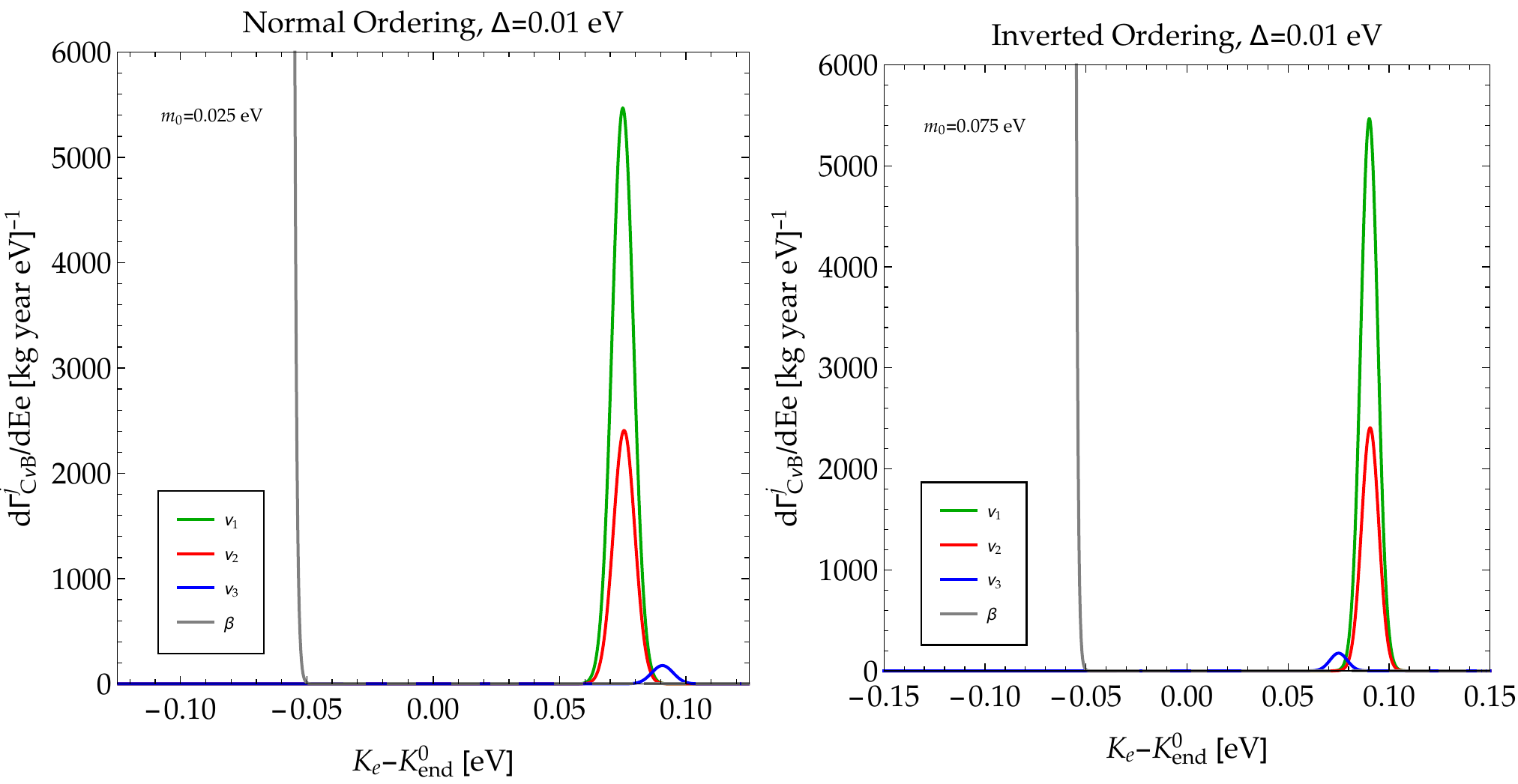}\label{fig:dGammadEe1em2a}}\\
        \subfloat[$m_0 = 0.025$ eV]{\includegraphics[width=0.9\columnwidth]{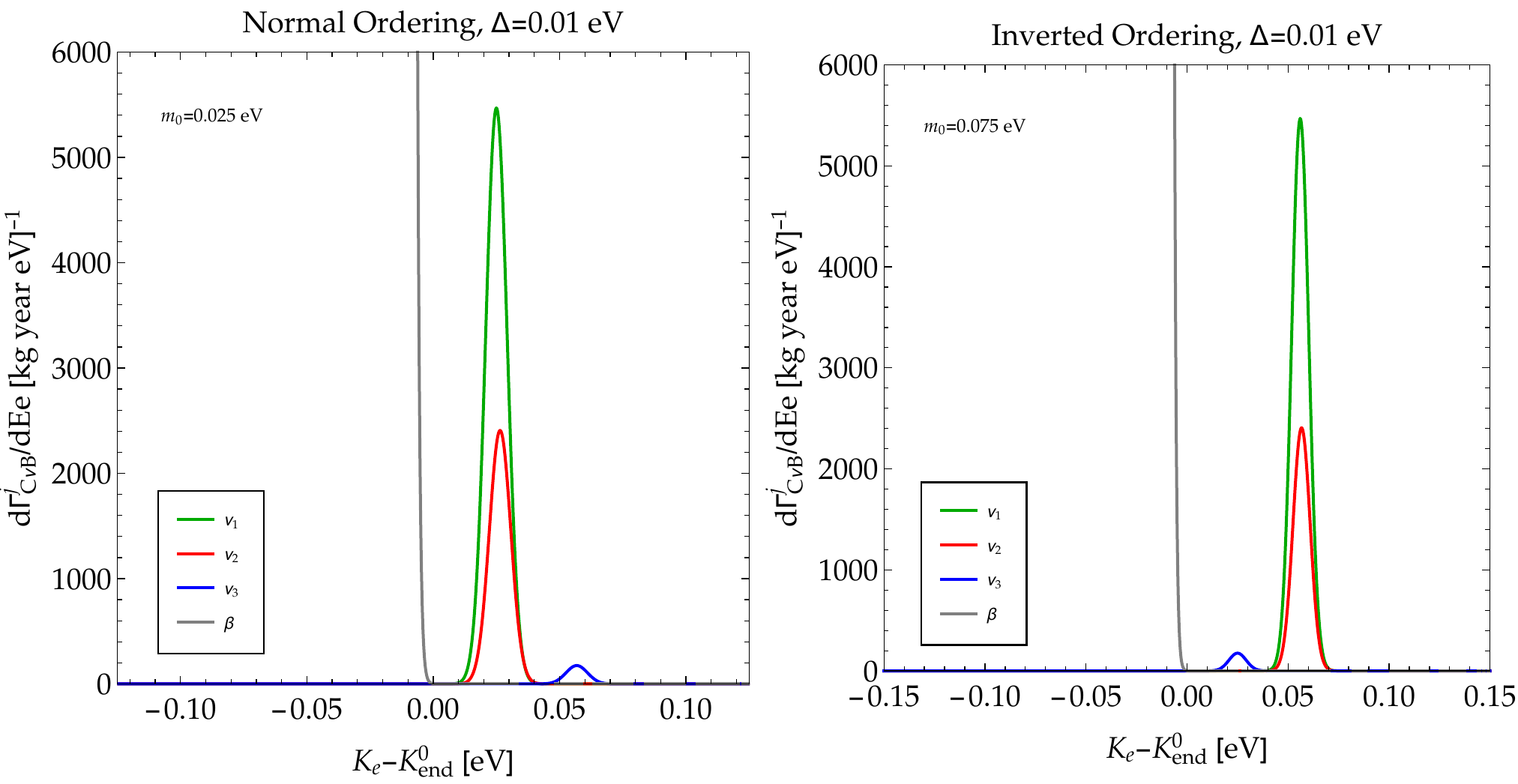}\label{fig:dGammadEe1em2b}}
        \caption{Simulated spectra of the electrons created by the relic neutrino capture for $\Delta=0.01$ eV for each mass eigenstate, $\nu_1$ (green), $\nu_2$ (red), $\nu_3$ (blue). The gray line corresponds to the beta decay background. We consider two values for the lightest neutrino mass. a) $m_0 = 0.075$ eV. b) $m_0 = 0.025$ eV.}
  	\label{fig:dGammadEe1em2}
\end{figure}

\begin{figure}[t!]
    \centering
        \subfloat[$m_0 = 0.025$ eV]{\includegraphics[width=0.9\columnwidth]{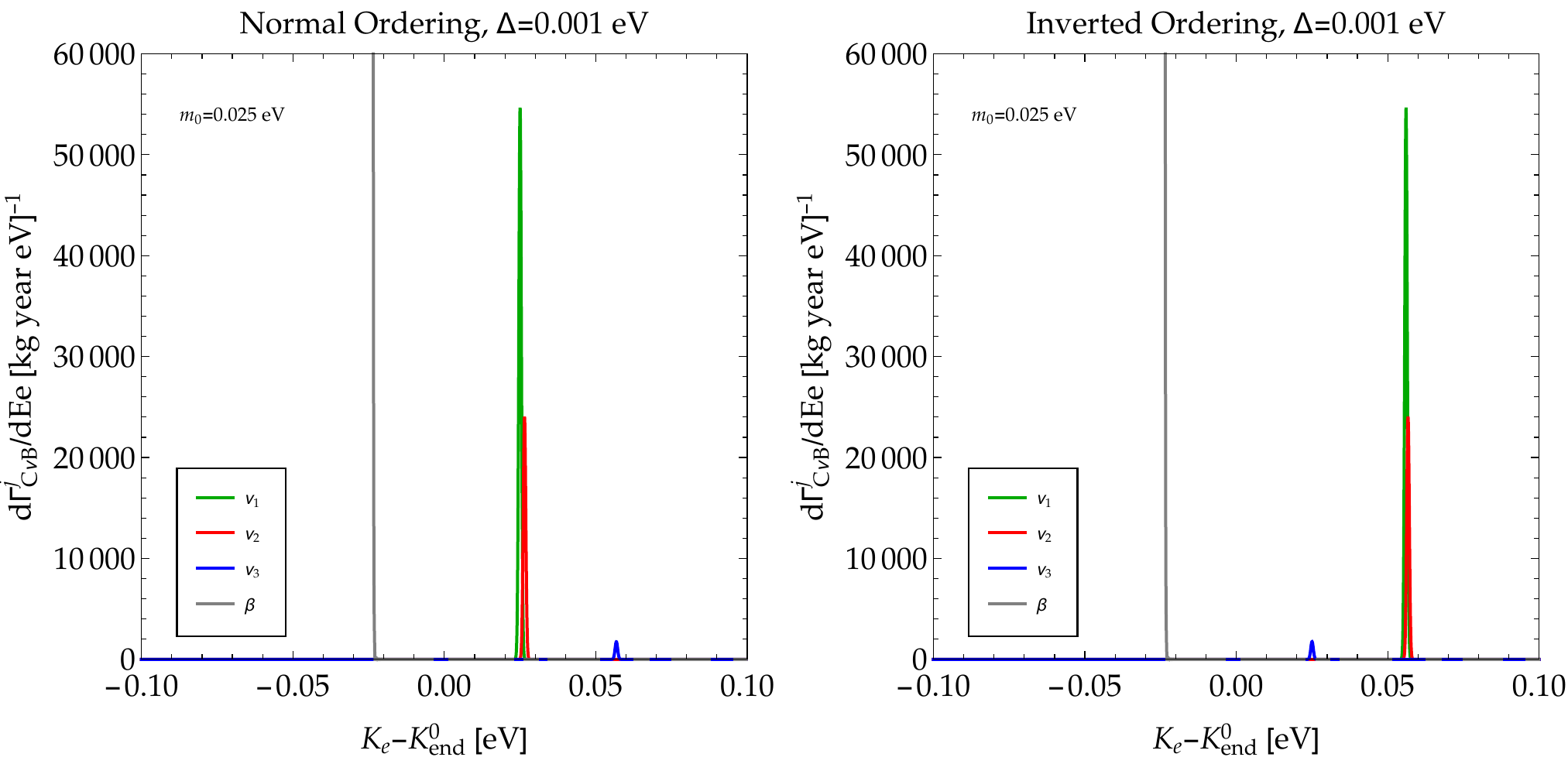}\label{fig:dGammadEe1em3a}}\\
        \subfloat[$m_0 = 0.005$ eV]{\includegraphics[width=0.9\columnwidth]{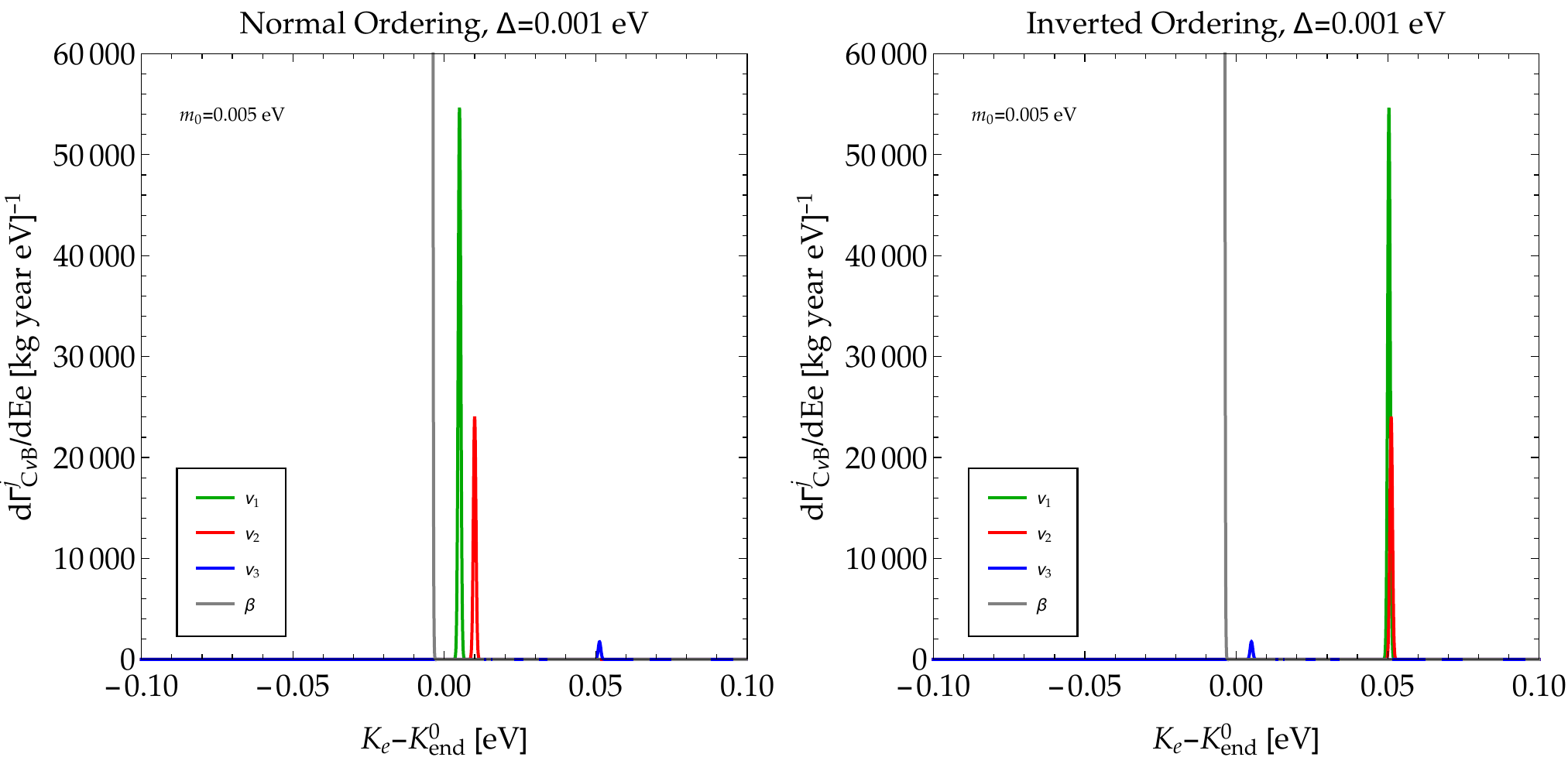}\label{fig:dGammadEe1em3b}}
        \caption{Simulated spectra of the electrons created by the relic neutrino capture for $\Delta=0.001$ eV for each mass eigenstate, $\nu_1$ (green), $\nu_2$ (red), $\nu_3$ (blue). The gray line corresponds to the beta decay background. We consider two values for the lightest neutrino mass. a) $m_0 = 0.075$ eV. b) $m_0 = 0.025$ eV.}
  	\label{fig:dGammadEe1em3}
\end{figure}

To illustrate how could be the spectra observed by PTOLEMY, we show two examples in figures \ref{fig:dGammadEe1em2} and \ref{fig:dGammadEe1em2} for the values of $\Delta$ considered. In figure \ref{fig:dGammadEe1em2} we show simulated spectra for two values of masses $m_0 = 0.075, 0.025$ eV and the beta decay spectrum (gray). In the first case, figure \ref{fig:dGammadEe1em2a}, we see that in both orderings the spectra for each mass eigenstate can not be distinguished as they are superimposed. Thus, in this case we expect a unique peak for the electrons produced by the $\CNB$. For the lightest neutrino mass of $m_0 = 0.025$ eV, figure \ref{fig:dGammadEe1em2b}, the peak related with the $\nu_3$ eigenstate can be in principle differentiated from the other two neutrinos. Nevertheless, such peak is tiny compared to the other two. This is due to the dependence of the neutrino capture with the PMNS matrix. Explicitly, we have
\begin{align*}
	\Gamma_{\CNB}^a\propto |\widetilde{U}_{ea}|^2,
\end{align*}
where \cite{Esteban:2016qun}
\begin{align*}
	\abs{\widetilde{U}_{e1}}^2&\approx 0.68,\\
	\abs{\widetilde{U}_{e2}}^2&\approx 0.3,\\
	\abs{\widetilde{U}_{e3}}^2&\approx 0.02.\\
\end{align*}
This shows that the neutrino capture rate for the third mass eigenstate is very small, and it can be lost in other background that could appear in PTOLEMY. This is further visualized in figure \ref{fig:dGammadEe1em3} for the extreme resolution $\Delta=0.001$ eV. For the mass of $m_0 = 0.025$ eV, figure \ref{fig:dGammadEe1em2a}, it is clear that $\nu_3$ neutrinos can be distinguished, but the size of the peak is extremely small. On the other hand, in the case $m_0 = 0.005$ eV, figure \ref{fig:dGammadEe1em2b}, we see that the three peaks can be separated in the Normal Ordering, but such separation is not possible in the Inverted Ordering. This is related to the degeneracy that appears in such ordering, as already pointed out.\\ 

Furthermore, as explained in \cite{Long:2014zva}, the detection in the Normal Ordering is harder than in the Inverted case, since in the Inverted case $\nu_1$ and $\nu_2$ have the largest rate and they are more separated from the beta decay background. While in the Normal Ordering case occurs the opposite, the $\nu_1$ and $\nu_2$ are always closer to the beta decay background. This also can be seen in figure \ref{fig:Xivsm0} since the region hidden in the background is different for each ordering; then, in the Inverted case, the gray points are only in the $\nu_3$ value of the discrimination function, while in the Normal one, they are in the largest contribution.

\newpage

The expected resolution of PTOLEMY is $\Delta =0.15$ eV \cite{Betts:2013uya}; thus, it will explore the initial part of the mass spectrum. In that case, the neutrinos are degenerated, and the experience will search a single peak. Accordingly, we will consider from now on the degenerated mass spectrum. Also, this implies that neutrinos should be non-relativistic in the $\CNB$. Next, we will consider the consequences of this fact in the presence of beyond SM physics.\\

\newpage

%%%%%%%%%%%%%%%%%%%%%%%%%%%%%%%%%%%%%%%%%%%%%%%%%%%%%%%%%%%%%%%%%%%%%%%%%%%%%%%%%%%%%%%%%%%%%%%%%%%%%%%%%%%%%%%%%%%%%%%%%%%%%%
\section{Parametrization of the Beyond SM Physics}
%%%%%%%%%%%%%%%%%%%%%%%%%%%%%%%%%%%%%%%%%%%%%%%%%%%%%%%%%%%%%%%%%%%%%%%%%%%%%%%%%%%%%%%%%%%%%%%%%%%%%%%%%%%%%%%%%%%%%%%%%%%%%%

%%%%%%%%%%%%%%%%%%%%%   table  %%%%%%%%%%%%%%%%%%%%%%%%%%%5
\begin{table}[t]
	\centering 
	\caption{Dimension-six operators considered in this work; $l_L,q_L$ are the fermion doublets, while $u_R,d_R,e_R,\nu_R$ are
			 the fermion singlets. The $\tau^A$ are the generators of SU(2)$_L$ and $\varepsilon_{ij}=(i\sigma_2)_{ij}$ is the 
			 totally antisymmetric symbol, with $\varepsilon_{12}=+1$.}
	\label{tab:d6Oper}
	\begin{tabular}{lll}
	\toprule\toprule
	\multicolumn{2}{c}{Four-fermion Operators} &  \multicolumn{1}{c}{Vertex Corrections}\\ 
	\cmidrule(lr){1-2}\cmidrule(lr){3-3}
	\multicolumn{1}{c}{$Q_{\nu_L}^{(6)}$} & \multicolumn{1}{c}{$Q_{\nu_R}^{(6)}$} &  \multicolumn{1}{c}{$Q_{\Phi}^{(6)}$} \\
	\midrule
		$Q_1=(\overline{l_L}e_R)(\overline{d_R}q_L)$ & $Q_5=(\overline{l_L}\nu_R)\varepsilon(\overline{q_L}d_R)$ & $Q_9=i(\Phi^T \varepsilon D_\mu\Phi)(\overline{u_R}\gamma_\mu d_R)$\\ 
		$Q_2=(\overline{l_L}e_R)\varepsilon(\overline{q_L}u_R)$ & $Q_6=(\overline{\nu_R}l_L)(\overline{q_L}u_R)$ & $Q_{10}=i(\Phi^T \varepsilon D_\mu\Phi)(\overline{\nu_R}\gamma^\mu e_R)$ \\
		$Q_3=(\overline{l_L}\gamma^\mu \tau^A l_L)(\overline{q_L}\gamma_\mu \tau^A q_L)$ & $Q_7=(\overline{e_R}\gamma^\mu \nu_R)(\overline{u_R}\gamma_\mu d_R)$ & $Q_{11}=(\Phi^\dagger i \overset{\text{\tiny$\longleftrightarrow$}}{D_\mu^a} \Phi)(\overline{q_L}\gamma_\mu \tau^A q_L)$\\ 
		$Q_4=(\overline{l_L}\sigma^{\mu\rho}e_R)\varepsilon (\overline{q_L}\sigma_{\mu\rho}u_R)$ & $Q_8=(\overline{l_L}\sigma^{\mu\rho}\nu_R)\varepsilon(\overline{q_L}\sigma_{\mu\rho}d_R)$  & $Q_{12}=(\Phi^\dagger i \overset{\text{\tiny$\longleftrightarrow$}}{D_\mu^a} \Phi)(\overline{l_L}\gamma^\mu \tau^A l_L)$\\ 
	\bottomrule
	\end{tabular}
\end{table}
%%%%%%%%%%%%%%%%%%%%%%%%%%%%%%%%%%%%%%%%%%%%%%%%%%%%%%%%%%%%

To obtain the previous relic capture rate we considered only the Fermi lagrangian, i.e. we considered only SM interactions. Nevertheless, it is possible that NSI play a role in neutrino capture since the $\CNB$ is composed by non-relativistic particles. Thus, we can expect contributions that are usually negligible for an ultrarelativistic neutrino to be sizable here. In order to maintain the discussion as general as possible, we will consider the Effective Field Theory approach. The first operator we could think about, the dimension-five \emph{Weinberg} operator, will give only mass to neutrinos, as we have seen in chapter \ref{cap:nuMaj}. So, we will start considering dimension-six operators which are invariant under the SM gauge group, SU($2$)$_L\times$U($1$)$_Y$, but also including right-handed neutrinos \cite{Grzadkowski:2010es}.\\ 

The general lagrangian we will consider is given by
\begin{align}\label{eq:d6OPer}
	\mathscr{L}_{\rm BSM}=\mathscr{L}_{\rm SM}^{(4)}+ \mathscr{L}_{m_\nu} + \frac{1}{\Lambda^2}\sum_{k = 1} ^{12} c_k^{(6)} Q_k^{(6)},
\end{align}
where $\Lambda$ is an energy scale, $c_k^{(6)}$ are the dimensionless Wilson coefficients, $\mathscr{L}_{m_\nu}$ is 
the neutrino mass lagrangian. The set of operators with left- and 
right-handed neutrinos $Q_k^{(6)}=\{Q_k^{(6)}(\nu_L),Q_k^{(6)}(\nu_R)\}$ is given in table \ref{tab:d6Oper}.
The relevant terms for the determination of the relic neutrino capture can be obtained from the complete lagrangian \ref{eq:d6OPer},
\begin{align}\label{eq:CNBL}
	\mathscr{L}_{\rm eff}^{\CNB}=-\frac{G_F}{\sqrt{2}} U_{ud} \widetilde{U}_{ea}\, \left\{[\bar{e}\gamma^\mu(1-\gamma^5)\nu_a][\bar{u}\gamma_\mu(1-\gamma^5)d] 
	+\sum_{l,q} \epsilon_{lq} [\bar{e}\mathscr{O}_l\nu][\bar{u}\mathscr{O}_q d]\right\}+{\rm h.c.},
\end{align}
where $\epsilon_{lq}$, related to the dimensionless couplings $c_k^{(6)}$, parametrize the NSI; $l$ corresponds to the 
coupling in the lepton sector while $q$ to the quark one, $\mathscr{O}_l$ and $\mathscr{O}_q$ are the Lorentz structures for each case. The couplings and structures are given in table \ref{tab:ParOPer} \citep{Cirigliano:2012ab,Cirigliano:2013xha,Ludl:2016ane}.
%%%%%%%%%%%%%%%%%%%%%   table  %%%%%%%%%%%%%%%%%%%%%%%%%%%5
\begin{table}[t]
	\centering
	\caption{Parameters and Lorentz structures for the BSM physics.}
	\label{tab:ParOPer}
	\begin{tabular}{ccc}
		\toprule\toprule
          $\epsilon_{lq}$ & $\mathscr{O}_l$ & $\mathscr{O}_q$ \\ \midrule
		$\epsilon_{LL}$ & $\gamma^\mu(1-\gamma^5)$ & $\gamma_\mu(1-\gamma^5)$ \\ 
		$\epsilon_{LR}$ & $\gamma^\mu(1-\gamma^5)$ & $\gamma_\mu(1+\gamma^5)$\\
		$\epsilon_{RL}$ & $\gamma^\mu(1+\gamma^5)$ & $\gamma_\mu(1-\gamma^5)$ \\ 
		$\epsilon_{RR}$ & $\gamma^\mu(1+\gamma^5)$ & $\gamma_\mu(1+\gamma^5)$\\
		$\epsilon_{LS}$ & $1-\gamma^5$ & $1$ \\ 
		$\epsilon_{RS}$ & $1+\gamma^5$ & $1$ \\ 
		$\epsilon_{LP}$ & $1-\gamma^5$ & $-\gamma^5$ \\ 
		$\epsilon_{RP}$ & $1+\gamma^5$ & $-\gamma^5$ \\ 
		$\epsilon_{LT}$ & $\sigma^{\mu\nu}(1-\gamma^5)$ & $\sigma_{\mu\nu}(1-\gamma^5)$ \\ 
		$\epsilon_{RT}$ & $\sigma^{\mu\nu}(1+\gamma^5)$ & $\sigma_{\mu\nu}(1+\gamma^5)$\\ \bottomrule
	\end{tabular}
\end{table}
%%%%%%%%%%%%%%%%%%%%%%%%%%%%%%%%%%%%%%%%%%%%%%%%%%%%%%%%%%%%
We need to translate these interactions to the hadron level by considering \cite{Ludl:2016ane}
\begin{align}
	\langle \mathscr{B}(p_{\mathscr{B}})|\bar{u} \mathscr{O}_q d |\mathscr{A}(p_{\mathscr{A}})\rangle,
\end{align}
where $|\mathscr{A}(p_{\mathscr{A}})\rangle$ and $|\mathscr{B}(p_{\mathscr{B}})\rangle$ are the hadronic initial and final states.
These matrix elements are calculated by matching the low-energy QCD effective theory to the quark-level lagrangian. For our case, 
we have that for protons and neutrons\footnote{We are not including the contribution of a weak-magnetic term as 
\[
\langle p(p_p)|\bar{u} \gamma_\mu d |n(p_n)\rangle_{\rm WM} = -i\frac{g_{\rm WM}}{2 M_N} \overline{u_p}(p_p)\sigma_{\mu\nu}(p_n-p_p)^\nu u_n(p_n),
\] 
since we found that it does not contribute to the $\CNB$ capture rate.}
\begin{subequations}\label{eq:hadmatelem}
	\begin{align}
		\langle p(p_p)|\bar{u}\gamma^\mu(1\pm\gamma^5)d |n(p_n)\rangle &=  \overline{u_p}(p_p)\gamma^\mu[g_V(q^2)\pm g_A(q^2)\gamma^5]u_n(p_n),\\
		\langle p(p_p)|\bar{u}d |n(p_n)\rangle &=  g_S(q^2)\,\overline{u_p}(p_p)\,u_n(p_n),\\
		\langle p(p_p)|\bar{u}\gamma^5 d |n(p_n)\rangle &= g_P(q^2)\, \overline{u_p}(p_p)\gamma^5 u_n(p_n),\\
		\langle p(p_p)|\bar{u}\sigma^{\mu\nu}(1\pm\gamma^5)d |n(p_n)\rangle &= g_T(q^2)\, \overline{u_p}(p_p)\sigma^{\mu\nu}(1\pm\gamma^5)u_n(p_n).
	\end{align}
\end{subequations}
We introduced the different hadronic form factors $g_h$, $h=\{V,A,S,P,T\}$ corresponding to the different vector, axial, scalar, pseudoscalar and tensor Lorentz structures, respectively. These couplings depend on the momentum transfer $q^2=(p_n -p_p)^2$; nonetheless, given the small momentum transfer for the capture rate, we can neglect safely this dependence. The values
of the couplings we are considering are in table \ref{tab:HadPar}. For the specific case of tritium and helium-3, we will suppose that the hadronic states are obtained by \eqref{eq:hadmatelem} with the substitutions $n \tto\! ^3{\rm H}$ and $p\tto\! ^3{\rm He}$
\cite{Ludl:2016ane}.

%%%%%%%%%%%%%%%%%%%%%   table  %%%%%%%%%%%%%%%%%%%%%%%%%%%5
\begin{table}[t]
	\centering
	\caption{Hadronic form factors we will consider in this chapter.}
	\label{tab:HadPar}
	\begin{tabular}{ccc}
		\toprule\toprule
          Form Factor & Value & Reference \\ \midrule
		$g_V(0)$ & $1$ & \protect\cite{Gershtein:1955fb,Feynman:1958ty}\\ 
		$g_A(0)/g_V(0)$ & $1.2646\pm 0.0035$ & \protect\cite{Akulov:2002gh}\\ 
		$g_S(0)$ & $1.02\pm 0.11$  & \protect\cite{Gonzalez-Alonso:2013ura}\\ 
		$g_P(0)$ & $349\pm 9$ & \protect\cite{Gonzalez-Alonso:2013ura}\\
		$g_T(0)$ & $1.020\pm 0.076$ & \protect\cite{Bhattacharya:2015esa}\\ 
		\bottomrule
	\end{tabular}
\end{table}
%%%%%%%%%%%%%%%%%%%%%%%%%%%%%%%%%%%%%%%%%%%%%%%%%%%%%%%%%%%%

\newpage

%%%%%%%%%%%%%%%%%%%%%%%%%%%%%%%%%%%%%%%%%%%%%%%%%%%%%%%%%%%%%%%%%%%%%%%%%%%%%%%%%%%%%%%%%%%%%%%%%%%%%%%%%%%%%%%%%%%%%%%%%%%%%%
\subsection{Specific cases to be considered }

The previous NSI lagrangian would not only affect the relic neutrino capture, but also other low energy processes, such as the beta decay \cite{Severijns:2006dr}, Cabbibo Universality~\cite{Hardy:2008gy}, radiative pion decay~\cite{Mateu:2007tr} 
and neutron decays~\cite{Bhattacharya:2011qm}. A complete compendium of the limits regarding all
low energy decays is given in refs.~\cite{Cirigliano:2012ab,Cirigliano:2013xha}. Thus, before considering the modification to the capture rate, we should analyse the limits on the $\epsilon_{lq}$ coefficients from experimental data. We will apply here the limits from the $\beta$-decay of several nuclei \citep{Severijns:2006dr}, obtaining the allowed parameter space for six specific $\epsilon_{lq}$ combinations. The bounds on those combinations come from a large set of experimental data on nuclear beta decay, considering pure Fermi and pure Gamow-Teller transitions plus data from neutron decay. In such work, the limits are imposed on the couplings among leptonic and hadronic currents, $C_h^{(\prime)}$, with $h=V,A,S,T$. The prime indicates when the coupling is related to a Lorentz structure containing the $\gamma^5$ matrix. This set of cases is relevant for our case since they include couplings 
with right-handed neutrinos. We first need to make a translation between the $C_h^{(\prime)}$ and the $\epsilon_{lq}$ coefficients. One finds that \cite{Cirigliano:2012ab}
\begin{align}
	\begin{aligned}
		C_V &= g_V(1+\epsilon_{LL}+\epsilon_{LR}+\epsilon_{RL}+\epsilon_{RR}),		& C_V^\prime &= g_V(1+\epsilon_{LL}+\epsilon_{LR}-\epsilon_{RL}-\epsilon_{RR}),\\
		C_A &= -g_A(1+\epsilon_{LL}-\epsilon_{LR}-\epsilon_{RL}+\epsilon_{RR}),		& C_A^\prime &= -g_A(1+\epsilon_{LL}-\epsilon_{LR}+\epsilon_{RL}-\epsilon_{RR}),\\
		C_S &= g_S(\epsilon_{LS}+\epsilon_{RS}),		& C_S^\prime &= g_S(\epsilon_{LS}-\epsilon_{RS}),\\
		C_T &= 4\,g_T(\epsilon_{LT}+\epsilon_{RT}),		&C_T^\prime &= 4\,g_T(\epsilon_{LT}-\epsilon_{RT}).
	\end{aligned}
\end{align}
Furthermore, the authors in \cite{Severijns:2006dr} constraint specific ratios of the $C_h^{(\prime)}$ parameters
\begin{align*}
	\frac{C_A}{C_V},\ \frac{C_S}{C_V},\ \frac{C_T}{C_A},\ \frac{C_V^\prime}{C_V},\ \frac{C_A^\prime}{C_A},\ \frac{C_S^\prime}{C_V},\ \frac{C_T^\prime}{C_A}.
\end{align*}
In each scenario, distinct combinations of these ratios are bounded. We performed a scan over the non zero $\epsilon_{lq}$ parameter, to find the allowed values of $\epsilon_{lq}$ at $3\sigma$ C.L. from the global fit for the beta decay data, in the ranges
\begin{align}
	\begin{aligned}
	-10^{-3}\le & \epsilon_{LL} \le 10^{-3}\, , & 	-10^{-3}\le & \epsilon_{LR} \le 10^{-3} \, ,\\
	-2.8\times 10^{-3}\le & \epsilon_{LS} \le 5\times 10^{-3}\, , & 
	-2\times 10^{-3}\le & \epsilon_{LT} \le 2.1\times 10^{-3} \, ,\\
	\end{aligned}
\end{align}
and
\begin{align}
	\abs{\epsilon_{Rq}}\le 10^{-1}.
\end{align} 
We have also scanned over the $g_A(0)/g_V(0)$ value given in table \ref{tab:HadPar} since 
such parameter is affected by the presence of BSM \cite{Gonzalez-Alonso:2013uqa}. 
The previous ranges in which the scan is performed have been chosen to include the constraints of 
refs.~\cite{Hardy:2008gy,Mateu:2007tr,Bhattacharya:2011qm} in the left-chiral 
coefficients at the $3\sigma$ level. Although stronger limits can be imposed 
on right-handed couplings using pion decay~\cite{Campbell:2003ir}, we will not 
include them as they are strongly dependent on the flavour structure of the 
model~\citep{Cirigliano:2012ab,Cirigliano:2013xha}. Other constraints, such as LHC bounds coming 
from $pp \to e + X + \slashed{E}_T$ have been considered in \cite{Bhattacharya:2011qm,Cirigliano:2012ab}. 
However, in such analysis is supposed that the interactions of eq.~\eqref{eq:CNBL} 
remain point-like up to the LHC energies, i.e.\ up to a few TeV. We will allow for the possibility of having
physics BSM at the electroweak scale; thus, we will use only low-energy constraints.
In all scenarios we found that the parameters $\epsilon_{LL}$ and $\epsilon_{LR}$ are unconstrained 
by the experimental data, as it has been previously noted in ref.~\cite{Gonzalez-Alonso:2013uqa}.\\

%\newpage

%%%%%%%%%%%%%%%%%%%%%%%%%%%%%%%%%%%%%%
\noindent {\bf Left-handed three parameter case.} The first case consists in considering only left-handed couplings in the effective lagrangian by imposing
\begin{align}
	\frac{C_V^\prime}{C_V}=\frac{C_A^\prime}{C_A}=1, \quad \frac{C_S}{C_V}=\frac{C_S^\prime}{C_V}, \frac{C_T}{C_A}=\frac{C_T^\prime}{C_A},
\end{align}
i.e., $\epsilon_{Rq}=0$, with $q=L,R,S,T$. The fit gives the following allowed ranges for the free parameters \citep{Severijns:2006dr}
\begin{align*}
	\frac{C_A}{C_V}&=-1.26994\pm 0.00246,\qquad
	\frac{C_S}{C_V}=0.0013\pm 0.0039,\qquad
	\frac{C_T}{C_A}= 0.0036\pm 0.0099.\\
\end{align*}

%%%%%%%%%%%%%%%%%%%%%%%%%%%%%%%%%%%%%%
\noindent{\bf Vector axial-vector three-parameter case.} In this case, scalar and tensor couplings are set to zero.
This is achieved by imposing $C_S=C_S^\prime=0$ and $C_T=C_T^\prime =0$. 
The free parameters, in terms of the $\epsilon_{lq}$ parameters, are
\begin{subequations}
	\begin{align}
		\frac{C_A}{C_V}&=-\frac{g_A}{g_V}\frac{1+\epsilon_{LL}-\epsilon_{LR}-\epsilon_{RL}+\epsilon_{RR}}{1+\epsilon_{LL}+\epsilon_{LR}+\epsilon_{RL}+\epsilon_{RR}},\\
		\frac{C_V^\prime}{C_V}&=\frac{1+\epsilon_{LL}+\epsilon_{LR}-\epsilon_{RL}-\epsilon_{RR}}{1+\epsilon_{LL}+\epsilon_{LR}+\epsilon_{RL}+\epsilon_{RR}},\qquad
		\frac{C_A^\prime}{C_A}=\frac{1+\epsilon_{LL}-\epsilon_{LR}+\epsilon_{RL}-\epsilon_{RR}}{1+\epsilon_{LL}-\epsilon_{LR}-\epsilon_{RL}+\epsilon_{RR}}.
	\end{align}
\end{subequations}
The fit obtained for the ratio $C_A^\prime/C_A$ is \citep{Severijns:2006dr}
\begin{align*}
	0.868\le \frac{C_A^\prime}{C_A} \le 1.153
\end{align*}
while for the other two free parameters the $3\sigma$ limits have a strong correlation \citep{Severijns:2006dr}. \\

%%%%%%%%%%%%%%%%%%%%%%%%%%%%%%%%%%%%%%
\noindent{\bf Right-handed scalar and tensor three-parameter case.} We can impose the constraints
\begin{align}
	\frac{C_V^\prime}{C_V}=\frac{C_A^\prime}{C_A}=1, \quad \frac{C_S}{C_V}=-\frac{C_S^\prime}{C_V},\quad \frac{C_T}{C_A}=-\frac{C_T^\prime}{C_A},
\end{align}
so, we are considering that the right-chiral scalar and tensor couplings $\epsilon_{RS},\epsilon_{RT}$ are different 
from zero.\\

%%%%%%%%%%%%%%%%%%%%%%%%%%%%%%%%%%%%%%
\noindent {\bf Five parameter case.} In this scenario, we will only impose that
\begin{align}
	\frac{C_V^\prime}{C_V}=\frac{C_A^\prime}{C_A}=1,
\end{align}
making a total of five free parameters. However, as noticed in the review \citep{Severijns:2006dr}, it is interesting to consider 
the limits of the difference and the sum of the scalar and tensor parameters
\begin{subequations}
	\begin{align}
		\frac{C_A}{C_V}&=-\frac{g_A}{g_V}\,\frac{1+\epsilon_{LL}-\epsilon_{LR}}{1+\epsilon_{LL}+\epsilon_{LR}},\\
		\frac{C_S + C_S^\prime}{C_V}&=\frac{g_S}{g_V}\,\frac{2\,\epsilon_{LS}}{1+\epsilon_{LL}+\epsilon_{LR}},\qquad
		\frac{C_S - C_S^\prime}{C_V}=\frac{g_S}{g_V}\,\frac{2\,\epsilon_{RS}}{1+\epsilon_{LL}+\epsilon_{LR}},\\
		\frac{C_T + C_T^\prime}{C_A}&=-\frac{g_T}{g_A}\,\frac{8\,\epsilon_{LT}}{1+\epsilon_{LL}-\epsilon_{LR}},\qquad
		\frac{C_T - C_T^\prime}{C_A}=-\frac{g_T}{g_A}\,\frac{8\,\epsilon_{RT}}{1+\epsilon_{LL}-\epsilon_{LR}}.
	\end{align}
\end{subequations}
The limits we will impose are
\begin{align*}
	-1.272\le \frac{C_A}{C_V} \le -1.265
\end{align*}
while, for the other parameters, we will consider the correlation bound at $3\sigma$ C.\ L.\ \citep{Severijns:2006dr}.

%%%%%%%%%%%%%%%%%%%%%%%%%%%%%%%%%%%%%%%%%%%%%%%%%%%%%%%%%%%%%%%%%%%%%%%%%%%%%%%%%%%%%%%%%%%%%%%%%%%%%%%%%%%%%%%%%%
\section{Capture rate of the $\CNB$ considering Beyond SM physics}

Having established the scenarios, we can compute the new contributions to the $\CNB$ capture rate at PTOLEMY. It is important to note here that Ludl and Rodejohann \cite{Ludl:2016ane} have shown that the endpoint of the beta decay is not modified significantly by the existence of NSI. Nonetheless, the spectrum has sizeable distortions which can improve the limits presented before. Thus, it could be possible to differentiate the $\CNB$ capture rate from the beta decay background even in the presence of NSI. We will consider neutrinos as Dirac particles since our purpose is to analyse the possible increase of the capture rate in such case. On the other hand, we will suppose here that the relic neutrino number density is not modified significantly by the NSI. We will consider the implications of modification in the Cosmology in the next section. Let us remember here that, in order to obtain the capture rate
\begin{align}
	\Gamma_{\CNB} = N_T\sum_{a=1}^3\,\esp{\sigma_{a} (+1/2)v_a n_{\nu_+^a}+\sigma_a(-1/2)v_a n_{\nu_-^a}},\tag{\ref{eq:caprate}}
\end{align}
we need to compute the cross section times the neutrino velocity. The procedure to obtain this quantity will be identical to the one showed in previous sections. Considering the complete effective lagrangian, the capture cross section for a neutrino mass eigenstate $a$, with helicity $h_a=\pm 1$ and velocity $v_a$ including BSM 
effects is given by
\begin{align}
	\sigma_a(h_\nu)v_{\nu_a}=\frac{G_F^2}{2\pi} \abs{U_{ud}}^2|\widetilde{U}_{ea}|^2 F_Z(E_e) \, \frac{m_{\rm ^3He}}{m_{\rm ^3H}} E_e \, p_e\, 	T_a(h_a,\epsilon_{lq}),
\label{eq:sigmaNSI}
\end{align}
where $m_{\rm ^3 He}$ and $m_{\rm ^3H}$ are the helium and tritium masses, and $E_e$, $m_e$, 
$p_e$ are the electron e\-ner\-gy, mass and momentum, respectively. The $T_a(h_a,\epsilon_{lq})$ 
function contains the dependence on the neutrino helicity and on the $\epsilon_{lq}$ parameters,
\begin{align}\label{eq:NSIepsilons}
	T_a(h_a,\epsilon_{lq})&={\cal A}(h_a)\left[g_V^2\left(\epsilon_{LL}+\epsilon_{LR}+1\right)^2+3 \, g_A^2\left(\epsilon_{LL}-\epsilon_{LR}+1\right)^2+g_S^2\, \epsilon_{LS}^2+48\, g_T^2\, \epsilon_{LT}^2\right.\notag\\
	&\qquad\qquad\left.+\frac{2m_e}{E_e}\,[g_S\, g_V\, \epsilon_{LS}\left(\epsilon_{LL}+\epsilon_{LR}+1\right)-12\, g_A\, g_T\, \epsilon_{LT}\left(\epsilon_{LL}-\epsilon_{LR}+1\right)]\right]\notag\\
	&\quad +{\cal A}(-h_a)\left[g_V^2\,(\epsilon_{RR}+\epsilon_{RL})^2+3 \, g_A^2\,(\epsilon_{RR}-\epsilon_{RL})^2+g_S^2\, \epsilon_{RS}^2+48\, g_T^2\, \epsilon_{RT}^2\right.\notag\\
	&\qquad\qquad\qquad\left.+\frac{2m_e}{E_e}\,[g_S\, g_V\, \epsilon_{RS}\,(\epsilon_{RR}+\epsilon_{RL})\,-12\, g_A\, g_T \,\epsilon_{RT}\,(\epsilon_{RR}-\epsilon_{RL})]\right]\notag\\
	&\quad+2\,\frac{m_a}{E_a}\left\{g_S\, g_V\,\epsilon_{RS}\,\left(\epsilon_{LL}+\epsilon_{LR}+1\right)+\epsilon_{LS}\,(\epsilon_{RR}+\epsilon_{RL}))\right.\notag\\
	&\qquad\qquad\ \left.-12\, g_A\, g_T(\epsilon_{RT}\,\left(\epsilon_{LL}-\epsilon_{LR}+1\right)+\epsilon_{LT}\,(\epsilon_{RR}-\epsilon_{RL}))\right\}\notag\\
	&\quad+2\,\frac{m_am_e}{E_aE_e}\left\{g_V^2\,(\epsilon_{LL}+\epsilon_{LR}+1)(\epsilon_{RR}+\epsilon_{RL})+3 \, g_A^2\,(\epsilon_{LL}-\epsilon_{LR}+1)(\epsilon_{RR}-\epsilon_{RL})\right.\notag\\
	&\qquad\qquad\qquad \left.+g_S^2\,\epsilon_{RS}\, \epsilon_{LS}+48\, g_T^2\,\epsilon_{RT}\, \epsilon_{LT}\right\},
\end{align}
with $m_a, E_a$ the mass and energy of the $a$-th neutrino mass. Let us study next each case independently.

\newpage

%%%%%%%%%%%%%%%%%%%%%%%%%%%%%%%%%%%%%%
\noindent {\bf Left-chiral three parameter case.} For the scenario in which only there are left-chiral couplings, we have
\begin{align}
	\sigma_a(h_\nu)v_{\nu_a}=\frac{G_F^2}{2\pi} \abs{U_{ud}}^2|\widetilde{U}_{ea}|^2 F_Z(E_e) \frac{m_{\rm ^3He}}{m_{\rm ^3H}} E_e p_e\,{\cal A}(h_{\nu }) 
	&\left[\left(\epsilon_{LL}+1\right)^2\left(3 g_A^2+g_V^2\right)\right.\notag\\
	&+2\frac{m_e}{E_e}\left(\epsilon_{LL}+1\right)\left(g_S\, g_V \epsilon _{LS}-12 g_A\, g_T \epsilon_{LT}\right)\notag\\
	&\left.+g_S^2\, \epsilon_{LS}^2+48g_T^2\, \epsilon_{LT}^2\right].
\end{align}
Evidently, when we take all the non-standard couplings to zero, we obtain the SM result. Let us also note that scalar and tensor parameters have distinct dependence on the electron energy and mass, because of the different Lorenz structure. Now, computing the total capture rate for relic neutrinos using the allowed points obtained in the scan previously performed, we found that the modification in this case is in the region $[-5.3,7]\%$ at $3\sigma$ level. Here, a minus percentage indicates a diminution of the rate compared to the SM value. Therefore, we find that the change is not very significant in this case. This is actually expected as we saw that the allowed values of the effective couplings are not large, especially, the values of the scalar and tensor parameters.\\

%%%%%%%%%%%%%%%%%%%%%%%%%%%%%%%%%%%%%%
\noindent{\bf Vector axial-vector three-parameter case.} In this scenario, we have,
\begin{align}
	\sigma_a(h_\nu)v_{\nu_a}=\frac{G_F^2}{2\pi} \abs{U_{ud}}^2|\widetilde{U}_{ea}|^2 F_Z(E_e)\frac{m_{\rm ^3He}}{m_{\rm ^3H}} E_e p_e \left(3 g_A^2+g_V^2\right)
	&\left[{\cal A}(h_{\nu})(\epsilon _{LL}+1)^2+{\cal A}(-h_{\nu})\epsilon _{RR}^2\right.\notag\\
	&\left.+\frac{2m_e m_a^\nu}{E_e E_a}\left(\epsilon_{LL}+1\right)\epsilon _{RR}\right].
\end{align}
Let us note that the term linear in the right-handed coupling is proportional to $m_a^\nu/E_\nu$; such term would be negligible
in the case of a ultrarelativistic neutrino.  Also, we can see that this term comes from the interference of the standard model with the right-handed neutrino current. The term proportional to $\epsilon _{RR}^2$ comes from the square of the right-handed currents, and it is proportional to the ${\cal A}(-h_{\nu})$. Using the allowed values at $3\sigma$ for the $\epsilon_{LL,RR}$ obtained from the beta decay data, we found that the modification on the $\CNB$ capture rate is in the range $[-15,15] \%$; such modification could be significant for a large enough set of data. \\

%%%%%%%%%%%%%%%%%%%%%%%%%%%%%%%%%%%%%%
\begin{figure}[t]
	\begin{center}
    			\includegraphics[width=0.6\textwidth]{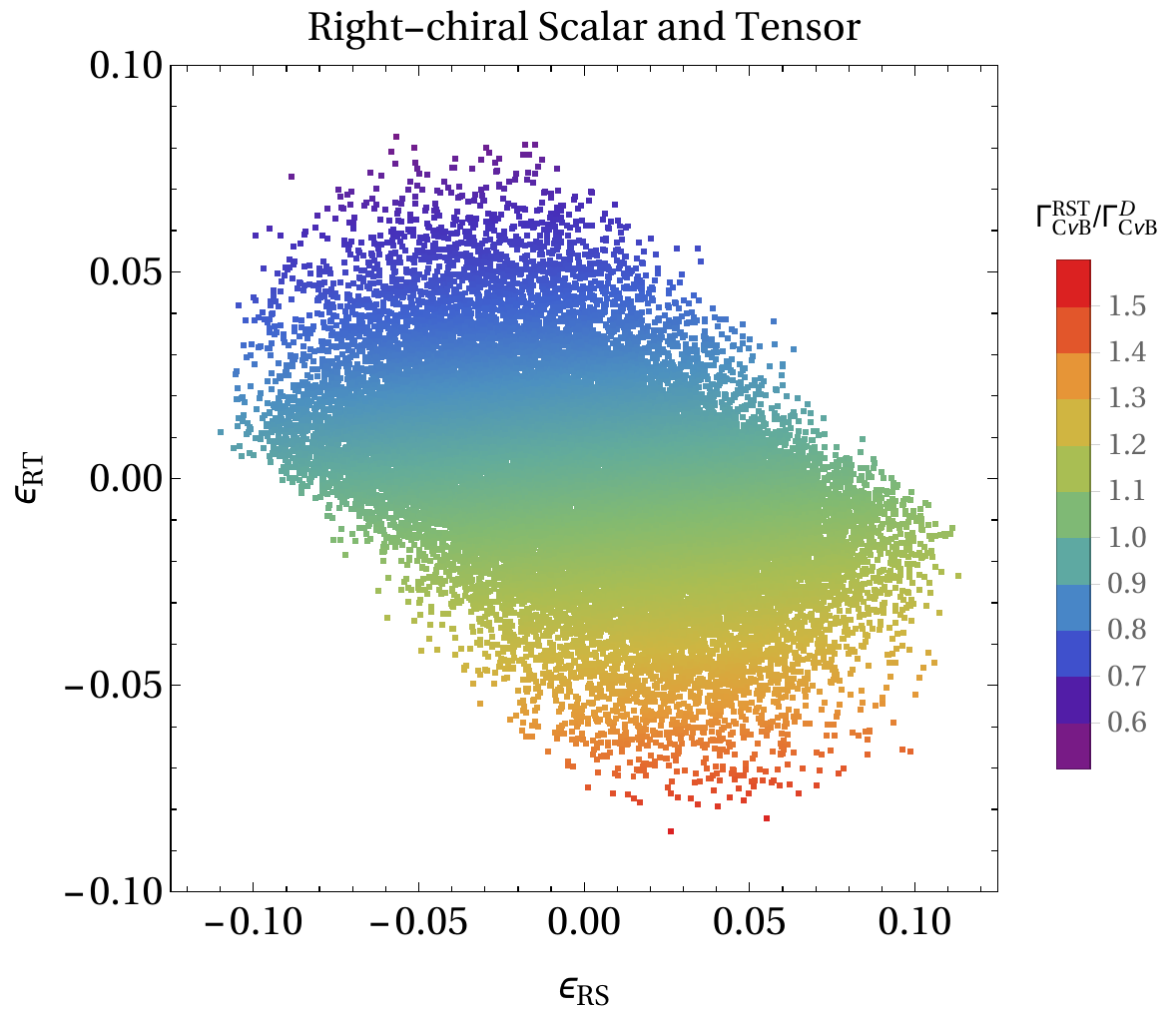}  
	\end{center}
  	\caption{Capture rate compared to the SM Dirac case in the $(\epsilon_{RS},\epsilon_{RT})$ plane for the case of right-handed scalar and tensor currents. The color code indicates the value of the ratio between the modified $\CNB$ capture rate and the SM value. In this case, such ratio can be in the interval of $[0.63,1.46]$.}
  	\label{fig:RSTCase}
\end{figure}

\noindent{\bf Right-handed scalar and tensor three-parameter case.} We have in this case
\begin{align}
	\sigma_a(h_\nu)v_{\nu_a}=\frac{G_F^2}{2\pi}\abs{U_{ud}}^2|\widetilde{U}_{ea}|^2 F_Z(E_e)\frac{m_{\rm ^3He}}{m_{\rm ^3H}}E_e p_e
		&\left[{\cal A}(h_{\nu})(\epsilon_{LL}+1)^2(3 g_A^2+g_V^2)\right.\notag\\
		&\left. +2\frac{m_a^\nu}{E_a}\left(\epsilon _{LL}+1\right) \left(g_S\, g_V\, \epsilon_{RS}-12 g_A\, g_T\, \epsilon_{RT}\right)\right.\notag\\
		&\left.+{\cal A}(-h_{\nu})(g_S^2\, \epsilon_{RS}^2+48g_T^2\, \epsilon_{RT}^2)\right],
\end{align}
again the term proportional to the neutrino mass comes from the interference between SM and right-handed currents. Furthermore,
we observe that this interference term is not dependent of the neutrino helicity, i.e. it does not depend on  ${\cal A}(h_\nu)$. This is due to the different Lorentz structures that appear in the BSM physics.\\ 

Considering the allowed parameter space, we obtained the possible modifications to the relic neutrino capture in this case. We found that the variation of the rate is in the range  $[-37,47]\%$ compared to the SM Dirac case. It is important to note here that the parameter space is highly correlated since the beta decay bounds impose such correlation. In figure \ref{fig:RSTCase} we present the correlation between the parameters $\epsilon_{RT}$ and $\epsilon_{RS}$ in this scenario. The color code indicates the ratio between the modified capture rate, denominated as $\Gamma_{\CNB}^{\rm RST}$ (RST from Right-handed Scalar Tensor), and the Dirac SM rate. We see that, for positive values of the couplings, the modified rate is lower than the SM one while for negative values the behaviour is the opposite. Therefore, the relic neutrino capture is sensitive to the sign of the couplings.\\

\begin{figure}[t!]
	\begin{center}
    			\includegraphics[width=0.6\textwidth]{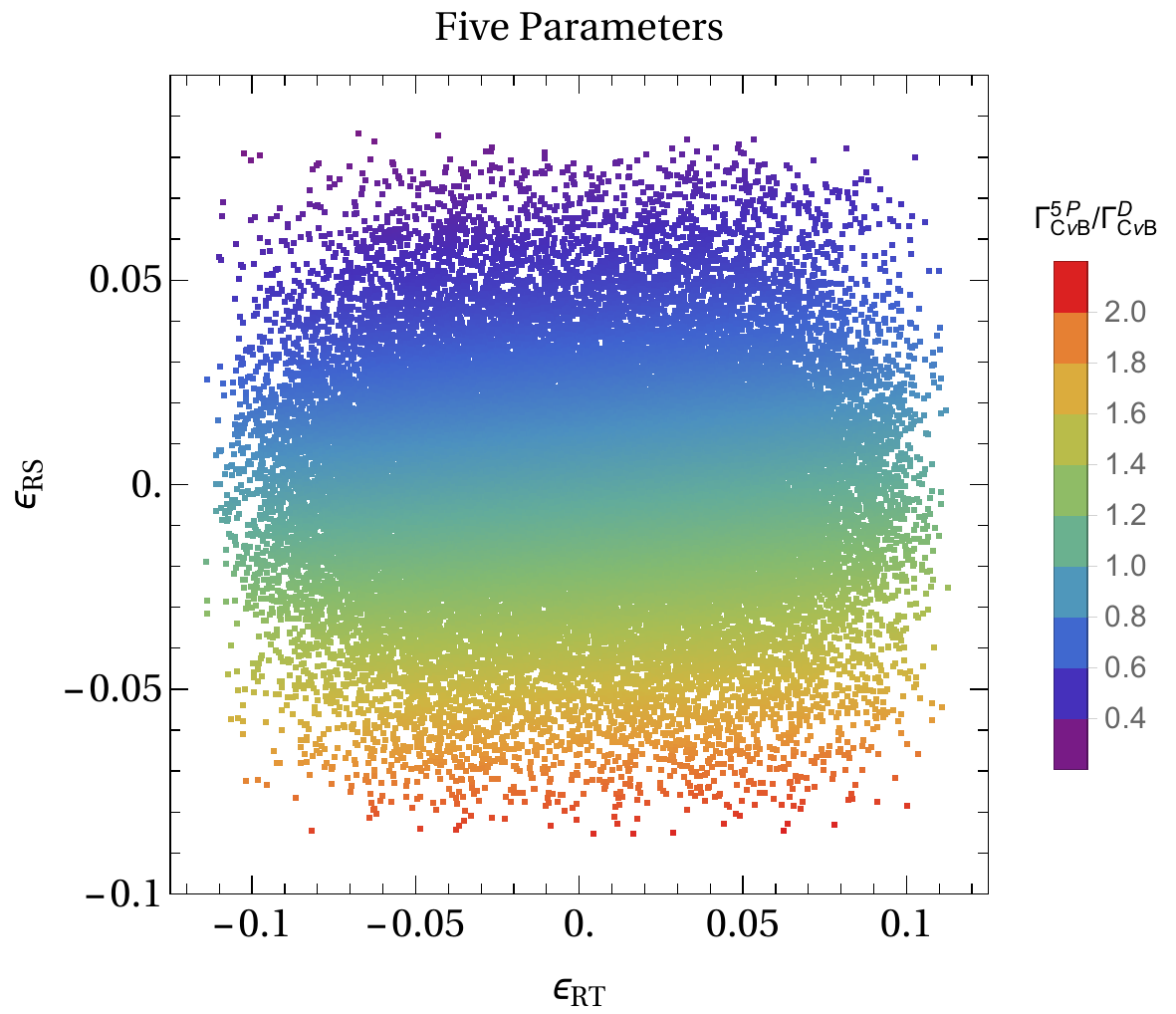}  
	\end{center}
  	\caption{Capture rate compared to the SM Dirac case in the $(\epsilon_{LS},\epsilon_{RT})$ (top left), $(\epsilon_{LT},\epsilon_{RT})$ (top right), $(\epsilon_{RS},\epsilon_{RT})$ (bottom) planes for the case of five simultaneously allowed NSI parameters. The color code indicates the value of the ratio between the modified $\CNB$ capture rate and the SM value. In this case, such ratio can be in the interval of $[0.30,2.19]$.}
  	\label{fig:5PCase}
\end{figure}

%%%%%%%%%%%%%%%%%%%%%%%%%%%%%%%%%%%%%%
\noindent {\bf Five parameter case.} In the fourth scenario, corresponding to the case in which we have five parameters, we find that the cross section times the velocity is basically the superposition of the left- and right-handed leptonic currents coupled with scalar and tensor quarks currents plus interference terms
\begin{align}
	\sigma_a(h_\nu)v_{\nu_a}=\frac{G_F^2}{2\pi} \abs{U_{ud}}^2|\widetilde{U}_{ea}|^2 F_Z(E_e) \frac{m_{\rm ^3He}}{m_{\rm ^3H}} E_e & p_e
	\left[{\cal A}(h_{\nu })\left(\epsilon_{LL}+1\right)^2\left(3 g_A^2+g_V^2\right)\right.\notag\\
	&+{\cal A}(h_{\nu })\left(g_S^2\, \epsilon_{LS}^2+48g_T^2\, \epsilon_{LT}^2\right)\notag\\
	&+{\cal A}(-h_{\nu })\left(g_S^2\, \epsilon_{RS}^2+48g_T^2\, \epsilon_{RT}^2\right)\notag\\
	&+2\frac{m_e}{E_e}{\cal A}(h_{\nu })\left(\epsilon_{LL}+1\right)\left(g_S\, g_V \epsilon _{LS}-12 g_A\, g_T \epsilon_{LT}\right)\notag\\
	&+2\frac{m_a^\nu}{E_e E_a}\left\{E_e\left(\epsilon _{LL}+1\right) \left(g_S\, g_V\, \epsilon_{RS}-12 g_A\, g_T\, \epsilon_{RT}\right)\right.\notag\\
	&\qquad\qquad+\left.\left.m_e\left(g_S^2\, \epsilon_{LS}\epsilon_{RS}+48g_T^2\, \epsilon_{LT}\epsilon_{RT}\right)\right\}\right]
\end{align}
where now the interference term proportional to the neutrino mass has mixing terms between $\epsilon_{LS,LT}$ with $\epsilon_{RS,RT}$. In figure \ref{fig:5PCase}, we show the ratio between the beyond SM capture rate with the SM Dirac rate for this scenario in the planes which present correlation among the couplings. Again, the color code indicates the value of the ratio of the modified capture rate with respect to the SM one. In this case, we find that the ratio can get the maximum value of $\sim 2.19$, which is very significant since it shows that NSI can alter the Dirac relic capture rate to a value close the Majorana one. Nevertheless, as observed in each plane, we see that the capture rate can also be diminished by the NSI. Therefore, if PTOLEMY finds data compatible with the $\CNB$ given the position of the peak, but the number of events are smaller than the predicted by the SM, this could suggest the contribution of NSI. Nevertheless, we should consider the impact of a modification in the relic neutrino density, and see if one can differentiate such modified cosmology from NSI.\\
%\begin{figure}[t]
%	\begin{center}
%    			\includegraphics[width=\textwidth]{P1n0}  
%	\end{center}
%  	\caption{Dependence on a BSM parameter $\epsilon$ of the Capture rate of the $\CNB$, in the case where is considered a 
%  	standard neutrino relic density $n_0$.}
%  	\label{fig:EpsDep}
%\end{figure}
%
%\begin{figure}[t]
%	\begin{center}
%    			\includegraphics[width=\textwidth]{P2n0pn0R}  
%	\end{center}
%  	\caption{Dependence on a BSM parameter $\epsilon$ of the Capture rate of the $\CNB$, in the case where is considered a 
%  	standard neutrino relic density $n_0$.}
%  	\label{fig:EpsDepn0R}
%\end{figure}

%%%%%%%%%%%%%%%%%%%%%%%%%%%%%%%%%%%%%%%%%%%%%%%%%%%%%%%%%%%%%%%%%%%%%%%%%%%%%%%%%%%%%%%%%%%%%%%%%%%%%%%%%%%%5
\section{Relic Right-Handed Neutrinos}

Since we are assuming that NSI are present, we can imagine that in the Early Universe those interactions may have been important in such a way that they modified the relic neutrino abundance. Thus, in the first place, we should consider the modifications in the left-handed Dirac neutrino abundance. Given that we have seen that these NSI can be at most $2$ orders of magnitude smaller than the weak interaction, as the $\epsilon_{lq}$ parameters are of such order, we see that active (left-handed) neutrinos were mainly maintained in equilibrium by the SM interactions. Therefore, we do not expect a significant change in the active neutrino abundance. On the other hand, an important right-handed Dirac neutrino abundance could have been created. The initial right-handed abundance can have a thermal or a non-thermal origin. Thus, in order to estimate such abundance, we will consider the cosmological constraints on right-handed neutrinos as previously presented in \cite{Anchordoqui:2012qu,SolagurenBeascoa:2012cz,Zhang:2015wua,Chen:2015dka} for these two different origins.\\ 

Let us begin with an initial thermal right-handed neutrino abundance present in the Universe and maintained in equilibrium by the NSI themselves or other interactions. As usual, when the expansion rate of the Universe becomes stronger than the interactions rate, the right-handed abundance will become decoupled from the plasma. At that point, the abundance of both left- and right-handed neutrinos per species are equal since they are maintained in equilibrium
\begin{align}\label{eq:rhlhab}
	n_{\nu_R^a} (T_R) = n_{\nu_L^a} (T_R),
\end{align}
being $T_R$ the right-handed neutrino freeze out temperature. The relationship between the temperature $T$ of a massless\footnote{Since neutrinos are Dirac by hypothesis, the right-handed component has the same mass as the active one; thus, as  $T_R>T_\nu$, the right-handed freeze out temperature should be larger than the left-handed one. Neutrinos can be considered massless at those temperatures.} species after the decoupling and $T_R$ is \cite{Kolb:1990vq}
\begin{align}
	a(T)\,T = a(T_R)\, T_R,
\end{align}
with $a(T)$ the scale factor at the epoch in which occurs the decoupling. Thus, recalling the relation between number density and temperature, equation \eqref{eq:nunumber}, we have \cite{Zhang:2015wua}
\begin{align}
	n_{\nu_R^a}(T) a(T)^3 = n_{\nu_R^a}(T_R) a(T_R)^3. 
\end{align}
To relate the right-handed neutrino abundance with the left-handed one, we should take into account the conservation of the entropy 
$S=g_{*S}T^3a(T)^3 = {\rm constant}$ \cite{Kolb:1990vq}, with the effective relativistic degrees of freedom for entropy $g_{*S}$ given by
\begin{align*}
	g_{*S}&=\sum_{i={\rm bosons}}g_i\left(\frac{T_i}{T}\right)^3+\frac{7}{8}\sum_{i={\rm fermions}}g_i\left(\frac{T_i}{T}\right)^3.
\end{align*}
Using this conservation, we can get the relation
\begin{align}
	\frac{n_{\nu_R^a}(T_\nu)}{n_{\nu_R^a}(T_R)}=\frac{g_{*S}(T_\nu)}{g_{*S}(T_R)}\left(\frac{T_\nu}{T_R}\right)^3
\end{align}
where we chose the final temperature as the left-handed neutrino decoupling temperature $T_\nu$. Furthermore, taking into account that left-handed neutrinos were in equilibrium in this period $n_{\nu_L^a}(T_\nu) T_{\nu}^3 = n_{\nu_L^a}(T_R) T_{R}^3$ and the equality at $T_R$, equation \eqref{eq:rhlhab}, one can relate the abundances as \cite{Zhang:2015wua},
\begin{align}
	n_{\nu_R^a}(T_\nu) = \frac{g_{*S}(T_\nu)}{g_{*S}(T_R)} n_{\nu_L^a}(T_\nu).
\end{align}
We need then to obtain the relation of the effective relativistic degrees of freedom for entropy at both temperatures. This can be done taking into account that the right-handed neutrino can modify the Big Bang Nucleosynthesis (BBN), i.e., the creation of light elements at the Early Universe. Therefore, we define the effective number of thermal neutrino species $N_{\rm eff}$ from the total energy density as \cite{Kolb:1990vq,Zhang:2015wua}
\begin{align}
	\rho=\left[1+N_{\rm eff}\,\frac{7}{8}\left(\frac{4}{11}\right)^\frac{4}{3}\right]\rho_\gamma,
\end{align}
with $\rho_\gamma$ the total radiation energy density. In the standard $\Lambda$CDM cosmological model, $N_{\rm eff}=3.046$. Actually, analysing the CMB spectrum it is possible to constrain this effective number. It has been shown that the deviation from the standard value $\Delta N_{\rm eff}=N_{\rm eff}^{\rm exp}-3.046$ is related to the ratio of the effective degrees of freedom as \cite{Anchordoqui:2012qu,SolagurenBeascoa:2012cz,Zhang:2015wua}
\begin{align}
	\Delta N_{\rm eff} = 3 \left(\frac{g_{*S}(T_\nu)}{g_{*S}(T_R)}\right)^\frac{4}{3},
\end{align}
so we have
\begin{align}
	n_{\nu_R^a}(T_\nu) = \left(\frac{1}{3}\Delta N_{\rm eff}\right)^\frac{3}{4} n_{\nu_L^a}(T_\nu).
\end{align}
From the Planck data, we will use the measured effective number of neutrino species as \cite{Ade:2015xua}
\begin{align*}
	N_{\rm eff}^{\rm exp} = 3.14^{+0.44}_{-0.43}\qquad \text{He + Planck TT + low P + BAO}\qquad \text{at $95\%$ C.L.},
\end{align*}
thus we have that this value can give us a limit on the right-handed neutrino density. Considering the upper bound, one gets the largest possible value of the right-helical neutrino and left-helical antineutrino number density per species at the present time \cite{Zhang:2015wua}
\begin{align}
	n_{\nu_+^a} &= n_{(\nu^a_-)^c} \approx 16\ {\rm cm}^{-3}.
\end{align}
As consequence, the relic neutrino capture needs to be modified to include the non zero right-handed abundance. In the pure SM case, we have that the capture rate is proportional to ${\cal A}(h_\nu)$ which for both left- and right-helical states is approximately $1$ in the case of non-relativistic neutrinos. Thus, the increase in the capture rate for Dirac neutrinos can be of $\sim 28\%$ \cite{Zhang:2015wua}. Furthermore, for all the cases in the presence of NSI we have considered, the modification to the $\CNB$ capture rate due to the additional right-handed abundance consists on increasing the rate. This is of course expected since we are including new neutrinos that can be captured. Nevertheless, the enlargement of the rate is dependent on the case considered. This can be seen from the definition on the capture rate, equation \eqref{eq:caprate}, and the value of the cross section times velocity in each scenario. We have checked that the presence of further neutrinos enlarges the capture rate on $30 \%$ in the vector axial-vector case to a value of $70\%$ in the five parameter case. In this last scenario, the $\CNB$ rate can be as large as $2.8\,\Gamma_{\CNB}^{\rm D}$, reinforcing our results on the possibility of having Dirac neutrinos with a relic capture rate numerically similar to the Majorana one.\\ 

The second possibility consists in having an initial non-thermal right handed neutrino abundance. We will follow the description given in \cite{Chen:2015dka}; we will suppose that right-handed Dirac neutrinos are initially a degenerated Fermi gas, decoupled from the thermal bath, with number and energy densities per species given by \cite{Chen:2015dka}
\begin{align}
	n_{\nu_R^a}=\frac{1}{3\pi^2}\varepsilon_F^3,\quad \rho_{\nu_R^a}=\frac{1}{4\pi^2}\varepsilon_F^4,
\end{align}
being $\varepsilon_F$ the Fermi energy. Since the entropy conservation is not spoiled by the presence of these right-handed neutrinos, the number density for any given temperature is 
\begin{align}
	n_{\nu_R^a}(T)=\frac{\vartheta}{3\pi^2}\frac{g_{*S}(T)}{g_{*S}(T_R)}T^3,
\end{align}
with $\vartheta=\varepsilon_F/T_R$. Thus, the relic right-handed neutrino density can be related to the photon number density as \cite{Chen:2015dka}
\begin{align}
	n_{\nu_R^a}(T_\gamma)=\frac{1}{6\zeta(3)}\frac{g_{*S}(T_\gamma)}{g_{*S}(T_R)}\vartheta\, n_\gamma.
\end{align}
It is necessary to determine the bound from observations on the value of $\vartheta$. Noticing that these additional degrees of freedom could also modify the BBN, we can get a value for $\Delta N_{\rm eff}$ using the energy density at a given temperature \cite{Chen:2015dka}. Adding over the three species, one has
\begin{align}
	\Delta N_{\rm eff} = 3\,\frac{8}{7}\frac{30}{8\pi^4}\zeta(3)\left(\frac{g_{*S}(T_{\rm BBN})}{g_{*S}(T_R)}\right)^\frac{4}{3};
\end{align}
thus, from the Planck value, we have the limit,
\begin{align}
	\vartheta \lesssim 3.26.
\end{align}
This constraint implies that a right-helical neutrino and the left-helical antineutrino number density per species today is given by \cite{Chen:2015dka}
\begin{align}
	 n_{\nu_+^a} &= n_{(\nu^a_-)^c} \approx 36\ {\rm cm}^{-3}.
\end{align}
Including these additional neutrinos, we find that the capture rates increase even more compared to the thermal right-handed neutrino abundance, which is completely expected. Furthermore, for the left chiral and vector axial-vector cases we found that the modification on the rate can be in the interval $[40,90]\%$, getting closer to the expected value for SM Majorana neutrinos. For the other three scenarios we found larger modifications. In the right-handed scalar-tensor case, the NSI capture rate has a maximum value of $\sim 2.46\times \Gamma_{\CNB}^{\rm D}$, while in the five-parameter case we obtained $\sim 3.51\times \Gamma_{\CNB}^{\rm D}$. This confirms that it is possible to have a capture rate for Dirac neutrinos identical to the Majorana case.\\

Nevertheless, we should notice an important fact. As previously stressed, the additional relic right-handed neutrinos can only increase the relic capture rate; thus, if PTOLEMY detects a {\it decrease} in the total capture rate, this may indicate the possible existence of NSI. On the other hand, let us suppose that PTOLEMY shows results compatible with an enhanced Dirac relic capture rate, yet lower than the Majorana value. Such result could be interpreted as a modification in the number of the relic neutrinos predicted by the Standard Cosmology, but it also can be described by the existence of NSI. This degeneracy can not be solved by PTOLEMY-like experiments only as they just measure the number of events compatible to a peak expected from relic neutrino capture. Thus, other studies need to be performed.\\

In this chapter we have considered the impact of NSI in the detection of the cosmic neutrino background. We have discussed briefly the origin of such background and its properties, such as number density, pressure, and the root mean square momentum. These characteristics indicate that relic neutrinos can be non-relativistic particles if their masses are of ${\cal O}(10^{-3})$ eV or bigger. This has many important consequences; the most crucial one dwells with the differentiation between Dirac and Majorana natures. This can be explained taking into account the difference between chirality and helicity for non-relativistic neutrinos. When the neutrino decouples from the thermal primordial bath, it is basically a left-handed particle. The free streaming after the freeze out imposes that helicity, not chirality, is conserved as neutrinos are massive.\\ 

Since Majorana and Dirac particles have different chiral and helical components, the final abundances for the helical states are different in both cases. Dirac particles and antiparticles are distinct, and each chiral state can have two possible helicity projections. Keeping in mind that left-handed neutrinos interact weakly and a left-handed neutrino becomes a left-helical state, we can see that only the abundances for left-helical particle and right-helical antiparticle states are different from zero. In the other cases, we expect their abundances to be negligible. On the other hand, if neutrinos are left-handed Majorana fermions, which have the two helical states, we see that both helical neutrino states will have a non zero abundance after the decoupling. The right-handed heavy neutrino will be decoupled way before the left-handed one, making its abundance zero at the present.\\

The detection of the cosmic neutrino background is of main interest for both Particle Physics and Cosmology as these neutrinos contain information from a time of about $1$ second after the Big Bang. Nonetheless, given its tiny energy, their detection is quite complicated. Several methods have been proposed, as the Stodolky effect, a Cavendish-like torsion balance and scattering with UHE cosmic rays. All these methods have extremely small rates, far away from the current sensitivity. The most promising candidate for the $\CNB$ detection is the capture by tritium. The PTOLEMY experiment has been designed as a first attempt for such detection. As result of the possible non-relativistic nature of relic neutrinos, we have seen that the capture rate for the Majorana case is twice the value for Dirac fermions. However, this is a result obtained considering only SM weak interactions. Therefore, we asked ourselves about the implications of having NSI for the specific supposition of neutrinos being Dirac particles. Considering an Effective Field Theory approach and using the limits from data on beta decays, we found that it is possible to increase the Dirac relic capture rate to values numerically identical to the Majorana rate.\\ 

This is a substantial result since it shows that a detection compatible with the Majorana capture rate does not exclude the Dirac nature. Nevertheless, if the NSI bounds are improved this result can change. We also found that NSI can decrease the value of the capture rate. A possible discovery of relic neutrinos with a smaller rate may suggest the existence of beyond SM physics. To confirm this affirmation, we included the largest possible relic right-handed neutrino abundance allowed by the Planck data. Evidently, such additional neutrinos can only increase the relic neutrino capture. Notwithstanding, if the neutrino abundance is diminished by some unknown process, the relic capture can also be reduced, but, a NSI interpretation is also possible in such case. This degeneracy may be solved by a significant improvement on the measurements on the leptonic-hadronic couplings. Let us also stress here that we have not considered other possible beyond SM physics, such as neutrino decay, sterile neutrinos, or clustering processes. Nevertheless, the extension for those cases should be straightforward. In the final chapter, we will deviate from the main study of the thesis, i.e, the analysis of the consequences of the neutrino nature. We will consider the implications of neutrinos in experiments trying to detect the Weakly Interacting Massive Particle, candidate for Dark Matter. %\newpage
			%%%%%%%%%%%%%%%%%%%%%%%%%%%%%%%%%%%%%%%%%%%%%%%%%%%%%%%%%%%%%%%%%%%%%%%%%%%%%%%%%%%%%%%%%%%%%%%%%
%%%%%%%%%%%%%%%%%%%%%%%%%%%%%%%%%%%%%%%%%%%%%%%%%%%%%%%%%%%%%%%%%%%%%%%%%%%%%%%%%%%%%%%%%%%%%%%%%%
%%%%%%%%%%%%%%%%%%%%%%%%%%%%%%%%%%%%%%%%%%%%%%%%%%%%%%%%%%%%%%%%%%%%%%%%%%%%%%%%%%%%%%%%%%%%%%%%%%
\chapter[Neutrino background in DM direct detection experiments]{Neutrino background in Dark Matter direct detection experiments}
\label{cha:NeutrinoFloor}
\chaptermark{Neutrino background in DM direct detection experiments}

\lettrine{M}{any} experimental evidences have shown that there exists in the Universe matter which can not be directly detected by the usual telescopes. Such matter, de\-no\-mi\-na\-ted Dark Matter (DM) as it does not couple with photons, composes approximately $25\%$ of the Universe. It is then a crucial task to unveil its nature and its relationship with the known particles. Several candidates exist to form the Dark Matter; among them, we have the Weakly Interacting Massive Particle (WIMP) \cite{Gelmini:2015zpa}. This candidate has called a lot of attention since it gives the correct cosmological abundances via thermal production with cross sections in the range expected by weak interactions and masses in the $100$ GeV scale; thus, coinciding with the expected region for beyond SM physics. Therefore, if a WIMP exists and if it was created in the Early Universe, it should couple with SM particles, making its detection possible. Hence, experiments have been searching these particles by the interaction with nucleus, which creates a detectable recoil. Nevertheless, confirmed evidences have not been found yet, so, new experiments have been planned.\\ 

Recently \cite{Gutlein:2010tq,Billard:2013qya,Ruppin:2014bra,O'Hare:2016ows}, it has been shown that neutrinos will turn into an irreducible background in those searches. This is due to the coherent scattering between a neutrino and a nucleus, predicted by the SM. We will concentrate ourselves in this chapter in the study of the impact of neutrinos in direct detection searches, considering the presence of NSI. We will consider first the characteristics of direct detection experiments and the parametrization of the neutrino background through the discovery limit. We will introduce the NSI in the form of simplified models, and then we will analyse the limits and consequences of their possible existence. This chapter contains original research mainly published in \cite{Bertuzzo:2017tuf}, and other new results, regarding the inclusion of the reactor antineutrino flux contribution to the discovery limit considering the total number of reactors on the Earth.

%\newpage

%%%%%%%%%%%%%%%%%%%%%%%%%%%%%%%%%%%%%%%%%%%%%%%%%%%%%%%%%%%%%%%%%%%%%%%%%%%%%%%%%%%%%%%%%%%%%%%%%%%%%%%%
\section{Dark Matter Problem and Weakly Interacting Massive Particle}
%%%%%%%%%%%%%%%%%%%%%%%%%%%%%%%%%%%%%%%%%%%%%%%%%%%%%%%%%%%%%%%%%%%%%%%%%%%%%%%%%%%%%%%%%%%%%%%%%%%%%%%%

%%%%%%%%%%%%%%%%%%%%%%%%%%%%%%%%%%%%%%%%%%%%%%%%%%%%%%%%%%%%%%%%%%%%%%%%%%%%%
\subsection{Dark Matter Evidences and The WIMP Miracle}
%%%%%%%%%%%%%%%%%%%%%%%%%%%%%%%%%%%%%%%%%%%%%%%%%%%%%%%%%%%%%%%%%%%%%%%%%%%%%

The DM problem is one of the oldest open problems in Particle Physics. The first evidences originated from the work
of Fritz Zwicky, who studied the Coma Cluster. He showed that the cluster can not be bound by the gravitational attraction of the matter observed. Therefore, one needs the existence of some invisible matter to explain it \cite{Zwicky:1933gu}. Later, in the 1970's Vera Rubin and others concluded that the rotation curves of disk galaxies indicate that the mass of the galaxy is bigger than what is actually observed \cite{Hoekstra:2013via}. Other evidences from gravitational lensing \cite{vanUitert:2012bj,Moustakas:2002iz} show that the mass in a galaxy is about 4 times greater than observed through light. In larger scales, there are additional evidences. At cluster level, the {\it Bullet Cluster} \cite{Clowe:2006eq} is considered a definitive proof of the existence of DM. In that system, in which two clusters are colliding, visible matter and DM behave different from each other. Visible matter suffers a modification in its trajectory while gas in the two clusters is emitting x-rays. But, more important, observations from gravitational lensing show that most of the matter is not in the places in which visible matter is. This is an indication of the existence of an invisible 
collisionless matter, i.e. DM. Finally, at cosmological scales, the observations indicate that visible matter is 5\% of the 
Universe content while an unknown invisible component is about $25\%$ \cite{Ade:2015xua}.\\ 

What do we know about DM? There are some basic properties we know about this invisible component \cite{Gelmini:2015zpa}:
\begin{enumerate}
	\item[1] \textbf{DM is stable or with a lifetime bigger than the age of the Universe.} Evidently, we see DM nowadays in several
	scales, and we know that it has existed since the Big-Bang.
	
	\item[2] \textbf{DM interacts gravitationally and it does not interact with photons.} The current evidences of the existence of 
	DM come	from gravitational observations only. We also know that DM does not couple with light in any observable way. DM can 
	interact	through other interactions, but there are still no evidences of it.
	
	\item[3] \textbf{The mass of DM has only been constrained in $\sim$ 80 orders of magnitude.} An upper bound of the DM mass
	of $m_{\rm DM} \lesssim 2\times 10^{-9} M_{\astrosun}=2\times 10^{48}$ GeV comes from the study of the gravitational 
	microlensing by the Kepler satellite \cite{Griest:2013aaa}. There is no lower experimental limit on the DM mass. For instance, 
	it has been proposed the so-called {\it Fuzzy} DM, which DM is a boson with a de Broglie wavelength of 1 kpc \cite{Hu:2000ke}. 
	
	\item[4] \textbf{DM requires physics beyond the SM.} In the early studies of DM, it was supposed the neutrino could actually be 
	perfect DM candidate. However, as we have seen, neutrino masses are of $\mathcal{O}({\rm eV})$ and they constitute a 
	{\it hot} component in the Universe. This basically means that neutrinos decoupled from the primordial thermal bath at 
	temperatures $T\sim\mathcal{O}({\rm MeV})$ as we previously showed. For DM we need for it to be warm or cold in order to explain 
	the structures that we see in the Universe \cite{Kolb:1990vq}.	Thus, there is no viable candidate of DM in the SM set of 
	particles; new particles are required.
\end{enumerate}

There are several possible DM candidates \cite{Gelmini:2015zpa}. We will concentrate ourselves in the denominated WIMP candidate. These particles are supposed to be produced in the early Universe by interactions present in the primordial bath. Then, due to the expansion of the Universe, the abundance of such particles {\it freezes out} in the same manner we studied in chapter \ref{cap:nuMaj} when we considered Leptogenesis. Thus, if we consider a Dirac stable particles $\chi$ with mass $m_\chi\gg T_{\rm FO}$, $T_{\rm FO}$ the freeze-out temperature, it can been shown that the relic density of those particles is given by \cite{Gelmini:2015zpa}
\begin{align}
	\Omega_\chi h^2 \simeq 0.1 \corc{\frac{x_{\rm FO}}{20}} \corc{\frac{60}{g_{\rm *}}}^{1/2}\frac{3\times 10^{-26}{\rm cm^3 s}}{a+3b/x_{\rm FO}}
\end{align}
where $x_{\rm FO}=m_\chi/T_{\rm FO}$, $g_{\rm *}$ are the relativistic degrees of freedom at freeze-out and $\langle\sigma v\rangle=a+b\langle v^2 \rangle$ is the annihilation cross section, with $a$ and $bv^2$ the s-wave and p-wave contributions. In the case of WIMP
masses in the GeV range, and imposing a correct DM relic density, one obtains that the cross sections should be of 
order $10^{-45}$ cm$^2$, which is of the same order of the weak interactions. This is the denominated WIMP {\it miracle};
the WIMP candidate suggests the existence of new physics above the $100$ GeV scale as indicated by other independent arguments
such as the hierarchy problem. This is the main reason why WIMPs have attracted the attention in the last years. Usually, beyond SM
physics has a stable neutral state which can have a mass of order $100$ GeV, making it a perfect DM candidate. Anyhow, we should keep in mind some assumptions that are made to obtain the previous relic density \cite{Gelmini:2015zpa}. 1) The DM decouples when the Universe is in its radiation-dominated epoch. 2) The WIMP is stable. 3) There is no asymmetry if the WIMP has an antiparticle. Any modification in these assumptions can modify the final result of the relic abundance. We will consider in the rest of the chapter the simplest possibility.\\

Therefore, it is of main interest the detection of WIMPs. There are three basic modes to detect them supposing the
existence of an interaction with SM particles; such interaction can be originated by some beyond SM physics. The three basic
modes are the {\it direct}, {\it indirect} and {\it collider} searches. The direct detection looks for elastic scatterings between
a WIMP and some particle, in such a way that it creates some energy that can be detected. The indirect search probe the
particles resulting from annihilations of WIMP in form of photons, neutrinos, or other particles. In the case of photons,
it is supposed that interactions are originated from higher order levels, suppressing the magnitude of those interactions.
Finally, in colliders, the existence of processes in which there is transverse missing energy are considered as evidence 
of an long lived particle, a WIMP. We will study the direct detection in the next section.

%%%%%%%%%%%%%%%%%%%%%%%%%%%%%%%%%%%%%%%%%%%%%%%%%%%%%%%%%%%%%%%%%%%%%%%%%%%%%
\subsection{Direct Detection Principle}
%%%%%%%%%%%%%%%%%%%%%%%%%%%%%%%%%%%%%%%%%%%%%%%%%%%%%%%%%%%%%%%%%%%%%%%%%%%%%

Since the Earth is moving around the Sun, and we suppose that DM is distributed in the Milky Way galaxy, we expect
that the Earth is under a shower of WIMPs. Furthermore, the flux of these particles is expected to be large, $\Phi\sim 10^7$ 
(GeV/m)/cm$^2$\ s \cite{Gelmini:2015zpa}. Therefore, if DM particles interact with nuclei, for instance, we can expect elastic
scatterings among them. Nonetheless, the energies and rates of such processes are expected to be small, given the order
of magnitude of cross sections involved. This is the reason why experiments are performed underground to avoid large
backgrounds as cosmic rays. Depending on the nature of the DM-SM interactions and the target material of the detector, 
two distinct kind of events can be probed: spin-independent and spin-dependent scatterings. The limits in the first
case are stronger than in the second one, but future experiments will prove both scenarios. Nevertheless, let us focus 
on the spin-independent case. The differential recoil rate for the scattering between a WIMP and a nucleus is given by 
\cite{Gelmini:2015zpa}
\begin{align}\label{eq:rate_DM}
	\left.\frac{dR}{dE_R}\right|_{\chi} &= M \,\ds{\frac{\rho_0}{m_N\,m_{\chi}}} \int_{v_{\rm min}}v f(v)\frac{d\sigma^{\chi}_{\rm SI}}{dE_R}d^3v
\end{align}
where $\rho_0=0.3$~GeV/$c^2$/cm$^3$ is the local DM density~\cite{Agrawal:2010fh,Lewin:1995rx}; $M$ is the total number of nuclei
in the detector; $m_N$ is the mass of the nucleus; $v$ and $m_\chi$ are the DM velocity and mass, respectively; $v_{\rm min}(E_R)$ 
is the minimum WIMP speed required to cause a nuclear recoil with energy $E_R$ for an elastic collision; and  $f(v)$ the WIMP velocity distribution in the Earth's frame of reference. Usually, the differential cross section is parametrized in terms of the total \textit{\textbf{nucleon}}-WIMP cross-section at zero momentum transfer $\sigma_{\chi n}$, defined by 
\cite{Agrawal:2010fh,Gelmini:2015zpa}
\begin{align}
\label{eq:DM_xsec_indep}
\sigma_{\chi n} &= \frac{\mu_n^2}{\mu_N^2} \frac{1}{(Z+N)^2} \int_0^{2 \mu_N^2 v^2/m_N} \left.\frac{d\sigma_{\rm SI}(E_R=0)}{d E_R}\right|_{\rm NP} dE_R,
\end{align}
being $Z,N$ the number of protons and neutrons in the nucleus and $\mu_{n(N)} = m_{n(N)} m_\chi / (m_{n(N)} + m_\chi)$ the 
reduced mass of the WIMP-nucleon(nucleus) system. We write
\begin{align}
	\frac{d\sigma_{\rm SI}}{dE_R}=\frac{m_N}{2\mu_n^2v^2}\sigma_{\chi n} (Z+N)^2{\cal F}(E_R)^2
\end{align}
where we introduced the nuclear form factor~\cite{Lewin:1995rx}
\begin{align}\label{eq:FormFactor}
{\cal F}(E_R)&=3\,\frac{j_1\left(q(E_R)r_N\right)}{q(E_R)r_N}\exp\left(-\frac{1}{2}[s \, q(E_R)]^2\right),
\end{align} 
with $j_1(x)$ is a spherical Bessel function, $q(E_R)=\sqrt{2 m_n (N+Z) E_R}$ the momentum exchanged during the scattering, 
$m_n \simeq 932$~MeV the nucleon mass, $s\sim 0.9$ the nuclear skin thickness and $r_N\simeq 1.14 \, (Z+N)^{1/3}$ the 
effective nuclear radius.
\newpage 
\noindent Thus, the WIMP recoil rate is given by
\begin{align}\label{eq:DMrecoil}
	\left.\frac{dR}{dE_R}\right|_{\chi} &= M \,\ds{\frac{\rho_0}{2\,\mu_n^2\,m_{\chi}}}\sigma_{\chi n} (Z+N)^2{\cal F}(E_R)^2 \int_{v_{min}}\frac{f(v)}{v}d^3v\notag\\
	&=M \, \ds{\frac{\rho_0}{\sqrt{2}\,\mu_n^2\,m_{\chi}}v_0^2}\sigma_{\chi n} (Z+N)^2{\cal F}(E_R)^2\, T(E_R).
\end{align}
The factor $T(E_R)$ is the integral
\begin{align}
	T(E_R)=\frac{\sqrt{\pi} v_0^2}{2}\int_{v_{min}}\frac{f(v)}{v}d^3v.
\end{align}
We will assume a Maxwell-Boltzmann distribution
\begin{align}
	f(v) = \begin{cases}
	\ds{\frac{1}{N_{\rm esc}\,(2\pi\,\sigma_v^2)^{3/2}}}\,\exp\Big[\ \ds{\frac{-(v + v_{\rm lab})^2}{2\sigma_v^2}}\Big] &  | v + v_{\rm lab}| < v_{\rm esc}, \\
 	0 &  |v + v_{\rm lab}| \geq v_{\rm esc},
		\end{cases}
\end{align}
where $v_{\rm esc}=544$ km s$^{-1}$, $v_{\rm lab}=232$ km s$^{-1}$ and $N_{\rm esc}=0.9934$ is a normalization factor taken from~\cite{Agrawal:2010fh}. Then, $T(E_R)$ is given by
\begin{align}
 	T(E_R) &= \ds{N_{\rm esc}}\left\{\ds{\frac{\sqrt{\pi}\,v_0}{4\,v_E}}\left[\text{erf}\left(\ds{\frac{v_{\rm min}+v_{\rm lab}}{v_0}}\right)-\text{erf}\left(\frac{v_{\rm min}-v_{\rm lab}}{v_0}\right)\right]-e^{-\ds{\frac{v^2_{\rm esc}}{v^2_0}}}\right\},\notag\\
 	&=N_{\rm esc}\left(c_1\exp\left[-c_2\frac{E_R}{E_\chi r}\right]-\exp\left[-\frac{v_{\rm esc}^2}{v_0^2}\right]\right),
\end{align}
being $v_0=220$ km s$^{-1}$ is the WIMP velocity and $E_\chi$ is the WIMP kinetic energy, $E_\chi=\frac{1}{2}m_\chi v_0^2$. The numerical factors $c_1,c_2$ are obtaining by fitting the last expression, and are given by $c_1=0.751,c_2=0.561$ \cite{Lewin:1995rx}. The number of DM events per ton-year can be obtained integrating the recoil rate in the energy recoil,
\begin{align}\label{eq:chi_events}
	{\cal N}^\chi &= \int_{E_{\rm th}}\, \left.\frac{dR}{dE_R}\right|_\chi \, \varepsilon(E_R)\, dE_R\,,
\end{align}
where $E_{\rm th}$ is the energy threshold and $\varepsilon(E_R)$ efficiency of the experiment.\\

%%%%%%%%%%    FIG CURRENT BOUNDS WIMP PLANE    %%%%%%%
\begin{figure}[t!]
	\centering
	\includegraphics[width=\textwidth]{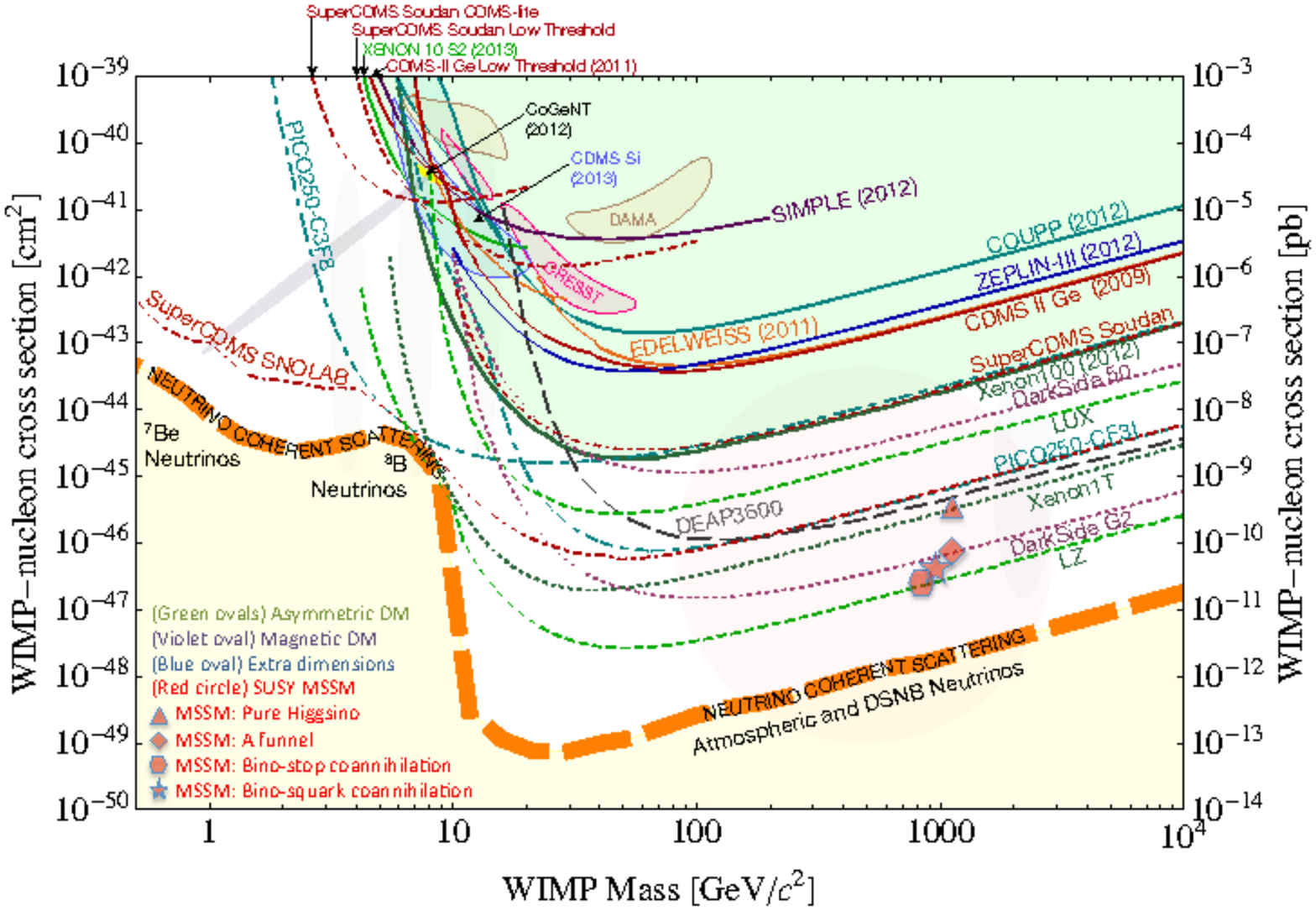} 
	\caption{Current limits (full lines) and future sensitivity (dashed lines) for the WIMP direct detection in the plane $(m_\chi,\sigma_{\chi n})$ assuming WIMPs as the whole DM. The lines correspondent to DAMA/LIBRA (light brown), CoGENT (Yellow), CRESST-II (pink) and CDMS-II-Si (light blue) show signal regions obtained by those experiments. The {\it neutrino floor} corresponds to the thick dashed orange line, computed for a Xe experiment. Regions of interest of some DM candidates are shown as points in the plane. Taken from \protect\cite{Gelmini:2015zpa}.}
	\label{fig:WIMP_lims}
\end{figure}
%%%%%%%%%%%%%%%%%%%%%%%%%%%%%%%%%%%%%%%

Several experiments have been performed in the last decade looking for evidences of WIMPs through direct detection.
The detectors are built with different materials, in order to take advantage of their properties. A first class consists 
of liquid noble gas detectors, using Argon or Xenon, which look for nuclear and electron recoils in form of photons and electrons
in photomultipliers. In this case, we have the experiments DarkSide \cite{Agnes:2014bvk}, Particle and Astrophysical Xenon Detector 
(PandaX) \cite{Tan:2016diz}, the Large Underground Xenon (LUX)~\cite{Akerib:2016vxi}, Xenon10 \cite{Angle:2008we}, 
Xenon100 \cite{Aprile:2012nq}, Xenon $1$T \cite{Aprile:2017iyp}, and the future LUX-ZonEd Proportional scintillation in LIquid Noble 
gases (LUX-ZEPLIN)~\cite{Akerib:2015cja}, DARk matter WImp search with liquid xenoN (DARWIN)~\cite{Aalbers:2016jon}. 
Other experiments, such as Cryogenic Dark Matter Search (CDMS) \cite{Agnese:2015nto}, Cryogenic Rare Event Search with 
Superconducting Thermometers (CRESST) \cite{Angloher:2015ewa}, CoGeNT Dark Matter Experiment \cite{Aalseth:2010vx}, Expérience pour 
DEtecter Les WIMPs En Site Souterrain (EDELWEISS) \cite{Armengaud:2012pfa} are cryogenic detectors in which a nuclear recoil is 
detected by ionization and phonons created in crystals. A third type of detector is a bubble chamber, such as PICO, union of the 
Project In Canada to Search for Supersymmetric Objects (PICASSO) and Chicagoland Observatory for Underground Particle Physics (COUP) 
collaborations \cite{Archambault:2009sm} and Superheated Instrument for Massive ParticLe Experiments (SIMPLE) 
\cite{Felizardo:2011uw}. So far, no strong evidence of WIMPs has been found. There are claims of WIMP detection in experiments such 
as DAMA/LIBRA \cite{Bernabei:2013xsa}, but these are in contradiction with other experiments' results. In figure 
\ref{fig:WIMP_lims}, we show the current limits and future sensitivities of WIMP searches through direct detection in the plane 
$m_\chi \times \sigma_{\chi n}$. The regions above the curves are excluded at $90$\% C.L. As we can see, large part of the parameter 
space has been excluded by experiments putting an stringent limit. Thus, new experiments have been proposed with large 
sensitivities and exposures. However, these future experiments will have an inconvenient. Neutrinos will become a source of 
irreducible background due to the existence of the Coherent Neutrino Scattering off Nuclei, which we will consider next.

%%%%%%%%%%%%%%%%%%%%%%%%%%%%%%%%%%%%%%%%%%%%%%%%%%%%%%%%%%%%%%%%%%%%%%%%%%%%%%%%%%%%%%%%%%%%%%%%%%%%%%%%
\section{Coherent Neutrino Scattering off Nuclei}
%%%%%%%%%%%%%%%%%%%%%%%%%%%%%%%%%%%%%%%%%%%%%%%%%%%%%%%%%%%%%%%%%%%%%%%%%%%%%%%%%%%%%%%%%%%%%%%%%%%%%%%%

A similar event to the one we expect from a WIMP scattering can be produced by neutrinos in the phenomenon called
Coherent Neutrino Scattering off Nuclei (CNSN)~\cite{Cabrera:1984rr,Drukier:1986tm,Strigari:2009bq}, constituting
a background for those searches. Differently from other known backgrounds, such low energy electron recoils,
neutron scatterings or cosmic rays, the CNSN background is irreducible~\cite{Billard:2013qya,Gutlein:2010tq}. An important fact
related to this process is that it has not been observed so far, due to the small cross section; several experiments 
are nevertheless trying to directly observe the CNSN~\cite{Moroni:2014wia,Akimov:2015nza} in the very near future.
Let us study in detail this process, and show the problems that it brings to DM direct detection experiments.

%%%%%%%%%%%%%%%%%%%%%%%%%%%%%%%%%%%%%%%%%%%%%%%%%%%%%%%%%%%%%%%%%%%%%%%%%%%%%
\subsection{A brief on Quantum Mechanical Coherence}
%%%%%%%%%%%%%%%%%%%%%%%%%%%%%%%%%%%%%%%%%%%%%%%%%%%%%%%%%%%%%%%%%%%%%%%%%%%%%

Let us start with a general and brief discussion about coherence in scattering theory. A cross section can be obtained 
from the transition amplitude $f(\vec{k}^\prime, \vec{k}\,)$ as
\begin{align*}
	\frac{d\sigma}{d\Omega}=\abs{f(\vec{k}^\prime, \vec{k}\,)}^2;
\end{align*}
such amplitude contains the information about the interaction between the incident particle and the target. In the
case in which the target is a composed system, the amplitude is the sum over each element, $N$ being the total number of 
constituents
\begin{align*}
	f(\vec{k}^\prime, \vec{k}\,)=\sum_{i=1}^N \tilde{f}_j(\vec{k}^\prime, \vec{k}\,)\exp\llav{i(\vec{k}^\prime-\vec{k})\cdot\vec{x_j}},
\end{align*}
where we introduced a phase factor related to the relative phase of the wave scatterings at $\vec{x_j}$ \cite{Freedman:1977xn}. 
Defining the momentum transfer as
\begin{align*}
	\vec{q}=\vec{k}^\prime-\vec{k},
\end{align*}
and the relative size of the target
\begin{align*}
	R=\max\abs{\vec{x_i}-\vec{x_j}},
\end{align*}
we have that the scattering cross section will depend on the value of $QR$, with $Q=\abs{\vec{q}}$.
If $QR\gg 1$, the phase factors can create cancellations among the different contributions, making the scattering
small, but it can give information about the spacial structure of the target system \cite{Freedman:1977xn}. On the other hand, 
when $QR\ll 1$, the relative phases are negligible and we can compute the cross section as \cite{Freedman:1977xn}
\begin{align}
	\frac{d\sigma}{d\Omega}=N^2\abs{\frac{1}{N}\sum_{i=1}^N \tilde{f}_j(\vec{k}^\prime, \vec{k}\,)}^2,
\end{align}
therefore we see that each contribution adds up {\it coherently}. The incident particle sees the target as a 
whole, and each constituent particle contributes to the cross section in the same form. Now, if the constituents
have a spacial density distribution $\rho(\vec{x})$, the amplitude is modified to \cite{Freedman:1977xn}
\begin{align*}
	f(\vec{k}^\prime, \vec{k}\,)&=\tilde{f}(\vec{k}^\prime, \vec{k}\,)\int\, d^3 x\, \exp\llav{i\vec{q}\cdot\vec{x}}\rho(\vec{x}),\\
	&=\tilde{f}(\vec{k}^\prime, \vec{k}\,){\cal F}(\vec{q}),
\end{align*}
where we defined the form factor ${\cal F}(\vec{q})$ as the Fourier transform of the spatial density distribution,
\begin{align*}
	{\cal F}(\vec{q})=\int\, d^3 x\, \exp\llav{i\vec{q}\cdot\vec{x}}\rho(\vec{x}).
\end{align*}
Let us note that we supposed that each constituent has the same amplitude $\tilde{f}(\vec{k}^\prime, \vec{k}\,)$.
This is true for the nucleus from the point of view of the neutrino; the difference among proton and neutron
is basically a coefficient as we will see in the next section. Again, in the case in which the energies are small,
i.e. $QR\ll 1$, the process is coherent, and the cross section will be proportional to the square of the number of
constituents.

%%%%%%%%%%%%%%%%%%%%%%%%%%%%%%%%%%%%%%%%%%%%%%%%%%%%%%%%%%%%%%%%%%%%%%%%%%%%%
\subsection{SM Cross section}
%%%%%%%%%%%%%%%%%%%%%%%%%%%%%%%%%%%%%%%%%%%%%%%%%%%%%%%%%%%%%%%%%%%%%%%%%%%%%

Now, we can determine the cross section for the CNSN. This is a process in which a neutrino scatters elastically 
from a nucleus $A$, creating a recoil
\begin{align*}
	\nu+A\tto\nu+A.
\end{align*}
Such process is mediated by neutral currents in the SM, so we can start with the effective lagrangian
\begin{align*}
	\mathscr{L}_{\rm eff}^Z=4\sqrt{2}G_F J^\mu_ZJ_{\mu\,Z}
\end{align*}
where the current is given by
\begin{align*}
	J^\mu_Z=\sum_{f=\nu,u,d} \left[\overline{f_L}\gamma^\mu \tau^3 f_L-s_W^2 \overline{f}\gamma^\mu Q f\right].
\end{align*}
Here, the sum is made over all the relevant fermions in the process, $f=\nu,u,d$. We explicitly
have that the relevant lagrangian for the CNSN is given by
\begin{align}
	\mathscr{L}_{\rm eff}^{\rm CNSN}=\frac{G_F}{\sqrt{2}}[\overline{\nu}\gamma^\mu(1-\gamma^5)\nu]\esp{\overline{u}\gamma_\mu\corc{1-\frac{8}{3}s_W^2-\gamma^5}u+\overline{d}\gamma_\mu\corc{-1+\frac{4}{3}s_W^2+\gamma^5}d}.
\end{align}
To compute the CNSN cross section, we need to go from quarks to hadrons. Accordingly, we consider the hadronic
matrix elements of quark currents as in WIMP computations \cite{Agrawal:2010fh}. The amplitude for the CNSN
in terms of initial and final nuclear states $\ket{A^\prime},\ket{A}$, respectively, is given by
\begin{align}
	i\mathcal{M}_{\rm CNSN}=\frac{i G_F}{\sqrt{2}}\sum_q \bra{\nu^\prime}|\overline{\nu}\gamma^\mu(1-\gamma^5)\nu\ket{\nu}\bra{A^\prime}|c_q \overline{q}\gamma_\mu q \ket{A}.
\end{align}
We use the matrix element \cite{Agrawal:2010fh}
\begin{align}
	\bra{A^\prime}|c_q \overline{q}\gamma_\mu q \ket{A}=c_q \widetilde{A}_q
\end{align}
where $\widetilde{A}_q$ is the number of quarks $q$ inside the nucleus $A$, and $c_q$ is the quark coupling 
with the Z boson. Then, to obtain the cross section, we just need to keep in mind that the spin average is 
only done over nucleus states since the neutrino is automatically left-handed in the sources we are considering. 
Finally, the differential cross section in terms of the nuclear recoil energy $E_R$ is \cite{Billard:2013qya}
\begin{align}\label{eq:recoil_SM}
 \left. \frac{d\sigma^{\nu}}{dE_R} \right|_{\rm SM} &= [{\cal Q}_V^{\rm SM}]^2 {\cal F}^2(E_R) \frac{G_F^2 m_N}{4\pi}   \left(1-\frac{m_N E_R}{2 E_\nu^2} \right) \, ,
\end{align}
with the SM coupling factor
\begin{align}
 {\cal Q}_V^{\rm SM} = N + (4 s_W^2 -1) Z.
\end{align}
Here, $N$ and $Z$ are the number of neutrons and protons in the target nucleus, respectively, $E_\nu$ the 
incident neutrino energy and $m_N$ the nucleus mass. Let us stress that this
cross section is proportional to the square of the number of constituents, such as expected in a coherent
scattering. The only difference is the $4 s_W^2 -1$ factor, which comes from the different effective weak coupling of
protons and neutrons. Also, notice the introduction of the nuclear form factor ${\cal F}^2(E_R)$, defined in an
analogous manner as in equation \eqref{eq:FormFactor}.

%%%%%%%%%%%%%%%%%%%%%%%%%%%%%%%%%%%%%%%%%%%%%%%%%%%%%%%%%%%%%%%%%%%%%%%%%%%%%
\subsection{CNSN in Direct Detection Experiments}\label{sec:direct_neutrino}
%%%%%%%%%%%%%%%%%%%%%%%%%%%%%%%%%%%%%%%%%%%%%%%%%%%%%%%%%%%%%%%%%%%%%%%%%%%%%

\begin{figure}[t]
	\begin{center}
    			\includegraphics[width=\textwidth]{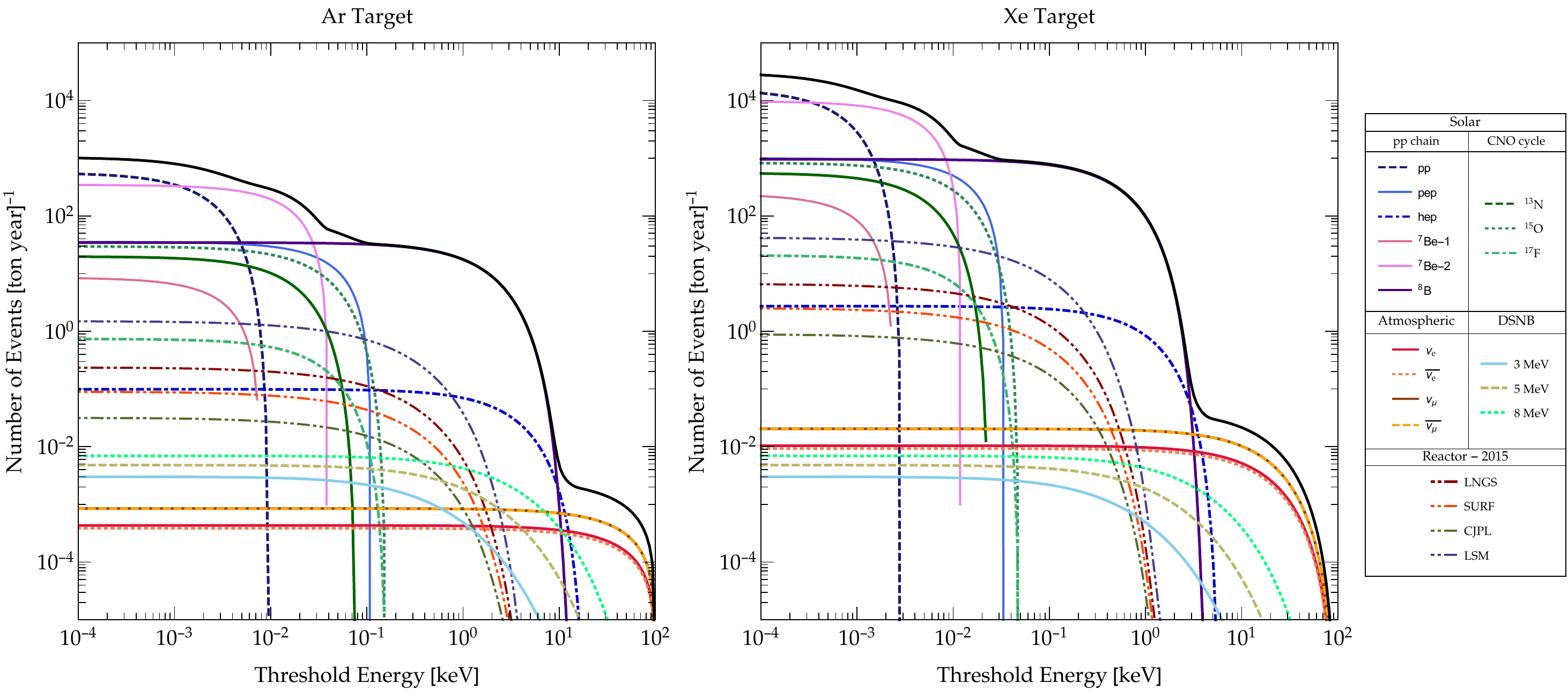}  
	\end{center}
  	\caption{Number of events as a function of the threshold energy for each neutrino sources considered in the present thesis.
  	We show for the cases of a Ar (left) and Xe (right) targets. The black full line corresponds to the total contribution of 
  	all neutrinos at the LSM location. For the cases of reactor antineutrinos, we present the event rates at four different laboratories, LNGS, SURF, CJPL and LSM, computed using the full reactor data from 2015.}
  	\label{fig:TotalNEv}
\end{figure}

After obtaining the differential cross section for the CNSN, we can determine the recoil event rates in terms of the
detector properties, such as exposure, efficiency and target material. The differential recoil rate will be
the integration over the neutrino e\-ner\-gy of the multiplication of the cross section and the incoming flux of neutrinos
\begin{align}
\ds{\left.\frac{dR}{dE_R}\right|_\nu} &= M\int_{E^{\nu}_{\rm min}}\ds{\frac{d\Phi}{dE_{\nu}}}\,\ds{\frac{d\sigma^{\nu}}{dE_R}}dE_{\nu}
\end{align}
where $M$ is the number of target nuclei per unit mass, $d\Phi/dE_{\nu}$ the incident neutrino flux and 
\begin{align*}
	E^{\nu}_{\rm min}=\sqrt{\frac{m_N\,E_R}{2}}
\end{align*}
is the minimum neutrino energy.
%-
\begin{figure}[t]
	\begin{center}
    			\includegraphics[width=\textwidth]{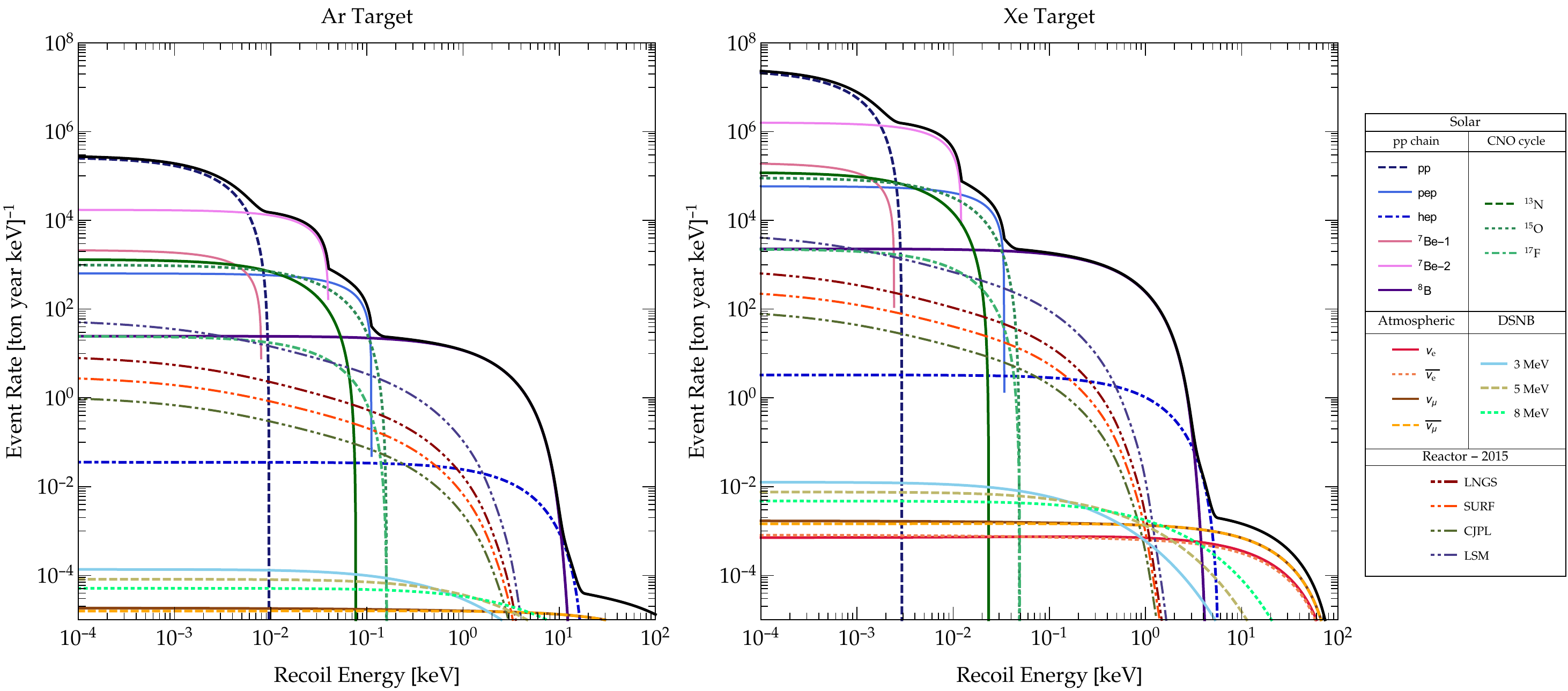}  
	\end{center}
  	\caption{Recoil event rate as a function of the recoil energy for each neutrino sources considered in the present thesis.
  	We show for the cases of a Ar (left) and Xe (right) targets. The black full line corresponds to the total contribution of 
  	all neutrinos at the LSM location. For the cases of reactor antineutrinos, we present the event rates at four different laboratories, LNGS, SURF, CJPL and LSM, computed using the full reactor data from 2015.}
  	\label{fig:TotalRR}
\end{figure}
%-
The fluxes we will consider in this chapter are those described
in the chapter \ref{cha:nu-MP}, i.e., neutrinos coming from solar, atmospheric, DSNB and reactor sources. In figure 
\ref{fig:TotalRR} we present the recoil rates in terms of the nuclear recoil energy for each source and Xe and Ar targets at the 
four different locations on Earth, the positions of the LNGS, SURF, CJPL and LSM laboratories. We can see that the recoil rates are large for small recoil energy, given the large low energy solar neutrino flux; however, such small recoil energies are extremely difficult to measure. On the other hand, $^8$B and $hep$ solar and atmospheric neutrinos have a smaller rate, but larger recoil energies, which in principle could be detected. The reactor antineutrino flux has been computed considering the 2015 data for all the reactors on the Earth, see subsection \ref{subsec:ReacFlux}. We also should notice that the contribution of the reactor antineutrinos to the total rate is indeed small for the laboratories we have chosen, even in the case in which the flux was large, as for the LSM case. Nonetheless, we will study the impact of the reactor antineutrinos flux in the discovery potential of WIMPs for detectors located in those laboratories\footnote{We should mention here the reason of choosing these laboratories. LNGS hosts the Xenon1T, DAMA, CRESST and future DARWIN experiments, SURF is the location of LUX and future LZ experiments,  EDELWEISS and future EURECA will be placed in the LSM site and CJPL is currently operating the PandaX experiment.}. Also, we should note the difference between the recoil rates for Xe and Ar targets. It mainly comes from the fact that Xe has more nucleons than Ar, and the cross section is proportional to the number of components squared.  Now, integrating the recoil rate from the experimental threshold $E_{\rm th}$ up to $100$~keV~\cite{Billard:2013qya}, one obtains the number of neutrino events 
\begin{align}\label{eq:nu_events}
 {\cal N} ^\nu &= \int_{E_{\rm th}}\, \left.\frac{dR}{dE_R}\right|_\nu \, \varepsilon(E_R)\, dE_R.
\end{align}
$E_{\rm th}$ is again the detector threshold energy and $\varepsilon(E_R)$ is the detector efficiency function.
In figure \ref{fig:TotalNEv} we present the number of events in terms of the energy threshold for each source considered,
and in the same locations for a $100\%$ efficiency. As expected by the previous results, we see that the number of events is larger for a smaller threshold, due to the increasing contribution of the solar neutrinos. Yet, from an experimental point of view, such tiny threshold are quite difficult to achieve. For instance, the LUX experiment have a threshold of 1.1 keV \cite{Akerib:2016vxi}, while CRESST has one of $\mathcal{O}(0.1)$eV \cite{Angloher:2015ewa}. Anyhow, we see that for those realistic thresholds, the number of events are of order $10^2$ per ton-year. Let us stress that the dominant contributions in that realistic case are the $^8$B and $hep$ solar neutrinos. Besides this, we see again that the contribution of reactor antineutrinos is smaller when compared with the solar neutrinos contribution, but they should not be neglected for thresholds smaller than 1 keV.\\ 

\begin{figure}[t]
	\begin{center}
    			\includegraphics[width=\textwidth]{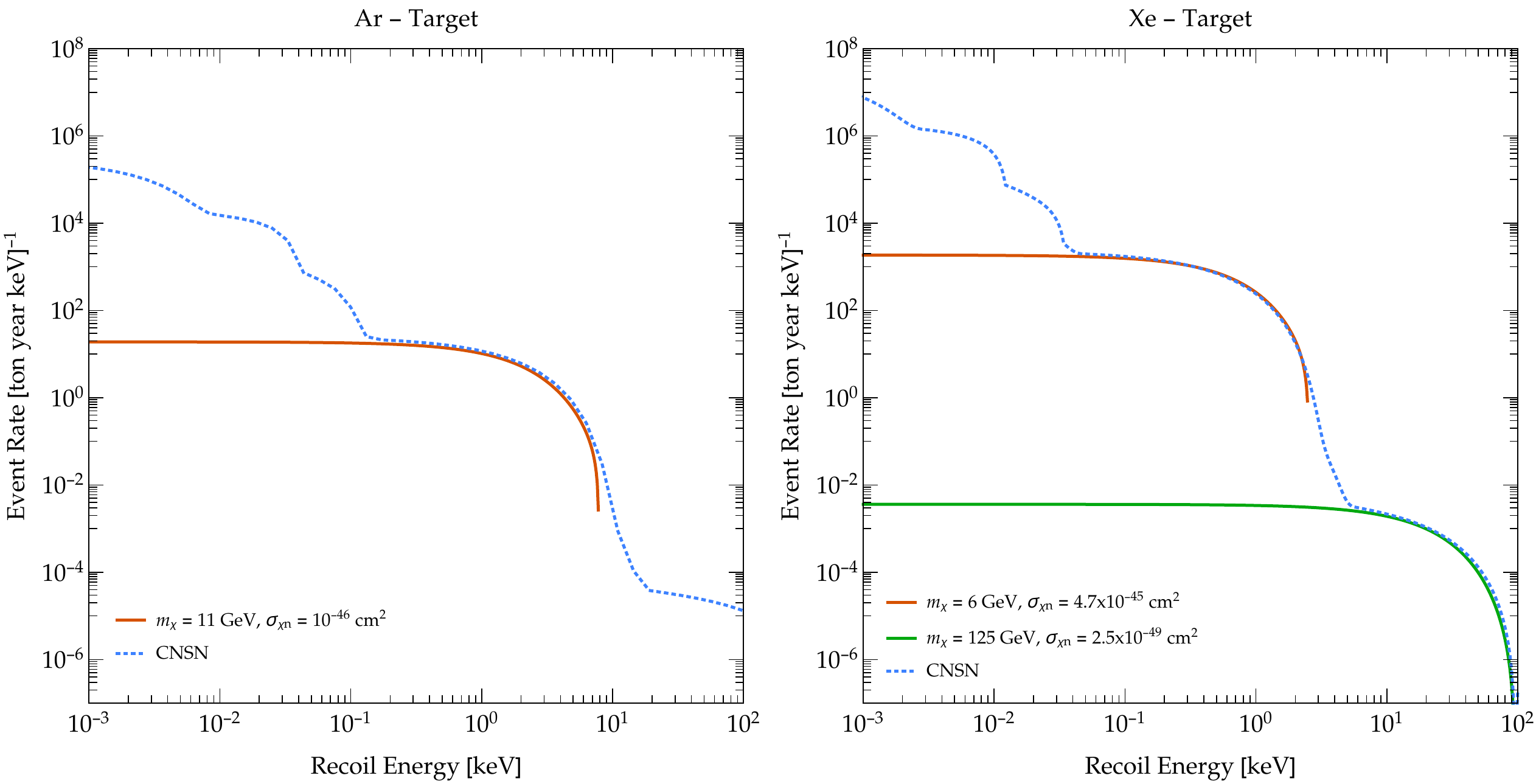}  
	\end{center}
  	\caption{Comparison between the recoil rate of the total CNSN at direct detection experiments with some benchmark WIMPs recoil 
  	rates for a Xe(Ar) target, on the right(left) panel. The benchmark cases correspond to WIMPs with $m_\chi=6$ GeV and 
  	$\sigma_{\chi n}=4.6\times 10^{-45}$ cm$^2$ (red line, right panel) and $m_\chi=125$ GeV and $\sigma_{\chi n}=2.5\times 
  	10^{-49}$ cm$^2$ (green line, right panel) in the Xe case, and $m_\chi=11$ GeV and $\sigma_{\chi n}=10^{-46}$ cm$^2$ (red line, 
  	left panel) for the Ar target.}
  	\label{fig:ExXeAr}
\end{figure}

To demonstrate the implications of the CNSN phenomenon in direct detection searches, we compare the recoil rate for two specific 
cases with the CNSN recoil rate in figure \ref{fig:ExXeAr} for Xe and Ar detectors. We see that in the Xe case the recoil rate for 
a WIMP with a mass of $m_\chi=6$ GeV and cross section of $\sigma_{\chi n}=4.6\times 10^{-45}$ cm$^2$ (red line, right panel) is 
completely mimicked by the CNSN rate (blue line), specifically by the contribution of the $^8$B neutrinos \cite{Billard:2013qya}. 
For a WIMP with $m_\chi=125$ GeV and $\sigma_{\chi n}=2.5\times 10^{-49}$ cm$^2$ (green line, right panel), the atmospheric 
contribution to the CNSN rate resembles the WIMP event rate. In the case of the Ar, we show the case of a WIMP with $m_\chi=11$ GeV 
and $\sigma_{\chi n}=10^{-46}$ cm$^2$ (red line, left panel). This shows that  neutrinos are an irreducible background for WIMP 
direct detection experiments, and such background is dependent of the specific target of the experiment. Therefore, it is important 
to perform a detailed analysis to understand in which point such background becomes significant. This has been done by the introduction of the denominated {\it neutrino floor}, which is basically a discovery limit for direct detection
searches. Let us notice that recently there have been several attempts to distinguish between WIMP and neutrino events in direct 
detection experiments \cite{Dent:2016iht, Cerdeno:2016sfi, Franarin:2016ppr, Grothaus:2014hja, Dent:2016wor}. Nonetheless, we
will consider only the basic direct detection experiments.
%-
%-
%%%%%%%%%%%%%%%%%%%%%%%%%%%%%%%%%%%%%%%%%%%%%%%%%%%%%%%%%%%%%%%%%%%%%%%%%%%%%%%%%%%%%%%%%%%%%%%%%%%%%%%%
\section{Discovery Limit in Direct Detection Experiments}
%%%%%%%%%%%%%%%%%%%%%%%%%%%%%%%%%%%%%%%%%%%%%%%%%%%%%%%%%%%%%%%%%%%%%%%%%%%%%%%%%%%%%%%%%%%%%%%%%%%%%%%%

To establish a minimal DM - nucleon cross section from which the neutrino background due to the CNSN can not be avoided,
we will consider first a background free approach by the introduction of the {\it One-neutrino event contour line}.
Nevertheless, it is important to perform a complete statistical analysis to understand deeply the influence of the neutrino
coherent scattering in direct detection experiments. This will be the task of a subsequent subsection.
%-
%-

%%%%%%%%%%%%%%%%%%%%%%%%%%%%%%%%%%%%%%%%%%%%%%%%%%%%%%%%%%%%%%%%%%%%%%%%%%%%%
\subsection{One-neutrino Event Contour line}\label{sec:dd}
%%%%%%%%%%%%%%%%%%%%%%%%%%%%%%%%%%%%%%%%%%%%%%%%%%%%%%%%%%%%%%%%%%%%%%%%%%%%%

We can start by considering a background free approach in the sense that we want to represent the CNSN in the plane $(m_\chi, \sigma_{\chi n})$. This is done by introducing the {\it one-neutrino event contour line}, line which defines the best background-free WIMP cross section that can be constrained supposing the background to be composed by {\it one} neutrino event. This contour line depends on the experiment performed, i.e. it relies upon its characteristics, such as exposure, threshold energy and target material. To determine the one-neutrino event contour line, we will adopt the procedure established in \cite{Billard:2013qya}. Let us consider a generic experiment. The exposure needed to detect a single neutrino event as a function of the energy threshold
is given by
\begin{align}
	{\cal E}_\nu(E_{\rm th}) &= \frac{{\cal N}^\nu = 1}{\int_{E_{\rm th}}\,  \left.\frac{dR}{dE_R}\right|_{\nu}\,dE_R} .
\end{align}
Then, for each threshold, we compute the background-free exclusion limit at $90\%$ C.L.; that is, the curve in which
we obtain $-\ln(1-0.9)\approx 2.3$ WIMP events employing the previous exposure function
\begin{align}\label{eq:xsec_nufloor}
	\sigma_{\chi n}^{1\nu} & = \frac{2.3}{{\cal E}_\nu(E_{\rm th})\,\int_{E_{\rm th}}\,  \left.\frac{dR}{dE_R}\right|_{\chi,\, \sigma_{\chi n}=1}\, dE_R}\, .
\end{align}
%
%%%%%%%%%%    FIG MODIFIED 1 NEUTRINO EVENT CONTOUR LINE     %%%%%%%
\begin{figure}[t!]
	\centering
	\includegraphics[width=\textwidth]{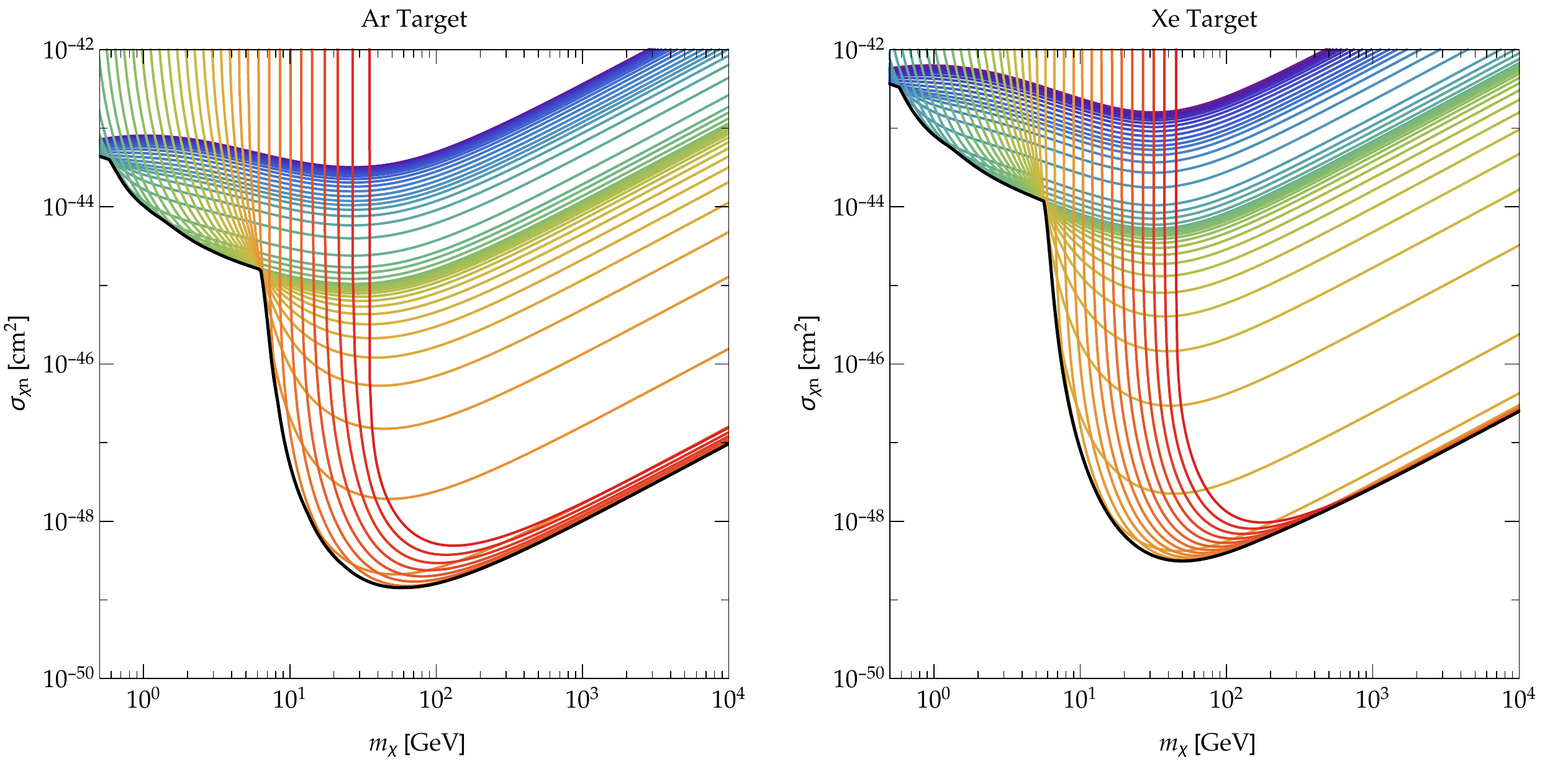}
	\caption{One-neutrino event contour lines (black lines) for Ar (left) and Xe (right) targets in the plane $(m_\chi,\sigma_{\chi n})$. The colored lines correspond to the background-free exclusion limits obtained determining the exposure needed to detect a single neutrino event, hence, the one-neutrino event contour line correspond to the minimum cross section for each WIMP mass.}
	\label{fig:1nuevCL}
\end{figure}
%%%%%%%%%%%%%%%%%%%%%%%%%%%%%%%%%%%%%%%
\newpage
\noindent Finally, we take the lowest excluded cross section for each WIMP mass. Such cross section as function of the mass will
be the one-neutrino event contour line. In other words, the one-neutrino event contour line describes the best background free
sensitivity achievable for each WIMP mass with one-neutrino event. In figure \ref{fig:1nuevCL} we present the one-neutrino event 
contour line (black line) for Ar (left panel) and Xe (right panel) targets computed with solar, atmospheric and DSNB 
neutrinos and reactor antineutrinos at LNGS; we also see that background-free exclusion limits computed considering the exposure 
needed to have a neutrino event for several thresholds.\\ 

We observe that the line corresponding to Xe is higher that the one for Ar. This is due basically to the 
dependence of the recoil rate with the number of nucleons in the WIMP case as well as for neutrinos. Thus, we expect the CNSN to be 
important first for Xe experiments. Furthermore, we see that there is a kink near to a WIMP mass of $6$ GeV; such change is due to 
the presence of the $^8$B neutrinos, and, as we will see later, the discovery limit is worsened for that specific mass.  Let us 
notice that these lines have been computed considering a $100\%$ efficiency in the detector, which is a crude approximation since 
each experiment has a definite efficiency. Anyhow, the one-neutrino contour event line is not a definitive limit for WIMP searches, 
but a preliminary estimate of the region in which CNSN effect becomes relevant. Thus, it is necessary to perform a complete 
statistical analysis which includes background fluctuations related to the uncertainty on the neutrino fluxes. 

%%%%%%%%%%%%%%%%%%%%%%%%%%%%%%%%%%%%%%%%%%%%%%%%%%%%%%%%%%%%%%%%%%%%%%%%%%%%%
\subsection{Discovery Limit}
%%%%%%%%%%%%%%%%%%%%%%%%%%%%%%%%%%%%%%%%%%%%%%%%%%%%%%%%%%%%%%%%%%%%%%%%%%%%%

Experiments constructed for a specific pursuit, such as direct detection experiments, perform detailed statistical
analysis to identify signal events over known backgrounds. Such statistical analysis are undertaken by considering 
the spectra of the possible signal compared to the background one; then, the statistical significance can be evaluated
using some test. The significance is a endorsement of the model predictions by the data obtained experimentally. 
In other words, it indicates if the model represents accurately the data. However, in general, there exists some parameters which are not known a priori when an experiment is carried out; instead, they are obtained by some other data set. Those parameters are known as {\it nuisance parameters}. Therefore, a procedure to determine the statistical significance needs to take into account those parameters in a fully determined way. A common frequentist approach uses a \textbf{likelihood ratio} as test statistics. 
It will be the basis to estimate the discovery limit in a direct detection experiment. We will follow the approach conceived 
from the general treatment by Cowan et.\ al.\ \cite{Cowan:1992xc} by Billard and collaborators \cite{Billard:2011zj}. It was later applied to the CNSN background in the references \cite{Ruppin:2014bra,O'Hare:2016ows}. Let us discuss the method to obtain the discovery limit with some detail. We start by defining a binned likelihood function as
\begin{align}\label{eq:likelihood}
	{\cal L}(\sigma_{\chi n},m_\chi,\phi_\nu,\Theta) &= \prod_{i=1}^{n_{\rm b}} {\cal P} 
	\left( {\cal N}^{\rm obs}_i \vert {\cal N}^\chi_i + \sum_{j=1}^{n_\nu} {\cal N}^{\nu}_i(\phi_\nu^j);\Theta \right) \times  		\prod_{j=1}^{n_\nu} {\cal L}(\phi_\nu^j).
\end{align}
This likelihood is built as the product of Poisson probability distribution functions ${\cal P}$ for each recoil energy bin $i$. We will consider $n_{\rm b}=50$ as the total number of bins. The likelihood functions ${\cal L}(\phi_\nu^j)$ correspond to Gaussian functions parametrizing the uncertainties on each neutrino flux parameter; thus, neutrino fluxes will be here the nuisance parameters. We will consider the fluxes and their uncertainties presented in the chapter \ref{cha:nu-MP}, section \ref{sec:NFl}. Besides, $j=1,\ldots, n_\nu$ correspond to each neutrino component so far considered in this thesis. The neutrino (${\cal N}^{\nu}_i$) and WIMP (${\cal N}^\chi_i$) number of events are computed as in equations \eqref{eq:nu_events} and \eqref{eq:chi_events}, but integrated in the intervals of the energy bin $[E_i,E_{i+1}]$. Finally, $\Theta$ will include the information on extra parameters that could be present in either neutrinos or WIMPs number of events. The test between the neutrino-only hypothesis $H_0$ and the neutrino+WIMP hypothesis $H_1$ consists in defining the ratio --for a fixed WIMP mass--
\begin{equation}
\lambda(0) = \frac{{\cal L}(\sigma_{\chi n}=0,\hat{\hat{\phi}}_\nu,\Theta)}{{\cal L}
(\hat{\sigma}_{\chi n},\hat{\phi}_\nu,\Theta)}
\end{equation}
where $\hat{\phi}_\nu$ and $\hat{\sigma}_{\chi n}$ are the fluxes and WIMP-nucleon cross section values which
maximize the likelihood ${\cal L} (\hat{\sigma}_{\chi n},\hat{\phi}_\nu,\Theta)$; meanwhile, $\hat{\hat{\phi}}_\nu$ 
is obtained by maximizing the likelihood function in the case of the null hypothesis, ${\cal L}(\sigma_{\chi n}=0,\hat{\hat{\phi}}_\nu,\Theta)$.
\newpage
\noindent Then, to assess the positive signal we compute the {\it test statistics} \cite{Cowan:1992xc,Billard:2011zj}
\begin{align*}
	q_0=\begin{cases}
		-2\ln \lambda(0) & \hat{\sigma}_{\chi n}\ge 0,\\
		0 & \hat{\sigma}_{\chi n}< 0;
		\end{cases}
\end{align*}
we have that $q_0$ measures the discrepancy between the null and the alternative (positive)
hypothesis. This test is specially outlined to determine the rejection of the background-only hypothesis 
to appraise a discovery. Hence, we need to compute the {\it p}-value $p_0$ from the probability density function
 of $q_0$ under the background-only hypothesis $H_0$, $f(q_0\vert H_0)$, as
\begin{align*} 
	p_0=\int_{q^{\rm obs}}^\infty\,f(q_0\vert H_0)\,dq_0,
\end{align*}
being $q^{\rm obs}$ the observed value of $q_0$ from the data. In other words, $p_0$ is the probability of 
a disagreement between $H_0$ and $H_1$ to be equal or larger than the value $q^{\rm obs}$. Thus, it is necessary to
know the probability density function (p.d.f.) $f$. Considering Wilk's theorem, it can be shown that such p.d.f.\
follows a half chi-square distribution for one degree of freedom, $\frac{1}{2}\chi_1^2$ \cite{Cowan:1992xc}. Therefore, the 
significance $Z_0$ will be simply
\begin{align*}
	Z_0=\sqrt{q_0},
\end{align*}
in units of $\sigma$.\\

Evidently, the previous procedure is done when a real experiment has data to be analysed. Nevertheless, we can perform
a simulation in order to estimate the discovery limit in direct detection searches. We will create simulated spectra for
both WIMP and neutrinos, and then we will compute the significance through the procedure previously presented. Clearly, such significance will not have a definite meaning since it will dependent on the details of the simulated spectrum. Thus, we will construct an ensemble of $500$ simulated experiments to obtain a statistical set of significances from which we can extract relevant information. Let us define $Z_{90}$ as the significance that can be obtained $90\%$ of the times in the statistical ensemble by computing the quantile function of $0.9$ as \cite{Billard:2011zj,Ruppin:2014bra,O'Hare:2016ows}.
\begin{align*}
	Z_{90}&=Q(0.90),
\end{align*}
or, equivalently,
\begin{align}
	\int_0^{Z_{90}} p(Z\vert H_0) \, dZ &=0.90 \, ,
\end{align}
where $p(Z\vert H_0)$ is the p.d.f.\ of the significances obtained from the simulated experiments ensemble. Hence, for
each WIMP mass and cross section, we will have a value of $Z_{90}$.
\newpage
\noindent We will define the discovery limit as {\it the minimum
value of the cross-section in which an experiments has a $90\%$ probability of making a 3$\sigma$ discovery for each WIMP 
mass} \cite{Billard:2011zj,Ruppin:2014bra,O'Hare:2016ows}. In other words, the {\bf neutrino floor} will correspond to the 
value of $\sigma_{\chi n}$ having $Z_{90}=3$.\\ 

\begin{figure}[t]
	\centering
	\includegraphics[width=\textwidth]{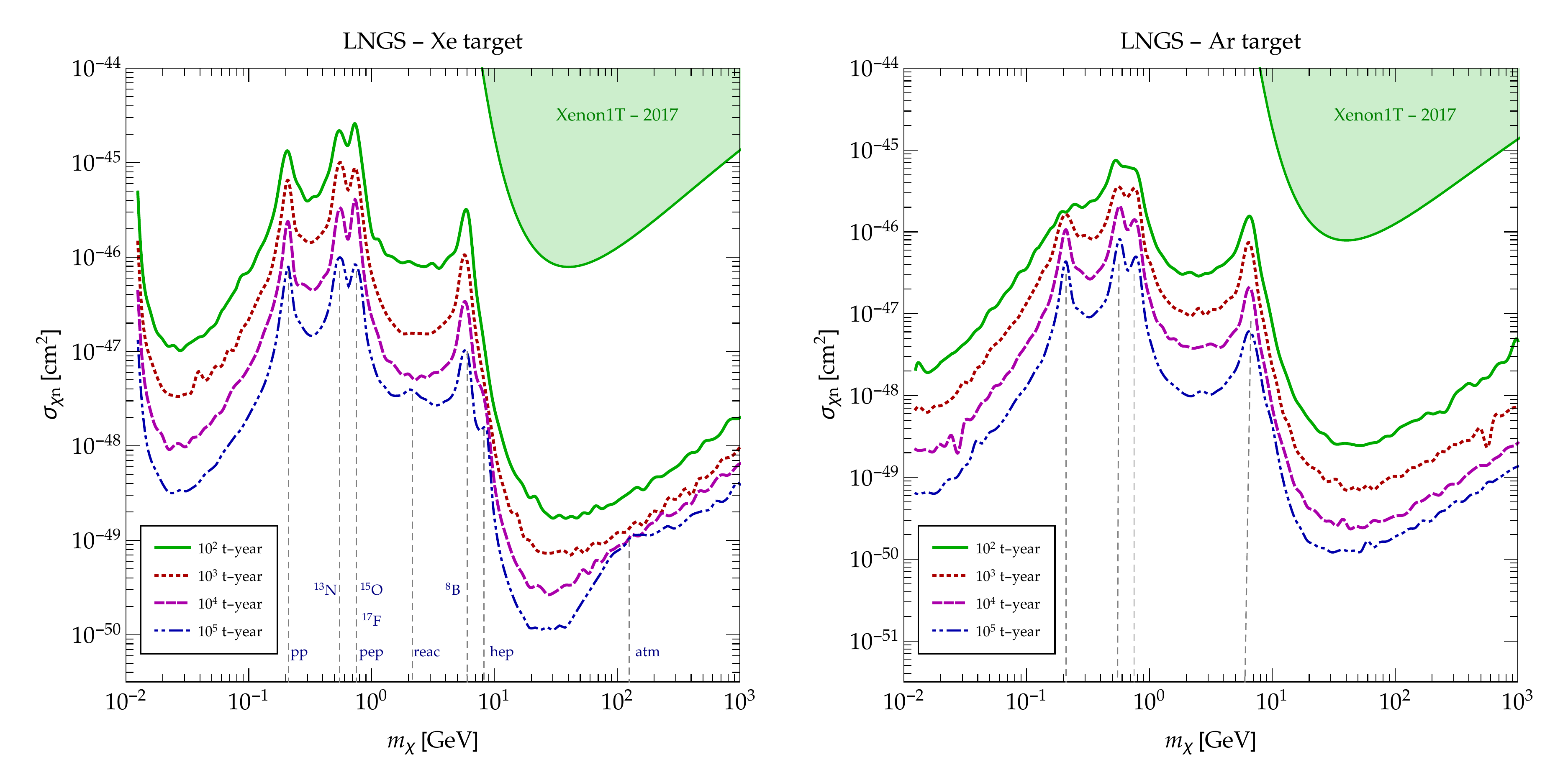}
	\caption{Dependence of the discovery limit on the exposure for Xe (left) and Ar (right) targets. The region in green has been excluded by Xenon1T experiment \protect\cite{Aprile:2017iyp}. The gray lines correspond to the masses which are most affected by the CNSN background; we show for each peak the relevant neutrino component related to it.}
	\label{fig:DLXeArExp}
\end{figure}

\begin{figure}[t]
	\centering
	\includegraphics[width=\textwidth]{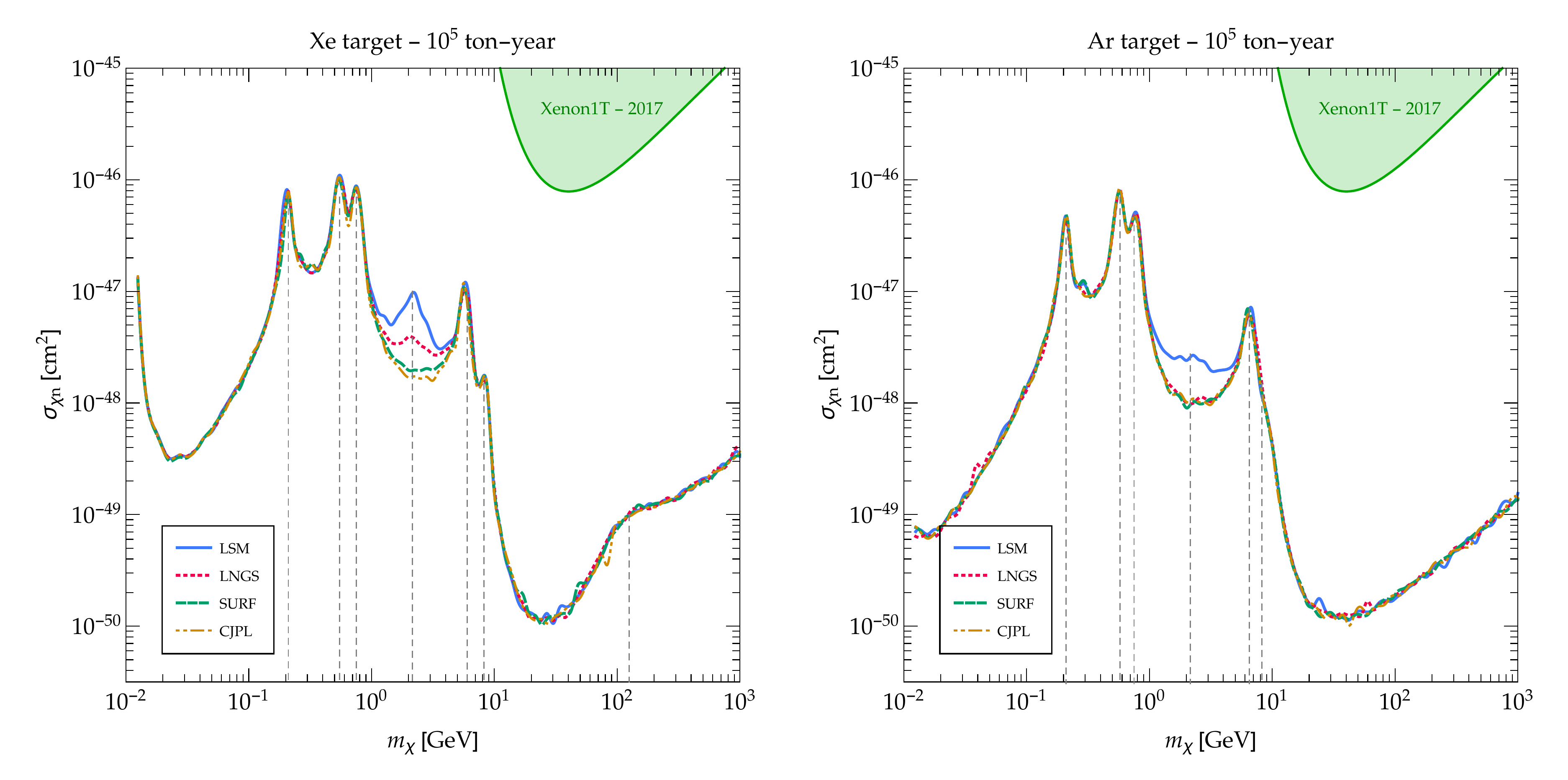}
	\caption{Dependence on the location of the discovery limit for a Xe (left) and Ar (right) targets. We considered an exposure of $10^5$ ton-year, and the four laboratories we have analysed so far. The LSM has the largest reactor contribution, as expected.}
	\label{fig:DLXeArLoc}
\end{figure}

In figure \ref{fig:DLXeArExp} and \ref{fig:DLXeArLoc}, we present the results of the 
WIMP discovery limit in the plane $(m_\chi,\sigma_{\chi n})$ for the same laboratories considered previously, using Xe and Ar as 
targets, and an artificial threshold of $0.01$ eV. Specifically, in figure \ref{fig:DLXeArExp}, we show the dependence of the 
discovery limit on the exposure of a simulated experiment at LNGS. Each peak appearing in the discovery limit corresponds to the 
region in which the WIMP spectrum is highly mimicked by some neutrino component. Explicitly, the peaks correspond to the $pp$,
$^{13}N$, $pep+\,^{15}O+\,^{17}F$, reactor, $^8B$, $hep$, and atmospheric neutrinos, respectively 
\cite{Ruppin:2014bra,O'Hare:2016ows}. We indicate the masses more affected by these neutrinos using gray dashed lines.  We also see 
that when increasing the exposure, different peaks start to appear.\\ 

Different from what has been considered in literature, we considered the contribution due to reactor antineutrinos to the discovery 
limit. Such reactor contribution at LNGS appears at large exposures ($10^5$ ton-year). Furthermore, we determined the discovery 
limit for different laboratories to understand the dependence on the location. Figure \ref{fig:DLXeArExp} shows the discovery limit 
for the four laboratories we have considered so far. As expected, the discovery limit only differs on the reactor associated peak, 
near to a WIMP mass of $\sim 2$ GeV. For the LSM, the contribution is the largest as such laboratory is relatively close to the 
reactors in France. Meanwhile, for the LNGS, CJPL and SURF, we see that the reactor contribution is small for this exposure.\\ 

Finally, we 
see in both figures the difference between the discovery limits for distinct target materials. The positions of the peaks are not 
modified significantly, since both WIMP and neutrino spectra are multiplied by the number of the constituents. Nonetheless, for a 
experiment whose target is composed by different atoms, the discovery limit should be different. Also, for different exposures, the 
lowest cross section that can be studied by the experiment is different for each target.

%%%%%%%%%%    
%%%%%%%%%%%%%%%%%%%%%%%%%%%%%%%%%%%%%%%

%%%%%%%%%%    
%%%%%%%%%%%%%%%%%%%%%%%%%%%%%%%%%%%%%%%
%%%%%%%%%%%%%%%%%%%%%%%%%%%%%%%%%%%%%%%%%%%%%%%%%%%%%%%%%%%%%%%%%%%%%%%%%%%%%%%%%%%%%%%%%%%%%%%%%%%%%%%%
\section{Non-Standard Interactions and the Discovery Limit}\label{sec:frame}
%%%%%%%%%%%%%%%%%%%%%%%%%%%%%%%%%%%%%%%%%%%%%%%%%%%%%%%%%%%%%%%%%%%%%%%%%%%%%%%%%%%%%%%%%%%%%%%%%%%%%%%%

We have considered so far the existence of a WIMP $\chi$ which couples to the SM particles. Such interaction is only parametrized by 
the WIMP-nucleon cross section, since further details about the couplings are actually unnecessary. On the other hand, we can 
speculate about the existence of NSI coupling with neutrinos in such a way that they modify the results 
obtained previously. Earlier works have shown that NSI affecting neutrino-nucleus scattering will change the CNSN at direct 
detection experiments \cite{Harnik:2012ni, Pospelov:2012gm, Cerdeno:2016sfi, Dent:2016wor}. Notwithstanding, these works suppose 
that the beyond SM physics is present in either the WIMP sector or the neutrino scattering. It is reasonable that NSI converse with both neutrino and WIMP sectors; furthermore, it is possible that the new physics is responsible of the smallness of neutrino masses 
\cite{Huang:2014bva}. We can then analyse the effects of an NSI coupling with both neutrinos and WIMPs on the discovery limit at 
future experiments. Accordingly, we will work within the framework of simplified models~\cite{Abdallah:2015ter,DeSimone:2016fbz}, 
models which consist in the addition of a WIMP and a mediator to the SM. In such frameworks, it is also supposed that the WIMP is odd 
under an additional $\Z_2$ symmetry, forbidding its decay on SM particles, but allowing scattering processes. We will concentrate 
ourselves on two cases for the mediator, a scalar and a vector boson \cite{Bertuzzo:2017tuf}. Let us start with the vector scenario.

%%%%%%%%%%%%%%%%%%%%%%%%%%%%%%%%%%%%%%%%%%%%%%%%%%%%%%%%%%%%%%%%%%%%%%%%%%%%%
\subsection{Vector Mediator}
%%%%%%%%%%%%%%%%%%%%%%%%%%%%%%%%%%%%%%%%%%%%%%%%%%%%%%%%%%%%%%%%%%%%%%%%%%%%%

%
In this case, our mediator will be a real boson $V_\mu$, with a mass $m_V$, described by the following lagrangian
\begin{align}
\label{eq:simp_model_vec}
{\cal L}_{\rm vec}  &= V_\mu (J_f^\mu + J_\chi^\mu)+ \displaystyle \frac{1}{2}\,m_V^2 V_\mu V^\mu \, , 
\end{align}
\newpage
\noindent coupling with the fermions $f=\{u,d,\nu, \chi\}$ as
\begin{align}
\label{eq:currents}
J_f^\mu = \sum_f  \overline{f} \gamma^\mu (g_V^f + g_A^f \gamma_5 ) f
\end{align}
where we have introduced the vector $g_V^f$ and axial-vector $g_A^f$ couplings. Let us also notice that we have written
down left- and right-handed currents, supposing the existence of right-handed neutrinos; thus, neutrinos can be Dirac or Majorana here, but this will be irrelevant for the purposes of the present chapter as the CNSN is independent of the neutrino nature. It is important to note that constraints coming from other experiments should be applied to these simplified models to obtain a consistent Ultraviolet completion. For instance, a $U(1)$ gauge scenario, studied in \cite{Kahlhoefer:2015bea}, has large regions of the parameter space that may be excluded depending on the nature of the couplings $g_A^f$.\\ 

Other constraints may be applied if there is isospin breaking or if the new physics affects the electroweak precision measurements. However, such analysis is model dependent, and we will just consider here limits in the {\it invisible} sector of neutrinos and WIMPS coming from direct detection experiments. Clearly, experimental limits constraint the neutral vector boson mass depending on the couplings with electrons and muons \cite{Abreu:1994ria,Agashe:2014kda,Carena:2004xs,Aad:2012hf,Chatrchyan:2012oaa,ATLAS:2012pu}; nevertheless, these limits can be avoided for specific fermion charges.

\newpage The differential cross section for the CNSN is indeed modified by the additional vector boson; we get
\begin{align}
\label{eq:recoil_vector}
{\displaystyle \left. \frac{d\sigma^{\nu}}{dE_R}\right|_{\rm V}} &= {\displaystyle {\cal G}_V^{2} \left. \frac{d\sigma^{\nu}}{dE_R}\right|_{\rm SM} \, , }\quad \text{with} \quad
{\displaystyle{\cal G}_V} = {\displaystyle  1 + \frac{\sqrt{2}}{G_F} \frac{{\cal Q}_V}{{\cal Q}_V^{\rm SM}} \frac{g_V^\nu - g_A^\nu}{q^2-m_V^2} \,,}
\end{align}
where $q^2 = - 2 \, m_N E_R$ is the square of the momentum transferred in the scattering process, and the coupling ${\cal Q}_V$ is obtained using the same matrix element as in the SM, but keeping in mind that couplings with quarks are now arbitrary \cite{Agrawal:2010fh}
\begin{align}
	{\cal Q}_V = (2Z+N) g_V^u + (2N+Z) g_V^d\,.
\end{align}
Let us stress here that we have assumed that neutrino production in the Sun is basically unaltered by the vector boson as 
only left-handed neutrinos interact with the target nuclei. Furthermore, if the new interaction couples only with right-handed 
neutrinos, there will not be any modification to the CNSN, so we see that the interference term, proportional to $g_V^\nu - 
g_A^\nu$, can increase or decrease the number of events expected in a detector. Therefore, ${\cal G}_V$ can be smaller (larger) than 
$1$ if $g_V^\nu < g_A^\nu$ ($g_V^\nu > g_A^\nu$) since $q^2 - m_V^2 = - (2 \, m_N E_R + m_V^2)$ is negative. Notice that we recover 
the SM result when ${\cal G}_V = 1$, for zero or completely right-handed interactions. In the WIMP sector, we will keep considering 
only spin-independent interactions by setting $g_A^\chi=0$. It is possible then to obtain the explicit dependence on the couplings 
of the differential cross section as
\begin{align}
 \left. \frac{d\sigma^{\chi}_{SI}}{dE_R} \right|_V &= 
 		{\cal F}^2(E_R) \frac{(g_V^\chi)^2 {\cal Q}_V^2}{4\pi}\frac{m_\chi m_N}{E_\chi (q^2-m_V^2)^2},
\end{align}
with $E_\chi$ the incident WIMP energy.

%%%%%%%%%%%%%%%%%%%%%%%%%%%%%%%%%%%%%%%%%%%%%%%%%%%%%%%%%%%%%%%%%%%%%%%%%%%%%
\subsection{Scalar Mediator}
%%%%%%%%%%%%%%%%%%%%%%%%%%%%%%%%%%%%%%%%%%%%%%%%%%%%%%%%%%%%%%%%%%%%%%%%%%%%%

The second scenario we will consider here consists in the introduction of a real scalar boson, $S$, with mass $m_S$, whose 
lagrangian is 
\begin{align}
\label{eq:simp_model_scal}
{\cal L}_{\rm sc} &= S\sum_f g_S^f\, \overline{f}\, f - \displaystyle \frac{1}{2}\,m_S^2 S^2,
\end{align}
where $g_S^f$ are the couplings between the scalar and the fermions. Notice that we are considering a CP even real scalar to forbid spin dependent interactions at direct detection experiments. Besides, a right-handed neutrino is also assumed to be present in the particles set. Analogous to the vector case, one can wonder about a UV-completion of this simplified model. Nonetheless, such discussion is not crucial for the results we will obtain later; hence, we will not enter in details about it here. See, for instance \cite{Bertuzzo:2017tuf}. The differential cross section for the CNSN is once more modified by the new physics, as expected. The modification is however different to the previous case as the Dirac structure is different. Therefore, the hadron matrix element needs to take into account that a scalar is coupling to the quarks. In such case, the matrix element is given by \cite{Agrawal:2010fh}
\begin{align}
	{\cal Q}_S = \sum_q \bra{A^\prime}|g_S^q \overline{q} q \ket{A} = Z f_p + (A-Z)f_n,
\end{align}
where $f_{p,n}$ are the effective couplings with protons and neutrons, which are given by
\begin{align}
	f_{p(n)}&=m_n\sum_{q=u,d,s} g_S^q \frac{f_{Tq}^{p(n)}}{m_q}  +\frac{2}{27} \left( 1- \sum_{q=u,d,s}  f_{Tq}^{p(n)} \right) \sum_{q=c,b,t} \frac{g_S^q}{m_q},
\end{align}
being $m_n$ is the average nucleon mass and $f_{Tq}^{p,n}$ correspond to the effective low energy coupling between a scalar mediator
and a quark inside a proton or neutron, respectively. These form factors can be obtained using chiral perturbation theory. We will 
use the values from the micrOMEGAs package \cite{Belanger:2014vza}, given by $f_{Tu}^p = 0.0153$, $f_{Td}^p=0.0191$, $f_{Tu}^n = 
0.011$, $f_{Td}^n = 0.0273$ and $f_{Ts}^{p,n} = 0.0447$. Let us stress that a more recent determination has been done in 
\cite{Alarcon:2011zs,Alarcon:2012nr}; we have concluded that such new determination will affect our results by $\sim 30\%$. The
differential cross section is then found to be
\begin{align}
\label{eq:recoil_scalar}
{\displaystyle\left. \frac{d\sigma^{\nu}}{d E_R} \right|_{\rm S}}  &= {\displaystyle\left. \frac{d\sigma^{\nu}}{d E_R} \right|_{\rm SM} + {\cal F}^2(E_R) \frac{{\cal G}_S^2 G_F^{2}}{4\pi} \, \frac{m_S^4 E_R m_N^2 }{E_\nu^2 (q^2-m_S^2)^2} \, ,} \quad \text{with} \quad
{\displaystyle{\cal G}_S }= {\displaystyle \frac{|g_S^\nu| {\cal Q}_S}{G_F \, m_S^2}.}
\end{align}
We see here that the modified differential cross section has an additional term coming from the different Dirac structure present
in the lagrangian. Therefore, we expect that the differential recoil rate and the number of events to be modified in shape.\\ 

On the 
other hand, the values of ${\cal Q}_S$ are quite large since it is a target dependent quantity. For instance, in a Xe experiment
we have that ${\cal Q}_S \approx 1400 \,g_S^q$, considering universal quark-scalar couplings. Hence, for $|g_S^\nu|\sim 1$, $|g_S^q|
\sim 1$, $m_S\sim 100$ GeV, values of ${\cal G}_S$ are $\sim 10^4$. Finally, we can compute the differential cross section for WIMP
scatterings at direct detection experiments
\begin{align}
 \left. \frac{d\sigma^{\chi}_{SI}}{dE_R} \right|_S &= 
 		{\cal F}^2(E_R) \frac{(g_S^\chi)^2 {\cal Q}_S^2}{4\pi} \frac{m_\chi m_N}{E_\chi (q^2-m_S^2)^2}.
\end{align}
Now, having established the models which will be our framework, we can analyse their impact in direct detection experiments.

\newpage

%%%%%%%%%%%%%%%%%%%%%%%%%%%%%%%%%%%%%%%%%%%%%%%%%%%%%%%%%%%%%%%%%%%%%%%%%%%%%
\subsection{Current Limits and Future Sensitivity}\label{sec:limits}
%%%%%%%%%%%%%%%%%%%%%%%%%%%%%%%%%%%%%%%%%%%%%%%%%%%%%%%%%%%%%%%%%%%%%%%%%%%%%

As a first step we should consider the limits that current and future experiments can impose on the simplified models. Direct 
detection experiments actually do not discriminate among neutrinos and WIMPs as they just compute the number of nuclear recoil 
events seen in the detector. Therefore, the constraints will be in the sum of scattering events from the two kinds of particles. For 
the present analysis we will adopt the 2016 results for the LUX experiment \cite{Akerib:2016vxi}, although Xenon $1$T results 
\cite{Aprile:2017iyp} are more recent and stronger. The main reason is that Xenon $1$T has a larger threshold than LUX, and the 
efficiency is smaller for lower recoil energies. Hence, we have checked that the limits in the neutrino sector are weaker than those 
from LUX, while in the WIMP sector bounds have not improved considerably. On the other hand, we will analyse the reach of two future 
proposed Xe experiments, LUX-ZEPLIN \cite{Akerib:2015cja} and DARWIN \cite{Aalbers:2016jon}.\\

\noindent{\bf Current bounds}. For the current limits we consider the LUX results obtained after a $3.35 \times 10^4$ kg-days run presented in 2016~\cite{Akerib:2016vxi}, with an energy threshold of $1.1$ keV. The energy efficiency $\varepsilon(E_R)$ is also 
acquired from the same published result. We compute the likelihood function built as a Poisson p.d.f.\ 
\begin{align}\label{eq:likeN}
	\mathcal{L}({\cal N}^{\rm t}(\hat{\theta})|{\cal N}^{\rm o})=\frac{(b+{\cal N}^{\rm t}(\hat{\theta}))^{{\cal N}^{\rm o}} e^{-(b+{\cal N}^{\rm t}(\hat{\theta}))}}{{\cal N}^{\rm o}!}
\end{align}
where ${\cal N}^{\rm t}$ correspond to the total number of nuclear recoil events as
\begin{align}
{\cal N}^{\rm t}(\hat{\theta}) & = {\cal N}^{\chi}(\hat{\theta}) + {\cal N}^{\nu}(\hat{\theta}),
\end{align}
being $\hat{\theta}$ the set of parameters of each model. In the likelihood function, we also have the observed number of events ${\cal N}^{\rm o}$ and the expected background $b$. We employ ${\cal N}^{\rm o}=2$ for the number of observed events and $b=1.9$ for the estimated background. Thus, to obtain limits in the different planes of the parameter space, we will maximize the likelihood function \eqref{eq:likeN}.
%%%%%%%%%%    FIG CURRENT BOUNDS VECTOR MODEL    %%%%%%%
\begin{figure}[t!]
\centering
\includegraphics[width=\textwidth]{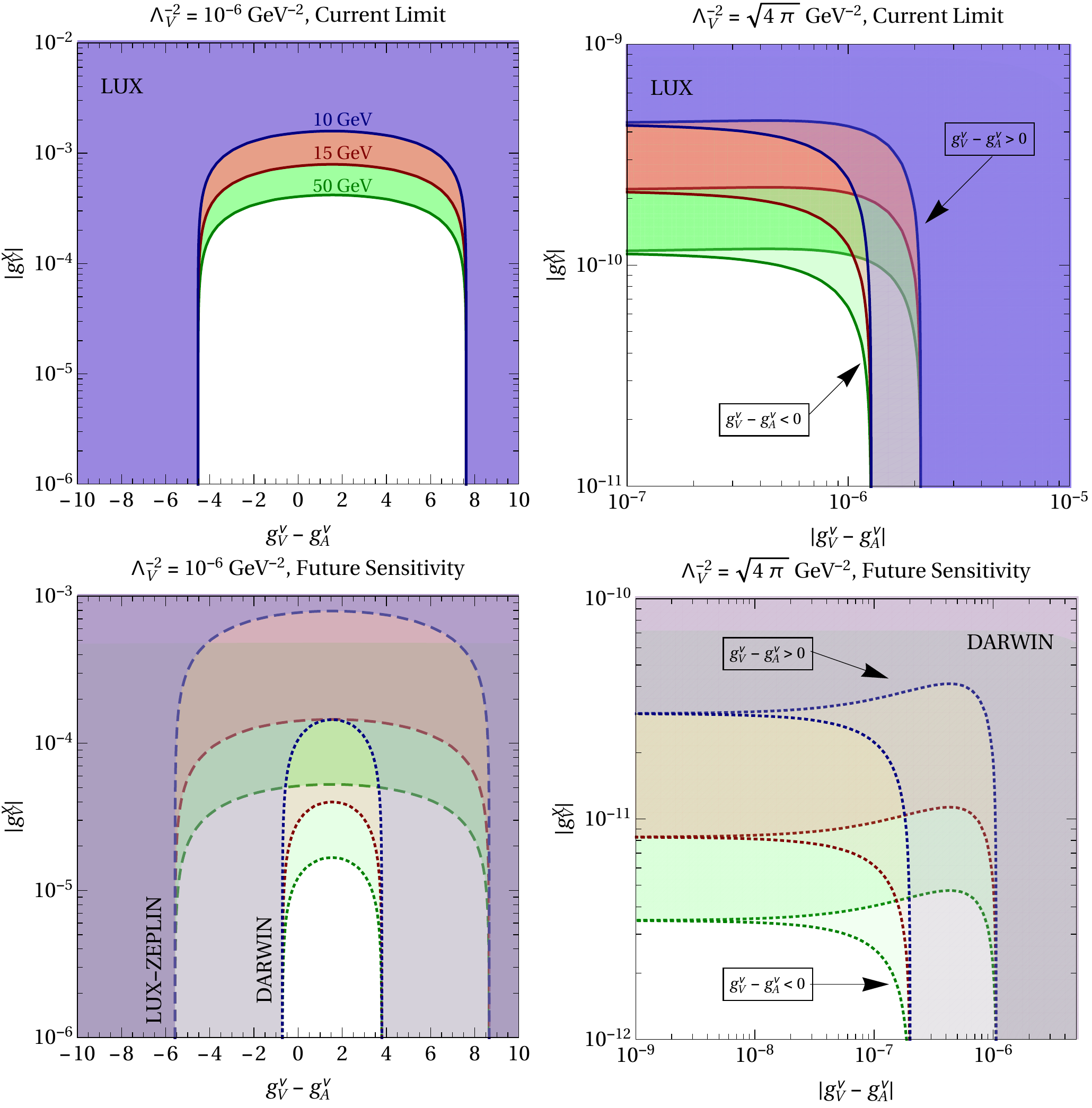} 
\caption{Current limits (top panels) and future sensitivity (bottom panels) on the parameters of the vector model. The coloured region can be excluded at $90\%$ C.L. by current LUX data \protect\cite{Akerib:2015rjg} (continuous lines) and by the future LUX-ZEPLIN \protect\cite{Akerib:2015cja} (dashed lines) and DARWIN \protect\cite{Aalbers:2016jon} experiments (dotted lines). The plots are for  $m_\chi = 10$ GeV  (violet), $15$ GeV (red) and $50$ GeV (green) for two different cases: $\Lambda^{-2}_V=10^{-6}$ GeV$^{-2}$ (left) and $\Lambda^{-2}_V=\sqrt{4\pi}$ GeV$^{-2}$ (right). For simplicity, in the latter case we only show the DARWIN future sensitivity, since the LUX-ZEPLIN results are qualitatively similar but a factor of $\sim 4-10$ less sensitive.}
\label{fig:V_lim}
\end{figure}
%%%%%%%%%%%%%%%%%%%%%%%%%%%%%%%%%%%%%%%
%%%%%%%%%%    FIG BOUNDS SCALAR MODEL    %%%%%%%
\begin{figure}[t!]
\centering
\includegraphics[width=\textwidth]{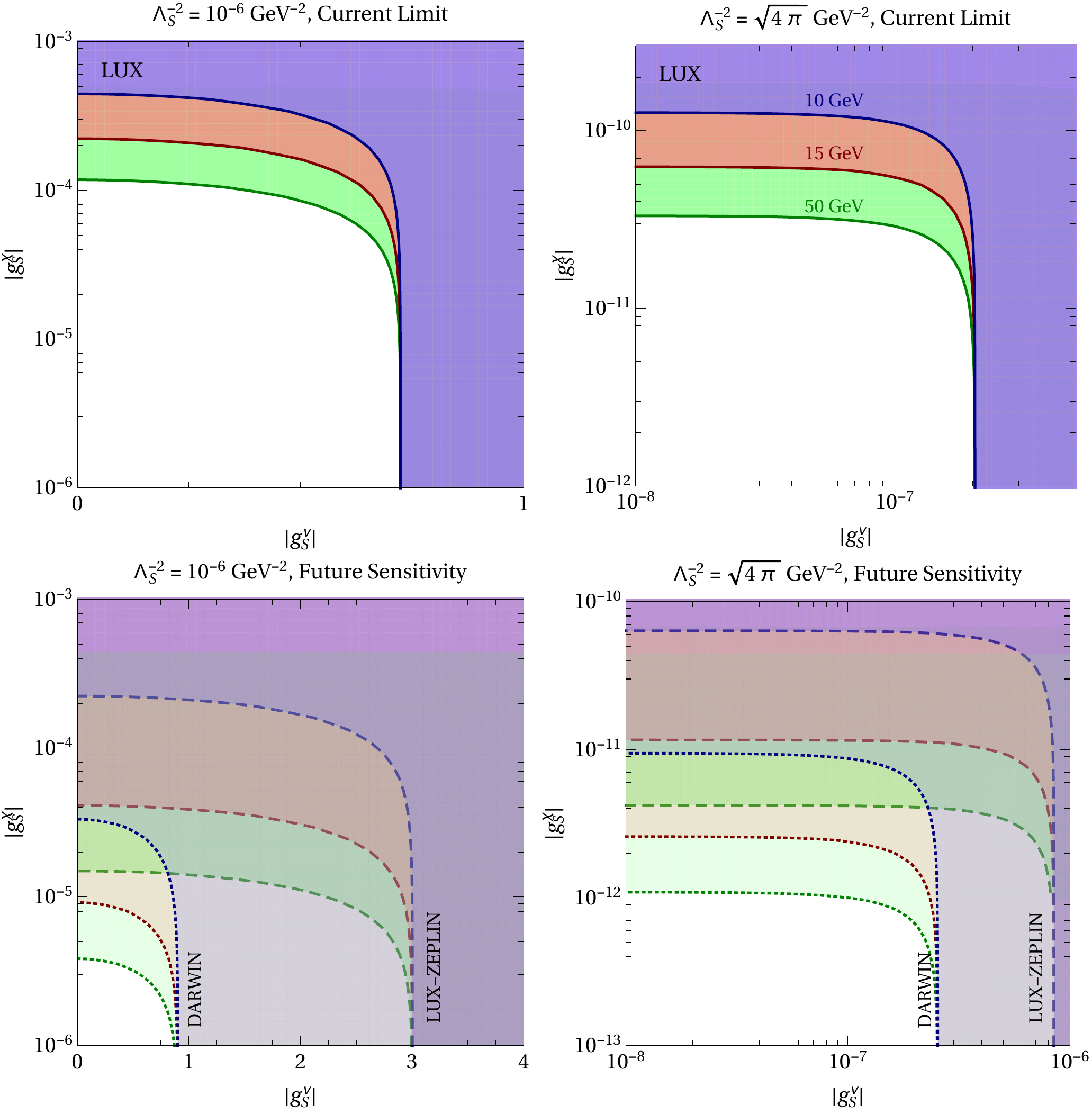} 
\caption{Current limits (top panels) and future sensitivity (bottom panels) on the parameters of the scalar model. The coloured region can be excluded at $90\%$ C.L. by current LUX data~\protect\cite{Akerib:2015rjg} (continuous lines) and by the future LUX-ZEPLIN~\protect\cite{Akerib:2015cja} (dashed lines) and DARWIN~\protect\cite{Aalbers:2016jon} experiments (dotted lines). The plots are for  $m_\chi = 10$ GeV  (violet), $15$ GeV (red) and $50$ GeV (green) for two different cases: $\Lambda^{-2}_S=10^{-6}$ GeV$^{-2}$ (left) and $\Lambda^{-2}_S=\sqrt{4\pi}$ GeV$^{-2}$ (right).}
\label{fig:S_lim}
\end{figure}
%}
%%%%%%%%%%%%%%%%%%%%%%%%%%%%%%%%%%%%%%%
\negthickspace Regarding the vector scenario, we scanned the parameter space in the ranges
\begin{equation}\label{eq:parameters_V}
  0\le \vert g_V^\nu - g_A^\nu \vert \le 10, \qquad
  0 \le \vert g_V^\chi\vert \le 1.
\end{equation}
Our results are shown in figure \ref{fig:V_lim} for the plane $(g_V^\nu-g_A^\nu,\abs{g_V^\chi})$ and the parameter $\displaystyle 
\Lambda^{-2}_V \equiv g^q_{V}/m_V^2=10^{-6} (4\pi)$ GeV$^{-2}$ on the left (right) panel. The $4\pi$ choice was done considering the 
couplings at the perturbativity limit, i.e. $g\sim 4\pi$ and a mass of the mediator of $m_V = 1$ GeV. For clarity, we present the 
limit for three different WIMP masses, $m_\chi = 10$ GeV (violet), $15$ GeV (red) and $50$ GeV (green). The bounds present two 
distinct regions; for $\Lambda^{-2}_V =10^{-6}$ GeV$^{-2}$, when $\vert g^\nu_V-g^\nu_A \vert \lesssim 3-4$, the WIMP contribution 
is dominant over the neutrino one, implying that $\vert g^\chi_V \vert \lesssim 2 \times 10^{-3}$ ($\lesssim 4 \times 10^{-4}$ ) for 
$m_\chi = 10 \,(50)$ GeV. When the neutrino couplings $\vert g^\nu_V-g^\nu_A \vert$ become large enough, the neutrino events 
increases and dominates the bounds. Let us also point out the asymmetry in the bounds of the neutrino couplings. Such asymmetry is 
related to the interference in the CNSN since, when $g_V^\nu-g^\nu_A<0$, the interference is positive making the number of events 
larger. Furthermore, the constraints on the neutrino couplings are independent of the WIMP mass because the mass becomes irrelevant when $g_V^\chi\to 0$. For the second case $\Lambda^{-2}_V =\sqrt{4\pi}$ GeV$^{-2}$, we can constrain the regions $\vert g^\chi_V \vert \lesssim 4.3 \times 10^{-10}$ ($\lesssim 1.2 \times 10^{-10}$ ) for $m_\chi = 10 \,(50)$ GeV and $\vert g_V^\nu - g_A^\nu \vert \lesssim$ few $10^{-6}$.
\newpage
\noindent
For the case of the scalar mediator, we analysed the constraints from LUX results, varying the ranges
\begin{equation}
\label{eq:parameters_S}
	0 \le |g_S^\nu| \le 5   \qquad  0 \le |g_S^\chi| \le 1\, .
\end{equation}
In figure \ref{fig:S_lim} we display the results concerning the scalar scenario for similar cases as in the vector case. We can constraint the region $\vert g^\chi_S \vert \lesssim 4.5 \times 10^{-4}$ ($\lesssim 1 \times 10^{-4}$) for $m_\chi = 10 \,(50)$ GeV if $\vert g^\nu_S \vert < 0.5$, when $\Lambda^{-2}_S=10^{-6}$ GeV$^{-2}$. For the extreme case of $\Lambda^{-2}_S=\sqrt{4\pi}$ GeV$^{-2}$, we get a bound of $\vert g^\chi_S \vert \lesssim 1.3 \times 10^{-10}$ ($\lesssim 3.2 \times 10^{-11}$) for $m_\chi = 10\,(50)$ GeV if $\vert g^\nu_S \vert < 10^{-7}$. When the neutrino couplings are strong enough to dominate the number of events, we can set a limit of $\vert g^\nu_S \vert \lesssim 0.7$ ($\vert g^\nu_S \vert \lesssim$ few $10^{-7}$) for $\Lambda^{-2}_S =10^{-6}$ ($=\sqrt{4\pi}$) GeV$^{-2}$. As in the previous case, such limit is independent of the WIMP mass.\\

\noindent{\bf Future sensitivity.} Given the future experiments being planned to discover, or constraint even more, the WIMP model, 
we can predict the future bounds on the simplified models we have studied so far. We will estimate the sensitivity of the future Xe 
based experiments LUX-ZEPLIN and DARWIN. In the design studies of those experiments, the collaborations present the initial 
parameters in which they will start operating. LUX-ZEPLIN collaboration assumes an energy threshold of $6$ keV, a maximum nuclear 
recoil energy of $30$ keV and a future exposure of $15.34$ ton-years~\cite{Akerib:2015cja}. Moreover, they suppose an energy-
independent efficiency of $50\%$. In the DARWIN case, it is designed for an aggressive exposure of $200$ ton-years, with a $30\%$ 
acceptance of recoils in the energy range of $5-35$ keV~\cite{Aalbers:2016jon}.\\

The future bounds for both models are shown on the bottom panels of figures \ref{fig:V_lim} and \ref{fig:S_lim} for LUX-ZEPLIN 
(dashed lines) and DARWIN (dotted lines). Certainly, to obtain those possible constraints we are supposing that these experiments do 
not find any evidence of WIMPs. Again, there are two distinct regions in the figures. First, in the region in which the WIMPs 
dominate, LUX-ZEPLIN will be able to improve the bounds on $\vert g_V^\chi \vert$ and $\vert g_S^\chi \vert$ by a factor of $2-10$. 
In the neutrino dominating region, we see nevertheless that the constraints are weaker than the current bounds obtained by LUX. This 
is an effect directly related to the energy threshold of the experiments. The higher LUX-ZEPLIN energy threshold diminishes the 
number of neutrino events since it does not allow for detection of solar ($^8$B) neutrinos. Actually, this also occurs for DARWIN, 
but, given the strong exposure, the effect is balanced.

%%%%%%%%%%%%%%%%%%%%%%%%%%%%%%%%%%%%%%%%%%%%%%%%%%%%%%%%%%%%%%%%%%%%%%%%%%%%%%%%%%%%%%%%%%%%%%%%%%%%%%%%
\subsection{Discovery Limit including Non-Standard Interactions}\label{sec:modified_spectra}
%%%%%%%%%%%%%%%%%%%%%%%%%%%%%%%%%%%%%%%%%%%%%%%%%%%%%%%%%%%%%%%%%%%%%%%%%%%%%%%%%%%%%%%%%%%%%%%%%%%%%%%%

\begin{figure}[t!]
	\centering
	\includegraphics[width=\textwidth]{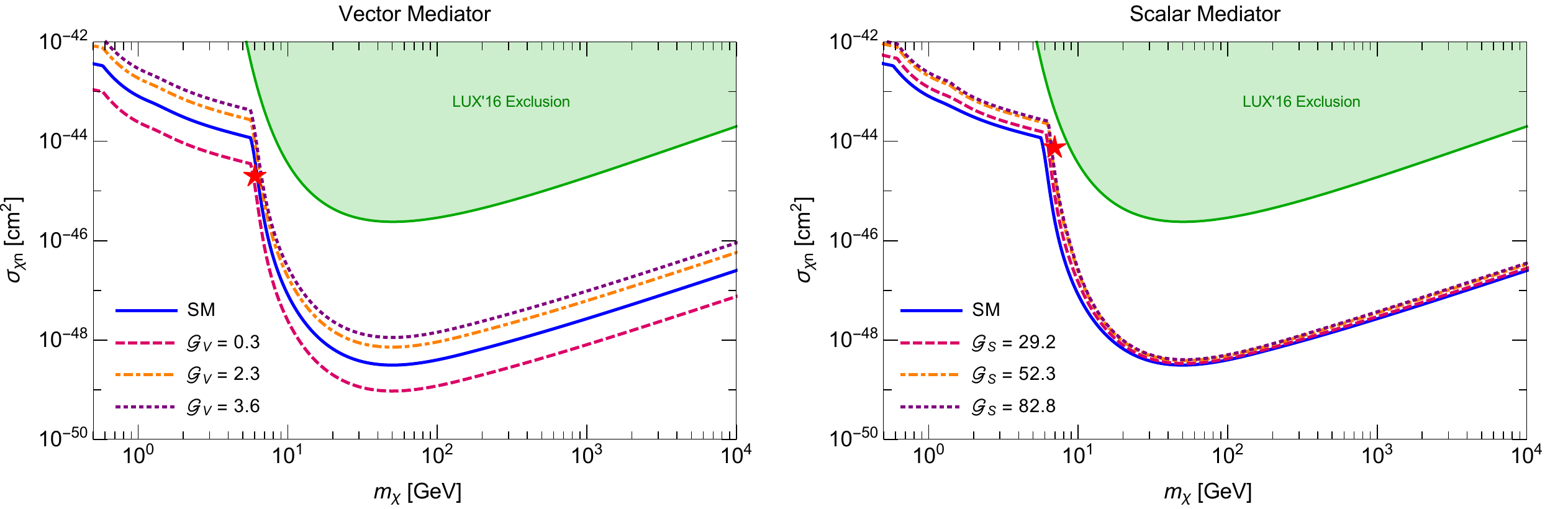}
	\caption{One-neutrino event contour lines for the two types of mediators, considering a Xe target detector. We show on the left 	(right) panel three examples for the vector (scalar) mediator. We also show the SM one-neutrino event contour line (in blue) for comparison. The red star is a point for which we will show the energy spectrum. The green region is excluded by LUX at $90\%$ of C.L.~\protect\cite{Akerib:2016vxi}.}
	\label{fig:MNF-V}
\end{figure}

\begin{figure}[t]
	\centering
	\includegraphics[width=\textwidth]{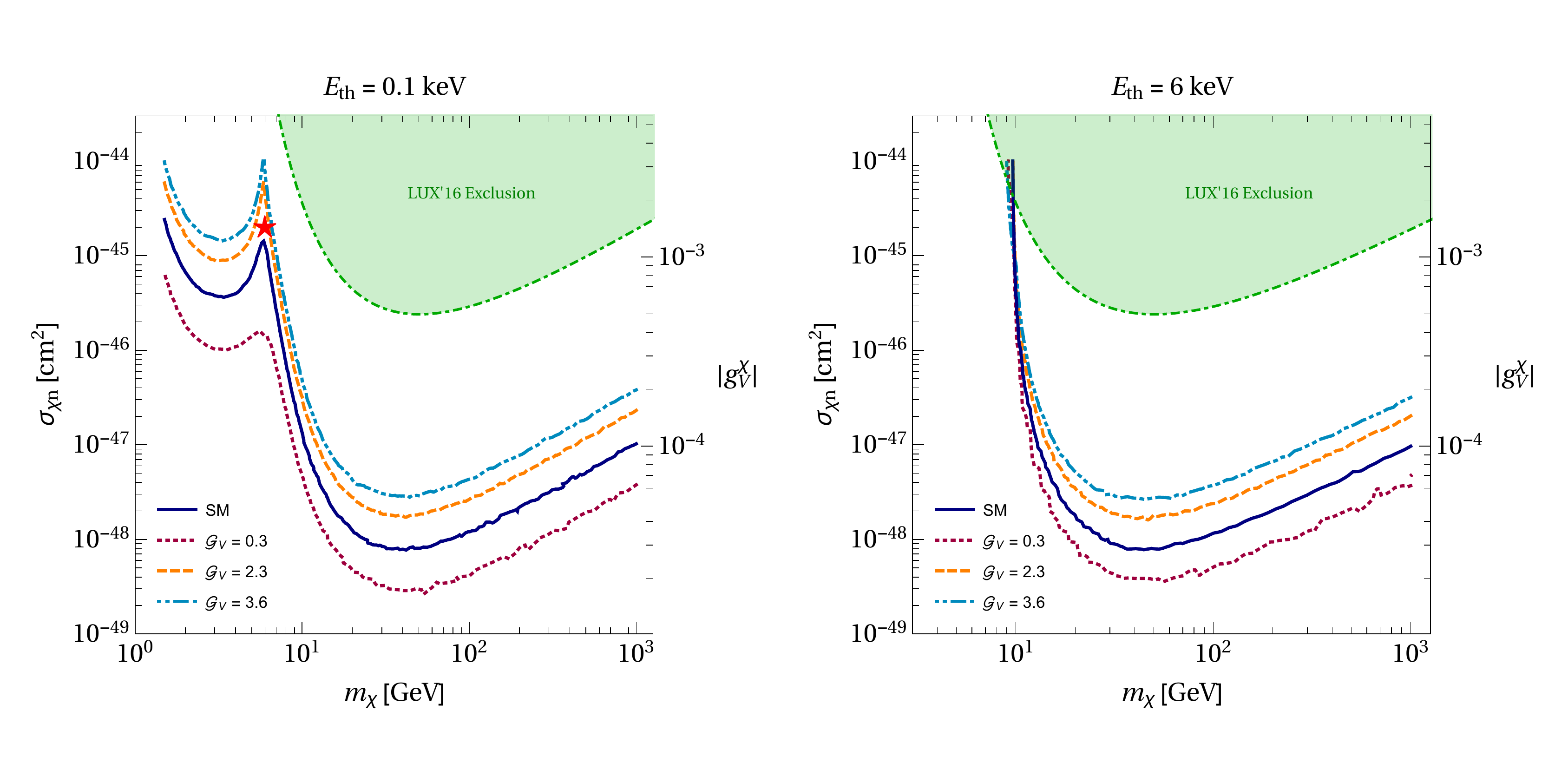}
	\caption{\label{fig:nufloor-v} Neutrino floor for the vector mediator case in the plane ($m_\chi, \sigma_{\chi n}$) and  ($m_\chi, \vert g^\chi_V \vert$). The results are for the LUX-ZEPLIN experiment with two different energy thresholds: a very low one, $E_{\rm th}$ =0.1 keV (left), and the nominal threshold used in proposal the experiment, $E_{\rm th}$ = 6 keV (right). The SM neutrino floor (dark blue) is shown, along with the most extreme case still allowed for the vector model (${\cal G}_V=3.6$, light blue), an intermediate case (${\cal G}_V=2.3$, orange), as well as a case where the neutrino floor can be smaller than the SM one (${\cal G}_V=0.3$, red). The axis corresponding to the value of the $\vert g^\chi_V \vert$ coupling was obtained 
considering $\Lambda_V^{-2}=10^{-6}$ GeV$^2$.}
\end{figure}

\begin{figure}[t!]
	\centering	
	\includegraphics[width=\textwidth]{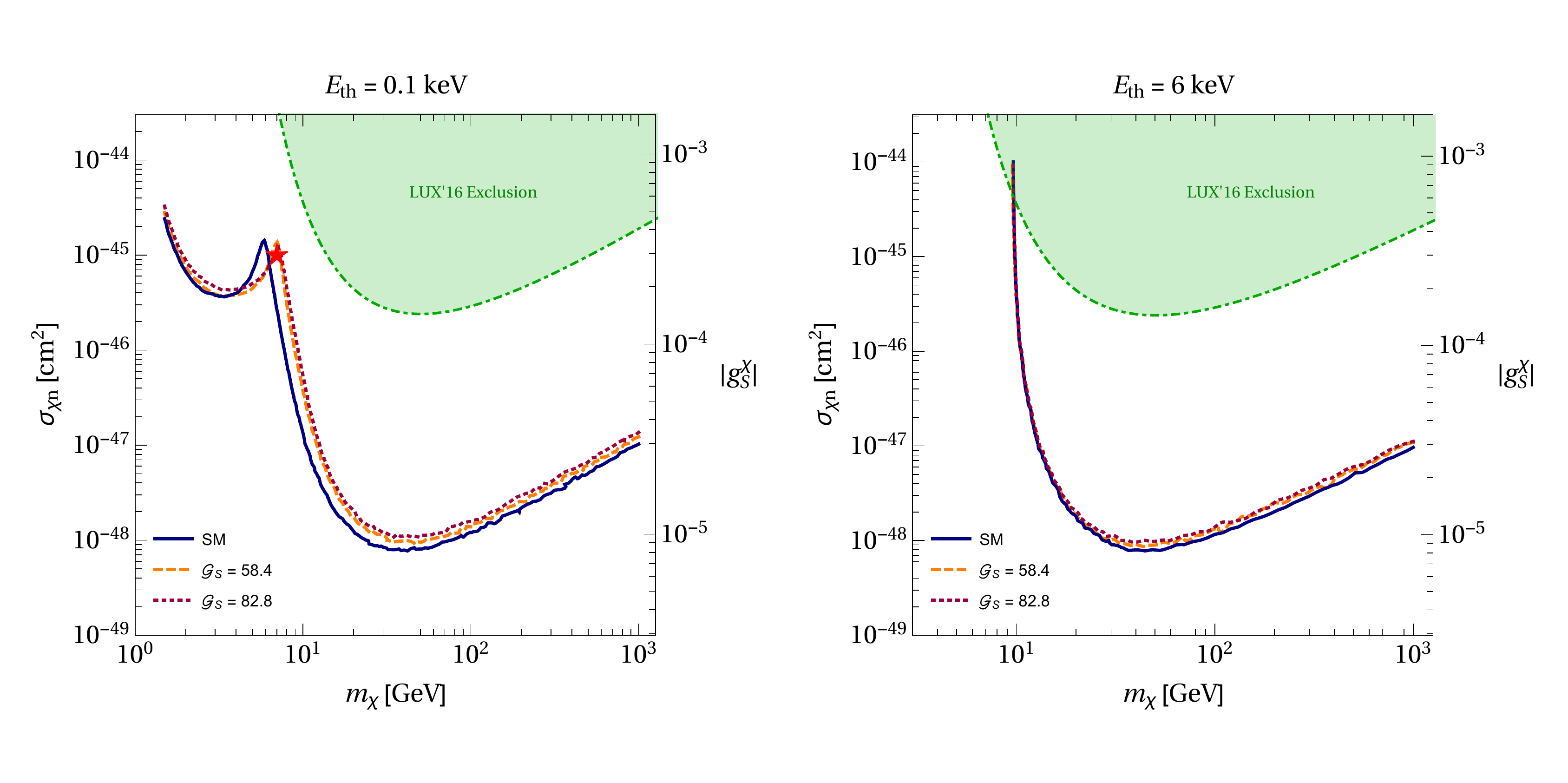}
	\caption{\label{fig:nufloor-s} 
As figure~\ref{fig:nufloor-v}, but for the scalar mediator case. The SM neutrino floor (dark blue) is shown, along with two cases still allowed by the scalar model (${\cal G}_S = 58.4$, orange; ${\cal G}_S = 82.8$, red).}
\end{figure}

A more crucial consequence of the possible existence of NSI is related to the dis\-co\-ve\-ry limit of direct detection experiments.
Let us start considering the modifications on the one-neutrino event contour line. In figure \ref{fig:MNF-V}. we present some 
examples of modified one-neutrino event contour lines, by fixing the values of the parameters $\mathcal{G}_V$ and $\mathcal{G}_S$ so 
that they are still allowed by the experimental constraints obtained in the previous section. The left (right) panel corresponds to the vector (scalar) scenarios. In the vector case, we see that the one-neutrino contour line is essentially a rescaling of the SM case, as expected from the modification of the cross section; the scalar scenario shows a deviation from the SM since the cross section in that case is altered by an additional factor, see equation \eqref{eq:recoil_scalar}.
%%%%%%%%%%    FIG MODIFIED 1 NEUTRINO EVENT CONTOUR LINE     %%%%%%%
%%%%%%%%%%%%%%%%%%%%%%%%%%%%%%%%%%%%%%%
As depicted in figure \ref{fig:MNF-V}, it is possible to lower the one-neutrino contour line in the vector case, for the value of 
${\cal G}_S=0.3$. Therefore, in principle, the CNSN contribution can be cancelled for a massive mediator case ($m_V \gtrsim 1 $ GeV) 
when
\begin{align}\label{eq:vlim}
	g^{\nu}_V  - g^{\nu}_A & = \frac{{\cal Q}_V^{\rm SM}}{{\cal Q}_V}\frac{G_F m_V^2}{\sqrt{2}}  = \frac{a^\nu_V}{g_V^q} \left( \frac{m_V}{\rm GeV}\right)^2 \,,
\end{align}
with the assumption of $g_V^{u}=g_V^{d}=g_V^{q}$, and $a^\nu_V$ is a target-related numerical value. We show in table \ref{tab:limits} the values of $a^\nu_V$ for some nuclei. On the other hand, in the scalar scenario, the cancelling of the one-neutrino event contour line is only partial. Examining the equations \eqref{eq:recoil_SM} and \eqref{eq:recoil_scalar} we observe that the positive scalar contribution can cancel merely the term proportional to $E_R/E_\nu^2$. This is achieved when 
\begin{align}
\label{eq:slim}
g^\nu_S &= \frac{{\cal Q}_V^{\rm SM}\,}{{\cal Q}_S} \frac{G_F m_S^2}{\sqrt{2}}= \frac{a^\nu_S}{g_S^q}\left( \frac{m_S}{\rm GeV}\right)^2 \,,
\end{align}
being $a^\nu_S$ a numerical value depending on the target, see table \ref{tab:limits}. In the right panel, orange line of figure \ref{fig:MNF-V} we present the case of a Xe experiment for $g_S^q = 1$ and $m_S = 100$ GeV, which corresponds to ${\cal G}_S = 52.3$.
\begin{table}[tb]
	\centering 
	\caption{Values of the coefficients $a^\nu_V$ and $a^\nu_S$ for various target nuclei, corresponding to strongest reduction of the CNSN cross session according  to equations \eqref{eq:vlim} and \eqref{eq:slim}.}
	\label{tab:limits} 
	\begin{tabular}{ccc}\toprule
		Nucleus & $a^\nu_V $  & $a^{\nu}_S$ \\ \midrule\midrule
	 	Xe & $1.0 \times 10^{-6}$ & $4.5 \times 10^{-7}$ \\
	 	Ar & $7.0 \times 10^{-7}$ & $3.7 \times 10^{-7}$ \\
	 	Ge & $9.4 \times 10^{-7}$ & $4.2 \times 10^{-7}$ \\
	 	Ca & $6.6 \times 10^{-7}$ & $3.7 \times 10^{-7}$ \\
	 	W  & $1.1 \times 10^{-6}$ & $4.5 \times 10^{-7}$ \\
	 	O  & $6.6 \times 10^{-7}$ & $3.7 \times 10^{-7}$ \\
		\bottomrule
	\end{tabular}
\end{table}
\newpage
%%%%%%%%%%%%%%%%%%%%%%%
Having determined the one-neutrino event contour line, we can proceed to compute the real discovery limit including NSI.
We will compute first the discovery limit for a more realistic case, the LUX-ZEPLIN experiment with two different energy 
thresholds. We show our results in figures ~\ref{fig:nufloor-v} and ~\ref{fig:nufloor-s}. In figure \ref{fig:nufloor-v}, we have the 
neutrino floor considering the vector mediator scenario. We have there the SM contribution (dark blue) and the case ${\cal  G}_V=3.6$ (light blue), which can be considered an extreme case since it corresponds to the current limit on $\vert g_V^\nu - g_A^\nu \vert$ ($\lesssim 10^{-6}$) for $\Lambda_V^{-2}= \sqrt{4 \pi}$ GeV$^{-2}$. We also show the cases ${\cal G}_V= 2.3, 0.3$ 
(orange, red). In the last case, the SM contribution is partially canceled by the vector contribution. Thus, the neutrino floor is
below the SM one. For the specific case of $E_{\rm th}=0.1$ keV, we see the peak corresponding to the $^8$B neutrinos. Let 
us stress that these and the other peaks are originated by the similarity of the WIMP spectra with the neutrino one, worsening the 
discovery limit. Furthermore, in the specific case of the vector mediator, we have that the modification on the neutrino floor only 
affects the value of the cross section in which the peaks appear, but not the WIMP mass correspondent to each peak. This is 
expected as the CNSN cross section is modified in a similar fashion.\\

On the other hand, in the scalar case, figure \ref{fig:nufloor-s} we considered the extreme value of ${\cal G}_S= 82.8$ (red) corresponding to the current limit on $\vert g_S^\nu \vert$ ($\lesssim 2\times 10^{-7}$) for $\Lambda_S^{-2}= \sqrt{4 \pi}$ GeV$^{-2}$. We also present the case of ${\cal G}_S= 58.4$ (orange). For the case of the smaller threshold, we have that the discovery limit is displaced to a mass of $\sim 7$ GeV. This shift is originated by the modification of the CNSN cross section as we have seen previously. In the case of the larger threshold, the scalar mediator does not affect the discovery limit significantly.\\ 

\begin{figure}[t!]
	\centering
	\includegraphics[width=0.6\linewidth]{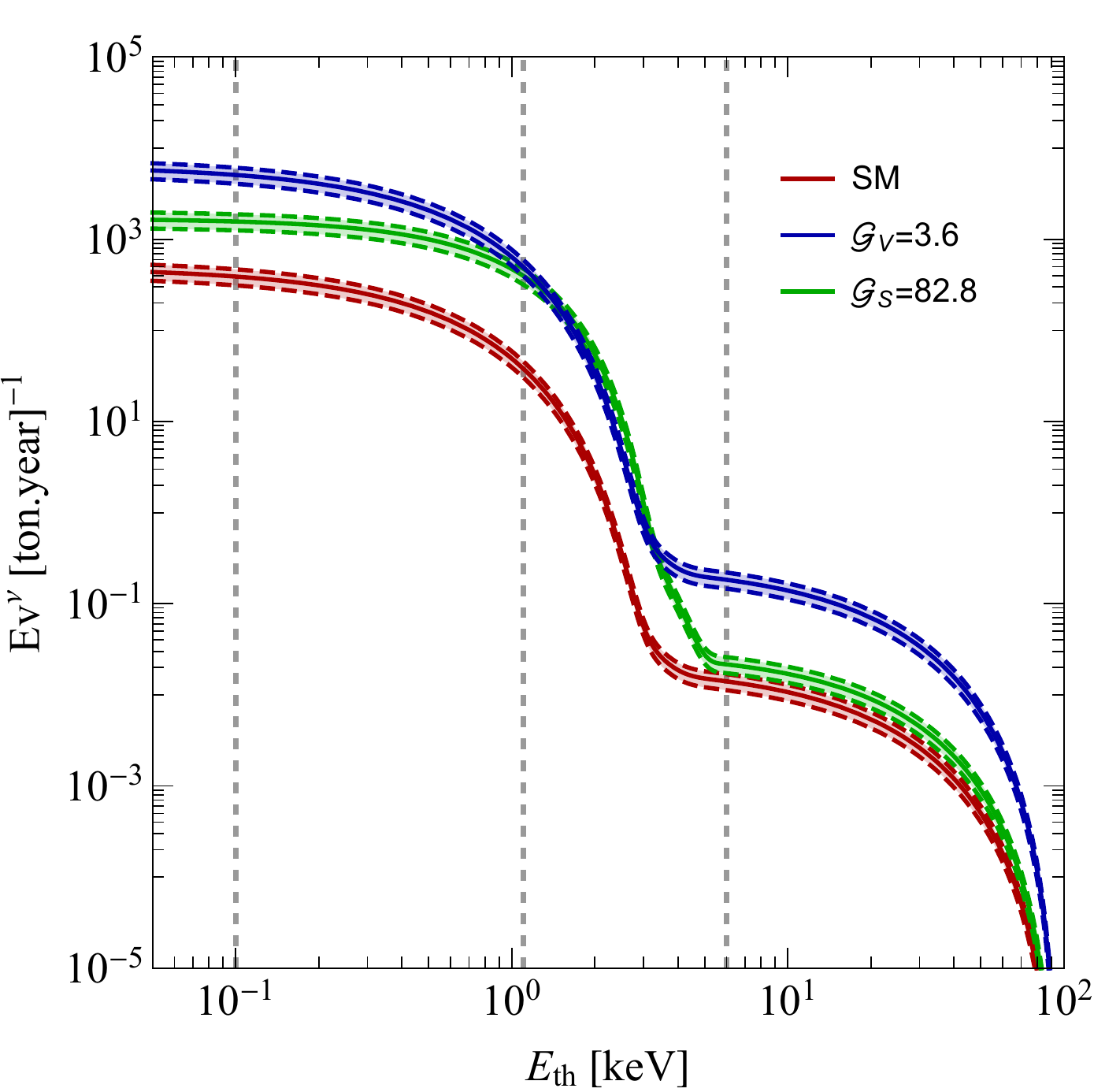}
	\caption{Number of CNSN events per ton-year for LUX-ZEPLIN as a functions of the energy threshold. In red we show the predictions for the SM and in blue (green)  for the vector (scalar) model with ${\cal G}_V=3.6$ (${\cal G}_S=82.8$).
The thickness of the curves represent a variation on the detector efficiency of $50\% \pm 10\%$.
}
	\label{fig:threshold}
\end{figure}

Taking into account that the scalar and vector limits are obtained from the LUX bound on the number of events, we can ask ourselves 
the reason of the difference among the discovery limits in each scenario. We determined the number of CNSN events as a function of 
the energy threshold, see figure \ref{fig:threshold} for the two extreme cases of ${\cal G}_V=3.6$ and ${\cal G}_S=82.8$. We can see 
there that for $E_{\rm th}\sim 1$ keV the contributions of both models are the same, as expected from the LUX limit. Nevertheless, 
due to the different behaviour of the rates in both scenarios, we see that the values of the number of events in the scalar scenario is smaller than in the vector one, approximately $4$ times different. This is explained by noting that the scalar scenario has the additional term  $E_R m_N/E_\nu^2$ (see equation \eqref{eq:recoil_scalar}) making the dependence on the threshold energy non-trivial. For a higher threshold, when only atmospheric neutrinos contribute, both SM and scalar contributions are of the same order. The thickness of the curves represent a variation on the efficiency of $10\%$. This variation on the detector efficiency does not modify significantly our results. In summary, for the LUX-ZEPLIN experiment the vector model only will affect the experimental sensitivity for $\sigma_{\chi n} \lesssim$ few $10^{-47}$ cm$^{2}$, while the scalar one does not.\\
%%%%%%%%%%%%%%%%%%%%%%%%%%%%%%%%%%%%%%%%%%%%%%%%%%%%%%%%%%%%%
%%%%%%%%%%%%%%%%%%%%%%%%%%%%%%%%%%%%%%%%%%%%%%%%%%%%%%%%%%%%

\begin{figure}[t]
	\centering
	\includegraphics[width=\textwidth]{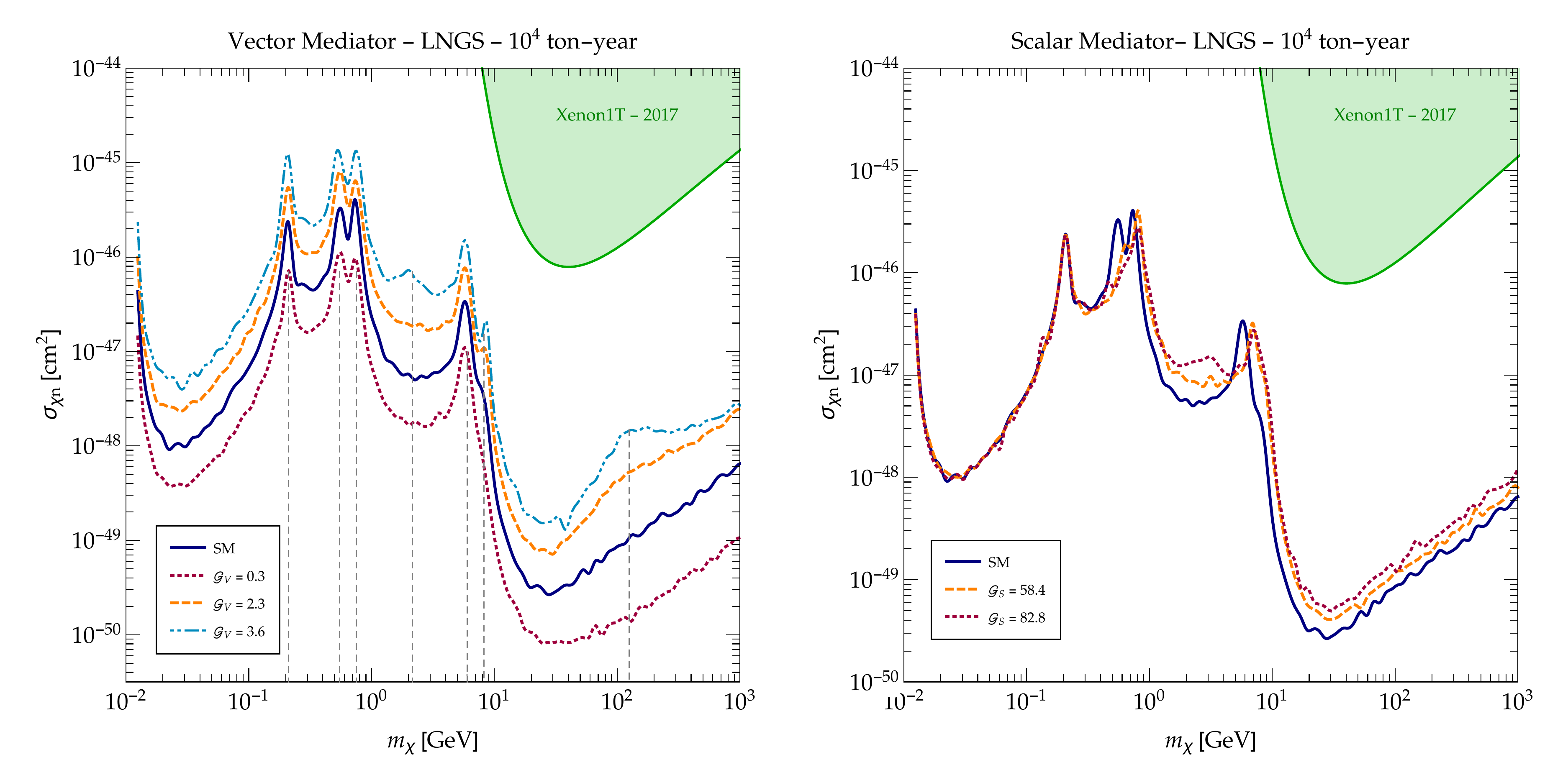}
	\caption{Neutrino floor for the vector (left) and scalar (right) mediator case in the plane ($m_\chi, \sigma_{\chi n}$) for a Xe experiment with an artificial threshold of $E_{\rm th} = 0.01$ eV and an exposure of $10^5$ ton-years. The dashed gray lines correspond to the WIMP mass more affected by neutrinos.}
	\label{fig:NFXeSV}
\end{figure}

In order to confirm our results for both scenarios, we determined the discovery limit for a unrealistic Xe experiment located at 
LNGS, with a extreme threshold of $0.01$ eV, exposure of $10^4$ ton-years and $100\%$ efficiency. 
In figure \eqref{fig:NFXeSV} we have the modification of the discovery limit for both vector (left) and scalar (right) scenarios.
In the vector scenario, the modification is manifested in a similar way to increasing the experimental exposure. Thus, 
for the extreme case of ${\cal G}_V=3.6$, the peaks corresponding to reactor antineutrinos, $hep$ and atmospheric neutrinos are now 
evident while in the SM case they are absent. This also happens for the value of ${\cal G}_V=2.3$, in which $hep$ and atmospheric 
neutrinos are noticeable, but reactor antineutrinos are not. In the case of ${\cal G}_V=0.3$, only the peaks of the largest neutrino 
contribution are present, due to the decrease of the number of events. In the scalar scenario, the behaviour is more involved as 
we have already stressed. For lower masses $m_\chi\lesssim 0.5$ GeV, the discovery limit coincides with the SM one since the 
scalar contribution is actually small. This may seem strange, as the $pp$ neutrinos are the least energetic neutrinos; however, let 
us remember that the recoil energy is in such case tiny. For the higher masses, the scalar contribution is important, even changing 
the structure of the discovery limit. Again, this is related to the additional contribution in this scenario.\\
%%%%%%%%%%%%%%%%%%%%%%%%%%%%%%%%%%%%%
%%%%%%%%%%%%%%%%%%%%%%%%%%%%%%%%%%%%%%%

\begin{figure}[t!]
\centering
\includegraphics[width=\linewidth]{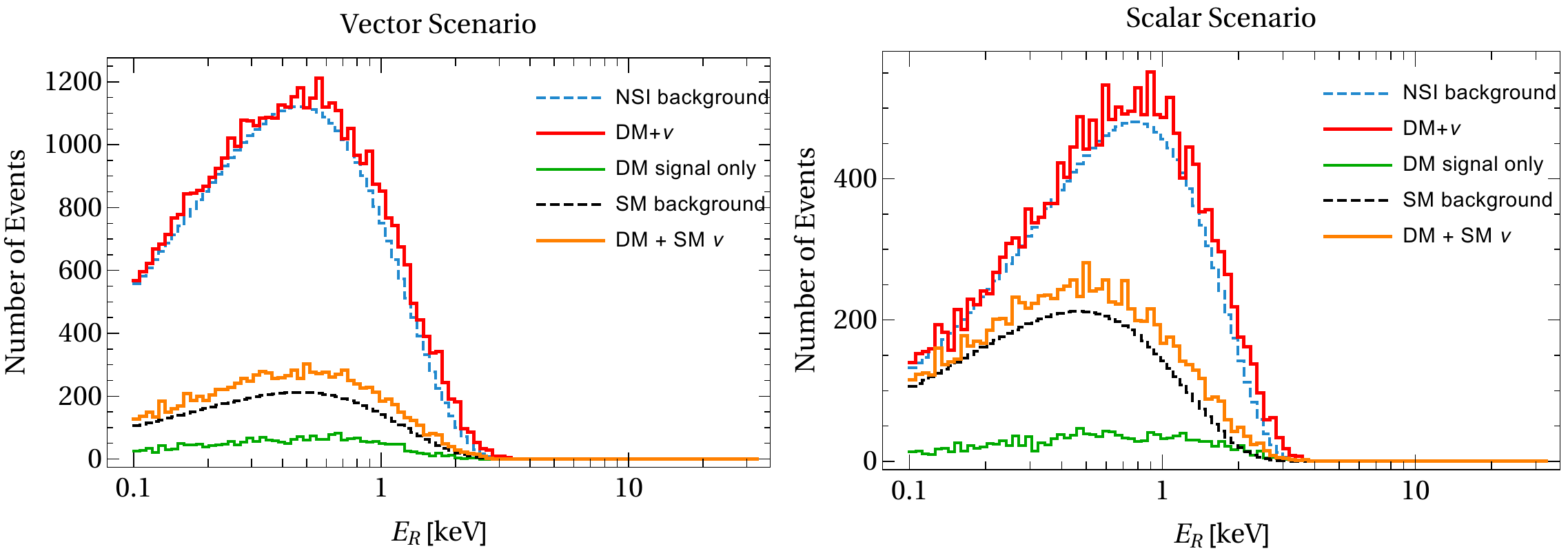}
\caption{Recoil spectrum in the vector (left) and scalar (right) case for the parameter point corresponding to the red star in figure~\ref{fig:nufloor-v} and figure~\ref{fig:nufloor-s}, respectively. The different contributions are shown separately: DM only (green), standard CNSN (black), non-standard CNSN (blue) as well as the combined spectrum (red).}
\label{fig:mod_spectra1}
\end{figure}

Finally, to show explicitly how NSI modify the detection of WIMPs at direct detection experiments, we present an example of two 
simulated energy spectra for the points in figure~\ref{fig:nufloor-v} (vector) and  figure~\ref{fig:nufloor-s} (scalar) in figure 
\ref{fig:mod_spectra1}. In such figure we present all possible contributions produced by WIMPs only (green), the SM CNSN (black) and 
the NSI CNSN (blue) for the vector (left) and scalar (right) scenarios. We present also in red the combined spectrum. For the cases 
we are considering, it would be possible to discriminate between WIMP and the SM neutrino background, using the test previously 
discussed. Nevertheless, the existence of NSI, which increases the neutrino background, does not allow to discriminate the WIMP in 
these cases.

\newpage 

In this chapter we have considered the influence of the neutrino background in the WIMP direct detection experiments. We presented 
briefly the properties of a WIMP as candidate to be the DM present in the Universe. Thus, in order to detect such particle, or 
particles, several experiments searching for nuclear recoil events that a WIMP can create have been performed. Although there 
have been some claims regarding a detection of a WIMP, more recent experiments have excluded a large part of the parameter space. 
Thus, more sensitive experiments have been proposed increasing the exposure and decreasing the energy threshold. Unfortunately, this 
creates an additional problem, neutrinos become an irreducible background for these searches. Such background is originated by the 
coherent neutrino scattering, process which is predicted by the SM. Since the neutrino energy is small enough to consider the 
nucleus components in a coherent way, the CNSN will depend on the square of the number of constituents.\\ 

We also introduced a 
simplified way to estimate when neutrinos become important through the definition of the one-neutrino event contour line, line 
corresponding to the lowest WIMP cross section which can be studied with a neutrino background of one event. Nevertheless, it is 
crucial to perform a complete statistical analysis to have certainty about the region affected by the neutrino background. This is 
done by introducing a discovery limit, corresponding to a curve from which a WIMP discovery can be achieved by an experiment with a 
significance of $3\sigma$ or higher.\\

We determined the discovery limit, or neutrino floor as it is also known, for a simulated Xe 
and Ar detectors with a minuscule energy threshold of $0.01$ eV to completely scan the limit. We obtained similar results of the 
position of the neutrino peaks appearing in the discovery limit as in other previous studies. However, we included the reactor 
antineutrinos to analyse their impact on the neutrino floor. Let us stress that this had not been done previously in the literature. 
We found that reactor antineutrinos will be important when experiments achieve exposures of order $10^5$ ton-year and thresholds 
which allow to study WIMP masses of order $1$ GeV, depending on the location.\\ 

%%%%%%%%%%%%%%%%    FIG MODIFIED SPECTRA 1,2   %%%%%%%%%%%%%
%%%%%%%%%%%%%%%%%%%%%%%%%%%%%%%%%%%%%%%%%%%%%%%%%%%%%%%%%%%%

In the final sections we devoted our study to 
understand the impact of NSI in direct detection experiments. Accordingly, we introduced two simplified models with a scalar and a 
vector mediator. First we analysed the limits on these scenarios coming from the latest LUX results. The more recent Xenon1T 
results give a worst limit since this experiment has a larger threshold. After we determined the current limits and future 
sensitivity which can be achieved by LUX-ZEPLIN and DARWIN experiments, we determined the modification of the discovery limit by the 
simplified models. We concluded that the vector mediator modifies significantly the neutrino floor, specially in the region $m_\chi< 
10$ GeV or $\sigma_{\chi n} \lesssim 10^{-47}$ cm$^2$. The scalar scenario does not modify in a significant way the discovery limit. 
We confirmed these results by scanning the whole mass range affected by the CNSN, including the new mediators. Therefore, we see 
that future WIMP searches with direct detection experiments will have to deal with the neutrinos in a very careful way, principally 
when considering small thresholds and large exposures. %\newpage

		%%%%%%%%%%%%%%%%%%%%%%%%%%%%%%%%%%%%%%%%%%%%%%%%%%%%%%%%%%%%%%%%%%%%%%%%%%%%%%%%%%%%%%%%%%%%%%%%%%
%%%%%%%%%%%%%%%%%%%%%%%%%%%%%%%%%%%%%%%%%%%%%%%%%%%%%%%%%%%%%%%%%%%%%%%%%%%%%%%%%%%%%%%%%%%%%%%%%%
%%%%%%%%%%%%%%%%%%%%%%%%%%%%%%%%%%%%%%%%%%%%%%%%%%%%%%%%%%%%%%%%%%%%%%%%%%%%%%%%%%%%%%%%%%%%%%%%%%
\chapter*{Conclusions}\label{cha:Conclu}\addcontentsline{toc}{chapter}{Conclusions}
\chaptermark{Conclusions}
%%%%%%%%%%%%%%%%%%%%%%%%%%%%%%%%%%%%%%%%%%%%%%%%%%%%%%%%%%%%%%%%%%%%%%%%%%%%%%%%%%%%%%%%%%%%%%%%%%
%%%%%%%%%%%%%%%%%%%%%%%%%%%%%%%%%%%%%%%%%%%%%%%%%%%%%%%%%%%%%%%%%%%%%%%%%%%%%%%%%%%%%%%%%%%%%%%%%%
%%%%%%%%%%%%%%%%%%%%%%%%%%%%%%%%%%%%%%%%%%%%%%%%%%%%%%%%%%%%%%%%%%%%%%%%%%%%%%%%%%%%%%%%%%%%%%%%%%

\lettrine{A} complete depiction of neutrino properties is still to be constructed. Experimental proposals pretend to unravel certain specific neutrino characteristics, such as mass ordering,  value of the CP violation phase, real scale of neutrino masses and, hopefully, the neutrino nature (Dirac or Majorana). The future of neutrino physics is indeed promissory of great days to come. In this thesis we have considered some phenomena which can be observed in the near future. We also have explored other more exotic processes that involve the possible existence of Beyond SM physics. The introduction of NSI is in fact justifiable as we know that the SM is not the final theory, if such theory even exists. There are unsolved problems related to the SM which could have an impact on our knowledge of neutrinos. For instance, it could be possible to discover a flavour theory that explains the families and the masses of all fermions. This, of course, is just speculation from the author. Physics evolves by an intricate combination of experimental observations and theoretical advances, and the theories which future physicists could create can be completely distinct from the ones we know. Anyhow, we must work within the framework we have in order to propel the scientific knowledge.\\

The present thesis was divided in two main parts. The first one intended to establish the basis for a subsequent understanding of the novel results obtained. In the \hyperref[cha:nu-MP]{first} chapter we described briefly the characteristics of the SM and the neutrino sources we used throughout the thesis. Specifically, we started by describing the basic properties of a Weyl fermion, and we showed that in this case {\it chirality}, i.e the type of fermion representation of the Lorentz group, coincides with {\it helicity}, which is the projection of the spin in the momentum direction. As we stressed all over the document, chirality and helicity only are identical for a massless fermion. After that, we described as succinct as possible the SM. It is not worth describing it further here as it has been outlined extensively in the literature. However, let us emphasize that neutrinos in the SM are massless particles by construction.

\newpage 

We considered next the neutrino sources relevant for our purposes: solar, atmospheric neutrinos and DSNB together with reactor antineutrinos. For each case we analysed their origin, spectra and the experiments which have detected or intend to detect them. One common result present in the neutrinos already detected was the divergence between the expected and measured number of events. Such discrepancy is explained by the existence of the neutrino oscillation phenomena, which occurs if neutrino are indeed massive and present mixing between mass and flavour eigenstates. Thus, in the final part of the chapter, we described oscillations, without explicitly considering the origin of the masses. We analysed briefly the neutrino propagation in vacuum and matter, and we quoted the current values of the quadratic mass differences and mixing angles. We also presented the parameters which are still unknown, as the CP phase and the mass ordering. In any case, oscillation experiments show that neutrinos are massive, but they do not clarify their Dirac or Majorana nature.\\

The \hyperref[cap:nuMaj]{second} chapter was constructed under the supposition that neutrinos are Majorana particles. Thus, we first described the properties of a Majorana fermion, showing two main differences. First, we saw that such fermions must be treated as quantum fields from the beginning since there are no travelling wave solutions. We also found that there are two distinct manners to define the Feynman propagator as the fermion number is not conserved by a Majorana field. The additional propagator, proportional to the fermion's mass, is the main ingredient for the neutrinoless double beta decay process, decay which could proof the Majorana nature of neutrinos. Afterwards, we considered the requirements for a neutrino to be Majorana within the SM framework. We saw that in the basic SM Majorana neutrinos are not possible as there is no SU$(2)_L$ triplet with hypercharge two.\\ 

Nevertheless, when one considers the dimension-five operator invariant under the SU$(2)_L\times$U$(1)_Y$ SM gauge group, the Weinberg operator, one finds that such term generates a Majorana mass term.Therefore, it can explain the smallness of neutrino masses as the non-renormalizable operator is suppressed by a scale larger than the electroweak one. Thus, one can analyse the possible extensions of the SM that can generate the Weinberg operator from tree-level interactions. We saw that there are three possibilities, including a set of right-handed singlets (type I), a scalar triplet (type II) or a fermion triplet (type III) to the SM. In all cases, we found that left-handed neutrino masses are suppressed by the large masses of the additional fields; a {\it see-saw} mechanism explains the smallness of neutrino masses. The main problem these models suffers is that finding the new states, in general, requires energies beyond the current technology. Finally, we outlined one of the most important consequences of the {\it see-saw} models, the explanation of the matter-antimatter asymmetry in the Universe by leptogenesis.

\newpage It is also possible to have models for Dirac neutrinos. Under this supposition, in the \hyperref[nu-2HDM]{third} chapter we analysed the two simplest SM extensions, the minimal introduction of right-handed neutrino singlets to the SM particle content and the neutrinophilic $2HDM$. In the first case, we saw how neutrinos can get masses from the electroweak symmetry breaking as the other SM particles. In such minimal scenario, the smallness of neutrinos is not explained at all; all fermion masses are not predicted by the SM. Nonetheless, inspired in the {\it see-saw} mechanism, we can suppose that neutrino masses are generated by a different physics. Thus, we considered the neutrinophilic 2HDM, in which a second scalar doublet is included to the set of SM particles. That second doublet is supposed to give mass to neutrinos by a spontaneous symmetry breaking. To avoid couplings between the new doublet and the charged fermions, we also need to add a new symmetry. The two minimal scenarios that have been considered in the literature correspond to a discrete $\Z_2$ or a continuous U$(1)$ symmetry. We presented the main features for both cases in the final sections of the chapter.\\
 
The second part of the thesis was focused on the novel results we achieved during the Doctorate. We divided them into three chapter, each one related to a different topic. The \hyperref[nu-2HDM-feno]{fourth} chapter contains the results related to the phenomenological constraints on the neutrinophilic $2HDM$. These constraints are divided in two classes, bounds on the scalar potential and flavour limits on the charged scalar sector. The constraints on the scalar potential are theoretical and phenomenological since such potential needs to fulfil certain properties, as stability, perturbative unitarity. The phenomenological limits correspond to bounds from the oblique parameters and Higgs, $Z^0$ decay widths. Applying these constraints we found that the $\Z_2$ symmetry model is basically excluded by precision measurements. Meanwhile, the U$(1)$ scenario is still allowed, but its spectrum needs to be highly degenerated. On the other hand, limits coming from flavour physics constrain the parameter space spanned by the VEV of the second doublet and the charged scalar mass. We found that bounds from $\mu\to e\gamma$ are the strongest ones in most of the region, but, for small VEV and large charged scalar mass, the constraint from $\mu\to 3e$ is more important. In the future, and if the proposed experimental sensitivities are achieved, the $\mu\to e$ conversion in nuclei could exclude a larger region on the parameter space. Hence, we have explored the possibility of having Dirac neutrinos with small masses and we a\-na\-ly\-sed the general bound on those models.\\

The \hyperref[cha:RelicNu]{fifth} chapter considered the possible consequences of having NSI on the detection of the cosmic neutrino background. We presented first the properties of the relic neutrinos, showing that those neutrinos can be non-relativistic, given the small momentum they have at the present time. This is crucial for the purpose of differentiating between neutrino natures. The reason is that the neutrinos belonging to the $\CNB$ are helicity eigenstates since the free hamiltonian does not conserve chirality. Thus, studying in detail the abundances for the helical states in the Dirac and Majorana cases, it was found that they are different. Moreover, we found that the Dirac $\CNB$ is composed by left-helical particle states and right-helical antiparticle states while it is composed by left- and right-helical particle states if neutrinos are Majorana. So, if we could detect the relic neutrinos it would be possible to shed some light on the real neutrino nature.\\ 

Nevertheless, such detection seems to be extremely difficult. There are several proposed methods to detect the $\CNB$. Among them, the most promising is the neutrino capture by tritium; such reaction creates a detectable electron. Previous works have computed the capture rate, showing that the expected value for Majorana neutrinos is twice the Dirac case result. This is basically related to the different helical composition of the $\CNB$ once the tritium can only capture particles but not antiparticles. Nonetheless, one should analyse the details of the detection processes. In fact, the $\CNB$ signal could be hidden by the background of the tritium beta decay. Also, it is important to see if any experiment can differentiate the mass eigenstates. We introduced a novel discrimination function to understand the requirements to distinguish the peaks related to the eigenstates. However, we found that it is necessary to have a extreme resolution for identifying at least two peaks, the one of the third mass eigenstate, and the other for the combination of $\nu_1$ and $\nu_2$ states. Furthermore, the capture rate is proportional to the PNMS mixing matrix element $|\widetilde{U}_{ea}|^2$, $a=\{1,2,3\}$, so the rate for $\nu_3$ is quite small, worsening the possibility of detection. The PTOLEMY proposal has a resolution which will not be able to differentiate the peaks, and only will be sensitive for masses $m_a^\nu\gtrsim 0.1$ eV. Thus, a detection could only discriminate the neutrino nature.\\

However, such statement should be considered carefully. Supposing that neutrinos are Dirac fermions, we considered the implications of having contribution of NSI in the $\CNB$ capture rate, which has not been considered previously. To do so, we wrote the NSI using the Effective Theory approach, by considering the SU$(2)_L\times$U$(1)_Y$ SM gauge invariant dimension-six operators relevant for the capture process. We computed the new contributions to the capture rate for four combinations of these operators. Such combinations were chosen from a work studying the bounds on NSI from nuclear beta decay of a set of isotopes and the neutron decay. An important result we obtained is that the rate has terms proportional to the ratio $m_a^\nu/E_a$, being $E_a$ the neutrino energy, coming from the interference between the SM and the NSI. Usually, such terms would be negligible as neutrinos are ultrarelativisitic. But this is not the case for the $\CNB$; furthermore, these contributions are as important as the others. Applying the existing limits, we found that it is possible to increase the Dirac capture rate to values close to the one expected for Majorana neutrinos. Therefore, a detection of the relic neutrinos compatible with the Majorana value is not a definite proof that neutrinos are their own antiparticles.
\newpage

On the other hand, we found that the capture rate for Dirac neutrinos can be decreased. If a $\CNB$ detection occurs with a smaller value than expected, it may indicate the existence of NSI. We also included the possibility of having relic right-handed neutrino with both thermal and non-thermal origins. We found that the capture rate is always increased since there are more relic neutrinos that could be captured.\\

Departing somewhat from the main subject of thesis, we considered in the \hyperref[cha:NeutrinoFloor]{final} chapter the impact on DM direct detections searches due to the neutrino background. For that purpose, we first gave the general properties of DM, and we described the direct detection principle of the WIMP candidate. In direct detection searches, it is supposed that WIMPs interact with nuclei in such a way that they create an experimentally detectable recoil. Unfortunately, the searches performed until now have presented negative results. Thus, more sensitive new experiments are being planned for execution in the near future. This however introduces an additional problem provided that neutrinos will become an irreducible background. To understand the origin of such background, we introduced the coherent neutrino scattering off nuclei, which is a process predicted by the SM. Such process is called  {\it coherent} due to the small energies involved, making the incoming neutrino not be able to differentiate the nuclear components. Then, a neutrino would ``see'' a nucleus as a whole, and the cross-section will be proportional to the number of constituents squared.\\ 

Taking into account the CNSN, we computed the recoil rate of neutrinos at direct detection experiments, and we found that it can mimic very well a recoil produced by a WIMP. Then, it is a central task to determine at which point neutrinos become unavoidable. A preliminary approach is done by introducing the {\it one-neutrino event contour line}, which is a contour line in the $\sigma_{\chi n}$ vs $m_\chi$ plane, $\sigma_{\chi n}$ the WIMP $\chi$-nucleon cross section and $m_\chi$ the WIMP mass, and it describes the best background-free $\sigma_{\chi n}$ that can be constrained considering one neutrino event background. We determined such contour line for two target materials, Xenon and Argon, and we found that the line for Xe is higher in $\sigma_{\chi n}$ than the one for Ar. This is simply explained recalling that Xe has more nucleons than Ar; therefore, the CNSN is larger for Xenon.\\ 

Nevertheless, a complete statistical computation is necessary to be performed to give a real estimative about the neutrino background. This is achieved by the introduction of the discovery limit of direct detection searches, or, as is also known in the literature, the {\it neutrino floor}. It is defined as the minimun value of the WIMP-nucleon cross section in which an experiment has a $90\%$ of probability of making a WIMP discovery with a significance of $3\sigma$. We also confirmed previous results in the literature regarding the peaks appearing in the discovery limit. Such peaks are related to the values of WIMP masses and cross sections whose spectrum is very similar to the spectrum of a neutrino. To compute the neutrino floor, we included the reactor antineutrino flux which has not been done previously. The reactor antineutrino flux is computed by considering {\it all} reactors on the Earth and the data available about their properties. We found that reactor antineutrinos contributions is highly dependent on the location of the experiment, which is completely expected. This was checked by computing the discovery limit for four different laboratories. We also obtained that the reactor fluxes will become important for experimental exposures of $10^5$ ton-years and thresholds small enough to detect WIMPs with masses of $\sim 1$ GeV.\\

All previous results were obtained by considering only the SM coherent neutrino scattering. We then asked ourselves what would be the impact of the existence of NSI coupling with both neutrinos and WIMPs at direct detection experiments. By considering the simplified model framework, we considered two scenarios regarding a vector and a scalar mediator. For these two cases, we computed the additional contributions to the CNSN cross sections. We found that, in the vector scenario, the modification comes in a form of a rescaling of the cross section while, in the scalar case, an additional term appears. Thus, to determine the influence of these simplified models, we first studied the constraints coming from the results of the LUX experiment. The limits show three different regions. One, where the WIMP contribution dominates, other, where neutrino contribution is the most important, and, finally an intermediate region. We also considered the future sensitivity of DARWIN and LUX-ZEPLIN experiments on the models; in this case, the future sensitivity seems weaker than the current limits. This is explained by noticing the larger energy threshold of the proposed experimental facilities.\\ 

Afterwards, we estimated the influence of the NSI in the WIMP searches. We first introduced the modifications on the one-neutrino event contour line, finding that it is possible to cancel the neutrino background in the vector scenario for some specific values of the couplings. For the scalar case, it is only possible to partially cancel the cross section. Another feature we found is that the contour line is only modified by an overall rescaling in the vector case while there is a shift in the position in the scalar scenario. Furthermore, we calculated the discovery limit in the presence of NSI. We found that the neutrino floor modification in the vector scenario behaves as a rescaling that would be produced by a increase or decrease of the experimental exposure. In the scalar scenario, we found that the alteration is dependent on the WIMP mass since the cross section has a different dependence on the recoil energy compared to the SM case. However, the modification is not as significant as in the vector case. For completeness, we showed simulated spectra for a Xe target, and we showed explicitly the difficulties that could appear if one considers NSI. So, direct detection searches need to be very careful to include the neutrino background in a proper manner.

\newpage

In this thesis we have investigated several aspects of neutrino physics, from mass models to non-standard neutrino interactions, and their implications in breakthrough experiments. As already mentioned, future experiments will hopefully unravel the missing neutrino properties expanding our knowledge of the extraordinary particles. Such properties can shed some light in other open problems in Particle Physics, and enlarge our comprehension of the Universe. From the point of view of the author, neutrinos have always boosted the advances of Particle Physics, and surely this will be similar in the future. We certainly live in exciting times in neutrino physics. 

	\appendix
	
		%%%%%%%%%%%%%%%%%%%%%%%%%%%%%%%%%%%%%%%%%%%%%%%%%%%%%%%%%%%%%%%%%%%%%%%%%%%%%%%%%%%%%%%%%%%%%%%%%%
%%%%%%%%%%%%%%%%%%%%%%%%%%%%%%%%%%%%%%%%%%%%%%%%%%%%%%%%%%%%%%%%%%%%%%%%%%%%%%%%%%%%%%%%%%%%%%%%%%
%%%%%%%%%%%%%%%%%%%%%%%%%%%%%%%%%%%%%%%%%%%%%%%%%%%%%%%%%%%%%%%%%%%%%%%%%%%%%%%%%%%%%%%%%%%%%%%%%%
\chapter{Conventions}\label{ap:Conv}

We present the conventions used in the development of the present thesis.

\begin{itemize}
\item  We use natural units in which reduced Planck, light speed and Boltzmann constants are equal to the unity. $\hbar=c=k_B=1$. This implies the conversion factors \cite{Agashe:2014kda}
\begin{align*}
	\hbar c &= 197.3269718(44)\ {\rm MeV\ fm} ,\\
	(\hbar c)^2 &=0.389379338(17)\ {\rm GeV^2\ mbarn}.
\end{align*}
\item We use the metric tensor with trace $-2$,
 \begin{align*}
	\eta_{\mu\nu}=\eta^{\mu\nu}=\begin{pmatrix}
			1&0&0&0\\
			0&-1&0&0\\
			0&0&-1&0\\
			0&0&0&-1
		 \end{pmatrix}.
\end{align*}
%%%%%%%%%%%%%%%%%%%%%%%%%%%%%%%%%%%%%%%%%%%%%%%%%%%%%%%%%
\item Spacetime coordinates and momenta are contravariant four-vectors while derivatives are covariant four-vectors.
\begin{align*}
	x^\mu&=(t,\vec{r}),\\
	p^\mu&=(E,\vec{p}),\\
	\partial_\mu\defm\parc{}{x^\mu}&=\corc{\parc{}{t},\nabla}.
\end{align*}
%%%%%%%%%%%%%%%%%%%%%%%%%%%%%%%%%%%%%%%%%%%%%%%%%%%%%%%%%%
\item The Dirac gamma matrices obey the algebra
\begin{align*}
	\llav{\gamma^\mu,\gamma^\nu}\defm \gamma^\mu\gamma^\nu+\gamma^\nu\gamma^\mu=2\eta^{\mu\nu},
\end{align*}
and the properties
\begin{align*}
	(\gamma^\mu)^\dag&=\gamma^0\gamma^\mu\gamma^0,\\
	\Tr \gamma^\mu &=0,\\ 
	\Tr \gamma^\mu\gamma^\nu &= 4g^{\mu\nu},\\
	\Tr \gamma^\mu \gamma^\nu \gamma^\tau \gamma^\rho& = 4(g^{\mu\nu}g^{\tau\rho}-g^{\mu\tau}g^{\nu\rho}+g^{\mu\rho}g^{\nu\tau}).
\end{align*}
%%%%%%%%%%%%%%%%%%%%%%%%%%%%%%%%%%%%%%%%%%%%%%%%%%%%%%%%%%
\item When needed, we will use the Dirac representation of the gamma matrices, given by
\begin{align*}
\gamma^\mu=\begin{pmatrix}
				 0 & \bar{\sigma}^\mu\\
				 \sigma^\mu & 0
				\end{pmatrix},
\end{align*}
with $\bar{\sigma}^\mu=(1_{2\times 2}, -\vec{\sigma})$ and $\sigma^\mu=(1_{2\times 2}, \vec{\sigma})$. Here, we have the $2\times 2$ identity matrix $1_{2\times 2}$ and the Pauli matrices $\vec{\sigma}=\{\sigma_1,\sigma_2,\sigma_3\}$,
\begin{align*}
	\sigma_1=\begin{pmatrix}
				0 & 1\\
				1 & 0
			\end{pmatrix}, \quad
	\sigma_2=\begin{pmatrix}
				0 & -i\\
				i & 0
			\end{pmatrix}, \quad
	\sigma_3=\begin{pmatrix}
				1 & 0\\
				0 & 1
			\end{pmatrix}.
\end{align*}
%%%%%%%%%%%%%%%%%%%%%%%%%%%%%%%%%%%%%%%%%%%%%%%%%%%%%%%%%%
\item  The chirality $\gamma^5$ matrix is defined as
\begin{align*}
\gamma_5\defm\gamma^5\defm i\gamma^0\gamma^1\gamma^2\gamma^3,
\end{align*}
and it obeys the following properties
\begin{align*}
	\left\{\gamma^5,\gamma^\nu\right\}=0,\\
	\gamma_5^2=I_{4\times4},\\
	(\gamma_5)^\dagger=\gamma_5.
\end{align*}

%%%%%%%%%%%%%%%%%%%%%%%%%%%%%%%%%%%%%%%%%%%%%%%%%%%%%%%%%%
\item Chirality projectors are defined as
\begin{align*}
	P_{R,L}\defm\frac{1}{2}\corc{1\pm\gamma^5},
\end{align*}
with
\begin{align*}
	P_{R,L}\gamma^\mu&=\gamma^\mu P_{L,R},\\
	P_R+P_L&=1,\\
	(P_R)^2=P_R,\quad(P_L)^2&=P_L,\\
	P_RP_L=P_LP_R&=0.
\end{align*}
%%%%%%%%%%%%%%%%%%%%%%%%%%%%%%%%%%%%%%%%%%%%%%%%%%%%%%%%%%%%%
\item  The charge conjugation matrix $\mathcal{\hat C}$ fulfils the properties
\begin{align*}
	\mathcal{\hat C}(\gamma^\mu)^T\mathcal{\hat C}^{-1}&=-\gamma^\mu,\quad\mathcal{\hat C}(\gamma^5)^T\mathcal{\hat C}^{-1}=\gamma^5,\\
	\mathcal{\hat C}^\dag=\mathcal{\hat C}^{-1}&=\mathcal{\hat C}^T=-\mathcal{\hat C}.
\end{align*}
obeying the following relations with the chiral projector, $P_{R,L}$ $\MCC$,
\begin{align*}
	P_{R,L}\MCC=\MCC(P_{R,L})^T
\end{align*}

\end{itemize}

		%%%%%%%%%%%%%%%%%%%%%%%%%%%%%%%%%%%%%%%%%%%%%%%%%%%%%%%%%%%%%%%%%%%%%%%%%%%%%%%%%%%%%%%%%%%%%%%%%%
%%%%%%%%%%%%%%%%%%%%%%%%%%%%%%%%%%%%%%%%%%%%%%%%%%%%%%%%%%%%%%%%%%%%%%%%%%%%%%%%%%%%%%%%%%%%%%%%%%
%%%%%%%%%%%%%%%%%%%%%%%%%%%%%%%%%%%%%%%%%%%%%%%%%%%%%%%%%%%%%%%%%%%%%%%%%%%%%%%%%%%%%%%%%%%%%%%%%%
\chapter{The Lorentz Group and Fermion Representations}\label{ap:Lgroup}

\lettrine{T}{he} two postulates in which the Special Relativity is based are sa\-tis\-fi\-ed 
by im\-po\-sing that the interval between two events in the {\it space-time} does not depend on the
observer. Namely, two inertial observers moving relatively one to each other will
measure the same infinitesimal interval
\begin{align*}
	ds^2=dt^2-d\vec{r\,}^2.
\end{align*}
Therefore, the transformations that keep this interval invariant will be the basis
for constructing a relativistically consistent quantum theory of fermions. 
In general, a transformation can be understood through two alternative interpretations. 
The {\it passive} interpretation, when the transformation is performed upon the coordinate 
system, and the {\it active} one, where the transformation acts on the physical state. 
Usually, the special relativity is studied in a passive way, analysing the transformations 
between two inertial observers. However, we can also consider an active approach. 
In this case, the change among inertial frames is accomplished by changing one coordinate 
in the space-time into another,
\begin{align*}
	\bar{x}^\mu=\Lambda^\mu_{\ \nu}x^\nu,
\end{align*}
keeping the norm of the quadrivector constant, i.e.\ $\bar{x}^2=x^2$,
with $x^2=x^\mu x_\mu$. 
This kind of transformations will be denominated as {\bf Lorentz transformations}.
The invariance of the interval imposes a constraint in the transformation matrix 
$\Lambda^\mu_{\ \nu}$ ($\eta_{\mu\nu}$ the metric tensor)
\begin{align}\label{eq:PropLT}
	\eta_{\mu\nu}\Lambda^\mu_{\ \rho}\Lambda^\mu_{\ \tau}=\eta_{\rho\tau}.
\end{align}
Let us note that these transformations include all possible rotations and boosts. 
Several comments can be made about these transformations. The most important one
for our purposes is that the Lorentz Transformations form a group, and all the fields
will have unambiguous properties upon the action of these transformations. Since
our intention is to study massless and massive fermions, let us examine in detail
the spinorial representations of the Lorentz group. We will restrict ourselves to the 
proper and orthochronous Lorentz group, given that the parity and 
time reversal cases are usually treated separately. A generic quantum field $f_{a}(x)$,
with ${a}$ a finite number of indexes, transforms under a Lorentz transformation,
represented as a unitary operator $U(\Lambda)$, as \cite{Srednicki:2007qs}
\begin{align*}
	U(\Lambda)^\dagger f_{a}(x) U(\Lambda)= D(\Lambda)_{\ a}^{b}f_{b}(\Lambda^{-1} x),
\end{align*}
where $D(\Lambda)_{\ a}^{b}$ are matrices that obey all the properties of the Lorentz group,
i.e.\ they belong to a {\it representation} of the group. In the case of an infinitesimal Lorentz 
transformation,
\begin{align*}
	\Lambda^\mu_{\ \nu}&=\delta^\mu_{\ \nu}+\delta\omega^\mu_{\ \nu},
\end{align*}
the unitary operator $U(\Lambda)$ is given by \cite{Srednicki:2007qs}
\begin{align}
	U(1+\delta\omega)=I+\frac{i}{2}\delta\omega_{\mu\nu}M^{\mu\nu},
\end{align}
being $M^{\mu\nu}=-M^{\nu\mu}$ a set of operators obeying the following commutation
relations
\begin{align}\label{eq:comrelLG}
	[M^{\mu\nu},M^{\rho\tau}]=i(\eta^{\mu\rho}M^{\nu\tau}-\eta^{\nu\rho}M^{\mu\tau}+\eta^{\mu\tau}M^{\nu\rho}-\eta^{\nu\tau}M^{\mu\rho}).
\end{align}
This operators are known as the {\bf generators} of the Lorentz group
and the relations \eqref{eq:comrelLG} are the {\it Lie algebra} of the
group. Whereas the generators $M^{\mu\nu}$ are antisymmetric, we will have 
only 6 independent operators. Those operators will correspond to the components
of the angular momentum $J_{i}$ and the components of the boosts $K_i$, defined as \cite{Srednicki:2007qs}
\begin{align*}
	J_i\equiv\frac{1}{2}\varepsilon_{ijk}M^{jk}, \qquad K_{i}\equiv M^{i0}.
\end{align*}
An important consequence of these definitions is that the commutation relations
for $\{J_i,K_i\}$ will be
\begin{subequations}\label{eq:ComRelJK}
	\begin{align}
		[J_i,J_j]&=i\varepsilon_{ijk}J_k,\\
		[J_i,K_j]&=i\varepsilon_{ijk}K_k,\\
		[K_i,K_j]&=-i\varepsilon_{ijk}J_k,
	\end{align}
\end{subequations}
with $\varepsilon_{ijk}$ the Levi-Civita symbol in three dimensions. The interpretation of these relations is straightforward. 
The first and the second ones are the standard relations among the components of the angular momentum and the commutator
between the angular momentum and any vector, respectively. The third one shows that a 
set of boosts can be equivalent to a rotation, as in the case of the Thomas precession \cite{Thomas:1927yu}. 
Although the operators $\{J_i,K_i\}$ have a definite interpretation, let us introduce the 
operators $A_i, B_i$\cite{Srednicki:2007qs}
\begin{align*}
	A_i&\equiv\frac{1}{2}(J_i+i K_i),\\
	B_i&\equiv\frac{1}{2}(J_i-i K_i).
\end{align*}
The commutation relations \eqref{eq:ComRelJK} take a simpler form,
\begin{subequations}\label{eq:ComRelAB}
	\begin{align}
		[A_i,A_j]&=i\varepsilon_{ijk}A_k,\\
		[B_i,B_j]&=i\varepsilon_{ijk}B_k,\\
		[A_i,B_j]&=0
	\end{align}
\end{subequations}
This shows that the $A_i,B_i$ form two independent commuting SU$(2)$ algebras\footnote{Technically speaking, 
the {\it complex} linear combinations of the Lorentz algebra are isomorphic to the 
{\it complex} linear combinations of the Lie algebra of SU$(2)\times$SU$(2)$.} related
by hermitian conjugation. This simplifies the construction of the representations
since the properties of SU$(2)$ are well known. Hence, a re\-pre\-sen\-ta\-tion of the Lorentz
group will be characterized by {\it two} angular momenta $(j, j^\prime)$, corresponding
to $\{A_i,B_i\}$ respectively, with $j,j^\prime=0,\frac{1}{2}, 1, \frac{3}{2},\ldots$ Each
representation will have $(2j+1)(2j^\prime+1)$ components. Explicitly, we will designate
the first four cases as \cite{Srednicki:2007qs}
\begin{description}[align=right,labelwidth=6.5cm]
	\item [$(0,0)\to$] scalar or singlet,
	\item [$(\frac{1}{2},0)\to$] left-handed spinor,
	\item [$(0,\frac{1}{2})\to$] right-handed spinor,
	\item [$(\frac{1}{2},\frac{1}{2})\to$] vector.
\end{description}

Therefore, we see here that there are two different representations containing the one-half 
angular momentum. Besides, those inequivalent representations are related by hermitian conjugation.
The first thing we could ask here is if we can have a fermion field that transforms 
only in one of those representations. To answer this, let us examine the properties of 
the left-handed and right-handed representations. Given our knowledge of the SU$(2)$, we know
that in both cases the fundamental representation will have two components.

\newpage

\noindent {\bf Left-handed representation.} Let us call the fundamental representation for the left-handed
spinor as $\psi_\alpha$, $\alpha$ taking two values. We can choose the angular momentum and boosts 
operators as \cite{Srednicki:2007qs}
\begin{align}
	J_i=\frac{1}{2}\sigma_i, \qquad K_i=-\frac{i}{2}\sigma_i,
\end{align}
being $\sigma_i$ the Pauli matrices, see appendix \ref{ap:Conv}. As consequence of this choosing,
we have that $A_i=J_i$ and $B_i=0$, as expected. The matrix $D(\Lambda)^\beta_{\ \alpha}$ can be obtained
considering the transformation of the left-handed spinor,
\begin{align}
	U(\Lambda)^\dagger \psi_\alpha (x) U(\Lambda)=D(\Lambda)^\beta_{\ \alpha} \psi_\beta (\Lambda^{-1} x).
\end{align}
Explicitly, one can find that
\begin{align*}
	D_L(\Lambda)=\exp\corc{-\frac{i}{2}\omega_{\mu\nu}S_L^{\mu\nu}},
\end{align*}
being $S^{\mu\nu}$ the generators of this representation, i.e.\ the $M^{\mu\nu}$
operators for the $(\frac{1}{2},0)$ representation,
\begin{align*}
	(S_L^{ij})^\beta_{\ \alpha}=\frac{1}{2}\varepsilon_{ijk}(\sigma_k)^\beta_{\ \alpha},\qquad
	(S_L^{i0})^\beta_{\ \alpha}=\frac{i}{2}(\sigma_i)^\beta_{\ \alpha},
\end{align*}
or, considering the ``four-vectors'' $\sigma^\mu$ and $\bar{\sigma}^\mu$ defined in the appendix \ref{ap:Conv}, \cite{Srednicki:2007qs}
\begin{align}
	(S_L^{\mu\nu})^\beta_{\enspace \alpha}=+\frac{i}{4}(\sigma^\mu\bar{\sigma}^\nu-\sigma^\nu\bar{\sigma}^\mu)^\beta_{\enspace \alpha}.
\end{align}

\noindent {\bf Right-handed representation.} As stated before, when we take the hermitian conjugate
of a left-handed field, we will obtain a field transforming under the $(0,\frac{1}{2})$ or 
right-handed representation. It is customary to put a dot on the indexes of the right-handed
field in order to distinguish them from a left-handed field, \cite{Srednicki:2007qs}
\begin{align}
	\bar{\psi}_{\dot{\alpha}}(x)=[\psi_\alpha (x)]^\dagger.
\end{align}
In this case, the angular momentum and boost operators are given by,
\begin{align}
	J_i=\frac{1}{2}\sigma_i, \qquad K_i=\frac{i}{2}\sigma_i,
\end{align}
and the Lorentz transformation of a right-handed field is \cite{Srednicki:2007qs}
\begin{align}
	U(\Lambda)^\dagger \bar{\psi}_{\dot{\alpha}} (x) U(\Lambda)=D_R(\Lambda)^{\dot{\beta}}_{\ \dot{\alpha}} \bar{\psi}_{\dot{\beta}} (\Lambda^{-1} x).
\end{align}
The transformation matrix is given by 
\begin{align*}
	D_R(\Lambda)=\exp\corc{-\frac{i}{2}\omega_{\mu\nu}S_L^{\mu\nu}},
	\qquad \qquad
	(S_R^{\mu\nu})^{\dot{\beta}}_{\enspace \dot{\alpha}}=-\frac{i}{4}(\bar{\sigma}^\mu \sigma^\nu-\bar{\sigma}^\nu \sigma^\mu)^{\dot{\beta}}_{\enspace \dot{\alpha}}.
\end{align*}

\section{Weyl Algebra}

To write terms which will be invariant under Lorentz transformations, we will introduce some definitions next. Let us first define the antisymmetric $\epsilon_{\alpha\beta}$ symbol \cite{Srednicki:2007qs}
\begin{align*}
	\epsilon^{\alpha\beta}=\epsilon^{\dot{\alpha}\dot{\beta}}&=(i\sigma_2)_{\alpha\beta},\\
	\epsilon_{\alpha\beta}=\epsilon_{\dot{\alpha}\dot{\beta}}&=(-i\sigma_2)_{\alpha\beta},
\end{align*}
The $\epsilon$ symbol will be useful to raise and lower spinorial indices,
\begin{align}
	\psi^\alpha&=\epsilon^{\alpha\beta}\psi_\beta,\quad \bar{\psi}^{\dot{\alpha}}=\epsilon^{\dot{\alpha}\dot{\beta}}\bar{\psi}_{\dot{\beta}}.
\end{align}
Thus, it will be possible to contract spinorial indices as is done with spacetime ones. However, it is important to note that the {\it ordering} in which the indices are contracted is crucial. For instance, using the antisymmetric property of $\epsilon_{\alpha\beta}$
\begin{align*}
	\chi^\alpha\xi_\alpha=\epsilon^{\alpha\beta}\chi_\beta\xi_\alpha=-\epsilon^{\beta\alpha}\chi_\beta\xi_\alpha = -\chi_\beta\xi^\beta. 
\end{align*}
It is necessary to establish a convention about the contracted indices. Following Srednicki \cite{Srednicki:2007qs}, we will identify a pair of undotted contracted indices always as $^\alpha_{\enspace \alpha}$ and for dotted indices we will have the opposite $_{\dot{\alpha}}^{\enspace \dot{\alpha}}$. Therefore, if we attempted to write invariant term $\psi^\alpha\psi_\alpha$, 
\begin{align*}
	\psi^\alpha\psi_\alpha = -\psi_\alpha\psi^\alpha,
\end{align*}
it should be identical to zero. Nevertheless, it is clear that the representations we are considering correspond to fermion representations. Then, we expect that such fields do {\bf anticommute}. Thus, the term will be
\begin{align*}
	\psi^\alpha\psi_\alpha = -\epsilon^{\beta\alpha}\psi_\beta\psi_\alpha = \epsilon^{\beta\alpha}\psi_\alpha\psi_\beta.
\end{align*}
For a general combination,
\begin{align*}
	\chi^\alpha\xi_\alpha = -\chi_\beta\xi^\beta = \xi^\beta\chi_\beta. 
\end{align*}
and we safely can ignore the spinorial indices keeping in mind the convention on the ordering
\begin{align*}
	\chi\xi = \xi\chi. 
\end{align*}
For right-handed fields, we will have $\bar{\chi}\bar{\xi} = \bar{\xi}\bar{\chi}$ \cite{Srednicki:2007qs}. We will define the indices of the $\bar{\sigma}^\mu$ and $\sigma^\mu$ as
\begin{align}
	\sigma^\mu_{\alpha\dot{\alpha}},\quad \bar{\sigma}^{\mu\dot{\alpha}\alpha};
\end{align}
notice the position and order of the indices. Such strange definitions are clarified by noting that \cite{Srednicki:2007qs}
\begin{align}\label{eq:barss}
	\bar{\sigma}^{\mu\dot{\alpha}\alpha}=\epsilon^{\alpha\beta}\epsilon^{\dot{\alpha}\dot{\beta}}\sigma^\mu_{\alpha\dot{\alpha}}. 
\end{align}
Other properties that $\sigma^\mu$ and $\bar{\sigma}^\mu$ obey are,
\begin{align*}
	\sigma^\mu+\bar{\sigma}^\nu+\sigma^\nu+\bar{\sigma}^\mu&=2\eta^{\mu\nu},\\
	\Tr \sigma^\mu\bar{\sigma}^\nu = \sigma^\mu_{\alpha\dot{\beta}}\bar{\sigma}^{\nu\dot{\beta}\alpha}&=2\eta^{\mu\nu},\\
	\sigma_{\mu\rho\dot{\tau}}\bar{\sigma}^{\mu\dot{\alpha}\beta}&=2\delta^\beta_\rho\delta^{\dot{\alpha}}_{\dot{\tau}}.
\end{align*}
Now, armed with these definitions, we can write invariant terms under Lorentz transformations. Evidently, we already have invariant $\chi\xi$ terms, but let us consider \cite{Srednicki:2007qs}
\begin{align}
	\bar{\xi}\bar{\sigma}^\mu \chi = \bar{\xi}_{\dot{\alpha}}\bar{\sigma}^{\mu\dot{\alpha}\beta}\chi_\beta.
\end{align}
Applying the transformation properties of left- and right-handed fields, we have that the previous term transforms as a vector
field,
\begin{align}
	U(\Lambda)^\dagger [\bar{\xi}\bar{\sigma}^\mu \chi] U(\Lambda)=\Lambda^\mu_{\enspace\nu} [\bar{\xi}\bar{\sigma}^\nu \chi].
\end{align}
Thus, a term like
\begin{align}
	i\bar{\xi}\bar{\sigma}^\mu \partial_\mu \chi,
\end{align}
is Lorentz invariant. Having the two main types of terms with and without derivatives, we can write invariant lagrangians.

\section{Weyl Fermions}

Let us start considering a fermion field $\chi$ in such a way that it has a conserved charged, that is, its lagrangian
has to be invariant under transformations \cite{Coleman:2011xi}
\begin{align}
	\chi \tto e^{i\theta} \chi;
\end{align}
this avoids terms as $\chi \chi$ or $\bar{\chi}\bar{\chi}$. Therefore, the lagrangian for such fields, which we will name as Weyl fermions, is
\begin{align}
	\mathscr{L}_{\rm Weyl}=i\chi^\dagger \bar{\sigma}^\mu \partial_\mu\chi.
\end{align}
As we can see here, this lagrangian is built from one type of field. Thus, a Weyl fermion does have a definite chirality. It is also possible to show that it has a definite helicity, as proven in the chapter \ref{cha:nu-MP}.

\section{Majorana Fermions}

Now, let us suppose that there is no conserved charge; we can then include bilinear terms in the fields \cite{Srednicki:2007qs}
\begin{align}
	\mathscr{L}_{\rm M}=i\chi^\dagger \bar{\sigma}^\mu \partial_\mu\chi +\frac{1}{2}m(\chi \chi+\bar{\chi}\bar{\chi}).
\end{align}
We have included both terms with the field in order to maintain our lagrangian hermitian. We can obtain the equations of motion for the $\chi$ field. The Euler-Lagrange equation for $\chi$ is
\begin{align}
	i\bar{\sigma}^{\mu\dot{\alpha}\beta}\partial_\mu\chi_\beta-m\bar{\chi}^{\dot{\alpha}}=0,
\end{align}
which coincides with the equation \eqref{eq:EMDC} appearing in chapter \ref{cap:nuMaj} by writing explicitly the $\epsilon$ matrix to lower the spinorial index of $\chi$. This fermion will have a definite chirality, as it is built from a unique Weyl field; but, it actually has two helicities, as presented in the chapter \ref{cap:nuMaj}. This kind of fermion representation is named after Ettore Majorana, i.e. this is a Majorana fermion.

\section{Dirac Fermions}

Now, let us consider a theory with two left-handed fermions, $\psi_i$ with $i=1,2$, having the same mass $m$ \cite{Srednicki:2007qs}
\begin{align}
	\mathscr{L}_{\rm D}=i\psi_i^\dagger \bar{\sigma}^\mu \partial_\mu\psi_i-\frac{1}{2}m(\psi_i\psi_i+\bar{\psi}_i\bar{\psi}_i).
\end{align}
Note that this lagrangian is invariant under the SO($2$) transformation,\cite{Srednicki:2007qs}
\begin{align}
	\begin{pmatrix}
		\psi_1\\
		\psi_2
	\end{pmatrix}
	\tto 
	\begin{pmatrix}
		\cos\theta & \sin\theta\\
		-\sin\theta & \cos\theta
	\end{pmatrix}
	\begin{pmatrix}
		\psi_1\\
		\psi_2
	\end{pmatrix},
\end{align}
so, we can define the following two combinations
\begin{subequations}
	\begin{align}
		\chi&=\frac{1}{\sqrt{2}}(\psi_1+i\psi_2),\\
		\xi&=\frac{1}{\sqrt{2}}(\psi_1-i\psi_2),
	\end{align}
\end{subequations}
in such a way that our lagrangian becomes
\begin{align}
	\mathscr{L}_{\rm D}=i\chi^\dagger \bar{\sigma}^\mu \partial_\mu\chi+i\xi^\dagger \bar{\sigma}^\mu \partial_\mu\xi-m(\chi\xi+\bar{\xi}\bar{\chi}).
\end{align}
We can now derive the equations of motion for the fields,
\begin{subequations}
	\begin{align}
		i\bar{\sigma}^{\mu\dot{\alpha}\beta}\partial_\mu\chi_\beta-m\bar{\xi}^{\dot{\alpha}}&=0,\\
		i\bar{\sigma}^{\mu\dot{\alpha}\beta}\partial_\mu\xi_\beta-m\bar{\chi}^{\dot{\alpha}}&=0,
	\end{align}
\end{subequations}
these equations are evidently coupled. We can write them in a better way using matrix properties,
\begin{align}
	\begin{pmatrix}
		-m\delta_\alpha^{\enspace\beta} & i\sigma^\mu_{\alpha\dot{\beta}}\,\partial_\mu\\
		i\bar{\sigma}^{\mu\dot{\alpha}\beta}\,\partial_\mu & -m\delta_{\enspace\dot{\alpha}}^{\dot{\beta}}
	\end{pmatrix}
	\begin{pmatrix}
		\chi_\beta\\
		\bar{\xi}^{\dot{\beta}}
	\end{pmatrix}=0,
\end{align}
where we have taken the hermitian conjugate of the second equation and we used the properties of the $\bar{\sigma}^\mu$ matrices, equation \eqref{eq:barss}. Thus, we can define a four-component field,
\begin{align}
	\Psi=\begin{pmatrix}
		\chi_\beta\\
		\bar{\xi}^{\dot{\beta}}
	\end{pmatrix},
\end{align}
and define new \emph{gamma} matrices as
\begin{align}
	\gamma^\mu=\begin{pmatrix}
		0 & \sigma^\mu_{\alpha\dot{\beta}}\\\
		\bar{\sigma}^{\mu\dot{\alpha}\beta} & 0
	\end{pmatrix}.
\end{align}
We obtain the acclaimed Dirac equation,
\begin{align}
	i\gamma^\mu\partial_\mu \Psi-m\Psi=0.
\end{align}
Therefore, we can conclude that a Dirac fermion is composed by two chiral fields $\chi_\beta,\bar{\xi}^{\dot{\beta}}$, having opposite chiralities. Hence, a Dirac field will not have a definite chirality. 

\newpage
\noindent It is possible to get back to the Weyl two-component notation. Let us define the chirality matrix as
\begin{align}
	\gamma^5=\begin{pmatrix}
		\delta_\alpha^{\enspace\beta} & 0\\
		0 & -\delta_{\enspace\dot{\alpha}}^{\dot{\beta}}
	\end{pmatrix},
\end{align}
and the left and right projectors $P_{L,R}=\frac{1}{2}\corc{1\pm\gamma^5}$
\begin{align}
	P_L=\begin{pmatrix}
		\delta_\alpha^{\enspace\beta} & 0\\
		0 & 0
	\end{pmatrix},\quad
	P_R=\begin{pmatrix}
		0 & 0\\
		0 & -\delta_{\enspace\dot{\alpha}}^{\dot{\beta}}
	\end{pmatrix}.
\end{align}
Projecting the Dirac field $\Psi$, we find
\begin{align}
	\Psi_L=P_L\Psi=\begin{pmatrix}
		\chi_\beta\\
		0
	\end{pmatrix}, \quad
	\Psi_R=P_R\Psi=\begin{pmatrix}
		0\\
		\bar{\xi}^{\dot{\beta}}
	\end{pmatrix}.
\end{align}
Therefore, we see that we can pass from one notation to the other by using the projector operators. We also can conclude that a Dirac fermion is composed by two Majorana fermions. Let us also stress that the Dirac fermion $\Psi$ is invariant under the trasnformation,
\begin{align*}
	\Psi\to e^{-i\theta}\Psi, 
\end{align*}
which comes from the SO($2$) original invariance.\\

In this appendix we have shown how the distinct types of fermions are generated in a systematic manner by considering the properties of the Lorentz trasformations and the Lorentz group. Considering that the Lie algebra of the Lorentz group can be written in the form of two independent SU($2$) algebras, we found that the representations have two different numbers, that is, two different values of the angular momenta. For the case of interest, fermion representations, there are two possible representations, called left- and right-handed representations. The type of representation of a fermion is usually called {\it chirality}. We explicitly considered each representation. After developing the notation usually applied to fermions, we wrote explicitly the lagrangians for Weyl, Majorana and Dirac fermions. We saw that a Weyl fermion corresponds to a field, with a well defined chirality, invariant under phase transformations. If one eliminates the restriccion of the invariance, it is possible to include mass terms for the fermion, and we obtain a Majorana field. Finally, if we write a lagrangian for two left-handed Majorana fermions invariant under a SO($2$) symmetry, it is possible to derive the Dirac equations in four-components; then, we can conclude that a Dirac fermion is composed by two Majorana fields.

	\backmatter
	
	\singlespacing

	\addcontentsline{toc}{chapter}{References}	
	\bibliographystyle{utphys}
	\bibliography{Ref}

\end{document}